\documentclass[letterpaper,final,5p]{elsarticle}
\usepackage{amsmath,amssymb,graphicx}
\usepackage{lineno}
\usepackage[usenames,dvipsnames]{color}
\usepackage{subfigure}
\usepackage{yhmath}
\usepackage{ulem}

\numberwithin{equation}{section}

\newcommand{\eps}{\varepsilon} 
\newcommand{\la}{\lambda}

\newcommand{\K}{\mathrm{K}}
\newcommand{\E}{\mathrm{E}}
\newcommand{\sn}{\mathrm{sn}}

\newcommand{\R}{\mathbb{R}}

\newcommand{\sgn}{\mathrm{sgn}\,}
\newcommand{\rmd}{\,\mathrm{d}}
\newcommand{\kt}{\tilde{k}}
\newcommand{\wt}{\tilde{\omega}}
\newcommand{\oA}{\overline{A}}
\newcommand{\at}{\tilde{\alpha}}
\newcommand{\rhob}{\overline{\rho}}
\newcommand{\ub}{\overline{u}}

\newcommand{\be}{\begin{equation}}
\newcommand{\ee}{\end{equation}}
\newcommand{\bea}{\begin{eqnarray}}
\newcommand{\eea}{\end{eqnarray}}
\newcommand{\bes}{\begin{equation*}}
\newcommand{\ees}{\end{equation*}}
\newcommand{\beas}{\begin{eqnarray*}}
\newcommand{\eeas}{\end{eqnarray*}}
\newcommand{\p}{\partial}

\newcommand{\caseI}{%
  \vrule height6pt width7pt depth-5.6pt %
  \vrule height6pt width0.4pt depth-2.1pt %
  \vrule height2.5pt width7pt depth-2.1pt %
  \vrule height2.5pt width0.4pt depth1pt %
  \vrule height-0.6pt width7pt depth1pt %
  }
\newcommand{\caseII}{%
  \vrule height6pt width7pt depth-5.6pt %
  \vrule height6pt width0.4pt depth1pt %
  \vrule height-0.6pt width7pt depth1pt %
  \vrule height2.5pt width0.4pt depth1pt %
  \vrule height2.5pt width7pt depth-2.1pt %
  }
\newcommand{\caseIII}{%
  \vrule height2.5pt width7pt depth-2.1pt %
  \vrule height6pt width0.4pt depth-2.1pt %
  \vrule height6pt width7pt depth-5.6pt %
  \vrule height6pt width0.4pt depth1pt %
  \vrule height-0.6pt width7pt depth1pt %
  }
\newcommand{\caseIV}{%
  \vrule height2.5pt width7pt depth-2.1pt %
  \vrule height2.5pt width0.4pt depth1pt %
  \vrule height-0.6pt width7pt depth1pt %
  \vrule height6pt width0.4pt depth1pt %
  \vrule height6pt width7pt depth-5.6pt %
  }
\newcommand{\caseV}{%
  \vrule height-0.6pt width7pt depth1pt %
  \vrule height6pt width0.4pt depth1pt %
  \vrule height6pt width7pt depth-5.6pt %
  \vrule height6pt width0.4pt depth-2.1pt %
  \vrule height2.5pt width7pt depth-2.1pt %
  }
\newcommand{\caseVI}{%
  \vrule height-0.6pt width7pt depth1pt %
  \vrule height2.5pt width0.4pt depth1pt %
  \vrule height2.5pt width7pt depth-2.1pt %
  \vrule height6pt width0.4pt depth-2.1pt %
  \vrule height6pt width7pt depth-5.6pt %
  }

\begin{document}
\date{}

\begin{frontmatter}
  \title{\bf Dispersive shock waves and modulation theory}
  \author[Loughborough]{G.~A.~El}
  \ead{g.el@lboro.ac.uk}

  \author[CUBoulder]{M.~A.~Hoefer}
  \ead{hoefer@colorado.edu}

  \address[Loughborough]{Department of Mathematical Sciences,
    Loughborough University, Loughborough, LE11 3TU, UK} 
  
  \address[CUBoulder]{Department of Applied Mathematics, University of
    Colorado, Boulder, CO 80309, USA} 
  
  \begin{abstract}
    There is growing physical and mathematical interest in the
    hydrodynamics of dissipationless/dispersive media.  Since
    G.~B.~Whitham's seminal publication fifty years ago that ushered
    in the mathematical study of dispersive hydrodynamics, there has
    been a significant body of work in this area.  However, there has
    been no comprehensive survey of the field of dispersive
    hydrodynamics.  Utilizing Whitham's averaging theory as the
    primary mathematical tool, we review the rich mathematical
    developments over the past fifty years with an emphasis on
    physical applications.  The fundamental, large scale, coherent
    excitation in dispersive hydrodynamic systems is an expanding,
    oscillatory dispersive shock wave or DSW.  Both the macroscopic
    and microscopic properties of DSWs are analyzed in detail within
    the context of the universal, integrable, and foundational models
    for uni-directional (Korteweg-de Vries equation) and bi-directional
    (Nonlinear Schr\"{o}dinger equation) dispersive hydrodynamics.  A
    DSW fitting procedure that does not rely upon integrable structure
    yet reveals important macroscopic DSW properties is described.
    DSW theory is then applied to a number of physical applications:
    superfluids, nonlinear optics, geophysics, and fluid dynamics.
    Finally, we survey some of the more recent developments including
    non-classical DSWs, DSW interactions, DSWs in perturbed and
    inhomogeneous environments, and two-dimensional, oblique DSWs.
  \end{abstract}

\end{frontmatter}

\tableofcontents

\section{Introduction} 
\label{sec:introduction}

Dispersive hydrodynamics is the domain concerned with fluid or
fluid-like motion in which dissipation, e.g., viscosity, is negligible
relative to wave dispersion.  In conservative media such as
superfluids, optical materials, and water waves, nonlinearity has the
tendency to engender wavebreaking that is mediated by dispersion.
Generically, the result of these processes is a multiscale, unsteady,
coherent wave structure called a dispersive shock wave or DSW.  This
review is concerned with the mathematical study of dispersive shock
waves and applications, principally via nonlinear wave modulation
theory, often referred to as Whitham averaging
\cite{whitham_linear_1974}.  

\subsection{Dispersive hydrodynamics}
\label{sec:disp-hydr}

In classical fluid dynamics, unphysical hydrodynamic singularities
(gradient catastrophe) are resolved by a transfer of energy from large
to small spatial scales accompanied by an increase in entropy.  This
results in a smooth but rapid transition in hydrodynamic and
thermodynamic quantities, a shock wave that moves faster than the
local speed of sound.  These irreversible processes are fundamental to
classical, dissipative hydrodynamics.  In contrast, when singularities
are resolved by a conservative, reversible process such as wave
dispersion, the resulting dynamics are quite different. Rather than
the generation of a turbulent energy cascade, gradient catastrophe in
dispersive hydrodynamic media leads to the generation of an expanding
nonlinear wavetrain, a dispersive shock wave.  Dispersive
hydrodynamics is therefore quite different from its dissipative
hydrodynamic analogue.  Nevertheless, the rich theory of classical
hydrodynamics provides a natural compass that is helpful in navigating
the relatively new field of dispersive hydrodynamics, e.g., Riemann
problems, characteristics, shock loci, admissibility, et cetera.

\begin{figure}
  \centering
  \includegraphics{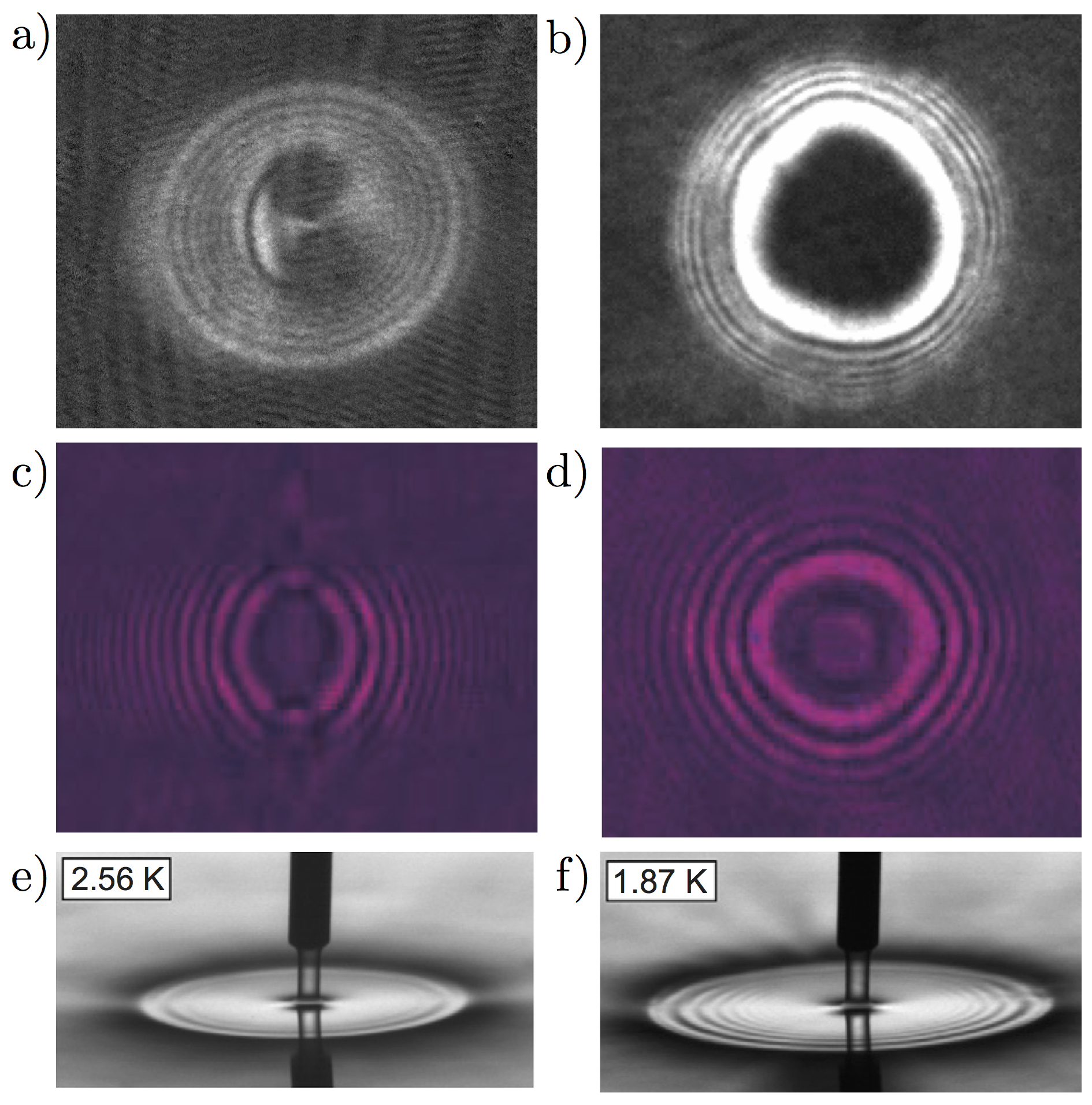}
  \caption{Dispersive shock waves. a,b) Blast waves in a Bose-Einstein
    condensate with expansion post (a) and during (b) laser pulse
    application \cite{hoefer_dispersive_2006}.  c,d) Nonlinear
    diffraction of an elliptical (c) and circular (d) Gaussian beam
    after propagation through a photorefractive crystal. Reprinted by
    permission from Macmillan Publishers Ltd: Nature Physics
    \cite{wan_dispersive_2007}, copyright 2007.  e,f) Hydraulic jump
    in liquid He II above (e) and below (f) the superfluid transition
    temperature 2.17 K.  Reprinted from \cite{rolley_hydraulic_2007}
    with permission from Elsevier.}
  \label{fig:dsw_expt_super_bec_optics}
\end{figure}

DSWs have recently garnered attention in connection with ground
breaking experiments on ultracold atoms
\cite{dutton_observation_2001,simula_observations_2005,hoefer_dispersive_2006,chang_formation_2008,hoefer_matter-wave_2009,meppelink_observation_2009}
(Figs.~\ref{fig:dsw_expt_super_bec_optics}a,b) and in optical media
\cite{rothenberg_observation_1989,couton_self-formation_2004,wan_dispersive_2007,jia_dispersive_2007,barsi_dispersive_2007,ghofraniha_shocks_2007,conti_observation_2009,ghofraniha_measurement_2012,fatome_observation_2014}
(Figs.~\ref{fig:dsw_expt_super_bec_optics}c,d), where they have been
referred to as quantum or optical shocks, respectively.  Figures
\ref{fig:dsw_expt_super_bec_optics}e,f show a Helium II surface wave
pattern above \ref{fig:dsw_expt_super_bec_optics}e and below
\ref{fig:dsw_expt_super_bec_optics}f the superfluid temperature
\cite{rolley_hydraulic_2007}.  The development of rank ordered,
stationary concentric rings in this hydraulic jump are indicative of
DSWs.  Dispersive shock waves have also been observed in intense
electron beams \cite{mo_experimental_2013} and rarefied plasma
\cite{taylor_observation_1970,tran_shocklike_1977} over sufficiently
short time scales.

\begin{figure}
  \centering
  \includegraphics{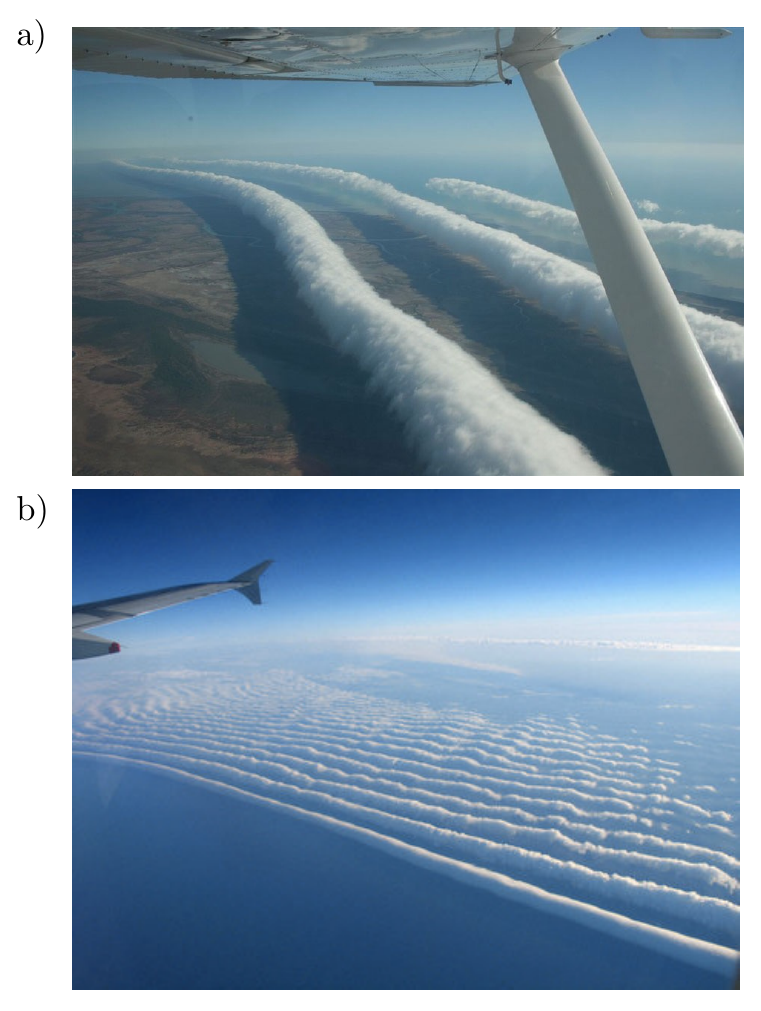}
  \caption{Atmospheric dispersive shock waves.  a) Morning glory roll
    cloud (copyright Mick Petroff, Creative Commons 3.0, 2009).
    b) Mountain waves (photo courtesy of the
    COMET$^{\textrm{\textregistered}}$ Program).}
  \label{fig:atmospheric_dsws}
\end{figure}

Dispersive hydrodynamics are not limited to exotic media but can also
occur in geophysics: stratified environments of the ocean and
atmosphere where the dispersive hydrodynamic medium is the interface
between two fluids, e.g., the ocean surface and air
\cite{bazin_recherches_1865,hammack_korteweg-vries_1978,treske_undular_1994,madsen_solitary_2008,chanson_current_2009,chanson_tidal_2011,trillo_observation_2016},
an internal ocean pycnocline
\cite{apel_oceanic_2002,scotti_observation_2004,wang_propagation_2011},
or lower and upper atmospheric layers
\cite{smyth_hydraulic_1988,christie_morning_1992,hills_nonstationary_2012}.
Here, the dispersive hydrodynamic medium is the interface itself.  The
conservation of mass enables approximately frictionless interfacial
dynamics.  DSWs in these systems are often referred to as undular
bores or roll clouds.  Striking examples of DSWs include Morning Glory
roll clouds and mountain waves (Fig.~\ref{fig:atmospheric_dsws}).
Shallow water undular bores or oscillatory hydraulic jumps are one of
the earliest observed DSWs (see Secs.~\ref{sec:shallow-water} and
\ref{sec:dsw_oblique}).  Internal ocean wave DSWs are quite prevalent,
especially in coastal regions during the summer months
\cite{apel_oceanic_2002} (see Sec.~\ref{sec:dsw-perturbed}).  An
additional system fitting into this category is the interfacial
dynamics of two Stokes fluids with high viscosity contrast, proposed
as a model of mantle magma dynamics
\cite{whitehead_magma_1990,lowman_dispersive_2013} (see
Sec.~\ref{sec:visc-fluid-cond}).  Indeed, any approximately
conservative medium exhibiting wave dispersion can be classified
dispersive hydrodynamic over sufficiently short spatio-temporal
scales.  All of these DSW examples exhibit a common, rank ordered
structure of oscillations.  

DSWs are believed to play an important role in a number of atmospheric
and oceanic events such as thunderstorm initiation
\cite{christie_morning_1992}, coastal tsunami propagation
\cite{madsen_solitary_2008}, \cite{grue_formation_2008} and internal
ocean transport \cite{scotti_observation_2004}.  Under appropriate
conditions, all the aforementioned media exhibit the requisite
dispersive hydrodynamic features of nonlinearity and dispersion.
These features can be modeled, in the simplest case, by a hyperbolic
system with dispersive corrections
\begin{equation}
  \label{eq:1}
  \mathbf{u}_t +  \nabla \cdot \mathbf{F}(\mathbf{u}) = 
  \nabla \cdot \mathbf{D}[\mathbf{u}], 
\end{equation}
where $\mathbf{u} = \mathbf{u}(\mathbf{x},t)$ is the state vector,
$\mathbf{F}$ is the nonlinear flux tensor, and $\mathbf{D}$ is an
integro-differential dispersive operator (in general, a 2-tensor),
distinguished by giving rise to a real-valued dispersion relation,
$\omega(\mathbf{k},\mathbf{u}_0)$--found by linearizing equation
\eqref{eq:1} according to $\mathbf{u}(\mathbf{x},t) = \mathbf{u}_0 +
\mathbf{v} e^{i(\mathbf{k}\cdot \mathbf{x} - \omega t)}$,
$|\mathbf{v}| \to 0$--such that the dispersion sign defined as
$\mathrm{sgn} [\det (\partial_{k_i k_j} \omega)] $ is not identically
zero.  Partial differential equations (PDE) of the type in
eq.~\eqref{eq:1} will be the focus of this review.  Suppose the evolution
described by (\ref{eq:1}) is characterized by two distinct
spatio-temporal scales: the hydrodynamic scale $\Delta x \sim l$,
$\Delta t \sim l/c_s$ ($c_s$ is the characteristic dispersionless,
hyperbolic speed in the system, i.e., the speed of sound), typically
specified by initial or boundary conditions, and the dispersive scale
$\Delta x \sim L$, $\Delta t \sim L/c_p$ ($c_p$ is the typical phase
velocity), where $L$ is the typical wavelength of dispersive waves,
the intrinsic coherence scale of the system. The motion in such a
system can be classified as dispersive-hydrodynamic if $\varepsilon =
L/l \ll 1$.

The scalar, one-dimensional, uni-directional version of
eq.~\eqref{eq:1} takes the form
\begin{equation}
  \label{eq:105}
  u_t + f(u)_x = D[u]_x.
\end{equation}
When $D \equiv 0$ and $f''(u) \ne 0$, smooth initial data for
\eqref{eq:105} can develop singularities in finite time
\cite{whitham_linear_1974}.  The inclusion of dispersion, $D \ne 0$,
regularizes the dynamics.  The equations of this type are abundant but
perhaps the most ubiquitous is the Korteweg-de Vries (KdV) equation
\cite{boussinesq_essai_1877,korteweg_change_1895} where $f(u) =
\frac{1}{2}u^2$ and $D[u] = -u_{xx}$.  The choice of convective
nonlinearity $f(u)$ and the dispersive operator $D[u]$ has a
significant impact on the resultant dispersive hydrodynamics.  In this
review, we will explore the dispersive hydrodynamics of several models
in the form \eqref{eq:105} with particular emphasis on KdV.

A pervasive example of a bi-directional system of two dispersive
hydrodynamic equations are the dispersive Euler equations
\begin{equation}
  \label{eq:110}
  \begin{split}
    \rho_t + (\rho u)_x &= D_1[\rho,u]_x \\
    (\rho u)_t + \left ( \rho u^2 + P(\rho) \right )_x &=
    D_2[\rho,u]_x ,
  \end{split}
\end{equation}
where $\rho \ge 0$ is interpreted as the dispersive fluid density, $u$
is the fluid's velocity, and $P(\rho)$ is a given constitutive law
for the pressure.  In the absence of dispersion $D_1 = D_2 \equiv 0$,
equations \eqref{eq:110} are known as the Euler $P$-system.
The $P$-system is hyperbolic so long as $P'(\rho) > 0$ and therefore
admits the sound speed $c(\rho) = \sqrt{P'(\rho)}$.  A number of
well-known model equations fit within this framework, including the
generalized Nonlinear Schr\"{o}dinger (gNLS) equation
\begin{equation}
  \label{eq:111}
  i\psi_t + \frac{1}{2} \psi_{xx} - f(|\psi|^2) \psi = 0,
\end{equation}
where $f$ is a smooth, positive, monotonically increasing function of
its argument.  In order to write equation \eqref{eq:111} in the
hydrodynamic form \eqref{eq:110}, we utilize the transformation to
magnitude and phase variables
\begin{equation}
  \label{eq:112}
  \psi = \sqrt{\rho} e^{i \phi}, \quad u = \phi_x ,
\end{equation}
known as the Madelung transformation.
After insertion of \eqref{eq:112} into \eqref{eq:111} and equating
real and imaginary parts, the gNLS equation takes the hydrodynamic
form \eqref{eq:110} with
\begin{equation}
  \label{eq:113}
  \begin{split}
    P(\rho) &= \int_0^\rho s f'(s) \rmd s, \\
    D_1[\rho,u] &\equiv 0,
    \quad D_2[\rho,u] = \frac{1}{4} \rho (\ln \rho)_{xx} .
  \end{split}
\end{equation}
When $f(\rho) = \rho$, we obtain the integrable Nonlinear
Schr\"{o}dinger (NLS) equation.

\subsection {Anatomy of a dispersive shock wave}
\label{sec:anat-disp-shock}

\begin{figure}
  \centering
  \includegraphics{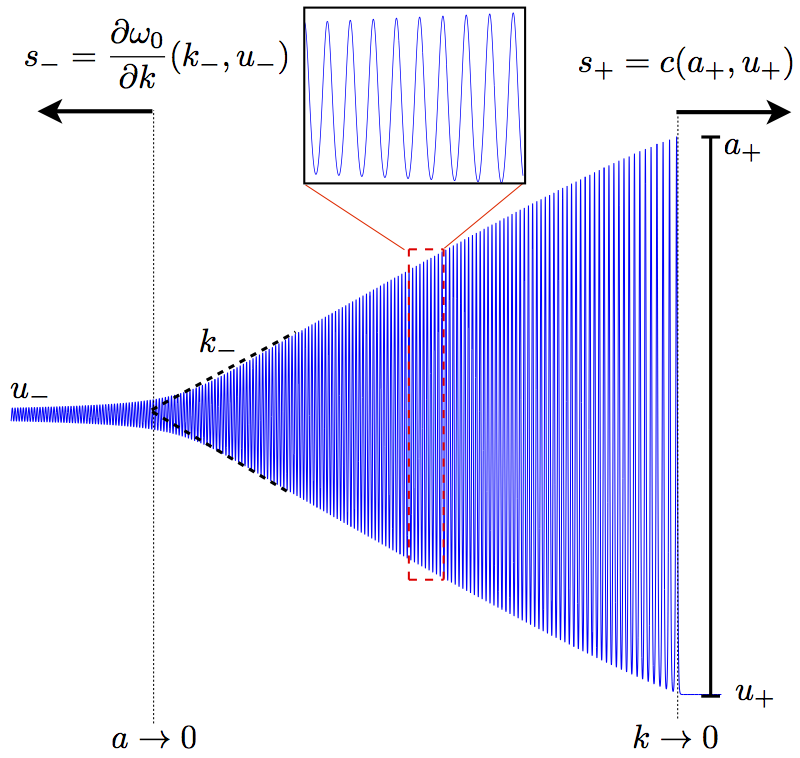}
  \caption{Schematic of a dispersive shock wave's macroscopic and
    microscopic (inset) structure.}
  \label{fig:dsw_schematic}
\end{figure}

In very general terms, DSWs are multi-scale, unsteady, nonlinear
coherent wave structures that are characterized by two complementary
identities.  When viewed locally, i.e., over a small region of space
and time, DSWs display periodic, or quasiperiodic structure (see
Fig.~\ref{fig:dsw_schematic} inset), forming due to the interplay
between nonlinear and dispersive effects.  Over a larger region
covering multiple wave oscillations, the DSW wavetrain exhibits slow
modulation of the wave's parameters (amplitude, frequency, mean), and
this modulation itself is a nonlinear hyperbolic wave (see
Fig.~\ref{fig:dsw_schematic}). This kind of ``dispersive-hyperbolic''
duality of modulated waves is not uncommon in linear wave theory but
DSWs prominently display it over the full range of nonlinearity, from
a weakly nonlinear regime to solitary waves realized as an integral
part of the modulated wavetrain.  A DSW connects two, disparate,
non-oscillatory states, which are either constant or slowly varying,
an important distinction from a generic nonlinear wavetrain \cite{ostrovsky_asymptotic_2014}.

The striking manifestation of
nonlinearity in a DSW is that it is characterized by at least two
distinct speeds of propagation, those of its leading $s_+$ and
trailing $s_-$ edges as shown in Fig.~\ref{fig:dsw_schematic}.  These
two DSW speeds bifurcate from the linear group velocity, a striking
realization of the splitting of the doubly degenerate characteristic
speed from linear wave theory \cite{whitham_linear_1974}. 
As depicted in the DSW schematic of Fig.~\ref{fig:dsw_schematic}, the
two distinct speeds of propagation correspond to a small amplitude,
harmonic limit $a \to 0$ of the modulated wavetrain and a large
amplitude solitary wave limit where the wavenumber $k$ vanishes.  As
described in Sec.~\ref{sec:dsw-non-integrable}, these distinguished
limits and an additional simple wave assumption enable the
determination of DSW edge speeds: the linear group velocity determines
the harmonic edge speed and the solitary wave amplitude-speed relation
determines the opposite edge speed.  Associated with the harmonic and
solitary wave speeds are a characteristic wavenumber of the weakly
nonlinear harmonic edge wavepacket, denoted $k_-$ in
Fig.~\ref{fig:dsw_schematic}, and the solitary wave amplitude, denoted
$a_+$ in Fig.~\ref{fig:dsw_schematic}.  The DSW speeds,
characteristic wavenumber, amplitude, and DSW envelope constitute the
macroscopic observables of a DSW.  The slowly modulated wave
connecting the leading and trailing edges describes the microscopic
DSW structure (see Fig.~\ref{fig:dsw_schematic} inset).

\begin{figure}
  \centering
  \includegraphics{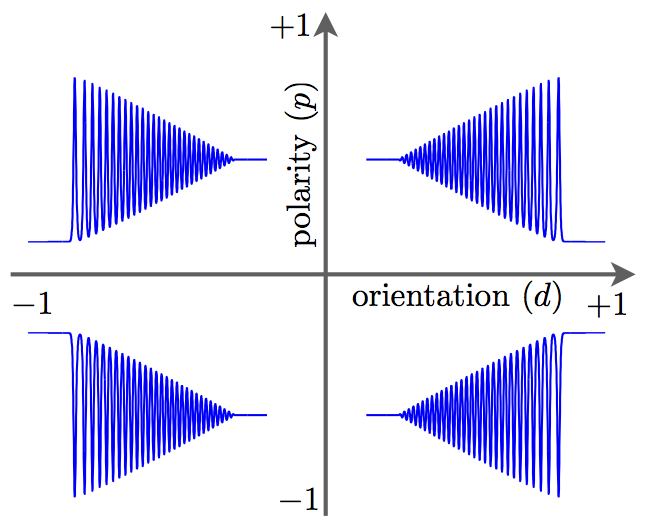}
  \caption{Orientation and polarity of a DSW.}
  \label{fig:dsw_classification}
\end{figure}

Because DSWs exhibit two distinct edges, it is natural to define the
DSW orientation $d$.  When the solitary wave is at the DSW leading
edge (rightmost), as in Fig.~\ref{fig:dsw_schematic}, $d = 1$; $d =
-1$ otherwise.  Associated with the solitary wave edge is the polarity
$p$, depending upon whether the edge is a wave of elevation ($p = 1$,
as in Fig.~\ref{fig:dsw_schematic}) or depression ($p = -1$).  Figure
\ref{fig:dsw_classification} depicts a DSW classification according to
orientation and polarity.  For scalar dispersive hydrodynamic
equations \eqref{eq:1}, the orientation and polarity of a DSW are
related to the curvature of the dispersion and the nonlinear flux
\cite{el_dispersive_2015}.  

\subsection{Dispersive vs.~dissipative-dispersive shocks}
\label{sec:disp-vs-diff}

The generation of DSWs represents a universal mechanism to resolve
unphysical hydrodynamic singularities in dispersive conservative
media, so their fundamental role in such media is similar to that of
viscous shock waves (VSWs) in classical gas and fluid dynamics. At the
same time, DSWs are sharply distinct from their well-studied viscous
counterpart both in terms of physical significance and mathematical
description. First of all, DSWs, unlike VSWs, do not dissipate energy.
The potential energy of the jump in hydrodynamic quantities across a
DSW leads to kinetic energy associated with nonlinear wave generation
and is not accompanied by an increase in entropy.  An increase in
entropy as a shock is traversed, on the other hand, is a defining
property of VSWs \cite{lax_hyperbolic_1973}.  Two salient features of
DSWs, their oscillatory microscopic structure and unsteady macroscopic
dynamics, make the distinction between DSWs and VSWs particularly
evident.  Indeed, the internal oscillatory structure of DSWs is in
sharp contrast with the monotone structure of VSWs.  But it is the
second unique feature of DSWs, their unsteady, expanding nature, that
makes DSWs so radically distinct from VSWs.  VSWs are characterized by
a fixed width and a single speed of propagation, the latter being
determined by a balance of physical integrals of motion (e.g., mass
and momentum) across the shock, not depending on the details of its
internal structure. The unsteady dynamics of DSWs have far-reaching
physical and mathematical implications, among which are the
inapplicability of classical Rankine-Hugoniot relations and the
inseparability of the description of macroscopic DSW dynamics from the
analysis of its nonlinear oscillatory microstructure.

In order to illustrate the distinction between DSWs and VSWs, we
consider a model equation incorporating nonlinearity, dispersion, and
dissipation in the Korteweg-de Vries-Burger's (KdVB) equation
\begin{equation}
  \label{eq:2}
  u_t + uu_x + \mu u_{xxx} = \nu u_{xx} ,
\end{equation}
where $\mu \in \R$ is the dispersion coefficient and $\nu \ge 0$ is
the dissipation coefficient.  Diffusive-dispersive dynamics
essentially reducible to the KdVB equation (\ref{eq:2}) arise in the
original theory of undular bores by Benjamin and Lighthill
\cite{benjamin_cnoidal_1954} (see also \cite{whitham_linear_1974}) and
in the theory of collisionless shocks in rarefied plasma due to
Sagdeev \cite{moiseev_collisionless_1963, sagdeev_cooperative_1966}
(see also the notable publication \cite{sagdeev_collisionless_1991} in
a popular magazine) and others (see, e.g., \cite{grad_unified_1967}).
The KdVB equation has also recently been invoked as a universal model
of cold atom hydrodynamics \cite{kulkarni_hydrodynamics_2012}.  When
$\mu = \nu = 0$ (Hopf equation), decreasing initial data leads to
singularity formation in finite time \cite{whitham_linear_1974}.  When
$\mu \ne 0$ and $\nu = 0$ (KdV equation), the singularity is resolved
by a DSW whose structure is described via a weak limit as $\mu \to
0^+$ or $\mu \to 0^-$ according to the Whitham modulation equations
\cite{whitham_non-linear_1965,gurevich_nonstationary_1974,lax_small_1983}
(see Sec.~\ref{sec: dsw-riemann}).  The direction $\mu$ approaches
zero determines the DSW orientation $d = \sgn \mu$ and polarity $p =
\sgn \mu$.  The purely dissipative regularization $\mu = 0$ and $\nu
\to 0^+$ (Burger's equation) strongly converges to a discontinuous
solution, a VSW, with shock speed determined by the Rankine-Hugoniot
jump conditions \cite{whitham_linear_1974}.

\begin{figure}
  \centering
  \includegraphics[width=\columnwidth]{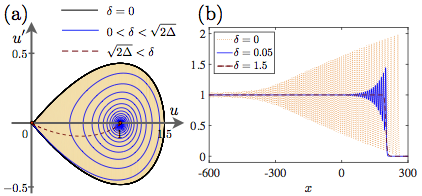}
  \caption{(a) Phase plane depicting KdVB TW solutions connecting
    $u_+=0 < u_-=1$ with $\mu=1$ via a monotone TW (dashed, strong
    dissipation) or an oscillatory TW (solid spiral, weak
    dissipation).  The KdV soliton (solid homoclinic orbit, zero
    dissipation) encloses the union of all KdV periodic orbits spanned
    across a DSW.  (b) Numerical solutions of KdVB at $t = 400$ with
    step-like initial data at the origin and the same $\delta$, $\mu$
    as in (a).  TWs ($\delta>0$) and DSWs ($\delta=0$) are attractors
    for the dynamics.}
  \label{fig:kdvbPhasePlane}
\end{figure}

The regularization where both dissipation and dispersion are included
(KdVB equation) crucially depends on the ratio $\delta =
\nu/\sqrt{|\mu|}$
\cite{grad_unified_1967,johnson_non-linear_1970,bona_travelling-wave_1985,gurevich_averaged_1987}.
When $\delta \to 0$ and $\mu \to 0^+$ or $\mu \to 0^-$, i.e., a
dispersion dominated regularization, the behavior is again described
by a DSW.  If $\delta > \delta_0 > 0$ and $\nu \to 0^+$, $\mu \to
0^\pm$, the regularization is dissipatively dominated and the solution
strongly converges to the same VSW solution as in Burger's equation.
The transition between these two limiting cases, DSW and VSW, can be
understood by taking $\delta > 0$ fixed (dissipation balancing
dispersion) and seeking traveling wave (TW) solutions in the form
$u(x,t) \to u(\xi)$, $\xi = (x-st)/\sqrt{|\mu|}$ with $u(\pm
\infty)=u_\pm$.  Inserting the TW ansatz into eq.~\eqref{eq:2},
integrating once and letting $v = u'$, we obtain the first order
system of ordinary differential equations (ODE)
\begin{equation}
  \label{eq:3}
  u'=v, \quad (\sgn \mu) v'=\delta v -\frac{1}{2}(u^2+u_-u_+)+su .
\end{equation}
The TW speed $s = \frac{1}{2}(u_+ + u_-)$, determined by the boundary
conditions, satisfies the Rankine-Hugoniot jump conditions.  The phase
portrait of the system, shown in Fig.~\ref{fig:kdvbPhasePlane}a for
$\mu = 1$, has two equilibria $(u,v)=(u_\pm, 0)$, one of which is a
saddle point, $(u_{\sgn\mu},0)$, and the other is a stable ($\mu < 0$)
or unstable ($\mu > 0$) node or spiral, depending on whether the
eigenvalues of the linearized system
\begin{equation*}
  \lambda_\pm(u)=\frac{1}{2} \left(\delta \, \sgn \mu \pm
    \sqrt{\delta^2+4(s-u)\sgn \mu}\right),
\end{equation*}
are real ($\delta \ge \sqrt{2\Delta}$) or complex ($\delta <
\sqrt{2\Delta}$), respectively.  The jump across the TW, $\Delta = u_-
- u_+ > 0$, corresponds to the Lax entropy condition
\cite{lax_hyperbolic_1973}.  The structure of the TW changes from a
monotonic to an oscillatory profile as $\delta$ is decreased across
the critical value $\sqrt{2\Delta}$.  Further analysis shows that for
$0 < \delta \ll 1$, the amplitude of the largest oscillation in the TW
profile is approximately $a = \frac{3}{2}\Delta$
\cite{johnson_non-linear_1970}.  This corresponds to the outermost
trajectory in the phase plane, intersecting $(u,v) = (1.5,0)$ in
Fig.~\ref{fig:kdvbPhasePlane}a.

For $\delta=0$, however, the (un)stable equilibrium becomes a center,
i.e., with imaginary eigenvalues, and a TW connecting $u_+$ and $u_-$
no longer exists. Instead, there is a homoclinic orbit connecting the
saddle point to itself, and the corresponding TW is a KdV soliton.
Enclosed within this homoclinic orbit are periodic orbits representing
KdV periodic TW solutions.  Numerical solutions to eq.~\eqref{eq:2}
with step-like initial data for the same three values $\delta >
\sqrt{2\Delta}$, $0 < \delta < \sqrt{2\Delta}$, and $\delta = 0$ used
in Fig.~\ref{fig:kdvbPhasePlane}a are shown in
Fig.~\ref{fig:kdvbPhasePlane}b.  We observe that the $\delta > 0$
solutions rapidly approach KdVB TWs and that the TWs with $0 < \delta
< \sqrt{2\Delta}$ exhibit oscillatory structure whose signature is
inherited by the orientation and polarity of the corresponding $\delta
= 0$ DSW.  The analysis in Sec.~\ref{sec: dsw-riemann} shows that a
DSW is described by a \textit{modulation} of the periodic TWs from the
zero amplitude, equilibrium solution to the infinite period, soliton,
homoclinic solution.  This allows us to identify the key distinction
between VSWs and DSWs:
\begin{itemize}
\item The VSW can be modeled by a  TW solution to an
ODE (a {\it single} heteroclinic orbit).
\item The DSW can be modeled by a PDE modulation of the periodic TW
  (effectively, a continuous \textit{family} of periodic orbits that
  is traversed across the wavetrain).
\end{itemize}
Note also that the DSW soliton edge amplitude is $a_{\sgn \mu} = 2
\Delta$, moving with speed $s_{\sgn \mu} =
\frac{1}{3}(u_-+u_++u_{(-\sgn\mu)})$ (see Sec.~\ref{sec:
  dsw-riemann}), different from the KdVB TW.

\subsection{Aim of this Review}
\label{sec:aim-this-review}

When writing a review article, there is always a competition between
breadth and depth.  We initially intended to survey a broad set of
problems related to DSWs in dispersive hydrodynamics, while touching
upon the theoretical tools involved.  However, while there are texts
\cite{whitham_linear_1974,kamchatnov_nonlinear_2000}, a shorter review
\cite{hoefer_dispersive_2009}, and articles with lengthy introductory
material
\cite{el_resolution_2005,hoefer_shock_2014,el_dispersive_2015}
covering some aspects of DSW theory, we felt that a more comprehensive
review of the fundamental mathematical tools would fill an important
void.  Dispersive hydrodynamics is experiencing significant
development in recent years as evidenced by this dedicated special
issue.  The bulk of this review surveys the fundamentals of modulation
theory (Sec.~\ref{sec:math-tools-modul}), the dispersive
regularization of shock waves in integrable systems (KdV and NLS,
Sec.~\ref{sec:disp-shock-waves}), and the construction of DSWs in
non-integrable systems (Sec.~\ref{sec:dsw-non-integrable}).  The
latter sections provide an overview of some more recent developments
including non-classical DSWs (Sec.~\ref{sec:non-classical-dsws}), DSW
interactions (Sec.~\ref{sec:dsw-interactions}), DSWs in inhomogeneous
environments (Secs.~\ref{sec:dsw-perturbed}, {\ref{sec:dsw_forced}}),
and multidimensional, steady, oblique DSWs
(Sec.~\ref{sec:dsw_oblique}).

\section{Mathematical tools of DSW modulation theory}
\label{sec:math-tools-modul}

The mathematical description of DSWs involves a synthesis of methods
from hyperbolic quasi-linear systems, asymptotics, and soliton
theory. One of the primary tools is nonlinear wave modulation theory,
introduced by Whitham in 1965 \cite{whitham_non-linear_1965}.  Whitham
theory and an additional technique, matched asymptotic analysis, are
approximate methods that have demonstrated practical impact in a wide
variety of problems despite their formal nature.  Rigorous, exact
solution methods such as the Inverse Scattering Transform (IST) and
the associated Riemann-Hilbert steepest descent method are quite
powerful, but only apply to a limited class of integrable systems and
problems. The PDE of dispersive hydrodynamics encompass a range of
integrable systems such as the KdV, NLS, and Kadomtsev-Petviashvili
(KP) equations as well as their nonintegrable generalizations.  In
this section, we review properties of quasi-linear systems and Whitham
theory.

\subsection{Hydrodynamic type systems}
\label{sec:basic-notions}

\subsubsection{Basic notions}

In this subsection, we present a brief account of some basic notions
and techniques from the theory of one-dimensional quasi-linear
systems.  A detailed description can be found in
\cite{rozhdestvenskii_systems_1983,dafermos_hyperbolic_2009}.

Systems of first-order, homogeneous, quasi-linear partial
differential equations (PDEs)
\begin{equation}\label{hyp1}
 {\bf u}_t +  \mathrm{A}({\bf u}){\bf u}_x   = 0\, ,
\end{equation}
where ${\bf u}(x,t) = \{u_1, u_2, \dots , u_N\}\in \mathbb{R}^N$ and
$\mathrm{A}(\mathbf u)$ is an $N \times N$ matrix, are often called
{\it hydrodynamic type systems} \cite{dubrovin_hydrodynamics_1989}.  Such systems arise in the modeling
of wave processes in continuum mechanics, plasma physics,
magnetohydrodynamics etc. In the context of DSW theory, hydrodynamic
type systems appear in two ways: (i) as the dispersionless
($\mathbf{D} \equiv 0$) limit of a dispersive
hydrodynamic system (\ref{eq:1}); (ii) as Whitham modulation systems
obtained by averaging the dispersive hydrodynamic system (\ref{eq:1})
over a family of periodic or quasi-periodic solutions.

If the equation
\begin{equation}\label{cl}
\frac{\partial }{\partial t} P ({\bf u}) + \frac{\partial }{\partial
  x} Q ({\bf u}) =0, 
\end{equation}
where $P ({\bf u})$ and $Q ({\bf u})$ are some scalar functions, is a
consequence of the hydrodynamic type system (\ref{hyp1}) for any
solution, then it is called a {\it hydrodynamic conservation law} of
system (\ref{hyp1}). We note that in models arising in continuum
physics, hydrodynamic type systems (\ref{hyp1}) are often {\it
  deduced} from a system of conservation laws expressing fundamental
physical principles such as conservation of mass, momentum, energy,
etc. In this case, one assumes that the relevant Jacobians are finite
and non-singular.  The number of independent conservation laws for the
system (\ref{hyp1}) could be less, equal or greater than
$N$. Sometimes hydrodynamic type systems have an infinite number of
conservation laws. This property is usually linked to integrability in
the sense of the generalized hodograph transform (see
Sec.~\ref{sec:gener-hodogr-meth} below).

The curve $x(t)$ is called a {\it characteristic} of the system
(\ref{hyp1}) if
\begin{equation}\label{char}
  \frac{dx}{dt}=\lambda, \qquad \det|\mathrm{A} - \lambda  I | =0, 
\end{equation}
where ${I}$ is the $N \times N$ identity matrix.  The eigenvalues
$\lambda=V_j$, $j =1, \dots, N$ of the matrix $\mathrm{A}$, are called
the {\it characteristic velocities}. Importantly, since
$\mathrm{A}=\mathrm{A}({\bf u})$, the characteristics of the
hydrodynamic type system (\ref{hyp1}) generally depend on the
solution ${\bf u}(x,t)$.

Let ${\bf l}^{(k)}({\bf u})$ be the left eigenvector of the matrix
$\mathrm{A}$ corresponding to the eigenvalue $V_k$.
The system (\ref{hyp1}) is called {\it hyperbolic} if all the
eigenvalues $V_k$ are real,
\begin{equation}
  \label{hyperb}
  V_1({\bf u}) \leqslant V_2 ({\bf u}) \leqslant \dots \leqslant
  V_N({\bf u})\, , 
\end{equation}
and the eigenvectors ${\bf l}^{(k)}$ form a basis in  $\mathbb{R}^N$. 
The system is called {\it strictly hyperbolic} if all eigenvalues
$V_k$ are real and distinct
\begin{equation}
  \label{strict_hyp}
  V_1({\bf u}) < V_2 ({\bf u}) < \dots < V_N({\bf u}) \, .
\end{equation}

A consequence of the system (\ref{hyp1})
\begin{equation}
  \label{charf}
  {\bf l}^{(k)} \cdot ({\bf u}_t + \mathrm{A}{\bf u}_x)={\bf l}^{(k)}
  \cdot ({\bf u}_t+ V_k {\bf u}_x)=0, 
\end{equation}
is called a {\it characteristic relation} of (\ref{hyp1}).  For a
strictly hyperbolic system (\ref{hyp1}), there are $N$ independent
characteristic relations (\ref{charf}), which form a system equivalent
to (\ref{hyp1}) called the characteristic form of (\ref{hyp1}).  Each
equation in the characteristic form contains differentiation only in a
single direction of the $(x,t)$-plane, $\left( \frac{d {\bf u}}{dt}
\right)_{k}=\left(\frac{\partial }{\partial t} + V_k \frac{\partial
  }{\partial x} \right){\bf u}$, so the system (\ref{hyp1}) transforms
into a system of $N$ ODEs along $N$ characteristic directions.

Each characteristic relation (\ref{charf}) introduces the differential
form
\begin{equation}\label{diff_form}
  {\bf l}^{(k)}  \cdot  d{\mathbf u} =l^{(k)}_1 du_1 +l^{(k)}_2du_2 +
  \dots + l^{(k)}_N du_N , 
\end{equation}
which vanishes on the $k$-th characteristic.
If this form is integrable, then it is possible to introduce a new
variable $r_k({\mathbf u})$ such that
\begin{equation}
  d r_k= \mu {\bf l}^{(k)}  \cdot  d{\mathbf u} , 
\end{equation}
for some $\mu({\bf u})$ called the integrating factor. Such a variable is
called a {\it Riemann invariant}.  If all $N$ characteristic forms
(\ref{charf}) are integrable, then the system (\ref{hyp1}) assumes
the diagonal or Riemann form
\begin{equation}
  \label{rim}
  \frac{\partial r_k} {\partial t} +V_k({\bf r}) \frac{\partial r_k}
  {\partial x} =0 \, , \qquad k=1, \dots, N, 
\end{equation}
where we have used the shorthand notation $V_k({\bf r}) \equiv
V_k({\bf u}(\bf r))$ and assumed invertibility of the mapping ${\bf u}
\mapsto {\bf r}$.  One can see that each Riemann invariant $r_k$ is
constant along the characteristic $dx/dt = V_k$.

Riemann invariants always exist if $N=2$ but generally do not exist if
$N >2$.  Each Riemann invariant $r_k$ is determined up to an arbitrary
function of a single variable so that $F(r_k)$ is also a Riemann
invariant for any differentiable function $F$ of a single variable.




There is no general method for the computation of Riemann invariants
(if they exist) for a given hydrodynamic type system
(\ref{hyp1}). However, for the important class of hydrodynamic type
systems obtained by Whitham averaging of dispersive-hydrodynamic
systems (\ref{eq:1}) (see Sec.~\ref{sec:whith-modul-theory} below),
the existence of Riemann invariants was shown to be intimately related
to the integrability of the original equations via the IST.  For such
systems, there is an effective method of finding Riemann invariants
using finite-gap integration theory (a periodic analogue of the IST
method) \cite{flaschka_multiphase_1980,kamchatnov_nonlinear_2000}.

The property of strict hyperbolicity (\ref{strict_hyp}) for Whitham
modulation systems can be maintained even when characteristic
velocities merge.  Strict hyperbolicity for such systems is retained
due to the reduction of the system order $N$ so that the merged
velocity defines a regular characteristic.  For diagonal Whitham
systems (\ref{rim}), this is taken into account by the following
modification of the definition of strict hyperbolicity:
\begin{equation}
  \label{strict_hyp_diag}
  V_j = V_k   \ \Longleftrightarrow \    r_j = r_k  \quad \forall \,j,~k
  =1, \dots, N. 
\end{equation}
For non-strictly hyperbolic diagonal systems, there could be two or
more distinct Riemann invariants associated with the same
characteristic velocity $V_k$.  In terms of the original system (\ref{hyp1}) (when it is diagonalizable) non-strict hyperbolicity is  related to
non-invertibility of the mapping ${\bf u} \mapsto {\bf r}$.




The $k$-th characteristic family of the diagonal system (\ref{rim})
is called {\it genuinely nonlinear} if
\begin{equation}
  \label{gnonl}
  \partial_k V_k  \ne 0 , \quad \hbox{where} \quad \partial_k \equiv
  \frac{\partial }{\partial r_k} . 
\end{equation}
If (\ref{gnonl}) holds for all $k$, then the system (\ref{rim}) is
called genuinely nonlinear.  For a scalar conservation law
$u_t+f(u)_x=0$, the condition of genuine nonlinearity is nonzero
curvature of the flux, $f''(u) \ne 0$ for all $u$.
The characteristic family $\frac{d x}{d t}=V_k$ is called {\it
  linearly degenerate} if $\partial_k V_k = 0$. The system (\ref{rim})
is called linearly degenerate if $\partial_k V_k= 0$ for all $k=1,
\dots, N$.

The notions of genuine nonlinearity and linear degeneracy can be
generalized to non-diagonal systems (\ref{hyp1}). The $k$-th
characteristic family of the system (\ref{hyp1}) is genuinely
nonlinear if \cite{lax_hyperbolic_1973}
\begin{equation}
  \label{laxnonlin}
  \nabla_{\bf u} V_{k} \cdot {\bf p}^{(k)} \ne 0 \, ,
\end{equation}
where ${\bf p}^{(k)}({\bf u})$ is the right eigenvector of the matrix
$\mathrm{A}$ corresponding to the eigenvalue $V_k$.

If the system (\ref{hyp1}) is non-strictly hyperbolic, it is not genuinely
nonlinear as well \cite{dafermos_hyperbolic_2009}. The converse is
generally not true.

Finally, we introduce the definition of the {\it integral curve} of the
vector field ${\bf p}^{(k)}$ as a solution of the system of $N$
ODE
\begin{equation}
  \label{int_curve}
  {\bf v}'(s) = {\bf p}^{(k)}({\bf v}(s)),
\end{equation}
where $s$ is a parameter.  One can see from (\ref{laxnonlin}),
(\ref{int_curve}) that the $k$-th characteristic field is genuinely
nonlinear if the characterisic velocity $V_k$ is monotone along the
integral curve.

\subsubsection{Simple waves}

\label{sec:simple_waves}

Closely related to the notion of a Riemann invariant is the notion of
a simple wave or Riemann wave, playing an especially important role in
the DSW theory.  We first assume that the system (\ref{hyp1}) is
strictly hyperbolic and genuinely nonlinear.

A simple wave is a particular solution of (\ref{hyp1}) such that all
components of the vector ${\bf u}$ depend on the same quantity
$\alpha(x,t)$,
\begin{equation}
  \label{simple1}
  {u_j} = u_j(\alpha(x,t)), \qquad j=1, \dots, N.
\end{equation}
Substituting (\ref{simple1}) into (\ref{hyp1}), we obtain the system of
ODE
\begin{equation}
  \label{simple2}
  (\mathrm{A} - \lambda I)\frac{d{\bf u}}{d\alpha} = 0\, , \quad
  \lambda = - \frac{\alpha_t}{\alpha_x} \, . 
\end{equation}
A nontrivial solution of (\ref{simple2}) for ${\bf u}(\alpha)$ is
possible only if $\lambda = V_k$, for some $k \in \{1, \dots,
N\}$. Then for each $k$, the solution of the algebraic system in
(\ref{simple2}) yields a system of $N$ integral curves ${\bf u}' =
{\bf p}^{(k)}({\bf u}(\alpha))$ (cf.~(\ref{int_curve})).  The
second equation in \eqref{simple2} can be written as a scalar PDE
\begin{equation} 
  \label{alpha_simple}
  \alpha_t + V_k({\bf u}(\alpha))\alpha_x=0.
\end{equation}
Because the system (\ref{hyp1}) was assumed strictly hyperbolic, it
has $N$ families of simple wave solutions.

Since $\alpha=const$ along the characteristic $dx/dt=V_k({\bf
  u}(\alpha))$, this characteristic represents a straight line in the
$x,t$ plane.  All the other the characteristics are curvilinear.  In
particular, if $\alpha(x,t)=f(x/t)$, where $f' \ne 0$, equation
(\ref{alpha_simple}) implies $V_k=x/t$. Such {\it similarity
  solutions} play a particularly important role in DSW theory.  Let
$s=x/t$. Then for the similarity solution, $\frac{\rmd}{\rmd s} V_k =
(\frac{\rmd}{\rmd \alpha} V_k) f'(s)=1$ and so $\frac{\rmd}{\rmd \alpha}
V_k = \nabla_{\bf u} V_k \cdot \frac{\rmd}{\rmd \alpha} {\bf u} =
\nabla_{\bf u} V_k \cdot \mathbf{p}^{(k)} \ne 0$
(cf. (\ref{laxnonlin})). Thus, the similarity solution of (\ref{hyp1})
exists only if (\ref{hyp1}) is genuinely nonlinear for all ${\bf u}$
involved.

The \textit{$k$-wave curve} associated with \eqref{hyp1} is the curve
$\mathbf{u}(\alpha)$ in $\R^N$ parametrized by $\alpha$ associated
with the $k^{\mathrm{th}}$ characteristic family where
$V_k(\mathbf{u}(\alpha)) = x/t$.  The curve identifies the states that
can be continuously connected by a simple wave in the
$k^{\mathrm{th}}$ characteristic family.  Suppose two states
$\mathbf{u}_\pm$ lie on the $k$-wave curve according to
$\mathbf{u}_\pm = \mathbf{u}(\alpha_\pm)$ where $\alpha_- < \alpha_+$.
Then a global, continuous solution to \eqref{alpha_simple} exists so
long as $V_k(\mathbf{u}_-) < V_k(\mathbf{u}_+)$.  This result imparts
a directionality to the $k$-wave curve expressing the
``connectability'' by a simple wave in the $k^{\mathrm{th}}$
characteristic family of two states on the curve.

In the construction of a simple wave one can use any of the
components, say $u_i$, of the vector ${\bf u}$ instead of the quantity
$\alpha(x,t)$ and seek the solution in the form ${\bf u}= \tilde {\bf
  u}(r)$, where $r=u_i$.  Then introducing $V(r)=V_k({\bf u}(r))$ one
obtains the simple wave equation of the form
\begin{equation}\label{simple0}
r_t+V(r)r_x=0,   \quad V'(r) \ne 0\, .
\end{equation}
If system (\ref{hyp1}) is diagonalizable then its {\it simple wave
  reduction} (\ref{simple0}) can be obtained directly from the
diagonal form (\ref{rim}) by setting constant all but one Riemann
invariant.  Indeed, let $r_j = \hbox{const}_j$, $\forall j \ne k$ in
(\ref{rim}). Then for $r=r_{k}(x,t)$ we obtain the {\it simple-wave}
equation (\ref{simple0}). There are $N$ different simple wave
reductions of system (\ref{rim}).

The general solution $r(x,t)$ of (\ref{simple0}) can be conveniently
represented by an implicit formula
\begin{equation}
  \label{swsol}
  x-V(r)t=W(r),
\end{equation}
where $W(r)$ is an arbitrary function. For an initial value problem,
the function $W(r)$ has the meaning of the inverse function to the
initial profile $r(x,0) = r_0(x)$, provided such an inverse exists.
Let $V'(r)>0$ and $r_0'(x)<0$ for some $x$. Then the solution
(\ref{swsol}) implies the occurrence of gradient catastrophe: $|r_x|
\to \infty$ for some $t=t_b$ and thus, does not exist globally.

\subsubsection{Generalized hodograph method}
\label{sec:gener-hodogr-meth}

Now we relax the simple-wave constraint and consider a more general
reduction of the diagonal system (\ref{rim}) in which all but two
Riemann invariants are constant. Without loss of generality, we assume
that the two changing invariants are $r_1$ and $r_2$ and consider the
system of two equations
\begin{equation}
  \label{2rim}
  \begin{split}
    (r_1)_t+V_1(r_1, r_2)(r_1)_x = 0, \\ 
    (r_2)_t+V_2(r_1, r_2)(r_2)_x = 0,
  \end{split}
\end{equation}
where $V_1 \ne V_2$ so that the system is strictly hyperbolic. 
To integrate the system (\ref{2rim}), we take advantage of the
classical {\it hodograph transform} (see
e.g. \cite{whitham_linear_1974,rozhdestvenskii_systems_1983}),
which is achieved by interchanging the role of dependent and
independent variables.

Here we present a somewhat modernized version of the hodograph method,
which admits, under a certain set of restrictions, a generalization to
systems (\ref{rim}) with $N>2$.  To this end, we consider $x=x(r_1,
r_2)$, $t=t(r_1, r_2)$ and, assumung that the Jacobian of the above
hodograph transformation $J=\partial_1 x\, \partial_2 t - \partial_2
x\, \partial_1 t$ is nonzero, we obtain a system of
two {\it linear} PDEs:
\begin{equation}
  \label{hodeq}
  \begin{split}
    \partial_1 x - V_2(r_1, r_2) \partial_1 t =0, \\ \partial_2 x -
    V_1(r_1, r_2) \partial_2 t =0, 
  \end{split}
\end{equation}
where, we recall $\partial_k=\partial / \partial r_k$. Thus we have
reduced the problem of integration of the quasilinear system
(\ref{rim}) for $N=2$ to integration of the system of two linear
PDEs (\ref{hodeq}). An essential part of the construction of the
solution to the original system (\ref{2rim}) using the hodograph
method is the inversion of the solution $x(r_1, r_2)$, $t(r_1, r_2)$
of system (\ref{hodeq}), which is not always possible, e.g., in the
vicinity of a wave-breaking point. Indeed, by solving (\ref{hodeq})
and inverting the hodograph solution $x(r_1, r_2)$, $t(r_1, r_2)$ one
generally obtains only a {\it local solution} of
(\ref{2rim}). However, the fact that any smooth, non-constant, local
solution of (\ref{2rim}) can be obtained in this way constitutes the
integrability of the system (\ref{2rim}) via the hodograph
transform. We note that the simple-wave solution cannot be obtained by
the classical hodograph transform due to degeneracy of the mapping
$(x,t) \mapsto (r_1,r_2)$ and thus, vanishing of the Jacobian $J$.

The hodograph method is known to be poorly compatible with the Cauchy
problem for (\ref{2rim}) (see, e.g., \cite{whitham_linear_1974}) so it
has not been often used in classical fluid dynamics. As we shall see,
however, it is ideally compatible with nonlinear free boundary
problems for the modulation equations arising in DSW theory.

We now introduce in (\ref{hodeq}) new characteristic dependent
variables $W_{1,2}(r_1, r_2)$ instead of $x$ and $t$:
\begin{equation}\label{W}
W_{1,2}({\bf r}) = x-V_{1,2}({\bf r})t\, ,
\end{equation}
so that (\ref{hodeq}) assumes a symmetric form
\begin{equation}\label{hodeq1}
\frac{\partial_1 W_2}{W_2 - W_1} = \frac{\partial_1 V_2}{V_2 - V_1}, \ \ \frac{\partial_2 W_1}{W_2 - W_1} = \frac{\partial_2 V_1}{V_2 - V_1}\, .
\end{equation}
Any smooth, non-constant {\it local} solution of system (\ref{2rim}) is
obtainable via (\ref{W}), (\ref{hodeq1}).  Note that the hodograph
solution in the form (\ref{W}) represents a natural generalization of
the simple-wave characteristic solution (\ref{swsol}), the crucial
difference being that in (\ref{W}), unlike in (\ref{swsol}), $W_{1,2}$
are not arbitrary functions but must satisfy the linear PDEs
(\ref{hodeq1}).

Remarkably, the hodograph method in the form (\ref{W}), (\ref{hodeq1})
can be extended to the multi-component case with $N > 2$. Such a
possibility is highly non-trivial since the classical hodograph
construction outlined above is not applicable due to the mapping
$(x,t) \mapsto (r_1, r_2, \dots , r_N)$ being no longer one-to-one.

A discovery made by Tsarev in 1985 \cite{tsarev_poisson_1985} (see
also \cite{dubrovin_hydrodynamics_1989}) was that the hodograph
construction in the symmetrized form \eqref{W}, (\ref{hodeq1}) is
still applicable to diagonal hydrodynamic type systems (\ref{rim})
with $N > 2$ if the characteristic velocities $V_j({\bf r})$ satisfy
the following set of conditions:
\begin{equation}\label{semih}
\partial_{j}\frac{\partial_{k}V_{i}}{V_{k}-V_{i}}=\partial_{k}\frac{
\partial_{j}V_{i}}{V_{j}-V_{i}}, \qquad i\neq j\neq k \, .
\end{equation}
Hydrodynamic type systems satisfying conditions (\ref{semih}) are
called {\it semi-Hamiltonian}.

A semi-Hamiltonian hydrodynamic type system possesses infinitely many
conservation laws parametrized by $N$ arbitrary functions of a single
variable. Its general local solution for $\partial _{x}r_{i}\neq 0$, $
i=1,\dots ,N$ is given by the {\it generalized hodograph} formula
\cite{tsarev_poisson_1985}
\begin{equation}\label{ghod}
  x-V_{i}(\mathbf{r})t=W_{i}(\mathbf{r})\,, \quad i=1, 2, \dots, N\, ,
\end{equation}
where the functions $W_{i}(\mathbf{r})$ are found from the linear system
of PDEs:
\begin{equation}
  \frac{\partial_{i}W_{j}}{W_{i} -W_{j}} = \frac{\partial _{i}V_{j}}{
    V_{i}-V_{j}} \,,\quad i,j=1,\dots ,N,\quad i\neq j.  \label{ts}
\end{equation}
One can see that the system of linear PDEs (\ref{ts}) is
overdetermined for $N \geqslant 3$. The semi-Hamiltonian property
(\ref{semih}) is nothing but the condition of compatibility of the
system (\ref{semih}) and thus, provides the criterion of {\it
  integrability } of the diagonal, hydrodynamic type system (\ref{rim}) in
the above generalized hodograph sense.
Note that for $N=2$ the conditions (\ref{semih}) do not exist so that
any $2 \times 2$ diagonal system is integrable.

\subsection{Whitham modulation theory} 
\label{sec:whith-modul-theory}

For the mathematical description of DSWs, we adopt Whitham's nonlinear
averaging principle \cite{whitham_non-linear_1965}, first applied to
KdV DSWs by Gurevich and Pitaevskii
\cite{gurevich_nonstationary_1974}.  The fundamental assumption is
that the DSW can be asymptotically represented as a slowly modulated
periodic traveling wave solution of the original nonlinear dispersive
equation with modulations of the wave's amplitude, wavelength, and
mean on a spatio-temporal scale that is much greater than the
wavelength and period of the traveling wave (see
Fig.~\ref{fig:dsw_schematic}).  This scale separation enables one to
effectively split the DSW problem into two separate tasks of differing
complexity: the relatively easy problem of finding a family of
periodic traveling wave solutions and the hard problem of finding an
appropriate modulation that matches the modulated wavetrain with a
smooth external flow.

\subsubsection{Whitham's method of slow modulations} 
\label{sec:whithams-method-slow}

The Whitham averaging method, a nonlinear generalization of long-time
asymptotic methods for linear waves or WKB, can also be viewed as a
field-theoretic analogue of the Krylov-Bogoliubov method of averaging
from the theory of ODE \cite{bogoliubov_asymptotic_1961}. The
modulation equations, originally derived by averaging conservation
laws \cite{whitham_non-linear_1965}, can also be determined by an
averaged Lagrangian procedure \cite{whitham_general_1965}, or
multiple-scale asymptotic methods \cite{luke_perturbation_1966}.
Descriptions of different versions of the Whitham method with various
degrees of completeness and rigor can be found in many monographs
(see, e.g.,
\cite{karpman_non-linear_1974,ostrovsky_modulated_1999,infeld_nonlinear_2000,kamchatnov_nonlinear_2000,ostrovsky_asymptotic_2014}).
A particularly succint exposition can be found in the review article
\cite{dubrovin_hydrodynamics_1989}.

Following Whitham \cite{whitham_non-linear_1965}, the equations
describing slow modulations of periodic nonlinear waves are obtained
by averaging conservation laws of the original dispersive equations
over a family of periodic traveling wave solutions. For simplicity we
shall consider the scalar case, the vector generalization is
straightforward. Given an $N^{\mathrm{th}}$ order nonlinear evolution
equation, $q_t = K(q, q_x, \dots, q^{(N)})$, where $q(x,t) \in
\mathbb{R}$ and $K(x_1, \dots, x_{N+1})$ is some function,
implementation of the Whitham method requires the existence of a
$N$-parameter family of periodic traveling wave solutions
$q(x,t)=\varphi(\theta;\mathbf{u})$, $\mathbf{u} \in \R^N$, with phase
$\theta = kx- \omega t$, wavenumber $k(\mathbf{u})$, and frequency
$\omega(\mathbf{u})$.  A natural parameterization of this solution is
to impose a fixed period, say $2\pi$: $\varphi(\theta+2\pi;\mathbf{u})
= \varphi(\theta;\mathbf{u})$.  Then the spatial $L(\mathbf{u}) =
2\pi/k(\mathbf{u})$ and temporal $\tau(\mathbf{u}) =
2\pi/\omega(\mathbf{u})$ periods are determined.  In order to carry
out Whitham's method, the evolution equation must admit at least $N -
1$ conserved densities $\mathcal{P}_i[q]$ and fluxes
$\mathcal{Q}_i[q]$, $i = 1,\ldots,N-1$ corresponding to local
conservation laws
\begin{equation}
  \label{eq:5}
  \frac{\partial}{\partial t} \mathcal{P}_i + \frac{\partial}{\partial
    x} \mathcal{Q}_i = 0, \quad i=1,\ldots,N-1 .
\end{equation}
The modulation equations are found by assuming slow spatio-temporal
evolution of the wave's parameters, $\mathbf{u} = \mathbf{u}(x,t)$,
relative to the periods $L$ and $\tau$: $|\mathbf{u}_x| \ll
|\mathbf{u}|/L$ and $|\mathbf{u}_t| \ll |\mathbf{u}|/\tau$.
Introducing $q(x,t)=\varphi(\theta; \mathbf{u}(x,t))$ in (\ref{eq:5}),
the conservation laws are then averaged, resulting in $N-1$ modulation
equations for $\mathbf{u}$ (see \cite{whitham_non-linear_1965},
\cite{kamchatnov_nonlinear_2000} for details)
\begin{equation}
  \label{eq:6}
  \frac{\partial}{\partial t}  \overline{\mathcal{P}_i[\varphi]} +
  \frac{\partial}{\partial x}   \overline{\mathcal{Q}_i[\varphi]} = 0, 
\end{equation}
where $
\overline{\mathcal{F}[\varphi]}(\mathbf{u}) \equiv
\frac{1}{2\pi}\int_0^{2\pi}
\mathcal{F}[\varphi(\theta;\mathbf{u})]\rmd \theta.
$
In order to
avoid secular growth at leading order, the reconstruction of the
modulated wave $\varphi(\theta;\mathbf{u})$ necessitates the
introduction of the generalized (modulated) wavenumber and frequency
via $\theta_x = k(\mathbf{u})$ and $\theta_t = -\omega(\mathbf{u})$.
The closure of the $N-1$ modulation equations \eqref{eq:6} is
achieved by equating mixed partials $\theta_{xt} = \theta_{tx}$,
resulting in the conservation of waves
\begin{equation}
  \label{eq:7}
  k_t + \omega_x = 0 .
\end{equation}
Note that the Whitham equations \eqref{eq:6}, \eqref{eq:7}, unlike
\eqref{eq:5}, are a system of non-dispersive conservation laws, which,
assuming non-vanishing of relevant Jacobians, can be represented in
standard form as a system of $N$ first order quasi-linear equations
\begin{equation}
  \label{eq:8}
  \mathbf{u}_t + A(\mathbf{u}) \mathbf{u}_x = 0 .
\end{equation}
The matrix $A(\mathbf{u})$ encodes information about both nonlinear
and dispersive properties of the original evolution equation.  A
subtle but important point to note is that a slowly varying phase
shift $\theta_0(x,t)$ of the modulated wave arising in the reconstruction of the phase $\theta$ from the 
dependencies $k(\mathbf{u})$ and $\omega(\mathbf{u})$ is an additional
parameter, undetermined by the modulation equations as written.  One
must appeal to higher order effects or, if available, utilize the
integrable structure of the Whitham equations.  An example of the
latter will be described in Sec.~\ref{gen_sol_kdv-whitham}.

Equations \eqref{eq:8} are a system of hydrodynamic type, as described
in Sec.~\ref{sec:basic-notions}.  Structural properties of the Whitham
equations \eqref{eq:8} provide practical information about the
nonlinear evolution of modulated waves.  For example, hyperbolicity of
the Whitham equations implies modulational stability of nonlinear
wavetrains and enables the use of the method of characteristics.  In
this case, the characteristic velocities can be viewed as nonlinear
group velocities.  In the limit of vanishing amplitude, two of these
characteristic velocities merge, becoming the linear group velocity,
the remaining characteristic velocity becoming the dispersionless
nonlinear wave speed.  Splitting of the linear group velocity for
modulated waves of finite amplitude is one of the fundamental
nonlinear effects predicted by Whitham theory \cite{whitham_linear_1974}. This effect plays a
decisive role in the unsteady, expanding structure of DSWs.

Although many dispersive hydrodynamic systems exhibit hyperbolic
Whitham equations, when the Whitham equations are of elliptic type,
the initial value problem is ill-posed and nonlinear wavetrains
exhibit modulational instability
\cite{lighthill_contributions_1965,benjamin_instability_1967,whitham_linear_1974,zakharov_modulation_2009}.
Additional structural properties such as genuine nonlinearity and
strict hyperbolicity play an important role in the structure of DSWs
and, in particular, the DSW fitting method
(Sec.~\ref{sec:dsw-fitting-method}).  When these properties are
relaxed, more complex wave structures such as double-waves,
undercompressive DSWs, and contact DSWs are possible
(Sec.~\ref{sec:non-classical-dsws}) \cite{el_dispersive_2015}.

After first proposing the averaged conservation law approach
\cite{whitham_non-linear_1965}, Whitham developed modulation theory
utilizing an equivalent, averaged Lagrangian
\cite{whitham_general_1965,whitham_linear_1974}, which we briefly
outline here.  Following \cite{whitham_linear_1974}, consider the
action functional
\begin{equation*}
  S[q] = \int \int
  L(q,q_t,q_x,q_{xx},q_{tt},q_{xt},\ldots) \rmd
  t \rmd x
\end{equation*}
with Lagrangian $L$.  The spatio-temporal evolution of the state
$q(x,t)$ can be characterized by a critical point of the
variational principle $\delta S[q] = 0$ yielding the
Euler-Lagrange equations 
\begin{equation*}
  L_{q} - \frac{\partial}{\partial t}
  L_{q_{t}} - \frac{\partial}{\partial x} L_{q_{x}} +
  \cdots = 0.
\end{equation*}
Whitham's key idea was to consider the average variational principal
\begin{equation}
  \label{eq:10}
  \delta \int \int \mathcal{L}(\mathbf{u}) \rmd t \rmd x = 0 ,
\end{equation}
determined by integrating the Lagrangian $L$ over the family of periodic
solutions $\varphi(\theta;\mathbf{u}) =
\varphi(\theta+2\pi;\mathbf{u})$ described earlier
\begin{equation*}
  \mathcal{L}(\mathbf{u}) \equiv \int_0^{2\pi} L(\varphi,\theta_t
  \varphi_\theta, \theta_x \varphi_\theta,\ldots) \rmd \theta .
\end{equation*}
For concreteness, we consider a scalar, second order PDE whose
traveling wave solution is characterized by its amplitude $a$,
wavenumber $k = \theta_x$, and frequency $\omega = -\theta_t$.  Then the
modulation equations are the Euler-Lagrange equations for the
average variational principal \eqref{eq:10}
\begin{align}
  \label{eq:13}
  &\delta a: \qquad \mathcal{L}_a = 0 \\
  \label{eq:14}
  &\delta \theta: \qquad \frac{\partial}{\partial t}
  \mathcal{L}_\omega - \frac{\partial}{\partial x} \mathcal{L}_k = 0,
\end{align}
and the consistency condition \eqref{eq:7}.  Equation \eqref{eq:13}
determines the nonlinear dispersion relation $\omega = \omega(k,a)$.
Equation \eqref{eq:14} is referred to as the conservation of wave
action and is the analogue of an adiabatic invariant of classical
mechanics.

A multiple scales, perturbative analysis of the variational method
enables its justification \cite{whitham_linear_1974}.  Yet another
approach to modulation theory, shown to be equivalent to the previous
two by Luke \cite{luke_perturbation_1966}, see also
\cite{ablowitz_evolution_1970}, is the multiple scales perturbation
method applied {\it directly to the governing PDE}.  As an example, we apply
this method to the KdV equation in the next section.

\subsubsection{KdV modulation system}
\label{sec:kdv-mod-sys}

The averaging method presented in the previous section does not make explicit distinction between ``slow'' and ``fast'' $x,t$-variables.
Here we present the construction of the KdV modulation system via
multiple scales perturbation theory, in which the small parameter  defining the separation of slow and fast scales in the modulated solution is introduced explicitly.  

We consider the KdV equation in the ``physical'' normalization
\begin{equation}
  \label{kdv1}
  u_t + uu_x + u_{xxx}=0 \, .  
\end{equation}
A slowly modulated wave is sought in the form
\begin{equation}
  \label{eq:17}
  u(x,t) = \varphi(\theta,X,T) + \varepsilon u_1(\theta,X,T) + \cdots,
  \quad 0 < \varepsilon \ll 1,
\end{equation}
with slow variables $X = \varepsilon x$, $T = \varepsilon t$ and rapid phase
$\theta = \varepsilon^{-1} S(X,T)$ satisfying $\theta_x = S_X=k(X,T)$, and $\theta_t = S_T=-\omega(X,T)$.  The
small parameter $\varepsilon$ characterizes the ratio of the wave's
typical wavelength (or period) to the typical modulation length (or
temporal) scale, ultimately determined by initial and or boundary
conditions.  Inserting this ansatz into eq.~\eqref{kdv1} and equating
like powers of $\varepsilon$, we obtain
\begin{align}
  \label{eq:16}
  &\mathcal{O}(1)\!: ~ -\omega \varphi_\theta + k \varphi
  \varphi_\theta + k^3 \varphi_{\theta\theta\theta} = 0, 
  \\
  \label{eq:18}
  &\mathcal{O}(\varepsilon^n)\!: ~ \mathcal{A} u_n = F_n , \quad n = 1,
  2, \ldots ,
\end{align}
where the linear operator $\mathcal{A}$ is defined according to
\begin{equation}
  \label{eq:21}
  \mathcal{A} = - \omega \partial_{\theta} + k
  (\varphi \partial_\theta + \varphi_\theta) +
  k^3 \partial_{\theta\theta\theta},
\end{equation}
and each inhomogeneity $F_n$ involves $\varphi$, $u_1$, $\ldots$,
$u_{n-1}$ and their derivatives.  The first two are
\begin{align}
  \label{eq:19}
  F_1 &= - \varphi_T - \varphi \varphi_X - 3k^2 \varphi_{\theta\theta X},\\
  \label{eq:20}
  F_2 &= -3k \varphi_{\theta XX} - \frac{k}{2} (u_1^2)_\theta - (u_1
  \varphi)_X - 3k^2 (u_1)_{\theta\theta X} .
\end{align}
The order one equation \eqref{eq:16} determines the family of periodic
traveling wave solutions.  We utilize the following normalization
\begin{equation*}
  \begin{split}
    \varphi(\theta + 2\pi) &= \varphi(\theta), \quad \theta \in \R, \\
    \varphi_\theta(0) &= 0, \quad \varphi_{\theta\theta}(0) < 0,
  \end{split}
\end{equation*}
so that $\theta = 0$ is the maximum of the wave and the invariance of
eq.~\eqref{eq:16} under the transformation $\theta \to -\theta$
implies that the periodic solution of interest is even in
$\theta$. The $2\pi$-periodicity of the wave determines the nonlinear
dispersion relation $\omega$ in terms of three other free parameters
(eq.~\eqref{eq:16} is third order).  It is necessary to fix the wave
period $p$ in order to avoid secularity, e.g., $\partial_X
\varphi(\theta, X, T) = \partial_X \varphi(\theta + np, X, T) = n
p_X \partial_\theta \varphi(\theta, X, T) + \partial_X
\varphi(\theta,X,T)$, for any integer $n$ holds if and only if $p_X =
0$.  The particular choice $p = 2\pi$ is for convenience.

The choice of parameterization is mainly driven by physical or
mathematical convenience.  A natural physical parameterization is to
utilize the wavenumber $k$, wave amplitude $a \equiv \varphi(0) -
\varphi(\pi)$, and wave mean $\overline{\varphi} \equiv \frac{1}{\pi}
\int_0^{\pi} \varphi(\theta) \rmd \theta$, all depending on the slow
variables $X$ and $T$.  We will first use this parameterization to
obtain the Whitham equations in their physical form, a procedure
readily adapted to other dispersive hydrodynamic equations.  Then, we
will consider mathematical parameterizations, of particular relevance
for the completely integrable KdV equation because they lead to a very
efficient, diagonalized representation of the Whitham equations.

According to the multiple scales procedure, we now seek solutions
$u_n$ to eqs.~\eqref{eq:18} that are bounded and $2\pi$-periodic in
$\theta$ in order to maintain the uniform asymptotic ordering of the
ansatz \eqref{eq:17}.  Necessary conditions for the existence of
periodic solutions of inhomogeneous, linear differential equations
such as \eqref{eq:18} can be formulated by considering the adjoint
operator
\begin{equation}
  \label{eq:25}
  \mathcal{A}^\dagger = (\omega -
  k \varphi) \partial_\theta - k^3 \partial_{\theta\theta\theta} .
\end{equation}
By direct verification, one can show that this operator admits the
following three linearly independent, homogeneous solutions
\begin{equation}
  \label{eq:26}
  \begin{split}
    w_1(\theta) &= 1, \\
    w_2(\theta) &= \varphi(\theta), \\
    w_3(\theta) &= 
    \int^{\theta}_{\theta_0} 
    \frac{\varphi(\theta) - \varphi(s)}{\varphi_s^2(s)} \rmd s  .
  \end{split}
\end{equation}
The dependence of $\varphi$ on $X$ and $T$ has been suppressed here
for simplicity.  Both $w_1$ and $w_2$ are $2\pi$-periodic and bounded
in $\theta$ but $w_3$ is not periodic as can be seen by the fact that
\begin{equation*}
  w_3(2\pi)-w_3(0) = \int_0^{2\pi} \frac{\varphi(0) -
    \varphi(s)}{\varphi'^2(s)} \rmd s > 0 .
\end{equation*}
Necessary conditions for the existence of a $2\pi$-periodic solution
$u_1$ of eq.~\eqref{eq:18} are therefore the two orthogonality
relations
\begin{align}
  \label{eq:28}
  \int_0^{2\pi} w_1 F_1 \rmd \theta = 0, \quad \int_0^{2\pi} w_2 F_1
  \rmd \theta = 0.
\end{align}
Substitution of \eqref{eq:19} into \eqref{eq:28} leads to two Whitham
equations
\begin{align}
  \label{eq:29}
  \overline{\varphi}_T + \frac{1}{2} \left ( \overline{\varphi^2}
  \right )_X &= 0 , \\
  \label{eq:30}
  \left ( \overline{\varphi^2} \right)_T +  \left ( \frac{2}{3}
    \overline{\varphi^3} - 6 k^2 \overline{\varphi_\theta^2} \right
  )_X &= 0 ,
\end{align}
also obtainable by averaging the first two KdV conservation laws per
Whitham's original approach.  Adding the conservation of waves ($S_{XT}=S_{TX}$),
\begin{equation}
  \label{eq:24}
  k_T + \omega_X = 0
\end{equation}
completes the Whitham modulation system. As a matter of fact,
(\ref{eq:24}) is equivalent to (\ref{eq:7}).

\begin{figure}
  \centering
  \includegraphics{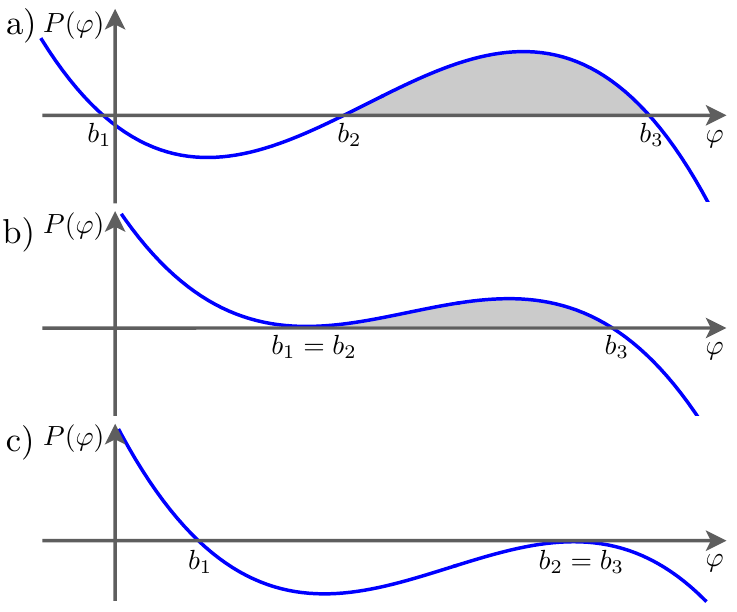}
  \caption{Example cubic potential curves for KdV traveling wave
    solutions: (a) periodic, (b) soliton, (c) constant.  When $0 < b_3
    - b_2 \ll 1$, the periodic solution is approximately sinusoidal.}
  \label{fig:kdv_potential}
\end{figure}
An alternative, mathematically motivated parameterization for
$\varphi$ is to integrate eq.~\eqref{eq:16} twice and obtain the ODE
\begin{equation}
  \label{kdv_ODE} 
  (\varphi_\theta)^2 = P(\varphi) \equiv \frac{1}{3k^2}
  (b_1-\varphi)(b_2-\varphi)(b_3-\varphi) \, ,
\end{equation}
where $b_1 \le b_2 \le b_3$, assumed real, are related to two
constants of integration and the wavenumber $k$.  As depicted in
Fig.~\ref{fig:kdv_potential}, the roots of the cubic potential
$P(\varphi)$ determine the behavior of the traveling wave. All real
solutions of interest satisfy $b_2 \le \varphi \le b_3$.  When two
roots coalesce, the periodic solution degenerates to either a soliton
when $b_1 = b_2$ or a constant when $b_2 = b_3$.  For $0< b_3-b_2 \ll 1$,
the periodic solution is approximately sinusoidal.  Averaging of
the quantity $\mathcal{F}[\varphi]$ is therefore achieved via
\begin{equation}\label{eq:aver}
  \overline{\mathcal{F}[\varphi]} = \frac{1}{\pi} \int_0^{\pi}
  \mathcal{F}[\varphi(\theta)] \rmd \theta = \frac{1}{\pi}
  \int_{b_2}^{b_3} \frac{\mathcal{F}[\varphi] \rmd \varphi}{\sqrt{P(\varphi)}} ,
\end{equation}
where derivatives of $\varphi(\theta)$ in $\mathcal{F}[\varphi]$ are
replaced by $\sqrt{P(\varphi)}$.  The $2\pi$-periodicity of $\varphi(\theta)$ implies on using (\ref{kdv_ODE}) that
\begin{equation}\label{eq:2pi_per}
2\pi = \int^{2\pi}_0 \rmd\theta = 2\sqrt{3}k \int^{b_3}_{b_2} \frac{\rmd \varphi}{\sqrt{(b_1-\varphi)(b_2-\varphi)(b_3-\varphi)}},
\end{equation}
which relates the wavenumber $k$ with the roots $\{b_i \}_{i=1}^3$ of the cubic \eqref{kdv_ODE}.
 
The roots $\{b_i\}_{i=1}^3$ of the
cubic \eqref{kdv_ODE} can be used as modulation variables in
eqs.~\eqref{eq:29}, \eqref{eq:30}, and \eqref{eq:24} rather than the
physical parameterization $\{k,a,\overline{\varphi}\}$.  The two are
related via
\begin{equation}
  \label{eq:33}
  \begin{split}
    a &= b_3 - b_2, \\
    k &= \frac{\pi\sqrt{b_3-b_1}}{2\sqrt{3}K(m)}, \quad m =
    \frac{b_3-b_2}{b_3-b_1}, \\ 
    \overline{\varphi} &= b_1 + (b_3 - b_1) \frac{E(m)}{K(m)}
    , 
  \end{split}
\end{equation}
where $K(m)$ and $E(m)$ are the complete elliptic integrals of the
first and second kind, respectively.  The modulation solution of
eqs.~\eqref{eq:29}--\eqref{eq:24} determine the evolution of the
slowly varying periodic solution of eq.~\eqref{kdv_ODE}, which is
\begin{equation}
  \label{eq:34}
  \varphi(\theta) = b_1 + (b_3 - b_1) \mathrm{dn}^2 \left (
    \frac{K(m)}{\pi} \theta; m 
  \right ),
\end{equation}
where $\mathrm{dn}$ is a Jacobi elliptic function and $m$ is defined
in eq.~\eqref{eq:33}.

In his 1965 paper \cite{whitham_non-linear_1965}, Whitham made an
important discovery: using ingenious algebraic manipulations (see
\cite{kamchatnov_nonlinear_2000} for detail), he showed that the
modulation system \eqref{eq:29}--\eqref{eq:24} for the KdV equation
can be represented in Riemann invariant form
\begin{equation}
  \label{eq:31}
  \frac{\partial r_i}{\partial T} + V_i \frac{\partial r_i}{\partial
    X} = 0 , \quad i = 1, 2, 3 \, .
\end{equation}
The Riemann invariants $r_1 \le r_2 \le r_3$ found by Whitham are
related to the roots of the potential curve according to
\begin{equation}
  \label{kdv_Riemann_inv} 
  r_1=\frac{b_1+b_2}{2}, \ \
  r_2=\frac{b_1+b_3}{2}, \ \ r_3=\frac{b_2+b_3}{2}, \ 
\end{equation}
while the characteristic velocities are expressed in terms of $r_j$ by
\begin{equation}
  \label{kdv_char_vel}
  \begin{split}
    V_1 &= V - \frac{2}{3}(r_2 - r_1)
    \frac{K(m)}{K(m) - E(m)}\\
    V_2 &= V - \frac{2}{3}(r_2 - r_1)
    \frac{(1-m)K(m)}{E(m) - (1-m)K(m)} \\
    V_3 &= V + \frac{2}{3}(r_3 - r_1) \frac{(1-m)K(m)}{E(m)},
  \end{split}
\end{equation}
where $m = (r_2-r_1)/(r_3-r_1)$.  The phase velocity is $V \equiv
\omega/k = \frac{1}{3}(r_1 + r_2 + r_3)$.

It will be convenient for the study of DSWs to identify the invertible
transformation between the physical para\-metrization
$(a,k,\overline{\varphi})$ and the Riemann invariants $(r_1,r_2,r_3)$:
\begin{equation}
  \label{eq:46}
  \begin{split}
    a &= 2(r_2 - r_1), \\
    k &= \frac{\pi \sqrt{r_3 - r_1}}{\sqrt{6} K(m)}, \quad m =
    \frac{r_2 - r_1}{r_3 - r_1}, \\
    \overline{\varphi} &= r_1 + r_2 - r_3 + 2(r_3 - r_1)
    \frac{E(m)}{K(m)} .
  \end{split}
\end{equation}
The periodic wave \eqref{eq:34} takes the form
\begin{equation}
  \label{eq:65}
  \varphi(\theta) = r_1 + r_2 - r_3 + 2(r_3 - r_1)\mathrm{dn}^2\left (
    \frac{K(m)}{\pi} \theta; m \right ) .
\end{equation}
In the limit $r_2 \to r_3$ ($m \to 1$) (cf.~Fig.~\ref{fig:kdv_potential}b), the
wave \eqref{eq:65} takes the form of a soliton
\begin{equation}
  \label{eq:66}
  \varphi = \overline{\varphi} + a_{\rm s} \mathrm{sech}^2 \left (
    \frac{\sqrt{a_{\rm s}}}{\sqrt{12}}(x - V_{\rm s} t - \theta_0) \right ) ,
\end{equation}
where $V_{\rm s} = \overline{\varphi} + a_{\rm s}/3$ is the soliton
amplitude-speed relation.  The background $\overline{\varphi}$,
soliton amplitude $a_{\rm s}$, and velocity $V_{\rm s}$ are expressed in terms of
$r_1$ and $r_3$ as
\begin{equation}
  \label{eq:67}
  \overline{\varphi} = r_1, \quad a_{\rm s} = 2(r_3 - r_1), \quad
  V_{\rm s} =
  \frac{1}{3}(r_1 + 2r_3) .
\end{equation}
In the limit $r_2 \to r_1$ ($m \to 0$) (cf.~Fig.~\ref{fig:kdv_potential}c), the
wave \eqref{eq:65} is a vanishing harmonic wave
\begin{equation}
  \label{eq:68}
\varphi =  \overline{\varphi} + \frac{a_{\rm h}}{2}\left ( \cos ( kx - \omega_0
    t) - 1 \right ) + 
  \mathcal{O}(a_{\rm h}^2),
\end{equation}
where $\omega_0(k, \overline \varphi) =  \overline \varphi k - k^3$ is the linear dispersion
relation.  The wave amplitude $a_{\rm h}$, wavenumber $k$, and background
$\overline{\varphi}$ are related to the Riemann invariants via
\begin{equation}
  \label{eq:69}
  \overline{\varphi} = r_3, \quad a_{\rm h} = 2(r_2 - r_1) \ll 1, \quad k^2 =
  \frac{2}{3}(r_3 - r_1) .
\end{equation}

A convenient and compact representation of the characteristic
velocities (\ref{kdv_char_vel}) can be obtained by considering the
wave conservation equation \eqref{eq:24} as a consequence of the
diagonal system \eqref{eq:31}
\cite{gurevich_breaking_1991,kudashev_wave-number_1991,kamchatnov_nonlinear_2000}.
Equations \eqref{eq:31} imply that $k_T + \omega_X = 0$ can be written as
(recall $\partial_j = \partial/\partial r_j$)
\begin{equation*}
  \sum_{j=1}^3 \frac{\partial r_j}{\partial X} \left [  - V_j
    \partial_j k + \partial_j \omega \right ] = 0 .
\end{equation*}
Generic variations of $r_j$ imply that each expression in brackets
must be zero so that the characteristic velocities satisfy
\begin{equation}
  \label{vscalar}
  V_j=\frac{\partial_j \omega}{\partial_j k}= \left (
    1-\frac{L}{\partial_j L} \partial_j \right ) V,  \quad i=1,2,3,
\end{equation}
where $L = 2\pi/k = 2 K(m) \sqrt{6/(r_3-r_1)}$, and the relation $\omega=kV$ was used.
We note that expressions (\ref{vscalar}) are universal for diagonalizable modulation systems and elucidate the significance of the characteristic velocities $V_j$ of the Whitham modulation system as nonlinear group velocities. 

We introduced the small parameter $\eps$ and the slow variables $X$
and $T$ in order to clearly implement the multiple scales perturbation
procedure.  This is a standard approach for limit process expansions
(see, e.g., \cite{ablowitz_nonlinear_2011}).  What is effectively
being considered is the \textit{long time behavior} of the modulated
periodic wave.  Now that we know the Whitham modulation equations
\eqref{eq:31}, rather than explicitly identifying the small parameter
$\eps$, we can take $\eps = 1$ with the caveat that the modulation
equations are valid over long time and spatial scales, $t,x \gg 1$.

An alternative derivation method and properties of the KdV Whitham
system in Riemann invariant form \eqref{eq:31} will be described in
subsequent sections.

\subsubsection{Whitham vs. NLS}
\label{sec:whitham-vs.-nls}

The Whitham equations, a system of quasi-linear equations of
hydrodynamic type, are applicable to modulations of large amplitude
dispersive waves.  The structure of slowly varying, weakly nonlinear, nearly
monochromatic wavetrains is governed by the NLS equation, a universal
model of this phenomenon \cite{benney_propagation_1967}.  It is
natural to ask how these two models are related.  Newell
\cite{newell_solitons_1985} explored this issue with a specific model
equation but the ideas are applicable more generally
\cite{ablowitz_evolution_1970}.

An obvious difference between the NLS and Whitham equations is that
the former is dispersive and the latter is not.  Therefore, taking a
straightforward, small amplitude reduction of the Whitham equations
will not result in NLS.  Consider the derivation via multiple scales
of the KdV Whitham equations in the previous section.  The two
secularity conditions at first order \eqref{eq:29}, \eqref{eq:30}
result from orthogonality to the two-dimensional kernel of the adjoint
linear operator $\mathcal{A}^\dagger$ \eqref{eq:25} (recall that only
$w_1$, $w_2$ in \eqref{eq:26} are $2\pi$-periodic, while $w_3$ is
not).  This enabled the maintenance of a well-ordered, uniform
asymptotic sequence as $\varepsilon \to 0$.  In the weakly nonlinear
regime, we have $\varphi = \overline{\varphi} + \frac{a}{2} \cos
(\theta) + \mathcal{O}(a^2)$ so that, with $\theta_0 = \pi/2$ in
eq.~\eqref{eq:26},
\begin{equation*}
  w_3(\theta) = \frac{2}{a}(-1+\sin \theta) + \mathcal{O}(1), \quad 0 < a
  \ll 1 ,
\end{equation*}
a $2\pi$-periodic function, linearly independent of $w_1$ and $w_2$.
Now the operator $\mathcal{A}^\dagger$ admits a three-dimensional
kernel so an additional orthogonality condition is required to remove
secularity.  Therefore, it is not possible to first take the
asymptotic limit process $\varepsilon \to 0$ and then $a \to 0$.  The
derivation of NLS relies on the maximal balance $\varepsilon =
\mathcal{O}(a)$ and a time scale $\varepsilon^2 t$ and spatial scale
$\varepsilon (x - c_{\rm g} t)$ where $c_{\rm g}$ is the group
velocity of linear waves.  Therefore, before taking the small
amplitude reduction, higher order corrections to the Whitham equations
\eqref{eq:29} and \eqref{eq:30} must be included.  They include
dispersive effects.  See \cite{whitham_linear_1974}, where Whitham
incorporated dispersive effects into the modulation equations for the
nonlinear Klein-Gordon equation.  More recently, the emergence of
dispersion in modulations of a general class of second order equations
has been investigated in \cite{bridges_breakdown_2015}.

\subsubsection{IST- integrability and Riemann invariants}
\label{sec:integr-riem-invar}

After Whitham's discovery of the diagonal structure of the KdV
modulation system, a similar set of Riemann invariants was found in
\cite{driscoll_modulational_1975} for the modified KdV (mKdV)
equation, connected with the KdV equation by the Miura transformation
\cite{miura_kortewegvries_1968}.  Later, Flaschka, Forest and
McLaughlin (FFM) \cite{flaschka_multiphase_1980} showed that the
availability of Riemann invariants for the KdV-Whitham system is
intimately linked to the integrable structure of the KdV equation via
the IST. Subsequently, Riemann invariants were found for other
modulation systems associated with integrable equations such as the
nonlinear Schr\"odinger equation
\cite{forest_geometry_1986,pavlov_nonlinear_1987}, the Kaup-Boussinesq
system \cite{el_integrable_2001} and others.

The method of FFM is based on finite-gap integration theory, a highly
non-trivial extension of the IST utilizing tools from
algebraic-geometry to the case of periodic boundary conditions (see
\cite{belokolos_algebro-geometric_1994}). Finite-gap theory was used
in \cite{flaschka_multiphase_1980} to prescribe and study equations
for the slow modulations of $N$-phase wavetrains via averaging of conservation laws
over the $N$-torus.  Note that Whitham's original
paper \cite{whitham_non-linear_1965} considered only $N=1$. The main
result of \cite{flaschka_multiphase_1980} is that the Riemann
invariants of the KdV-Whitham modulation system are the endpoints of
spectral bands of $N$-phase or, alternatively, $N$-gap KdV
solutions. Krichever \cite{krichever_method_1989} generalized the FFM
construction to the two-dimensional case in the KP equation and also
derived a family of exact solutions to the modulation equations.

The FFM construction of multiphase averaging is quite technical and
makes prominent use of the theory of hyperelliptic Riemann surfaces
and associated Abelian differentials. For the single-phase case,
playing a key role in DSW theory, Kamchatnov
\cite{kamchatnov_new_1997,kamchatnov_nonlinear_2000} developed a
reduced version of the finite-gap averaging technique, which does not
require the use of algebro-geometric tools of FFM theory.
Kamchatnov's method applies to integrable equations belonging to the
Ablowitz-Kaup-Newell-Segur (AKNS) hierarchy
\cite{ablowitz_inverse_1974} and has the advantage of delivering both
the traveling wave solution and the Whitham equations in the Riemann
invariant parametrization. We now apply this method to recover the
Riemann invariant form (\ref{eq:31}) of the KdV-Whitham system.

We shall be using here the ``IST-friendly'' normalization of the KdV equation
\begin{equation} 
\label{kdv6}
 u_t+6uu_x+u_{xxx}=0.
\end{equation}
The integrability of the KdV equation is based on the possibility of
representing it as a compatibility condition of two linear
differential equations, called the Lax pair, for the same complex
function $\psi(x,t)$ \cite{lax_integrals_1968}. It is convenient to
represent the KdV Lax pair in the form
\begin{align}
  \psi_{xx} &= \mathcal{A}\psi,\label{laxeqn1}\\
  \psi_t &=
  -\frac{1}{2}\mathcal{B}_x\psi+\mathcal{B}\psi_x,\label{laxeqn2} 
\end{align}
where
\begin{equation} 
  \mathcal{A}=-(u+\lambda),\quad
  \mathcal{B}=4\lambda-2u, \label{laxkdv} 
\end{equation}
and $\la$ is a complex parameter. The first equation (\ref{laxeqn1})
is the quantum-mechanical Schr\"odinger equation \\ $-\psi_{xx}
+V(x)\psi= \lambda \psi$, which specifies for a given potential
$V(x)=-u(x,t)$ a stationary spectral problem with $t$ being a
parameter. The second equation (\ref{laxeqn2}) of the Lax pair
constitutes the evolution problem for $\psi$.  A calculation shows
that the compatibility condition $(\psi_{xx})_t=(\psi_t)_{xx}$
yields the KdV equation (\ref{kdv6}) provided the isospectrality
condition $\la_t=0$ holds.

We now take two basis solutions $\psi^+$ and $\psi^-$ of
(\ref{laxeqn1}) with the asymptotic behaviors $\psi^{\pm} \sim e^{\pm
  i\sqrt{\lambda} x}$  for $\lambda \gg 1$ and
construct the squared basis function
\begin{equation*}
  g=\psi^+\psi^-.
\end{equation*}
It is possible to show that the function $g(\lambda, x, t)$
satisfies the equations
\begin{gather}
  \label{gx} 
  g_{xxx}-2\mathcal{A}_xg-4\mathcal{A}g_x = 0, \\
  \label{gt} 
  g_t=\mathcal{B}g_x-\mathcal{B}_xg  \, , 
\end{gather}
equivalent to the Lax pair (\ref{laxeqn1}), (\ref{laxeqn2}).  Equation
(\ref{gt}) can be rewritten in the conservative form
\begin{equation} 
  \label{gener-conserv}
  \left(\frac{1}{g}\right)_t=\left(\frac{\mathcal{B}}{g}\right)_x
\end{equation}
provided $g \ne 0$. This expression generates an infinite series, each
term a distinct KdV conservation law, by expansion of
(\ref{gener-conserv}) in powers of $1/\la$ (see
\cite{kamchatnov_nonlinear_2000} for details).

Multiplying (\ref{gx}) by $g$ and integrating once yields 
\begin{equation} \label{gx1}
  \frac{1}{2}gg_{xx}-\frac{1}{4}g_x^2-\mathcal{A}g^2=R(\lambda), 
\end{equation}
where $R(\lambda)$ is the integration constant, which can be a
function of the spectral parameter $\lambda$ and, in principle, a
function of $t$.

The crucial step in Kamchatnov's method is the identification of the
function $R(\la)$ with the third degree polynomial defining, via the
radical $\sqrt{R(\lambda)}$, the elliptic Riemann surface on which
periodic KdV solutions exist. To this end, we assume
\begin{equation} 
  \label{Plam}
  \begin{split}
    R(\lambda)=(\lambda-\lambda_1)(\lambda-\lambda_2)(\lambda-\lambda_3) \\
    = \lambda^3-s_1\lambda^2+s_2\lambda-s_3,
  \end{split}
\end{equation}
where $\la_1 \le \la_2 \le \la_3$ are constants, and
\begin{equation}
  \left.
    \begin{split}
      s_1&=\lambda_1+\lambda_2+\lambda_3,\\
      s_2&=\lambda_1\lambda_2+\lambda_1\lambda_3+\lambda_2\lambda_3,\\
      s_3&=\lambda_1\lambda_2\lambda_3
    \end{split}
    \qquad \right\}\label{s}
\end{equation}
are elementary symmetric polynomials. In what follows, we will show
that the dependence (\ref{Plam}) indeed delivers the periodic solution
$u(x,t)$. We also note that in the general framework of finite-gap
theory (see, e.g., \cite{novikov_theory_1984}) the roots $\la_j$ of the
polynomial (\ref{Plam}) represent the endpoints of the spectral bands
of the Schr\"odinger operator (\ref{laxeqn1}) with the potential given
by the periodic KdV solution.

The structure of equation (\ref{gx1}) suggests that its solution
$g(\lambda, x,t)$ can be sought in the form of a first degree
polynomial in $\lambda$,
\begin{equation}
  g=\lambda-\mu(x,t),\label{gsol}
\end{equation}
where $\mu(x,t)$ is a new unknown function called the auxiliary
spectrum. Substituting (\ref{gsol}), (\ref{Plam}) into (\ref{gx1}), we
obtain
\begin{equation} 
  \label{mux}
  \begin{split}
    -\frac{1}{2}(\lambda-\mu)\mu_{xx}-\frac{1}{4}\mu_x^2+(u+\lambda)(\lambda-\mu)^2=
    \\  
    \lambda^3-s_1\lambda^2+s_2\lambda-s_3.
  \end{split}
\end{equation}
Equating the coefficients of $\lambda^2$ on both sides yields 
\begin{equation} 
  \label{u}
  u(x,t)=2\mu-s_1 \, .
\end{equation}
Next, after substitution of (\ref{gsol}) into (\ref{gt}), we obtain
\begin{equation}
\mu_t=(4\lambda-2u)\mu_x+2(\mu-\lambda)u_x.\label{mut}
\end{equation}
Setting the free spectral parameter $\lambda=\mu$ in (\ref{mux}) and
(\ref{mut}), which can be done at any point $(x, t)= (x_0, t_0)$, we
obtain:
\begin{equation} \label{muxt}
  \mu_x=2\sqrt{-R(\mu)}, \quad \mu_t=-2s_1\mu_x.
\end{equation}

The second equation (\ref{muxt}) implies $\mu= \mu(\xi)$, where
$\xi=x-Vt+\xi_0$ is the traveling phase with the phase velocity
\begin{equation}
  \label{phasevelocityriemann}
  V=-2s_1=-2(\lambda_1+\lambda_2+\lambda_3), 
\end{equation}
and $\xi_0$ is an arbitrary initial phase.

Now, the first ODE (\ref{muxt}) becomes
\begin{equation} \label{muODE}
  \mu_\xi = 2 \sqrt{-R( \mu)}
\end{equation}
implying that real-valued $\mu$ oscillates between $\la_2$ and
$\la_3$. Integrating (\ref{muODE}), we obtain the elliptic (cnoidal
wave) solution of the KdV equation, parametrized by the spectral
branch points $\lambda_1, \lambda_2, \lambda_3$:
\begin{equation}\label{cnoidalriemann}
  u(x,t)=\lambda_3-\lambda_1-\lambda_2-2(\lambda_3-\lambda_2)
  \textrm{sn}^2(\sqrt{\lambda_3-\lambda_1}\xi,m), 
\end{equation}
where the modulus $m=(\la_3-\la_2)/(\la_3 - \la_1)$.

The wavelength of (\ref{cnoidalriemann}), which is the period of the
nonlinear oscillator (\ref{muODE}), is
\begin{equation}
  L= \int \limits^L_0 d\xi = \int \limits_{\la_2}^{\la_3} \frac{\rmd
    \mu }{\sqrt{-R(\mu)}}. \label{wavelengthriemann} 
\end{equation}
We now recall that the direct computation of the periodic solution via
substituting the ansatz $u = \phi(\xi)$ into the KdV equation
(\ref{kdv6}) yields (see (\ref{kdv_ODE}), where $\theta = k\xi$ with
$k=2\pi/L$)
\begin{equation} \label{phiQ}
  \begin{split}
    \phi_\xi^2= 2(b_1- \phi)(b_2 - \phi) (b_3 - \phi) \equiv - Q(
    \phi),  \\ V=2(b_1 + b_2 + b_3), \quad  b_1 \le b_2 \le b_3 \, . 
  \end{split}
\end{equation}
Using (\ref{u}), (\ref{muxt}) we obtain $ \phi (\xi)= 2 \mu (\xi) -
s_1$. Substituting this into (\ref{phiQ}) we obtain $Q(2 \mu + V/2)=
16 R(\mu)$, which after some algebra yields the relations between the
roots $\lambda_1, \lambda_2, \lambda_3$ of the spectral polynomial
$R(\lambda)$ and the roots $ b_1, b_2, b_3$ of the potential curve
$Q(\phi)$ (cf.~\eqref{kdv_Riemann_inv})
\begin{equation} \label{lab}
  \lambda_1=-\frac{b_2 +  b_3}2; \ \  \lambda_2=-\frac{b_1 +  b_3}2,
  \ \  \lambda_3=-\frac{b_1 +  b_2}2 . 
\end{equation} 


We now introduce slow modulations of the periodic solution
(\ref{cnoidalriemann}) via $\lambda_j(X,T)$, $X=\varepsilon x$,
$T=\varepsilon t$, $\varepsilon \ll 1$, and derive modulation equations for
$\{\la_j\}_{j=1}^3$ using the Kamchatnov adaptation of the FFM-Whitham
averaging procedure, which, instead of averaging the necessary number
of conservation laws (\ref{eq:29}), (\ref{eq:30}), one averages all
KdV conservation laws via the generating equation
(\ref{gener-conserv}).

Before deriving the modulation equations, we represent the generating
equation (\ref{gener-conserv}) in a form suitable for averaging. From
(\ref{gx1}), it follows that $1/g$ is singular when $\la=\la_j$,
$j=1,2,3$, which becomes apparent if one introduces the normalization
\begin{equation} \label{gnorm}
  \hat g = g/\sqrt{R(\la)},
\end{equation}
which reduces equation (\ref{gx1}) to
\begin{equation} \label{gtilde}
  \hat g \hat g_{xx} - (\hat g_x)^2 + (u+\la) \hat g ^2 =1,
\end{equation}
so that $\hat g$ does not have  singularities at $\la=\la_j$.
Thus, to remove the singularities at $\la =\la_j$ in
(\ref{gener-conserv}) we multiply it by $\sqrt{R(\la)}$ and use
(\ref{gsol}), (\ref{u}) to obtain the normalized generating equation
\begin{equation} \label{gener-norm}
  \left[\sqrt{R(\lambda)}\cdot\frac{1}{\lambda-\mu}\right]_t+
  \left[\sqrt{R(\lambda)}\cdot\left(-4-\frac{2s_1}{\lambda-\mu}
    \right)\right]_x=0.  
\end{equation}
We now consider the slowly modulated cnoidal wave
(\ref{cnoidalriemann}) by assuming $\la_j=\la_j(X,T)$. Introducing a
period-average of (\ref{muODE}) by (cf. (\ref{eq:aver}))
\begin{equation*}
  \langle {\mathcal{F}[ \mu ]} \rangle = \frac{1}{L} \int_0^{L}
  \mathcal{F}[ \mu(\xi)] \rmd \xi = \frac{1}{L}
  \int_{\la_2}^{\la_3} \frac{\mathcal{F}[\mu] \rmd \mu}{\sqrt{-R(\mu)}} ,
\end{equation*}
we apply it to (\ref{gener-norm}) to obtain the generating equation
for the modulation equations
\begin{align}  
  & \left[\sqrt{R(\lambda)} \Big \langle \frac{1}{\lambda-\mu} \Big
    \rangle \right]_T  \nonumber \\ 
  & + \left[\sqrt{R(\lambda)} \left(-4 -  2s_1\Big \langle
      \frac{1}{\lambda-\mu} \Big \rangle \right )
  \right]_X=0. \label{aver} 
\end{align}
We now observe that differentiation of $\sqrt{R({\la})}$ in
(\ref{aver}) yields the factors
\begin{equation*}
  \frac{1}{\sqrt{\lambda-\lambda_j}}\frac{\p\lambda_j}{\p
    T}\qquad\textrm{and}\qquad\frac{1}{\sqrt{\lambda-\lambda_j}}
  \frac{\p\lambda_j}{\p X}, 
\end{equation*}
which are singular as $\lambda\rightarrow\lambda_j$.  We then multiply
(\ref{aver}) by $\sqrt{\la-\la_j}$ and take the limit $\la \to \la_j$
to obtain
\begin{equation}  \label{whithamr}
  \frac{\p\lambda_j}{\p T}+v_j\frac{\p \lambda_j}{\p X}=0,    \ \
  j=1,2,3, 
\end{equation}
where
\begin{equation} \label{v123}
  v_j= -2 s_1 -\frac{4}{\big \langle  \frac{1}{\la_j - \mu}\big
    \rangle} \, . 
\end{equation}
Thus $\la_j$ are the Riemann invariants of the modulation system
(\ref{whithamr}).  Computing the characteristic velocities
(\ref{v123}), we arrive at the expressions (\ref{kdv_char_vel}), where
$V_j({\bf r})=v_j(\boldsymbol{\lambda}(\bf r))$ and
\begin{equation}
  r_1 = - 6 \la_3, \quad  r_2 = - 6 \la_2,  \quad r_3 = -6 \la_1 \, . 
\end{equation}

\subsubsection{Properties of the KdV-Whitham system}
\label{sec:prop-kdv-whith}

Direct verification shows that the KdV-Whitham system (\ref{eq:31}),
(\ref{kdv_char_vel}) is: (i) strictly hyperbolic
(\ref{strict_hyp_diag}); (ii) genuinely nonlinear (\ref{gnonl}); (iii)
semi-Hamiltonian (\ref{semih}).  The first two properties imply the
existence of simple wave solutions (see Sec.~\ref{sec:basic-notions}),
and the third one implies integrability of the KdV-Whitham system via
the generalized hodograph transform (\ref{ghod}).  We note that the
general proof of strict hyperbolicity for the multiphase averaged
KdV-Whitham system was carried out by Levermore
\cite{levermore_hyperbolic_1988}. Integrability of the KdV-Whitham
system was proved by Tsarev \cite{tsarev_poisson_1985}.  Generally,
one can talk about the ``preservation of integrability when
averaging'' principle \cite{dubrovin_hydrodynamics_1989}.


Of particular interest are two limits of the KdV-Whitham system
(\ref{eq:31}), (\ref{kdv_char_vel}) corresponding to the harmonic and
soliton limits of the traveling wave \eqref{eq:34}. In the harmonic
limit, the modulus $m=0$ (i.e.~$r_2=r_1$) and the characteristic
velocities $V_2$ and $V_1$ merge together
\begin{equation}
  \label{v2=v1}
  m=0: \ \  V_2(r_1, r_1, r_3)=V_1(r_1, r_1, r_3)= - r_3 + 2r_1  \, . 
\end{equation}
One can also show that in this limit, the velocity $V_3(r_1, r_1,
r_3)=r_3$. Thus, in the harmonic limit $m=0$, the Whitham system
(\ref{eq:31}), (\ref{kdv_char_vel}) reduces to a system of two
equations,
\begin{equation}\label{whitham_harm_lim}
  \frac{\p r_1}{\p T}+(2r_1-r_3)\frac{\p r_1}{\p X}=0, \ \ 
  \frac{\p r_3}{\p T}+ r_3\frac{\p r_3}{\p X}=0.
\end{equation}
The second equation in (\ref{whitham_harm_lim}) is the dispersionless
KdV or Hopf equation $u_T+uu_X=0$ for $u = r_3$.

In the opposite, soliton limit $m=1$ (i.e. $r_2=r_3$), there is a
similar degeneracy, but now the merged characteristic velocities are
$V_2$ and $V_3$
\begin{equation}
  \label{v2=v3}
  m=1: \ \  V_2(r_1, r_3, r_3)  = V_3(r_1, r_3, r_3)=  \frac{1}{3}
  (2r_3 + r_1) \, , 
\end{equation}
while the remaining velocity $V_1(r_1, r_3, r_3)=r_1$ yields the Hopf
equation for $r_1$. Therefore, the soliton reduction of the
KdV-Whitham system is
\begin{equation}\label{whitham_sol_lim}
  \frac{\p r_3}{\p T}+\frac{1}{3}(2r_3+r_1)\frac{\p r_3}{\p X}=0, \ \  
  \frac{\p r_1}{\p T}+ r_1\frac{\p r_1}{\p X}=0.
\end{equation}
It is instructive to note that the merged characteristic velocities
(\ref{v2=v1}) and (\ref{v2=v3}) in the harmonic and soliton limits,
respectively, define a regular characteristic because the order of the
modulation system reduces from three to two in both limits.  Strict
hyperbolicity and genuine nonlinearity of the modulation system are
preserved in these limits.

We conclude this section with one more important property of the
KdV-Whitham equations: the characteristic velocities
(\ref{kdv_char_vel}) are homogeneous functions of ${\bf r}$ (i.e. $V_j(C{\bf r})=C^{\alpha}V_j({\bf r}) \ \hbox{for all}  \ C  > 0$) with 
homogeneity degree $\alpha=1$. 
This property defines the symmetry
\begin{equation} 
  \label{eq:scaling1}
  {\bf r } \to C {\bf r}, \quad T \to C^2 T, \quad X \to C^3 X ,  
\end{equation}
where $C=\hbox{const}$. The invariant scaling (\ref{eq:scaling1}) of
\eqref{eq:31}, \eqref{kdv_char_vel} along with the hydrodynamic
symmetry
\begin{equation}\label{scaling2}
  X \to CX, \quad T \to CT
\end{equation}
of all quasi-linear equations give rise to two special families of
similarity modulation solutions.  These families describe two
fundamental classes of DSWs described in Secs.~\ref{sec: dsw-riemann}
and \ref{sec:cubic_breaking} below.

\subsubsection{General solution of the KdV-Whitham equations}

\label{gen_sol_kdv-whitham}

The strict hyperbolicity and genuine nonlinearity properties of the
KdV-Whitham system guarantee the existence of three families of simple
wave solutions, which can be readily constructed using
characteristics. In these solutions, only one Riemann invariant is
changing while the other two are constant (see
Sec.~\ref{sec:simple_waves}).  More general solutions are given by the
generalized hodograph formulae (\ref{W}), which requires solving an
overdetermined system of linear PDEs (\ref{hodeq1}) due to Tsarev
\cite{tsarev_poisson_1985} in order to find the functions $W_j$,
$j=1,2,3$.  Despite linearity, the coefficients in (\ref{hodeq1})
involve rather complex combinations of complete elliptic integrals
(see (\ref{kdv_char_vel})), so their integration using standard
methods is not feasible.

We shall take advantage of the nonlinear group velocity
representation (\ref{vscalar}) for the characteristic velocities $V_j
= (1 - \frac{L}{\partial_j L} \partial_j)V$
where $V = \frac{1}{3}(r_1 + r_2 + r_3)$ is the phase velocity and the
wavelength $L$ can be conveniently represented as a loop
integral
\begin{equation} 
  \label{Loint}
  L = \sqrt{\tfrac{3}{2}}\oint \frac{\rmd  \lambda}{\sqrt{(\la -
      r_1)(r_2 - \la)(r_3 - \la)}} .
\end{equation}
The contour of integration surrounds clockwise the branchcut between
$r_1$ and $r_2$ on the upper sheet of the elliptic Riemann surface of
the radical $\sqrt{ (\la - r_1)(\la - r_2)(\la - r_3)}$.
 
Substituting (\ref{vscalar}) into the Tsarev equations \eqref{ts}
and using $\partial_{ij} \omega = \partial_{ji}\omega$, we obtain the
relationship
\begin{equation}
  \partial_j(\partial_i k W_i) = \partial_i(\partial_j k W_j) \, .
\end{equation}
This analogue of a curl-free condition implies the existence of a
scalar function $q(r_1, r_2, r_3)$ so that
(cf. (\ref{vscalar}))
\begin{equation} 
  \label{wscalar}
  W_j=\frac{\partial_j (kq)}{\partial_j k}= \left ( 1-\frac{L}{\partial_j
    L} \partial_j \right ) q,  \quad j=1,2,3. 
\end{equation}
The integration of the Tsarev equations \eqref{ts} for the vector
$(W_1, W_2,W_3)$ is then reduced to finding a single scalar potential function
$q({\bf r})$.  To obtain the equation for $q$, we substitute
(\ref{vscalar}) and (\ref{wscalar}) into \eqref{ts} and, using the
loop integral representation (\ref{Loint}), obtain the overdetermined
system of six {\it Euler-Poisson-Darboux} (EPD) equations in the
three-dimensional space of Riemann invariants
\begin{equation}
  \label{EPD}
  \begin{split}
    2(r_i-r_j)\frac{\partial^2 q}{\p r_i  \p r_j}
    = \frac{\partial q}{\partial r_i} - \frac{\partial q }{\partial r_j}\, , \\  \\
    i, j = 1,2,3; \quad i \ne j \, .
  \end{split}
\end{equation}
Each of the equations in (\ref{EPD}) for a given pair $(r_i, r_j)$
represents a particular case of the classical EPD equation appearing,
in particular, in the theory of surfaces \cite{darboux_sur_1972}, gas
dynamics \cite{copson_partial_1975} and the theory of colliding
gravitational waves \cite{hauser_initial_1989}.  The system
(\ref{EPD}) appears in the classical differential geometry study
\cite{eisenhart_triply_1919} by Eisenhart.

A calculation of the mixed third derivatives shows that the system
(\ref{EPD}) is compatible, which also proves the semi-Hamiltonian
property (\ref{semih}) and integrability of the KdV-Whitham system
\eqref{eq:31}.

Summarizing, the mapping between EPD system solutions $q(r_1, r_2, r_3)$ and
solutions of the KdV-Whitham system \eqref{eq:31} has the form
\begin{equation}
  \label{mapping}
  x-V_j t = \left( 1-\frac{L}{\partial_j L} \partial_j \right ) q,
  \quad j=1,2,3.  
\end{equation}
Importantly, all non-constant and non-singular local 
solutions of the modulation system \eqref{eq:31} can be obtained in
this way. The transformation (\ref{mapping}) was found independently
in
\cite{gurevich_breaking_1991,kudashev_inheritance_1991,tian_oscillations_1993}.

Direct verification shows that the function
\begin{equation} 
  \label{Generf}
  G(\la; r_1, r_2, r_3) = \frac{\Phi(\la)}{\sqrt{(\la - r_1)(\la
      -r_2)(\la-r_3)}} ,
\end{equation}
where $\Phi (\la)$ is an arbitrary function, satisfies the EPD system
for any $\la$, i.e.~$G$ represents the generating function for
solutions of the EPD system \eqref{EPD} and, consequently via
\eqref{mapping}, for the solutions to the Whitham-KdV equations
\eqref{eq:31}.  In particular, choosing $ \Phi(\lambda) =
\lambda^{3/2}$ and expanding the generating function $ G(\la, r_1,r_2,
r_3)$ for $ \lambda \gg 1$, we obtain
\begin{equation}\label{expan}
  G=  1+\frac{q_1}{\lambda} + \frac{q_2}{\lambda^2} + \dots,
\end{equation}
where
\begin{equation}
  \label{gi}
  \begin{split}
    q_1 &=  2 s_1\, , \ \   q_2= 6 s_1^2 -8 s_2, \\
    q_3 &= 5s_1^3 -12 s_1s_2 + 8s_3, \ \dots,
  \end{split}
\end{equation}
and $s_1, s_2, s_3$ are the elementary symmetric polynomials of $r_1,
r_2, r_3$ (recall (\ref{s})). Each $ q_\alpha({\bf r})$, $ \alpha =
1,2,3, \dots$ is a symmetric homogeneous function of $r_1, r_2, r_3$
with homogeneity degree $\alpha$ satisfying the EPD system \eqref{EPD}.
This family of homogeneous modulation solutions of the EPD system was
first obtained by Krichever \cite{krichever_method_1989} utilizing an
algebro-geometric approach to the integration of the multiphase
KdV-Whitham equations. Since the characteristic velocities $V_j$ are
homogeneous functions of ${\bf r}$ with the homogeneity degree $1$, it
is not difficult to show that the homogeneous solutions $q_\alpha({\bf
  r})$ of the EPD system give rise, via the mapping (\ref{mapping}),
to the generalized similarity (scaling) modulation solutions
\cite{gurevich_nonstationary_1974,krichever_method_1989},
\be 
\label{scaling_krich} r_j= t^{\gamma}
R_j(\frac{x}{t^{\gamma+1}})\, , \quad \gamma=\frac{1}{\alpha - 1}\, 
\quad j=1,2,3  
\ee 
for $\alpha \ne 1$.
The general solution of (\ref{EPD}) is parametrized by three functions
of a single variable, as is the solution to the Tsarev equations
\eqref{ts} for $(W_1,W_2,W_3)$. Then, using the generating function
(\ref{Generf}), one obtains the general solution to the EPD system
(\ref{EPD}) \cite{eisenhart_triply_1919}
 \begin{equation} \label{general_g}
   q(r_1, r_2, r_3) = \sum \limits_{j=1}^3 \int \limits^{r_j}_0
   \frac{\Phi_j(\la)}{\sqrt{(\la - r_1)(\la-r_2)(\la-r_3)}} \rmd
   \la ,
 \end{equation}
 where $\Phi_j(\la), \ j=1,2,3$ are arbitrary complex functions.
 \subsubsection{Modulation phase shift}
 
 \label{sec:mod_phase_shift}
  
 We now describe an important connection between solutions of the EPD
 system (\ref{EPD}) and the phase of the slowly modulated periodic
 solution of the KdV equation.  The phase appears in the modulation
 construction in two ways. In the derivation of the modulation
 equations via multiple-scale WKB-type expansions described in
 Sec.~\ref{sec:kdv-mod-sys}, the generalized phase $\theta$ is
 determined by the modulation solution via the local wavenumber and
 local frequency, which are defined as $k= \theta_x = \eps \theta_X$
 and $\omega = - \theta_t = - \eps \theta_T$ respectively
 \cite{whitham_linear_1974}. On the other hand, the IST based
 finite-gap approach to the derivation of the modulation equations
 described in Sec.~\ref{sec:integr-riem-invar} yields the explicit
 expression $\theta=k\xi= kx- \omega t + \theta_0$ for the
 phase. These two definitions of the phase are equivalent for the
 non-modulated periodic solution, but the presence of slow modulations
 imposes some constraints on the initial phase $\theta_0$, which can
 be viewed as a phase shift due to modulation.  To find this
 phase correction, we first represent the phase in the form $\theta
 =\eps^{-1}(kX - \omega T + \Theta_0)$ where $\Theta_0=\varepsilon
 \theta_0$. In the modulated wave, $\Theta_0$ undergoes slow
 spatiotemporal variations, so it is natural to assume that
 $\Theta_0(X,T)= \Upsilon({\bf r}(X,T))$. The modulation phase shift function $\Upsilon({\bf
   r})$ is then found from the definition of the local wavenumber
\begin{equation} \label{consistency}
  k = \varepsilon \theta_X = k+ \sum \limits_{j=1}^3 \left \{
    \frac{\partial k}{\partial r_j} X -  \frac{\partial
      \omega}{\partial r_j} T + \frac{\partial \Upsilon}{\partial r_j}
  \right \}\frac{\partial r_j}{\p X},
\end{equation}
which yields 
\begin{equation} 
  \label{eq:106} 
  \sum \limits_{j=1}^3 \left \{
    \frac{\partial k}{\partial r_j} X- \frac{\partial
      \omega}{\partial r_j} T+ \frac{\partial \Upsilon}{\partial r_j}
  \right \}\frac{\partial r_j}{\p X}=0\, .
\end{equation}
Since equation (\ref{eq:106}) must be valid for all non-constant
solutions $r_j(X,T)$, each expression in brackets must vanish and we have
\begin{equation} 
  \label{eq:107}
  \frac{\partial k}{\partial r_j} X -  \frac{\partial \omega}{\partial
    r_j} T + \frac{\partial \Upsilon}{\partial r_j} = 0, \quad j=1,2,3,
\end{equation}
provided $(r_j)_X \ne 0$ for all $j$. We note that using the local
frequency definition instead of the wavenumber for the determination
of the phase correction leads to the same expression (\ref{eq:107}).
Dividing (\ref{eq:107}) by $\frac{\partial k}{\partial r_j} $ and
using (\ref{vscalar}), we obtain
\begin{equation} 
  \label{eq:108}
  X-V_jT= \frac{\partial_j \Upsilon}{\partial_j k}, \quad j=1,2,3. 
\end{equation}
Comparing (\ref{eq:108}) with the generalized hodograph solution
(\ref{mapping}), we find that $\Upsilon = kq({\bf r})+C$, where $C$ is
an arbitrary constant, which can be set to zero without loss of
generality. Thus, the phase correction $\theta_0$ in the modulated KdV
solution is determined by the solution of the EPD equation.  A
particular form of this result was obtained in
\cite{grava_numerical_2007} by analyzing the expression for the phase
in the rigorous asymptotic solution of the initial value problem for
the semi-classical KdV equation obtained in
\cite{deift_extension_1998}. Here, it is derived as a general property
inherent within the modulation theory. We stress that the existence of the function $\Upsilon(\mathbf{r})$  is not obvious {\it a priori}.
It is guaranteed only due to the semi-Hamiltonian structure of the
KdV-Whitham system (i.e., to the consistency of the associated Tsarev
system \eqref{ts}), which is inherited from the integrability of the
KdV equation. 
Thus, one can expect that the representation of the phase $\theta(x,t)$ of the modulated solution in the form $\theta=kx - \omega t+ \varepsilon^{-1}\Theta_0(X,T)$ is only possible for integrable equations.

We now note that the phase compatibility condition (\ref{eq:107}) can
be written as the stationary phase condition
\begin{equation} \label{eq:109}
\nabla_{\bf r} \theta =0 \, .
\end{equation}
Modulations of the periodic KdV solution are defined by the stationary
point of the phase in the space of Riemann invariants of the Whitham
system.  Equation (\ref{eq:109}) has been derived for the general case
of multiphase modulated KdV solutions in \cite{el_unified_2001}.
Along with the nonlinear extension (\ref{vscalar}) of the definition
of the group velocity, the stationary phase condition (\ref{eq:109})
provides further striking parallels between modulation theories for
linear and nonlinear waves \cite{whitham_linear_1974}.  We note that
the precise meaning of the equivalence between the stationary phase
condition and the modulation solution has been revealed by Deift,
Venakides and Zhou \cite{deift_extension_1998} in the framework of the
Riemann-Hilbert problem approach to the semi-classical IST for the KdV
equation.

\subsection{NLS-Whitham equations}

\subsubsection{NLS dispersive hydrodynamics}

\label{sec:NLS_disp_hyd}

The cubic nonlinear Schr\"odinger equation
\begin{equation}\label{eq:NLS}
    i \psi_t+\frac{1}{2}\psi_{xx} - \sigma |\psi|^2\psi=0 , \quad
    \sigma = \pm 1\, ,
\end{equation}
is a particular case of the gNLS equation \eqref{eq:111}.  Equation
(\ref{eq:NLS}) models slow evolution of the normalized complex valued
envelope of a nearly monochromatic, weakly nonlinear wave
$\psi(x,t)e^{i(k_0 x' - \omega_0 t')}$, where $k_0$ and
$\omega_0=\omega (k_0)$ are the wavenumber and the frequency of the
short-wavelength ``carrier'' wave respectively, $c_{\rm
  g}=\omega'(k_0)$ is the group velocity; the independent variables
$x,t$ in the NLS equation \eqref{eq:NLS} are related to physical space
and time $x', t'$ as $x= \delta [x' - c_{\rm g} t' ]$, $t=\delta^2
t'$, where $\delta \ll 1$ is the amplitude parameter.

Some of the prominent physical contexts for the NLS equation
\eqref{eq:NLS} are the dynamics of deep water waves ($\sigma =-1$),
Langmuir waves in plasma ($\sigma=-1$), spin waves in a ferromagnet
($\sigma = \pm 1$), and electromagnetic waves in a nonlinear,
defocusing ($\sigma= 1$) or focusing ($\sigma= -1$) medium, notably in
nonlinear optics. It also models the dynamics of a geometrically
constrained Bose-Einstein condensate (BEC) in the mean-field
approximation (both signs of $\sigma$ are relevant).  Note that in the
BEC context, the derivation of the NLS equation (termed the
Gross-Pitaevskii equation, see \eqref{eq:75} below) is not a slowly
varying envelope approximation but rather a direct approximation of
the condensate order parameter.

From the mathematical point of view, the NLS equation
(\ref{eq:NLS}), similar to the KdV equation \eqref{kdv1}, is a
universal \cite{benney_propagation_1967}, integrable nonlinear dispersive
equation.  Using the Madelung
transformation $\psi \mapsto (\rho,u)$ \eqref{eq:112},
where $\rho(x,t)>0$ and $u(x,t)$ are real-valued functions,
we can exactly represent the NLS equation (\ref{eq:NLS})
in the dispersive-hydrodynamic form (\ref{eq:1}) 
\begin{equation}
  \label{eq:NLS_disp_hydro}
  \begin{split}
    \rho_t+(\rho u)_x &=0,\\
    (\rho u)_t+(\rho u^2 + \frac{\sigma}{2} \rho^2)_x &=
    \frac{1}{4}(\rho(\log \rho)_{xx})_x \, ,
  \end{split}
\end{equation}
with the fluid-like density $\rho$, velocity $u$, and pressure
\begin{equation}
  \label{eq:123}
   P(\rho) = \frac{\sigma}{2} \rho^2.
\end{equation}

The dispersionless, long-wave limit of the system
(\ref{eq:NLS_disp_hydro}) is obtained by neglecting the right-hand
side of the momentum equation (\ref{eq:NLS_disp_hydro}).  With some
manipulation, the dispersionless equations can be written
\begin{equation}
  \label{eq:sw}
 \rho_t+(\rho u)_x=0,\quad u_t+(\rho u^2+ \frac{\sigma}{2} \rho^2)_x=0\, .
\end{equation}
The system (\ref{eq:sw}) is hyperbolic for $\sigma=1$ and elliptic for
$\sigma = -1$.  The latter implies ill-posedness of the Cauchy problem
for all but analytic initial data.  This can be interpreted in the
context of the pressure law \eqref{eq:123} where $\sigma = -1$
coincides with a negative pressure and an attractive interaction.  In
nonlinear optics, this effect corresponds to focusing of the optical
signal.  When $\sigma = -1$, a slightly perturbed, uniform density
evolving according to \eqref{eq:NLS_disp_hydro} will experience
modulational instability \cite{zakharov_modulation_2009}, thus the
hydrodynamic background is unstable.  When $\sigma = 1$, the pressure
\eqref{eq:123} is positive, resulting in a repulsive or defocusing
interaction.  In this case, a uniform density solution to
\eqref{eq:NLS_disp_hydro} is stable and the dispersionless system
(\ref{eq:sw}) coincides with the ideal shallow water equations or,
equivalently, the isentropic gas dynamic equations with the
(unphysical) heat capacity ratio $\gamma = 2$
\cite{whitham_linear_1974}.

The system (\ref{eq:sw}) can be cast in the diagonal form
\begin{equation}\label{er}
  \frac{\partial r_{\pm}}{\partial t} + V_{\pm}(r_+, r_-)
  \frac{\partial r_{\pm}}{\partial x}=0\, ,
\end{equation}
with Riemann invariants and characteristic velocities satisfying
\begin{equation}
  \label{eq20}
  r_\pm=\frac12{u}\pm\sqrt{\sigma \rho}\, \qquad V_\pm = u  \pm
  \sqrt{\sigma \rho},
\end{equation}
respectively. For $\sigma = -1$ (focusing), both Riemann
invariants and the characteristic velocities form complex conjugate
pairs.  For both defocusing and focusing cases, the characteristic
velocities are expressed in terms of the Riemann invariants by the
formulae:
\begin{equation}
  \label{V}
  V_+=\frac{3}{2} r_+ + \frac12{r_-} \, , \qquad V_-=\frac32 
  {r_-}+ \frac12{r_+} \, .
\end{equation}
Thus, as a model for propagation of weakly nonlinear,
quasi-monochromatic,  modulated  waves (see
Sec.~\ref{sec:whitham-vs.-nls}), the NLS equation describes the
evolution of two dispersive envelope wave families relative to the
dominant propagation of the carrier wavetrain moving with the group
velocity $c_{\rm g}$. This is why it is classified as bi-directional
dispersive hydrodynamics.

\subsubsection{NLS-Whitham equations and their properties}

\label{sec:NLS_Whitham_prop}

Here we present an outline of the modulation theory results for the
defocusing NLS equation \eqref{eq:NLS}, $\sigma = 1$, although the
modulation equations will be equally applicable to the focusing case.
Similar to the KdV equation, the NLS equation is integrable via the
IST \cite{shabat_exact_1972,zakharov_interaction_1973}, which ensures
Riemann invariant structure and integrability of the associated
Whitham equations via the generalized hodograph transform.  The
Riemann invariants for the NLS-Whitham equations were found by Forest
and Lee \cite{forest_geometry_1986} and Pavlov
\cite{pavlov_nonlinear_1987} for the general multiphase case. In the
single-phase case, the application of Kamchatnov's technique
\cite{kamchatnov_nonlinear_2000} yields all the necessary results
without the need to invoke the more sophisticated tools of algebraic
geometry. The derivation is based on the NLS matrix analogue of the
Lax pair for the NLS equation found by Zakharov and Shabat
\cite{zakharov_interaction_1973} and then further developed by
Ablowitz, Kaup, Newell and Segur \cite{ablowitz_inverse_1974}.  It
follows similar lines as the derivation of the KdV periodic
solution described in Sec.~\ref{sec:integr-riem-invar}. The details
can be found in \cite{kamchatnov_nonlinear_2000}.

\smallskip

{\it Periodic solution}

\smallskip

The periodic traveling wave solution of the defocusing NLS equation
(\ref{eq:NLS}) with $\sigma = 1$ is parametrized by four integrals of
motion $r_1 \le r_2 \le r_3 \le r_4$ and can be expressed in terms of
the Jacobi elliptic $\sn$ function:
\begin{align}
  \rho  &=\frac14(r_4- r_3- r_2+r_1)^2+ (r_4-r_3)
  (r_2-r_1)  \nonumber \\
  & \times \,  {\rm sn}^2\left(\sqrt{(r_4-r_2)(r_3-r_1)}  \, \xi ;
    m\right) \, ,  \label{eq013} \\ 
  u=&V - \frac{C}{\rho} \, ,  \nonumber
\end{align}
where
\begin{equation*} 
  \begin{split}
    C=\frac{1}{8} (-r_1 - r_2 + r_3 +
    r_4) (-r_1 + r_2 - r_3 + r_4) \\
    \times \ (r_1 - r_2 - r_3 + r_4),
  \end{split}
\end{equation*}
and
\begin{equation}
  \label{eq016}
  \xi= x-Vt-\xi_0,\qquad V=\frac12 \sum_{i=1}^4 r_i,
\end{equation}
$V$ being the phase velocity of the nonlinear wave and $\xi_0$ the
initial phase.

The modulus $0 \le m \le 1$ of the elliptic solution (\ref{eq013}) is
defined as
\begin{equation}\label{eq015}
m=\frac{(r_2- r_1)(r_4- r_3)}{(r_4-r_2)(r_3-r_1)},
\end{equation}
and the wave amplitude is
\begin{equation}\label{amp}
a= (r_4- r_3)
(r_2-r_1) \, .
\end{equation}
The wavelength of the periodic wave (\ref{eq013})  is given by the loop integral with the contour of integration surrounding either the branchcut between
$r_1$ and $r_2$ or between $r_3$ and $r_4$, the result in both cases being
\begin{equation}
  \label{eq017}
  \begin{split}
    L =  \frac{1}{2}\oint \limits
    \frac{\rmd\lambda}{\sqrt{(\la-r_1)(\la-r_2)(\la-r_3)(r_4-\la)}} \\
    =\frac{2{ \K}(m)}{\sqrt{(r_4- r_2)(r_3- r_1)}}.
  \end{split}
\end{equation}
In the soliton limit   $m \to 1$ (i.e.  $r_3  \to r_2$) the traveling wave
solution (\ref{eq013}) turns into a dark soliton
\begin{equation}\label{sol}
  \rho = \overline{\rho} - \frac{a_{\rm s}}{\hbox{cosh}^2 (\sqrt{a_{\rm s}}(x-V_{\rm s} t -
    \xi_0))}\, , 
\end{equation}
where the background density $\rhob$, the soliton amplitude $a_{\rm s}$, and
its velocity $V_{\rm s}$ are expressed in terms of $r_1, r_2, r_4$ as
\begin{equation}\label{22}
  \begin{split}
    \rhob &= \frac{1}{4}(r_4 -  r_1)^2,  \quad a_{\rm s}=(r_4 - r_2)(r_2 -
    r_1) , \\   
    V_{\rm s} &=\frac{1}{2}(r_1+2 r_2+ r_4)\, . 
  \end{split}
\end{equation}

\smallskip
{\it Modulation equations} 

\smallskip The NLS-Whitham system for the $r_i$'s has diagonal
form 
\begin{equation}
  \label{eq18} 
  \frac{\partial r_i}{\partial
    T}+V_i({\bf r})\frac{\partial r_i}{\partial X}=0, \qquad
  i=1,2,3,4 \, ,
\end{equation}
where $X = \eps x$, $T = \eps t$ are the slow variables.

The characteristic velocities $V_i$ can be computed using the general
formula (\ref{vscalar})
\begin{equation}
  \label{eq019}
  V_i({\bf r})=\left(1-\frac{L}{\partial_i L}\partial_i \right)V ,
  \quad i=1,2,3,4 \, . 
\end{equation}
Substitution of eqs.~(\ref{eq016}), (\ref{eq017}) into
eq.~(\ref{eq019}) yields the explicit expressions
\begin{equation}\label{vi}
\begin{split}
V_1&=\tfrac12 \sum r_i
-\frac{(r_4-r_1)(r_2-r_1)}{(r_4- r_1)-(r_4-r_2)\mu(m)},\\
V_2&=\tfrac12 \sum r_i
+\frac{(r_3-r_2)(r_2-r_1)}{(r_3-r_2)-(r_3-r_1)\mu(m)},\\
V_3&=\tfrac12 \sum r_i
-\frac{(r_4-r_3)(r_3-r_2)}{(r_3-r_2)-(r_4-r_2)\mu(m)},\\
V_4&=\tfrac12 \sum r_i
+\frac{(r_4-r_3)(r_4- r_1)}{(r_4-r_1)-(r_3-r_1)\mu(m)},
\end{split}
\end{equation}
where  $\mu(m)=\E(m)/ \K(m)$.  

It is not difficult to show using the representation (\ref{eq019})
that the NLS-Whitham system (\ref{eq18}), (\ref{vi}) is genuinely
nonlinear and strictly hyperbolic (see \cite{el_refraction_2012}).
The general proof of these properties for the multiphase case can be
found in \cite{kodama_whitham_1999} (see also
\cite{jin_semiclassical_1999}) and is analogous to the original proof
in \cite{levermore_hyperbolic_1988} for the KdV-Whitham system.

Now we consider the harmonic $m=0$ and soliton $m=1$ limits.

We first note, that in contrast to the KdV case, the harmonic limit
$m=0$ can be achieved in one of two possible ways: either via
$r_2=r_1$ or $r_3=r_4$.  Using standard asymptotics for elliptic
integrals \cite{abramowitz_handbook_1972}, we obtain
\begin{equation}
  \label{m01}
  \begin{split}
    &\hbox{when} \ \  r_2=r_1:   \\   
    & V_2=V_1=r_1 + \frac{r_3 + r_4}{2}  + \frac{2(r_3- r_1)(r_4 -
      r_1)}{2r_1 - r_3-r_4} \, , \\ 
    &V_3= \frac{3}{2} r_{3} + \frac12{r_{4}}=V_-(r_3, r_4)\, , \\
    &V_4= \frac{3}{2} r_{4} + \frac12{r_{3}} = V_+(r_3, r_4)\, .
  \end{split}
\end{equation}
where $V_{\pm}$ are the characteristic velocities of the NLS
dispersionless limit (shallow water) system (\ref{er}).
\begin{equation} \label{m02}
\begin{split}
&\hbox{When} \ \  r_3= r_4:  \\    
 &V_3=V_4=r_4 + \frac{r_1 + r_2}{2} +
\frac{2(r_4-r_2)(r_4 - r_1)}{2r_4 - r_2-r_1} \, , \\
&V_1= \frac{3}{2} r_{1} + \frac12{r_{2}}=V_-(r_1, r_2)\, , \\ 
&V_2= \frac{3}{2} r_{2} + \frac12{r_{1}}=V_+(r_1, r_2)\, . 
\end{split}
\end{equation}
In the soliton limit, $m=1$ occurs only if $r_2=r_3$, so we obtain:
\begin{equation}
  \label{m1}
  \begin{split}
    &\hbox{when} \ \  r_2=r_3:   \\  
    & V_2=V_3=\frac{1}{2}(r_1+ 2r_2 + r_4)=V_{\rm s}\, , \\
    &V_1= \frac{3}{2} r_{1} + \frac12{r_{4}}=V_-(r_1, r_4)\, , \\ 
    &V_4= \frac{3}{2} r_{4} + \frac12{r_{1}}=V_+(r_1, r_4)\, . 
  \end{split}
\end{equation}

In both harmonic and soliton limits, the fourth-order modulation
system (\ref{eq18}), (\ref{vi}) reduces to a system of three
equations, two of which are decoupled.  Moreover, one can see that in
all considered limiting cases, the decoupled equations agree with the
dispersionless limit of the NLS equation (\ref{er}).

\smallskip

{\it General solution}

\smallskip

Similar to the KdV-Whitham system, the NLS-Whitham system is
semi-Hamiltonian (\ref{semih}), i.e., integrable via the generalized
hodograph transform.  Using the transformation (\ref{wscalar}) (with
$j=1, \dots, 4$, $L$ defined by (\ref{eq017})), and the
representation (\ref{eq019}) for the NLS-Whitham characteristic
velocities, one can show that the remarkable mapping between
solutions of the Whitham modulation equations and solutions of the EPD
system (\ref{EPD}), consisting now of nine equations, takes place for
the defocusing NLS equation as well. This fact was first established
in \cite{gurevich_evolution_1992}. Then one can construct the
appropriate generating function (see (\ref{Generf})), and the general
solution of the EPD system (see (\ref{general_g})).
 
\section{DSWs in integrable systems}
\label{sec:disp-shock-waves}

The primary application of Whitham modulation theory is to the long
time description of wave breaking in dispersive hydrodynamic systems.
Integrable systems afford a detailed analytical description of the
resultant dispersive shock waves.  In what follows, we focus upon the
key results from Whitham theory as applied to DSWs in the integrable
KdV and NLS equations.  We also briefly touch upon the relationship of
Whitham theory to the Inverse Scattering Transform.

In summary, the Whitham method effectively reduces the asymptotic
description of a DSW to the integration of a quasi-linear modulation
system of hydrodynamic type \eqref{eq:8} with a free boundary for the
leading and trailing DSW edges \cite{gurevich_nonstationary_1974}.
The boundary conditions are the continuous matching of the wave mean
in the DSW region with the smooth, dispersionless external flow along
double characteristics of the modulation system.  This nonlinear, free
boundary problem and its extensions are now commonly known as the
Gurevich-Pitaevskii (GP) matching regularization. Importantly, the
Whitham equations subject to GP matching conditions admit a global
solution describing the modulations in an expanding DSW.  The
simplest, yet very important, example of such a solution was obtained
in the original paper \cite{gurevich_nonstationary_1974} for the
prototypical problem of dispersive regularization of an initial step,
the Riemann problem for the KdV equation.  The DSW modulation solution
is a rarefaction wave solution of the Whitham equations.  We consider
this problem now.  As already was mentioned, unless explicitly
specified, we won't be making the formal disctinction between slow
($X,T$) and fast ($x, t$) variables by simply assuming that the small
parameter --- the ratio of fast and slow scales --- naturally arises
in the long time asymptotic solution.
 
\subsection{KdV Riemann problem} 
\label{sec: dsw-riemann}

The Riemann problem \cite{riemann_uber_1860} classically refers to the
initial value problem for a system of (1+1)D hyperbolic equations
consisting of two constant states with a step at the origin, see,
e.g., \cite{lax_hyperbolic_1973,dafermos_hyperbolic_2009}.  Inherent
to this formulation is an underlying regularization mechanism,
classically dissipative. The seminal paper of Gurevich and Pitaevskii
\cite{gurevich_nonstationary_1974} adapted the Riemann problem to
consider the long time behavior of step initial data
\begin{equation}
  \label{eq:36}
  u(x,0) = u_\pm, \quad \pm x > 0,
\end{equation}
for the KdV equation \eqref{kdv1}, a dispersively regularized
hyperbolic equation.  The solution to this problem is so fundamental
to most all of DSW theory, we refer to the KdV Riemann problem as the
GP problem.

\subsubsection{Case $u_- > u_+$:} 
\label{sec:case-u_-}

\begin{figure}
  \centering
  \includegraphics[width=0.8\columnwidth]{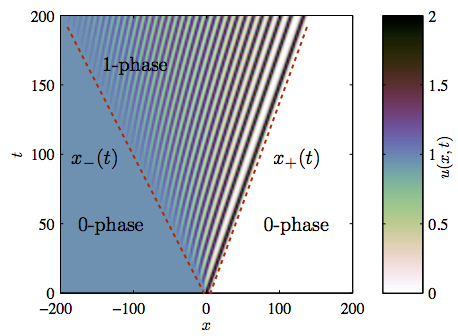}
  \caption{Contour plot of the KdV DSW solution found by GP
    \cite{gurevich_nonstationary_1974} for $u_- = 1$, $u_+ = 0$.}
  \label{fig:kdv_dsw_contour}
\end{figure}

We first examine the case where $u_- > u_+$ in \eqref{eq:36}.  For
this, the solution to the dispersionless Hopf equation
\begin{equation}
  \label{eq:37}
  r_t + r r_x = 0
\end{equation}
via the method of characteristics immediately becomes multivalued,
thus the dispersive term in the KdV equation \eqref{kdv1} is needed to
enable a single-valued solution.  The key insight of GP was to
represent the DSW solution by two separate regions in the $x$-$t$
plane, a region of expanding, rapid oscillations and a slowly varying,
dispersionless region, matched together appropriately at their
boundary, see Fig.~\ref{fig:kdv_dsw_contour}.  The \textit{1-phase}
region, $S_1$, associated with the slowly modulated periodic wave
$\varphi$ \eqref{eq:65}, emanates from the origin according to
\begin{equation*}
  S_1 = \{(x,t) ~ | ~ x_-(t) < x < x_+(t), ~ t > 0\} ,
\end{equation*}
where $x_-(0) = x_+(0) = 0$.  In the region $S_1$, the solution is
described by the modulation variables $r_i$, $i=1,2,3$ of the Whitham
equations (cf.~\eqref{eq:31})
\begin{equation}
  \label{eq:124}
  \frac{\partial r_i}{\partial t} + V_i \frac{\partial r_i}{\partial
    x} = 0, \quad i = 1, 2, 3,
\end{equation}
with characteristic velocities $V_i$ in eq.~\eqref{kdv_char_vel}.  We
emphasize that the solutions to eqs.~\eqref{eq:37} and \eqref{eq:124}
describe the long time behavior of the evolution of the KdV initial
data \eqref{eq:36}.  

The \textit{0-phase} region, $S_0$, is the complement in the upper
half plane, $S_0 = (\overline{S_1})^c$, where the solution is
described by $r$, the solution to the dispersionless Hopf equation
\eqref{eq:37}.  The boundary between $S_0$ and $S_1$,
\begin{equation}
  \label{eq:40}
  \begin{split}
    \partial S &= \partial S_- \cup \partial S_+, \\
    \partial S_\pm
    &= \{(x,t) ~|~ x = x_\pm(t), ~ t > 0 \} ,
  \end{split}
\end{equation}
is a \textit{free boundary} and must be determined along with the
solution in each region.  It is this existence of a free boundary that
make DSWs a challenge to analyze.

The step initial data \eqref{eq:36} is used to initialize the Hopf
equation \eqref{eq:37} for the region $S_0$
\begin{equation*}
  r(x,0) = u_\pm, \quad \pm x > 0 .
\end{equation*}
The Hopf equation then admits the piecewise, spatially uniform solution
\begin{equation*}
  r(x,t) = u_\pm, \quad x  \gtrless x_\pm(t), \quad t > 0 .
\end{equation*}

The GP matching conditions at the boundary $\partial S$ must determine
both the location of the boundary as well as the values of the
modulation variables there.  One natural condition is achieved by
equating the average $\overline{\varphi}$ of the modulated wave in
$S_1$ to the dispersionless solution in $S_0$:
\begin{equation}
  \label{eq:39}
  \begin{split}
    \overline{\varphi} |_{\partial S} = r |_{\partial S} ~
    \Rightarrow ~ \overline{\varphi}(x_\pm(t),t) = u_\pm .
  \end{split}
\end{equation}
Two additional conditions are achieved by continuous matching of $S_0$
to $S_1$.  The rapidly varying periodic wave in $S_1$ can terminate in
a slowly varying, in this case spatially uniform, region of $S_0$ in
one of two ways, either the wave amplitude goes to zero (harmonic
edge) or the wavenumber goes to zero (soliton edge).  A choice must be
made as to which boundary, $\partial S_-$ or $\partial S_+$, is
associated with which edge.  This choice is determined by an
admissibility condition, the dispersive hydrodynamic analogue of an entropy
condition.  In order to identify the harmonic and soliton edges, GP
appealed to the 2-wave curve (recall Sec.~\ref{sec:simple_waves}) of
the Whitham modulation system.

\begin{figure}
  \centering
  \includegraphics[scale=0.25]{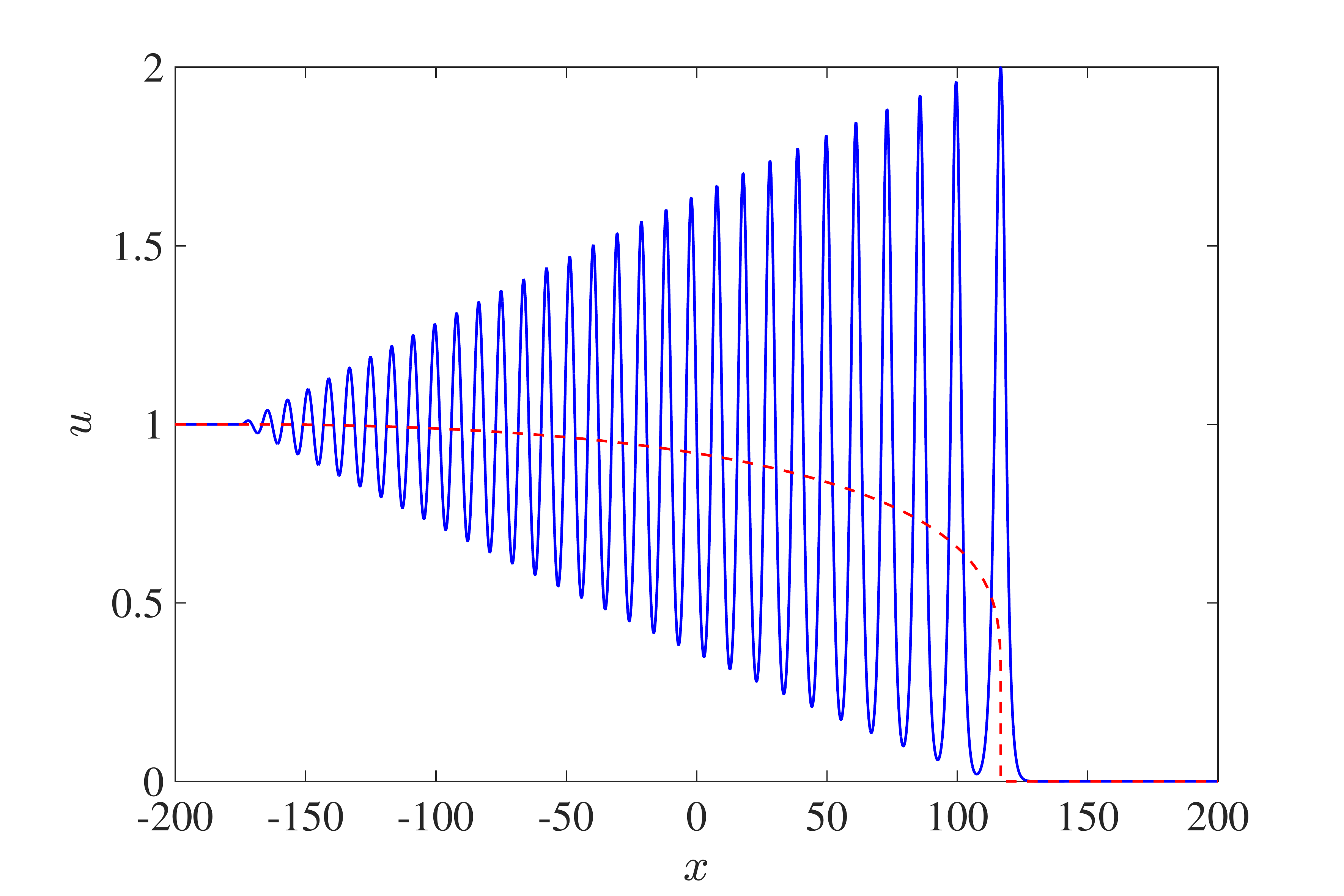}
  \caption{KdV DSW modulation solution (solid) and mean $\overline{u}$
    (dashed) for $t = 175$, $u_- = 1$, $u_+ = 0$.}
  \label{fig:kdv_dsw}
\end{figure}
As noted in Sec.~\ref{sec:prop-kdv-whith}, the harmonic edge limit
occurs when $\mathbf{r} \to \mathbf{r}_{\mathrm{h}} = (r_1,r_1,r_3)$
and the soliton edge limit occurs when $\mathbf{r} \to
\mathbf{r}_{\mathrm{s}} = (r_1,r_3,r_3)$.  The 2-wave curve of
eqs.~\eqref{eq:124} is parametrized according to
\begin{equation}
  \label{eq:43}
  \begin{split}
    s &= V_2(r_1,r_2(s),r_3), \quad s = \frac{x}{t},
    \quad r_1 < r_3,
  \end{split}
\end{equation}
where $r_1$ and $r_3$ are constant.  By the ordering
$V_2(\mathbf{r}_{\mathrm{h}}) < V_2(\mathbf{r}_{\mathrm{s}})$ (recall
eqs.~\eqref{v2=v1}, \eqref{v2=v3} for the merged characteristic velocities in the soliton and harmonic  limits respectively), we therefore observe the
admissibility criterion
\begin{equation*}
  \begin{split}
    -r_3 + 2r_1 < \frac{1}{3}(2r_3 +r_1) ~ \iff ~ r_1 < r_3 .
  \end{split}
\end{equation*}
Simply stated, the harmonic edge is on the left and the soliton edge
is on the right.  This determines the DSW's \textit{orientation} $d =
1$ (recall Fig.~\ref{fig:dsw_classification}).  The DSW boundaries
$x_\pm(t)$ are  merged characteristics of the Whitham system 
\begin{align}
  \label{eq:125}
  x_-(t) &= s_- t, \qquad s_- = V_2(\mathbf{r}_{\mathrm{h}}) \\
  \label{eq:126}
  x_+(t) &= s_+ t, \qquad s_+ = V_2(\mathbf{r}_{\mathrm{s}}) ,
\end{align}
straight lines in the $x$-$t$ plane.

With the DSW boundaries and its orientation determined, we can now
identify the final two GP matching conditions
\begin{align}
  \label{eq:45}
  a_- = a |_{\partial S_-} = 0, \quad k_+ = k |_{\partial S_+} = 0 .
\end{align}
Utilizing the expressions in eqs.~\eqref{eq:66} and \eqref{eq:68}  for  $\bar \varphi$  in the soliton and harmonic limit respectively,  we can relate the
matching conditions \eqref{eq:39} and \eqref{eq:45} to the following
conditions on the Riemann invariants
\begin{align}
  \label{eq:47}
  r_2(s_-) = r_1 = u_+, \quad r_2(s_+) = r_3 = u_- .
\end{align}
Now, using (\ref{kdv_char_vel}), (\ref{eq:47}) the modulation solution
(\ref{eq:43}) can be explicitly written as
\begin{equation}\label{eq:explicit}
\frac{x}{t} = u_+ + \frac{1}{3}\Delta \left(1+m - 
    \frac{2m(1-m)\K(m)}{\E(m) - (1-m)\K(m)}\right),
  \end{equation}
where  $\Delta = u_- - u_+$ is the initial jump height and
$m=(r_2-u_+)/\Delta$.

The macroscopic properties of the DSW include the motion of
the harmonic and soliton edges \eqref{eq:125}, \eqref{eq:126} with speeds found from (\ref{eq:explicit})  by computing the limits
$m \to 0$ and $m \to 1$ respectively,
\begin{equation}
  \label{eq:52}
  s_- = -\Delta + u_+ , \quad s_+ = \frac{2}{3}\Delta + u_+.
\end{equation}
 The expansion speed of the DSW is $s_+-s_-=\tfrac{5}{3}\Delta$.
Recalling the amplitude and wavenumber of the modulated wave
\eqref{eq:46}, we can now determine the soliton edge amplitude $a_+$
and the harmonic edge wavenumber $k_-$
\begin{equation}
  \label{eq:9}
  a_+ = 2\Delta , \quad k_- = \frac{1}{3} \sqrt{6\Delta}\, .
\end{equation}
One can directly verify that the harmonic edge moves with the linear
group velocity $s_- = \partial_k \omega_0(k_-,u_-)$ (recall
eq.~\eqref{eq:68}).  Similarly, the soliton edge moves according to the
KdV soliton amplitude-speed relation $s_+ = V_{\rm s} = u_+ + a_+/3$ (recall
eq.~\eqref{eq:66}).  The soliton edge rises above $u_+$ so the
\textit{polarity} of this DSW is $p = 1$ (recall
Fig.~\ref{fig:dsw_classification}).

So far, we have determined the macroscopic behavior of the DSW, its
leading and trailing edge properties.  The asymptotic behavior of the
quantities of physical interest (the amplitude, the mean value and the
wavenumber) in the vicinities of the trailing and leading edges is
readily obtained by expanding expressions (\ref{eq:46}) along with the
modulation solution (\ref{eq:47}), (\ref{eq:explicit}) for $m \ll 1$
and $1-m \ll 1$ respectively (see \cite{gurevich_nonstationary_1974}).
In particular, one has $a \simeq \tfrac{2}{3}(s-s_-)$ near the
trailing edge and $k \simeq \pi (2/3)^{1/2}/\ln({1}/(s_+-s))$ near the
leading edge.

The internal DSW structure is recovered by inserting the 2-rarefaction
modulation solution for $r_1, r_2, r_3 $ given by eqs.~\eqref{eq:47},
\eqref{eq:explicit} into the traveling wave \eqref{eq:65}.  An example
is shown in Fig.~\ref{fig:kdv_dsw} along with the weak limit
$\overline{\varphi}$ where we used a zero phase shift.

A numerical comparison to the GP problem was undertaken in the work by
Fornberg and Whitham \cite{fornberg_numerical_1978}, where they
demonstrated excellent agreement between the numerics and the leading
edge speed \eqref{eq:52}.  The trailing edge exhibits decaying waves
that propagate slower (more negative) than the trailing edge velocity
\eqref{eq:52}.  The structure of these waves has been studied by use
of matched asymptotic expansions \cite{ablowitz_interactions_2013},
where it was shown that they follow typical dispersive decay with
amplitude proportional to $t^{-1/2}$.  It is worth noting that there
is a logarithmic correction to the motion of the soliton edge, leading
Khruslov, who studied the GP problem using IST, to call the leading
oscillation a \textit{quasi-soliton} \cite{khruslov_asymptotics_1976}.

In summary, GP identified two matching conditions at each 1-phase,
0-phase boundary:
\begin{enumerate}
\item Matching the 1-phase wave mean to the 0-phase flow, eq.~\eqref{eq:39}.
\item The zero amplitude and zero wavenumber reductions at the
  harmonic and soliton edges, respectively, eq.~\eqref{eq:45}.
\end{enumerate}
The location of the harmonic and soliton edges are determined by the
existence of a self-similar, centered simple wave solution connecting
them.  These DSW ``jump conditions'' are the dispersive analogues of the
Rankine-Hugoniot jump conditions of classical shock theory.

We have been considering the strictly negative or concave dispersion
KdV equation \eqref{kdv1}.  The dispersion sign is defined according
to $\mathrm{sgn}\, [\partial_{kk}\omega_0(k,\overline{u})]$, $k > 0$.
The sign of dispersion provides an interpretation of the DSW
orientation $d = 1$.  For concave dispersion, the group velocity
decreases with increasing wavenumber so that the harmonic edge (larger
wavenumber) must move slower than the soliton edge (zero wavenumber),
hence the requisite orientation \cite{gurevich_nonlinear_1990}.

The transformation $x \to -x$, $u \to -u$ of the KdV equation
\eqref{kdv1} results in $u_t + uu_x - u_{xxx} = 0$, exhibiting
strictly positive or convex dispersion.  The transformation causes
the DSW to switch both its orientation and polarity to $d = -1$, $p =
-1$.

\subsubsection{Case $u_- < u_+$:}
\label{sec:case-u_--1}

Implicit in the previous analysis of the GP problem is the existence of a
modulated 1-phase region, which, due to wavebreaking of the
dispersionless equation \eqref{eq:37}, is guaranteed when $u_- > u_+$.
For the case $u_- < u_+$, a leading order description solely in terms
of the dispersionless equation \eqref{eq:37} is possible.  This
involves a centered rarefaction wave that takes the form
\begin{equation*}
  r(x,t) =
  \begin{cases}
    u_- & x < u_- t \\
    \frac{x}{t} & u_-t \le x < u_+t \\
    u_+ & u_+t < x
  \end{cases}.
\end{equation*}
Higher order corrections to this leading order solution have been
investigated using matched asymptotic methods
\cite{leach_large-time_2008}.  It was demonstrated that the left edge
of the rarefaction wave exhibits small oscillations with typical
dispersive long time amplitude decay $\mathcal{O}(t^{-1/2})$ and the right
edge decays exponentially to $u_+$.

\begin{figure}
  \centering
  \includegraphics{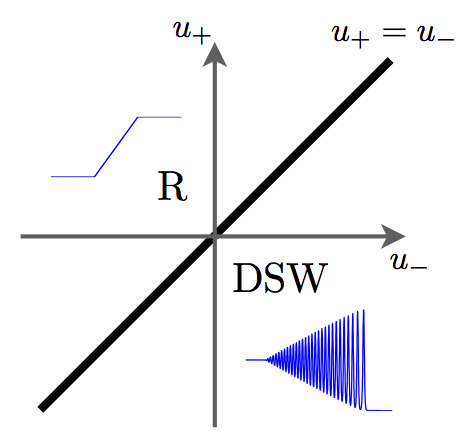}
  \caption{Classification of the GP problem including rarefaction waves (R)
    and DSWs.}
  \label{fig:gp_classification}
\end{figure}

Combining the two cases $u_- \gtrless u_+$, we obtain the complete
classification of the GP problem, represented in
Fig.~\ref{fig:gp_classification}.

\subsubsection{DSW characteristics}
\label{sec:dsw-characteristics}

\begin{figure}
  \centering
  \includegraphics{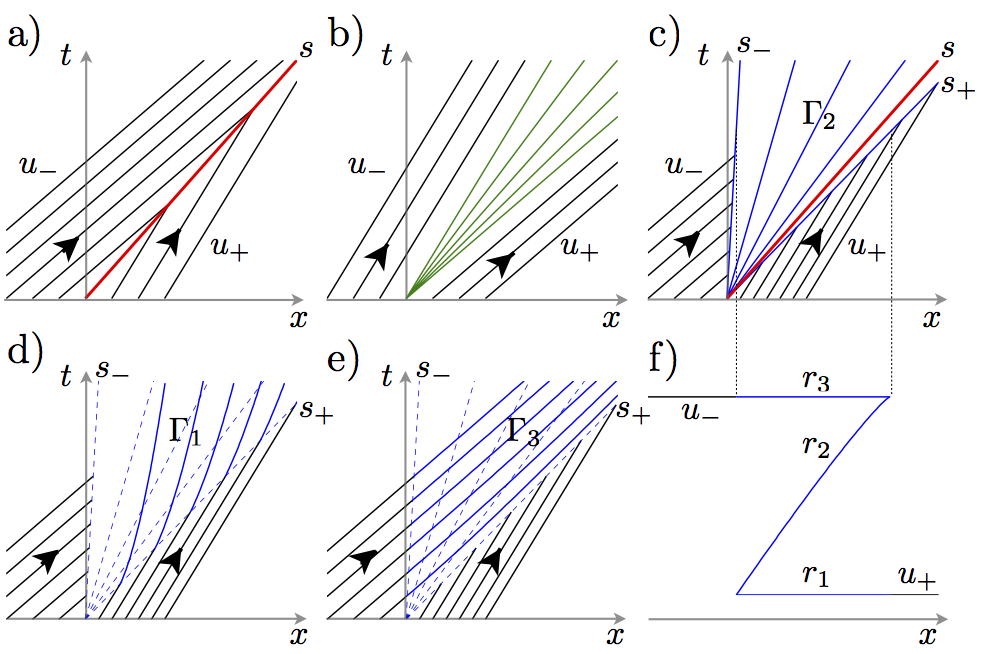}
  \caption{Characteristics for the classical shock (a), rarefaction (b)
    solutions of the Hopf equation \eqref{eq:37} and the (c-e)
    KdV DSW solution along with the (f) Whitham modulation solution in
    Riemann invariants.}
  \label{fig:kdv_characteristics}
\end{figure}

Adopting tools from the theory of hyperbolic conservation laws, we
consider the characteristic families associated with the KdV DSW in
Fig.~\ref{fig:kdv_characteristics}.  External to the 1-phase region,
the characteristics are straight lines $x = u_\pm t + x_0$.  Internal
to the DSW, there are three characteristic families of the Whitham
equations given by
\begin{equation*}
  \Gamma_i = \left \{ x(t) ~ \Big | ~  \frac{\rmd x}{\rmd t} = V_i \right
  \}, \quad i = 1, 2, 3 .
\end{equation*}
The second characteristic family in
Fig.~\ref{fig:kdv_characteristics}c consists of the expansion fan
solution in Fig.~\ref{fig:kdv_characteristics}f.  The first (slow) and
third (fast) characteristic families in
Figs.~\ref{fig:kdv_characteristics}d,e are curved and emanate from the
soliton, $x = s_+t$, and harmonic, $x = s_-t$, DSW boundaries,
respectively.  The slow and fast characteristics transfer information
into the DSW.  In long time, these characteristics become parallel to
the opposite DSW boundaries $x= s_-t$ and $x = s_+ t$, respectively.
Thus, merger of the first and second characteristic families occurs at
the trailing, harmonic edge of the DSW, coinciding with the group
velocity of linear waves.  The second and third characteristic
families merge at the leading, soliton edge of the DSW, corresponding
to propagation with the soliton phase velocity.

We also include the characteristic description of classical shocks
Fig.~\ref{fig:kdv_characteristics}a and rarefactions
Fig.~\ref{fig:kdv_characteristics}b for comparison.

An equivalent but alternative approach to dispersive regularization of
the Riemann problem can be achieved by choosing degenerate initial
conditions for the 1-phase modulation equations that match the initial
conditions \eqref{eq:36} and yield a global solution.  The degeneracy
relates to the reduction in order of the Whitham equations when two
Riemann invariants coincide.  This procedure was first utilized in
\cite{bloch_dispersive_1992} and later applied to NLS
\cite{kodama_whitham_1999}, \cite{biondini_whitham_2006} and the KdV Riemann problem
\cite{hoefer_dispersive_2006}.

\subsection{KdV: dispersive regularization of wavebreaking}
\label{sec:dsws-wave-breaking}

The Riemann problem solution for the KdV equation provides a major
insight into properties of DSWs. It is also of primary interest for
applications since it utilizes the simplest, centered simple wave
solutions of the associated modulation system.  In many situations,
however, this solution is not sufficient for the description of DSWs
as the general scenario of simple wave evolution can involve a
``spontaneous'' formation of gradient catastrophe at some point $(x_b,
t_b)$, whose location is determined by initial conditions
\cite{whitham_linear_1974}. To describe the dispersive regularization
of wavebreaking, a more general solution of the Whitham equations is
required.

We consider the KdV equation (\ref{kdv1}) with a ``large and smooth''
initial condition $u(x,0) =u_0(\varepsilon x)$ where $\varepsilon \ll 1$. It
is convenient to rescale the independent variables in (\ref{kdv1}) by
introducing $\tilde x = \varepsilon x$, $\tilde t = \varepsilon t$ (i.e. the
new variables are the slow variables $X$ and $T$ from
Sec.~\ref{sec:kdv-mod-sys}). Then, on dropping tildes, the initial
value problem of interest is:
\begin{equation}\label{kdv_eps}
 u_t + uu_x + \varepsilon^2 u_{xxx}=0,  \quad u(x,0)=u_0(x),
\end{equation}
where $ u_0 (x)$ is a sufficiently smooth function with $u_0'(x)<0$ on
some interval $\Delta \subset \mathbb{R}$ with a single inflection
point on $\Delta$.

Initially, if $\varepsilon$ is sufficiently small, one can neglect the
dispersive term in (\ref{kdv_eps}) and approximate the solution
$u=u(x,t; \varepsilon)$ of the small-dispersion KdV equation
(\ref{kdv_eps}) by the solution $r(x,t)$ of the dispersionless KdV or
Hopf equation (\ref{eq:37}) with the same initial condition $u_0(x)$.
This is, actually, a rigorous result due to Lax and Levermore \cite{lax_small_1983},
stating that $\lim \limits_{\varepsilon \to 0} u(x, t; \varepsilon)= r(x,t)$
for $0 < t < t_b$.  The solution of the Hopf equation can be written
in the implicit form
\begin{equation}
  \label{hs}
  x-rt=W(r),
\end{equation}
where $W(r)$ is the inverse function of $r=u_0(x)$ (see (\ref{swsol})).

Nonlinear evolution of $r(x,t)$ specified by the simple wave solution
(\ref{hs}) leads to the steepening of the profile for $r(x)$ and
eventually, to gradient catastrophe $ |r_x| \to \infty$ at some
point $(x_b,t_b, r_b)$, found from the conditions $W''(r_b)=0$, $t_b=-
W'(r_b)$ and $x_b=r_bt_b + W(r_b)$ (see,
e.g., \cite{whitham_linear_1974}).

Let the function $u_0(x)$ be analytic near the breaking point. Using
symmetries of the KdV equation, which are also symmetries of the Hopf
equation (\ref{eq:37}), one can always set the wave breaking point at
$x_b=t_b=r_b=0$ so that, without loss of generality, we can assume
that $W(0)=W'(0)=W''(0)=0$. Expanding the solution (\ref{hs}) near this
point assuming that $W'''(0) \ne 0$ and upon retaining the first
nonvanishing term, we obtain the universal ``cubic'' behavior of the
simple wave near the point of gradient catastrophe
\cite{gurevich_nonstationary_1974}:
\begin{equation}
  \label{cubic}
  x- rt = - \mu r^3 \, ,  \quad \mu >0.
\end{equation}
Using symmetries of the KdV equation, one can always set $\mu =1$. One
can show that the curve $ r(x,t)$ defined by (\ref{cubic}) is
single-valued for $ t<0$ (pre-breaking), and triple-valued for $t>0$
(post-breaking). The latter behavior is unphysical and dispersion has
to be taken into account for $t>0$ in order to regularize the
singularity and prevent formation of the multivalued solution.

We now observe that the function $r(x,t)$ implicitly specified by
equation (\ref{cubic}) is the solution to the Cauchy problem for the
Hopf equation (\ref{eq:37}) with the initial condition
\begin{equation}\label{cubic_ic}
r(x,0)=-x^{1/3} ,
\end{equation}
where we assumed $\mu=1$ in (\ref{cubic}).  This establishes our
interest in considering the KdV equation (\ref{kdv_eps}) with initial
data in the form of an inverse cubic (\ref{cubic_ic}) as a
universal model for the dispersive regularization of wavebreaking.

In what follows, we construct modulation solutions regularizing the
wavebreaking for two types of initial data: (i) a monotone decreasing
function with a single inflection point; (ii) a decaying, localized profile. The
last case is of particular interest as it establishes a connection
between Whitham theory and the semi-classical, Bohr-Sommerfeld
approximation in quantum mechanics \cite{landau_quantum_1965}.

\subsubsection{GP matching conditions}

\label{sec:gp_matching}

We now formulate free-boundary matching conditions for the modulation
regularization of the general wavebreaking problem. These are
essentially analogous to the GP conditions (\ref{eq:39}) for the
Riemann problem but now, instead of matching the average $\overline
\varphi$ to constant boundary values $u_{\pm}$ at the DSW edges, we
require that $u_{\pm}= r(x_{\pm}(t), t)$, where $r(x,t)$ is the
solution of the Hopf equation given by (\ref{hs}).  Assuming the same
DSW orientation $d = 1$ established in the solution of the Riemann
problem in Sec.~\ref{sec: dsw-riemann}, we identify the trailing
$x_-(t)$ and leading $x_+(t)$ edges as harmonic ($m=0$) and soliton
($m=1$), respectively, and formulate general GP matching conditions
for the KdV-Whitham system \eqref{eq:124}.  We first note that it
follows from \eqref{eq:46} that in the harmonic and soliton limits,
the mean of the cnoidal wave solution $\overline {\varphi}$ becomes
\begin{equation} 
  \label{phi_r}
  \begin{split}
    m=0: \quad \overline {\varphi} = r_3 ,  \\
    m=1: \quad \overline {\varphi} =  r_1 .
  \end{split}
\end{equation}
The generalized GP matching conditions are then formulated as 
\begin{equation}
  \begin{array}{l}
    x=x_-(t):\qquad r_2=r_1\, , \ \ r_3 = r \, , \\
    x=x_+(t):\qquad r_2=r_3\, , \ \ r_1 = r \, ,
  \end{array}
\label{GP_match}
\end{equation}
where $r(x,t)$ is given by (\ref{hs}).  The DSW boundary $\partial S$
(\ref{eq:40}) is defined by the merged characteristics of the Whitham
system for $m=0$, where $x=x_-(t)$, and $m=1$, where $x=x_+(t)$ (see
(\ref{v2=v1}), (\ref{v2=v3})).  Then $x_{\pm}(t)$ are found from the
ordinary differential equations
\begin{equation}
  \label{edges_general}
  \frac{ dx_-}{dt}=- r_- + 2r_{12}\, , \ \  \frac{dx_+}{dt}=
  \frac{1}{3} (2r_{23} + r_+),  
\end{equation}
where $r_-(t) \equiv r(x_-(t),t)$ and $r_{12}(t) \equiv r_1(x_-(t),t)
= r_2(x_-(t),t)$ is the value of the merged Riemann invariant at
$x=x_-(t)$. Analogously $r_+(t) \equiv r(x_+(t),t)$ and $r_{23}(t)
\equiv r_2(x_+(t),t) = r_3(x_+(t),t)$.  Note that, unlike in the
solution of the Riemann problem, where the DSW edges are defined by
{\it regular characteristics} of the Whitham modulation system, the
general matching regularization (\ref{GP_match}) implies that the DSW
edges $x_{\pm}(t)$ specified by \eqref{edges_general} are {\it
  characteristic envelopes}, i.e., singular solutions of the general
characteristic ODEs defined on the solution of the GP problem. As a
result, the corresponding merged Riemann invariants $r_{12}$ and
$r_{23}$ are allowed to vary along the edge curves
\cite{gurevich_nonstationary_1974}.

One can show using \eqref{eq:67} -- \eqref{eq:69} that the edge
definitions \eqref{edges_general} can be re-written in a physically
transparent form
\begin{equation}\label{kinematic_conds}
  \frac{d x_-}{dt}= c_{\rm g}(k_-, \overline \varphi_-); \qquad
  \frac{dx_+}{dt} =V_{\rm s}( a_+, \overline \varphi_+), 
\end{equation}
where $c_{\rm g}( k, \overline \varphi )= \partial_k\omega_0(k,
\overline \varphi)$ is the linear group velocity and $V_{\rm s}( a_s,
\overline \varphi )$ is the speed of the KdV soliton of amplitude
$a_{\rm s}$ riding on the backround $\overline \varphi$. It follows
from \eqref{phi_r}, \eqref{GP_match} that at the edges $\overline
\varphi_\pm=r_{\pm}$.  Also $a_+=a_{\rm s}(x_+, t)$, $k_-=k(x_-, t)$.

The definitions \eqref{kinematic_conds} of the trailing and leading
edges have a natural interpretation as {\it DSW kinematic boundary
  conditions} while the matching conditions \eqref{GP_match} are {\it
  dynamic boundary conditions}, using standard terminology from
water wave theory \cite{whitham_linear_1974}. We also note that the
edge definitions in the form (\ref{kinematic_conds}) admit a natural
extension to DSW theory in non-integrable dispersive hydrodynamics,
for which modulation systems do not possess Riemann invariant
structure.

\subsubsection{Hodograph solution: the scalar potential}

\label{sec:hod_phase_funct}

The generalized GP problem \eqref{eq:124}, (\ref{GP_match}) is a
nonlinear free-boundary problem, which is generally very difficult to
solve directly. However, it admits a remarkable simplification when
reformulated in hodograph space with Riemann invariants $r_1, r_2,
r_3$ as independent variables.  First of all, one can notice that the
DSW edges $x_{\pm}(t)$ in the $x$-$t$ plane map to planes in hodograph
space. These are given by $r_1=r_2$ (the trailing edge) and $r_2=r_3$
(the leading edge). Then, using that $V_3(r_1, r_1, r_3) = r_3$ and
$V_1(r_1, r_3, r_3)=r_1$ (see Sec.~\ref{sec:prop-kdv-whith}) and
comparing the generalized hodograph solution (\ref{ghod}) with the
simple-wave solution (\ref{hs}) at the above boundaries, we obtain the
boundary conditions for the linear Tsarev system (\ref{ts})
\cite{gurevich_breaking_1991,gurevich_evolution_1992}
\begin{equation}
  \label{bc_w}
  \begin{split}
    W_3(r_1, r_1, r_3) = W(r_3), \\  W_1(r_1, r_3, r_3)= W(r_1)\, .
  \end{split}
\end{equation}
Further, one can show that $\partial_1 W_3(r_1, r_1, r_3) = \partial_3
W_1(r_1, r_3, r_3)=0$ \cite{gurevich_evolution_1992} and, since $r =
r_3$ along $x = x_-(t)$ and $r = r_1$ along $x = x_+(t)$, the conditions
(\ref{bc_w}) can be replaced by
\begin{equation}
  \label{bc_w_1}
  \begin{split}
    W_3(0, 0, r_-) =W(r_-); \\
    W_1(r_+, 0, 0)=W(r_+)\, .
  \end{split}
\end{equation}
Note that the boundary conditions (\ref{bc_w_1}) are linear and 
specified on known boundaries: the characteristics of the Tsarev
system (\ref{ts}). Thus the problem (\ref{ts}), (\ref{bc_w_1}) is a
linear Goursat (Darboux) problem \cite{courant_methods_1989-1}.  This
remarkable {\it full linearization} of the nonlinear free-boundary,
generalized GP problem in hodograph space is in sharp contrast
with the well known poor compatibility of initial-value problems with
the classical hodograph transform \cite{whitham_linear_1974}.

We now use the Goursat data (\ref{bc_w_1}) for $W_j$ to formulate the
counterpart boundary conditions for the EPD system (\ref{EPD})
defining the $q$-function, a scalar potential for the solutions $W_j$
to the Tsarev equations (\ref{ts}).  The boundary conditions for
$q(r_1, r_2, r_3)$ are obtained by substituting (\ref{bc_w_1}) in the
transformation (\ref{wscalar}) and then integrating the resulting
ODEs. As a result we obtain
\begin{equation}
  \label{bc_g}
  q(r, 0, 0) = q(0, 0, r) = \frac{1}{2}r^{-1/2}
  \int^{r}_{0}y^{-1/2}W(y)\rmd y \, . 
\end{equation}
 We
note that $r_2=r_3 = r_{23}$ and $r_2=r_1 = r_{12}$
corresponding to the DSW leading and trailing edges, respectively, are
singular surfaces for the EPD system (\ref{EPD}).  For the elliptic
parameter $m=(r_2-r_1)/(r_3-r_1)$ in the DSW to be defined in the entire
interval $0 \le m \le 1$ for all $t$, one must impose additional
conditions of non-singularity of $q$ at these surfaces,
\begin{equation}\label{nonsing}
  q(r_+, r_{23}, r_{23}) < \infty, \quad q(r_{12},r_{12}, r_-) < \infty \, .
\end{equation} 

If one of the Riemann invariants is constant, the function $g$ becomes
a function of two variables and only one of the conditions
(\ref{bc_g}) should be used. This is the case, e.g., if one considers
the GP problem with initial function $u_0(x)$, which monotonically
decreases for $x<0$ and is a constant for $x>0$, leading to DSW
propagation into a uniform ``gas'', a constant equilibrium state.  For
smooth, monotonically decreasing initial conditions $u_0(x)$, causing
all three Riemann invariants in the modulation solution to be
non-constant, an alternative form of the boundary condition for the
EPD system can be used \cite{tian_oscillations_1993}: $q(r,r,r) =
W(r)$.

Using the boundary conditions (\ref{bc_g}), we can identify the
functions $\Phi_j(\lambda)$ in the general solution (\ref{general_g})
of the EPD equation.  These are shown in
\cite{gurevich_evolution_1992} to be expressed in terms of the Abel
transform of the inverse function $W(r)$.  Finally, the solution for
the  function $q({\bf r})$ can be represented in compact form
\cite{gurevich_evolution_1992}
\begin{equation}
  \label{fsol}
  \begin{split}
    q=\frac{1}{\pi (r_3-r_2)^{1/2}}\int \limits
    ^{r_3}_{r_2}\frac{W(\tau)}{\sqrt{\tau-r_1}}K(z)\rmd\tau  \\
    +
    \frac{1}{\pi (r_2-r_1)^{1/2}}\int \limits
    ^{r_2}_{r_1}\frac{W(\tau)}{\sqrt{r_3-\tau}}K\left ( z^{-1} \right
    ) \rmd\tau\, , 
  \end{split}
\end{equation}
where $K(z)$ is the complete elliptic integral of the first kind and
\begin{equation}
  \label{z} 
  z=\left[
    \frac{(r_2-r_1)(r_3-\tau)}{(r_3-r_2)(\tau-r_1)}\right]^{1/2}.
\end{equation}
A different, double integral form of the solution $q({\bf r})$ was
obtained in \cite{tian_oscillations_1993}, where also the
invertibility of the hodograph solution (\ref{mapping}) for $r_j(x,t)$
was proved.

\subsubsection{Regularization of cubic wave breaking}
\label{sec:cubic_breaking}

As was shown in Sec.~\ref{sec:dsws-wave-breaking}, the regularization
of ``cubic'' wavebreaking is of particular interest, so we shall
construct the corresponding modulation solution explicitly.  A general
way to approach this problem would be to substitute the inverse
function $W(r)= -r^3$ of the initial condition (\ref{cubic_ic}) into
the solution (\ref{fsol}) for the  $q$-function and then generate
the modulation solution $\{ r_j\}_{j=1}^3$ in an implicit hodograph form
(\ref{mapping}). Instead, we notice that the dispersionless limit
solution (\ref{hs}) has scaling properties, namely
$r=t^{1/2}R(x/t^{3/2})$, where $R(\zeta)$ satisfies the algebraic
equation $ R^3 - R +\zeta=0$. Then one can directly construct the
modulation solution having the same scaling properties and require  that
the two solutions continuously match according to the GP conditions
(\ref{GP_match}) at the phase transition boundaries $x_{\pm}(t)$, the
DSW edges. The family of modulation solutions having the necessary
scaling properties was shown in Sec.~\ref{gen_sol_kdv-whitham} to be
generated by homogeneous solutions of the EPD system.  To this end, we
identify the scaling parameter $\gamma=1/2$ in (\ref{scaling_krich})
and find the corresponding homegeneity degree $\alpha=1+
1/\gamma=3$. Thus the necessary homogeneous solution is $q_3 = 5s_1^3
-12 s_1s_2 + 8s_3$, where $s_j({\bf r})$ are the elementary symmetric
polynomials (see \eqref{gi} and (\ref{s})). We now set $q=Cq_3$, where
$C$ is a constant to be found from the matching conditions
(\ref{GP_match}), or their equivalent $q(r, r, r)=W(r)=-r^3$
\cite{tian_oscillations_1993} for strictly monotone initial
functions. We find that $C=-1/35$.  As a result, the generalized
similarity modulation solution has the form
\begin{equation}
  \label{rsim}
  r_j(x,t)=t^{1/2} R_j(\zeta), \ \  \zeta=x/t^{3/2},
\end{equation}
where $R_j$, $j=1,2,3$ are found from the system
\begin{equation}\label{potemin}
\begin{split}
  \zeta -V_i(R_1, R_2, R_3)= -\tfrac{1}{35} \left (
    1-\frac{L}{\partial_i L} \partial_i \right ) q_3,   \\
  i=1,2,3, 
\end{split}
\end{equation} 
with $L\equiv L({\bf R})$, $q_3 \equiv q_3({\bf R})$, and $\partial_i
\equiv \partial_{R_i}$.

\begin{figure}[h]
  \centering
  \includegraphics[scale=0.25]{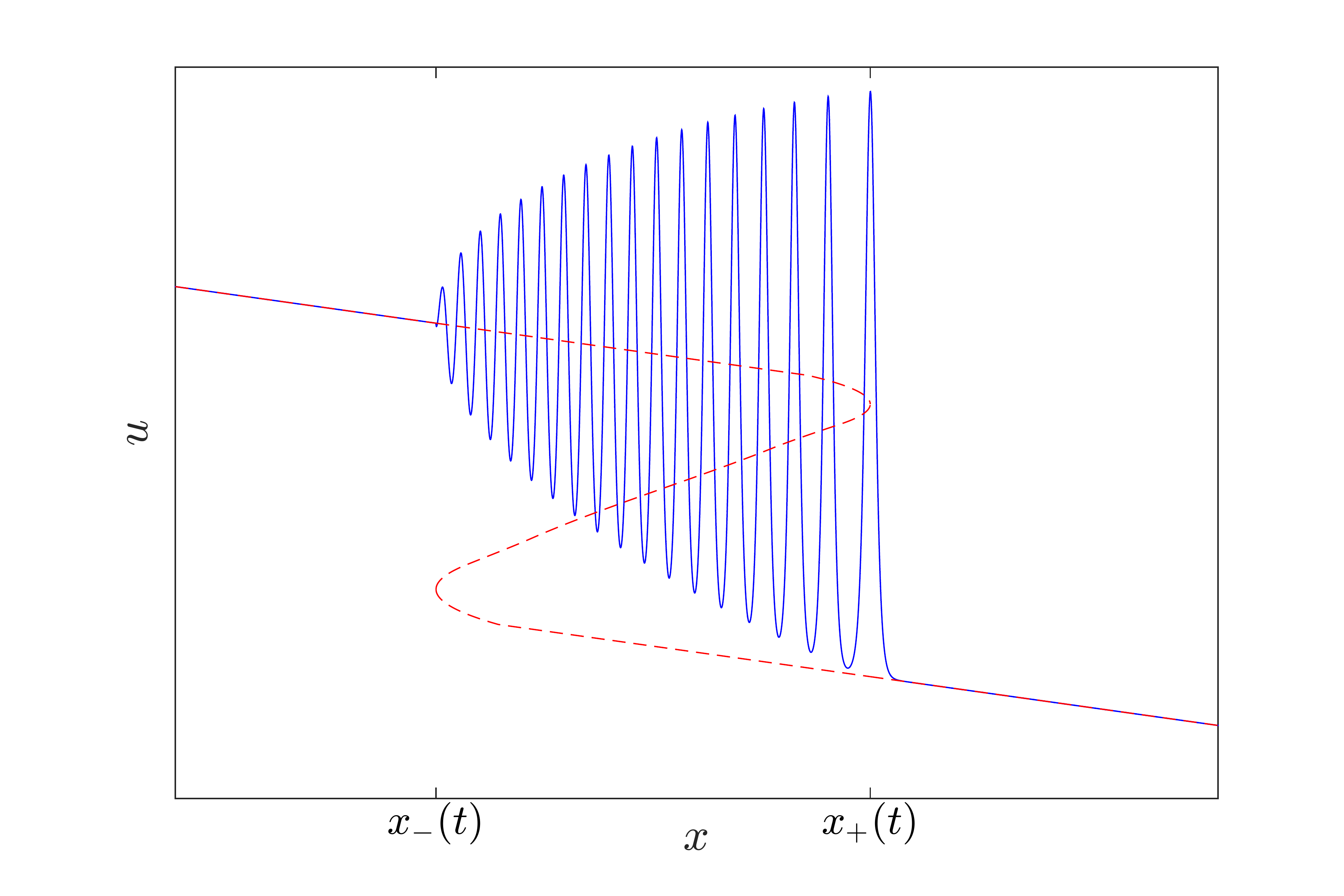}
  \caption{Sketch of the dispersive regularization of cubic
    wavebreaking in the KdV equation (solid) along with the
    modulation solution $r_{3}(x,t) \geqslant r_2(x,t) \geqslant
    r_1(x,t)$ (dashed).}
  \label{fig:wavebreaking}
\end{figure}
The constructed similarity solution describes the evolution of the
triple-valued function \cite{dubrovin_hydrodynamics_1989}
$\{r_1(x,t), r_2(x,t), r_3(x,t)\}$ (see Fig.~\ref{fig:wavebreaking}),
defined for all $t >0$ within the expanding interval $[x_-, x_+]$,
where
\begin{equation}\label{edges}
  x_-(t) = -2 \sqrt{3} t^{3/2}, \quad x_+(t) = \frac{\sqrt{15}}{27}
  t^{3/2} \, .
\end{equation}
The latter expressions are obtained by integrating
\eqref{edges_general}. We stress that the obtained modulation solution
{\it does not} coincide with the formal triple-valued solution
(\ref{cubic}) of the Hopf equation for $t>0$.  Substituting
$r_1(x,t)$, $r_2(x,t)$, $r_3(x,t)$ specified by (\ref{rsim}),
(\ref{potemin}) into the ``carrier'' traveling wave solution
(\ref{eq:65}), one obtains the full asymptotic description of the DSW.
The phase of the traveling solution is $ \theta= k(x - V t - \xi_0)$
with the modulation phase correction $\xi_0(x,t) = q({\bf r}) =
-\tfrac{1}{35}q_3({\bf r})$ (see Sec.~\ref{gen_sol_kdv-whitham}). A
schematic of dispersive regularization of cubic wavebreaking is shown
in Fig.~\ref{fig:wavebreaking}.

The amplitude of the lead soliton is found from the leading edge
definition in eq.~(\ref{kinematic_conds}), yielding $\dot x_+=
\overline \varphi_+ +a_+/3$ (see (\ref{eq:66})), whereas the soliton
background satisfies $\overline \varphi_+ = r_+=t^{1/2}R_+$, $R_+
\approx -1.065$ being the smallest root of the equation
$R^3-R+\sqrt{15}/27=0$.  As a result, using \eqref{edges} we obtain
\begin{equation}
  a_+(t) =   A t^{1/2}, \quad    A =  \tfrac{\sqrt{15}}{6} - 3 R_+
  \approx 3.840 
\end{equation}
The trailing edge wavenumber $k_-$ is found from the trailing edge
definition in eq.~\eqref{kinematic_conds} yielding $\dot x_-=
\overline \varphi_- -3k_-^2$ where the vanishing amplitude wave
background satisfies $\overline \varphi_- = r_- = t^{1/2}R_-$, $R_-
=\sqrt{3} $ being the greatest root of $R^3-R-2\sqrt{3}=0$. As a
result,
\begin{equation*}
  k_-(t) = (3t)^{1/4} .
\end{equation*}

Concluding, we note that the modulation solution for the cubic GP
problem was first studied numerically in the original GP paper
\cite{gurevich_nonstationary_1974}, where one can also find the
asymptotic behavior of the Riemann invariants near the trailing and
leading edges.  One qualitative difference to the GP solution for the
Riemann step problem is that in the step regularizing DSW, the wave
amplitude grows linearly near the trailing edge, $a \sim (x-x_-)$,
which results in a ``Martini glass'' DSW envelope shape, while in the
case of cubic wavebreaking, one has $a \sim \sqrt{x-x_-}$, a
``Bordeaux glass'' shape (cf.~Fig.~\ref{fig:kdv_dsw} and
Fig.~\ref{fig:wavebreaking}). The latter asymptotic behavior is, in
fact, generic for the dispersive wavebreaking regularization. Another
difference is that the lead amplitude $a_+$ as well as the trailing
edge wavenumber $k_-$ grow as $ t^{1/2}$ and $t^{1/4}$ respectively,
in the KdV cubic wavebreaking problem while they remain constant in
the Riemann problem. 

 
The modulation solution (\ref{potemin}) was first obtained by Pot\"emin
\cite{potemin_algebro-geometric_1988} by analyzing the relevant
Krichever algebro-geometric solution \cite{krichever_method_1989}.
This solution was also obtained in \cite{wright_exact_1999} within the
framework of the IST based semi-classical Lax-Levermore approach
\cite{lax_small_1983}. Stability of the modulation solution
regularizing cubic wave breaking was investigated numerically in
\cite{avilov_evolution_1987}.

\subsubsection{Evolution of  decaying profiles}

\label{sec:evolution_decaying_profiles}

We now consider KdV evolution (\ref{kdv_eps}) of decaying initial
profiles $u_0(x)$: $u_0(x) \to 0$ as $|x| \to \infty$. Since such an
evoution, under some mild restricions imposed on the decay rate of
$u_0(x)$, can be described using the IST, it will be particularly
instructive to see how the IST results compare with the results of the
modulation approach employed here.  We shall separately consider the
elevation and depression cases and then combine them into a dispersive
counterpart of the classical problem of $N$-wave evolution modeled by
the solution of Burger's equation \cite{whitham_linear_1974}.

\smallskip
{\it Localized elevation}

\begin{figure}[h]
  \centering
    \includegraphics{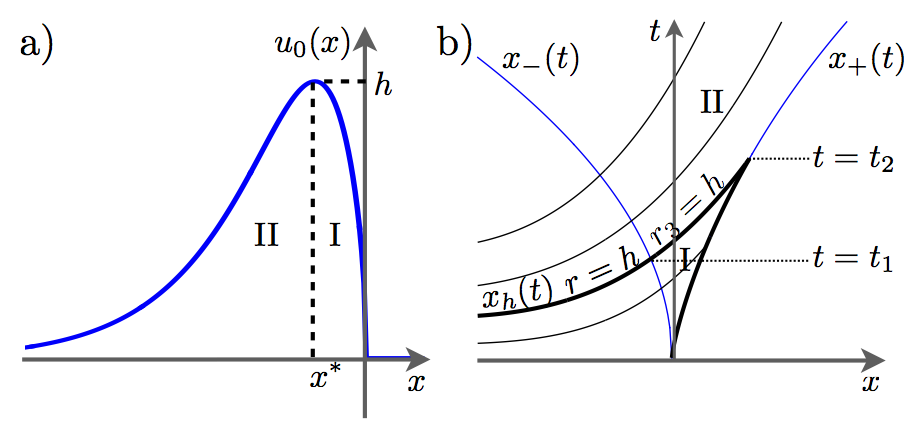}
    \caption{a) initial localized elevation $u_0(x)$; b)
      characteristics of the modulation system for the localized
      elevation problem.} 
    \label{fig:elevation}
\end{figure}

\smallskip With a mild violation of generality, we assume that the
initial hump $u_0(x) = u_0^+(x)\ge 0$, has semi-infinite support so
that $u_0^+(x) \equiv 0$ for $x>0$ and $u_0^+(x)>0$ for $x<0$.  This
corresponds to the aforementioned propagation of a DSW into a
uniform ``gas''.  We also assume that $u_0(x)$ has a single maximum at
some $x=x^*<0$: $u_0^+(x^*)=h>0$, $(u_0^+)'(x^*)=0$ and $u_0^+(x) \to
0$ as $x \to - \infty$. Now one can assume, without a further abuse of
generality, that the wavebreaking point is at $x=0, t=0$. A typical
form of the initial profile is shown in Fig.~\ref{fig:elevation}a.

Since $u_0^+(x)=0$ for $x \ge 0$, the GP matching condition at the
leading edge (\ref{GP_match}) is $r_1(x_+, t)=0$ which, together with
the modulation equation for $r_1$, implies that $r_1=0$ throughout the
modulation solution and there are only two varying Riemann invariants:
$r_2$ and $r_3$.  To handle non-monotonicity of the initial function
$u_0^+(x)$, we introduce a ``two-sheeted'' hodograph transform
\cite{gurevich_quasisimple_1989,gurevich_evolution_1992} so
that we have the {\it hodograph surface} consisting of two copies of
the hodograph plane $(0, r_2, r_3)$ which are glued together along the
line $r_3=h$. The construction is analogous to the Riemann surface
covering of the complex plane. On each sheet of the hodograph surface,
we define the mapping (\ref{mapping}) $(r_2,r_3) \leftrightarrow (x,
t)$ -- bijectivity of this mapping is not automatic and must later be proven
 for the particular solution -- reducing the modulation system
(\ref{eq:31}) for $r_2$ and $r_3$ to a single EPD equation (\ref{EPD})
for the  function $q_{k}(r_2,r_3)$ with $k=I, II$ labeling the
sheets of the hodograph surface.  Each function $g_{I, II}$ is defined
inside the triangle $0 \le r_2 \le r_3 \le h$ of the respective sheet.
We now introduce two monotone functions $W_I(r)$ and $W_{II}(r)$ so
that $W_I(r)$, $r \in [0,h]$, is the inverse of the decreasing part of
the initial profile $u_0^+(x)$ and $W_{II}(r)$, $r \in (0, h]$, is the
inverse of the increasing part with $W_I(x^*)= W_{II}(x^*)=h$.  Then the
boundary conditions for the functions $q_I(r_2, r_3)$ and
$q_{II}(r_2, r_3)$ assume the form
\begin{equation*}
  \begin{split}
    q_{I, II}(0, r_3) = \frac{1}{2} r_3^{-1/2} \int \limits^{r_3}_0
    y^{-1/2} W_{I, II}(y) \rmd y , \\ 
    q_I(r_2, h) = g_{II}(r_2, h).
  \end{split}
\end{equation*}
One must also add the non-singularity conditions (\ref{nonsing}) for
$q_I(r,r)$, but not for $q_{II}$, which is defined uniquely by 
continuous matching with $q_I$.

The conditions at $r_2=0$ follow from the GP matching (\ref{GP_match})
at the trailing edge and are analogous to (\ref{bc_g}) while the
condition at $r_3=h$ is new and provides a continuous matching of the
hodograph solutions defined on the sheets $I$ and $II$ along the
characteristic $r_3=h$ which is common for both sheets.

The resulting solution of the EPD equation can be compactly
represented in the form \cite{gurevich_evolution_1992}:
\begin{equation}\label{sheet_I}
\begin{split}
q_I (r_2, r_3) = \int^{r_2}_0 \kappa(r, r_2, r_3) W_I(r) \rmd r \\
+ \int^{r_{3}}_{r_2} \kappa(r_2, r, r_3)  W_{I}( r) \rmd r , 
\end{split}
\end{equation}
\begin{equation}\label{sheet_II}
q_{II} (r_2, r_3) = g_I + \int^h_{r_3}  \kappa(r_2, r_3, r) D( r) \rmd r, 
\end{equation}
where $D(r)=W_I(r) - W_{II}(r)$ and
\begin{equation*}
  \kappa(x,y,z) = \frac{1}{\pi\sqrt{(z- x) y}} \ \K\left(
    \frac{z-y}{z-x} \frac{x}{y}\right). 
\end{equation*}
Note that the solution (\ref{sheet_I}) on the first sheet is
equivalent to $q(0, r_2, r_3)$ given by (\ref{fsol}) with $W(r) \equiv
W_I(r)$. One can readily show that $\partial_3 q_I(r_2, h) =
\partial_3 q_{II}(r_2, h)$ thus ensuring the continuity of the
modulation solution (\ref{mapping}) across the characteristic
$r_3=h$. 

The analysis of the mapping $ (r_2, r_3) \mapsto (x,t)$ specified by
(\ref{mapping}) with $r_1 \equiv 0$, $q=q_{I, II}$ shows that it maps
the triangle $0 \le r_2 \le r_3$ of the sheet $I$ to some finite
region of the $x,t$ plane, bounded by the edge curves $x_-(t), x_+(t)$
defined by (\ref{edges_general}), and the characteristic $x_h(t)$
transferring the Riemann invariant $r_3=h$ (see
Fig.~\ref{fig:elevation}b).  Within the interval $t \in [0, t_1]$,
where $t_1$ is found from the condition $r_3(x_-(t_1), t_1)=h$, the
modulation solution is fully specified by $q_I$ and describes a DSW
regularizing wavebreaking of the monotonically decreasing part of the
initial function $u_0(x)$. For $t \in [t_1, t_2]$, where $t_2$ is
found from the condition $r_3(x_+(t_2), t_2)=h$, the solution
$(r_2(x,t), r_3(x,t))$ consists of two parts generated by $q_I$ and
$q_{II}$ respectively, each part being defined within its own domain
in the ($x,t$) plane. The domains have common boundary defined by the
characteristic transferring the maximum value $r_3=h$. Finally, for $t
> t_2$, the modulation solution is fully defined by the  function
$q_{II}$, which maps the triangle $0 \le r_2 \le r_3$ on the second
sheet of the hodograph surface onto the infinite region in the
$x,t$-plane bounded by the characteristics $[x_-(t), x_+(t)]$, $t >
t_2$ with the leading edge characteristic being a straight line
$x=\tfrac23 r_3 t$. The modulation solution generated by $g_{II}$
describes the interaction of the DSW with the rarefaction wave
developing due to the evolution of the increasing part of the initial
profile $u_0(x)$. This intepretation becomes particularly lucid when
the initial profile has the shape of a rectangular barrier
\cite{el_generation_2002}.

Strictly speaking, one should now show that the obtained solution is
global, i.e., the mapping $(r_2, r_3) \mapsto (x,t)$ is invertible for
all $t$. Partial proof of this invertibility can be found in
\cite{tian_initial_1994}.  Here, instead of giving a full mathematical
proof, we shall show that the obtained modulation solution implies a
long-time asymptotic behavior for $r_j(x,t)$ consistent with the
predictions of IST analysis for the case of a localized profile. This
agreement is a strong indication that the obtained modulation solution
is valid for all $t$.

The analysis of the mapping $ (r_2, r_3) \mapsto (x,t)$ specified by (\ref{mapping}) with $r_1 \equiv 0$,  $q=q_{I, II}$ shows that it maps the triangle $0 \le r_2 \le r_3$ 
of the sheet $I$ to some finite region  of the $x,t$ plane,  bounded by the edge curves  $x_-(t), x_+(t)$  defined by (\ref{edges_general}),  and the characteristic $x_h(t)$ transferring  the 
Riemann invariant $r_3=h$ (see Fig.~\ref{fig:elevation} b)).   Within the interval  $t \in [0, t_1]$, where $t_1$ is found from the condition $r_3(x_-(t_1), t_1)=h$,  the modulation solution is 
fully specified by $q_I$ and describes a DSW regularizing the wavebreaking of the monotonically decreasing part of the initial function $u_0(x)$. For $t \in [t_1, t_2]$, where $t_2$ is found from the condition $r_3(x_+(t_2), t_2)=h$, the solution $(r_2(x,t), r_3(x,t))$   consists of two parts generated by $q_I$ and $q_{II}$ respectively, each part  being defined within its own   domain in the ($x,t$) -plane. The domains have  common boundary defined by the characteristic  transferring the maximum value $r_3=h$. Finally, for $t>t_2$ the modulation solution is fully defined by the   function $q_{II}$ which maps the triangle $0 \le r_2 \le r_3$ on the second sheet of the hodograph surface onto the infinite region in the $x,t$-plane  bounded by the characteristics $[x_-(t), x_+(t)]$, $t > t_2$ with the leading edge characteristic being a straight line $x=\tfrac23 r_3 t$. The  modulation solution generated by $q_{II}$  describes the interaction of the DSW with the rarefaction wave developing  due to the evolution the increasing part of the initial profile $u_0(x)$. This interpretation becomes particularly lucid when the initial profile has the shape of a rectangular barrier \cite{el_generation_2002}. 

 Strictly speaking, one should now show that the obtained
solution is global, i.e.  the mapping $(r_2, r_3) \mapsto (x,t)$  is invertible for all $t$. Partial proof of this invertibility
can be found in \cite{tian_initial_1994}.
Here, instead of giving a full mathematical proof,  we shall show that the obtained
modulation solution implies  a long-time asymptotic
behavior for  $r_j(x,t)$ consistent with the predictions of the IST analysis applicable in the case of localized profile. This agreement is  a strong
indication that the obtained modulation solution is valid for all $t$.

The asymptotic behavior of the modulation solution  can be conveniently studied by representing the hodograph formulae (\ref{W}) in the form
\begin{equation}\label{hod_x_t}
t=\frac{W_3-W_2}{V_3 - V_2} ;  \quad x=V_3(r_2, r_3)t +W_3(r_2, r_3) \, . 
\end{equation}
The asymptotic behavior of $m=r_2/r_3$ for $t \gg 1$, following from
(\ref{hod_x_t}), is then found as \cite{gurevich_evolution_1992}
\begin{equation}\label{m_asym}
m \simeq 1 - e^{-t/T(r_3)} ,
\end{equation}
where 
\begin{equation}\label{T}
T(r_3)=  - \frac{3}{4 \pi} r_3^{-1/2} \int \limits^h_{r_3} \frac{D'(r)}{\sqrt{r-r_3}} \rmd\tau \, .
\end{equation}
Since $D'(r)<0$, one has $T(r_3)>0$, implying $m \to 1$ as $t \to
\infty$ at any fixed point of the $x$-domain of $r_3$. This has a
clear interpretation as the asymptotic transformation of a localized
elevation into a solitary wavetrain, which is consistent with IST
predictions.

For $x$ in (\ref{hod_x_t}) we obtain
\begin{equation}\label{x_soliton}
t \gg 1: \quad x=\frac{2}{3}r_3 t + x_0(r_3) + o(1),
\end{equation}
where
\begin{equation*}
  \begin{split}
    x_0(r_3) = \frac{1}{2 \pi (r_3)^{1/2}}  \int \limits^{h}_{r_3}  \ln
    \left (\frac{r-r_3}{r} \right) \frac{D'(r)}{\sqrt{r_3 - r}} \rmd r  \\ 
    + \frac{1}{2 (r_3)^{1/2}} \int \limits ^{r_3}_{r_0}
    \frac{W_I(r)}{\sqrt{r_3 - r}} \rmd r \, . 
  \end{split}
\end{equation*}
Invoking the traveling wave amplitude $a=2(r_2-r_1)=2r_2$, and using
the asymptotics (\ref{m_asym}), implying $r_2 \to r_3$ as $t \to
\infty$, we identify the leading order behavior in (\ref{x_soliton})
with the universal ``triangular'' asymptotic distribution of the
amplitude $a = 3 {x}/{t}$ in a solitary wavetrain consisting of
non-interacting solitons \cite{whitham_linear_1974}.  The amplitude of
the lead soliton in the train $a_{\max} = 2h$ is determined by the
maximum value $(r_3)_{\max}=h$, and is the same as for the lead
soliton in a DSW resulting from the Riemann problem with an initial
step of magnitude $h$, see Sec.~\ref{sec: dsw-riemann}.  Thus, the
range of amplitudes in the train is $0 < a <2h$, and the spatial
extent of the train is $0<x<\tfrac{2}{3}ht$. The correction $x_0(r_3)$
in (\ref{x_soliton}) is due to weak soliton interactions in the train
evolving from the initial localized elevation of the particular form
$u_0(x)$.

The asymptotic solitary wavetrain is also characterized by a
wavenumber $k$, which can be interpreted as the soliton density
distribution. From (\ref{eq:46}), (\ref{m_asym}), (\ref{T}),
(\ref{x_soliton}) we obtain, with account of the normalization of $x$
and $t$ in (\ref{kdv_eps}), the long-time asymptotic behavior
\begin{equation}\label{karpman0}
k  \simeq \frac{1}{t} F(a), 
\end{equation}
where 
\begin{equation}\label{karpman1}
F(a) = -\frac{1}{\varepsilon} \frac{\sqrt{6}}{4}  \int \limits^h_{a/2} \frac{D'(r)}{\sqrt{r-a/2}} \rmd r.
\end{equation}
From the balance relation $ k dx = f(a)da$,  where $\tfrac{1}{2\pi}f(a)$  is the distribution function of solitons over amplitude \cite{whitham_linear_1974},
we find, on using $a=3x/t$ and comparing with (\ref{karpman0}), that   $3f(a)= F(a)$.

 Each soliton corresponds to a discrete
eigenvalue $\la_n = - \eta_n^2$ of the quantum-mechanical
Schr\"odinger operator with the potential $-u_0(x)/6$, which is the
Lax operator associated with the KdV equation (\ref{kdv_eps}).  By
making the substitution $r=u_0(x)$, $a=2 \eta^2$ in (\ref{karpman1}),
we obtain the Weyl formula for the semi-classical density of the
(transformed) eigenvalues $\eta_n$ on the interval $[0, \sqrt{h}]$ (see, e.g., \cite{lax_generation_1994})
\begin{equation}\label{karpman2}
  \varphi(\eta)  \sim
  \frac{1}{ \pi \varepsilon} \oint  \frac{ \eta
  }{\sqrt{u_0(x) - \eta^2}} \rmd x. 
\end{equation}
In quantum mechanics the distribution (\ref{karpman2}) is a direct consequence of the
famous Bohr-Sommerfeld semi-classical quantization rule
\cite{landau_quantum_1965,whitham_linear_1974}.  In the framework of
soliton theory, it was first used by Karpman
\cite{karpman_asymptotic_1967} (see also \cite{karpman_non-linear_1974}).

The total number of solitons  ``contained'' in
the initial elevation $u_0(x)$ and released as $t \to \infty$ can be estimated by evaluating 
the integral 
\begin{equation}\label{total_N}
    N \simeq \frac{1}{2 \pi} \int \limits ^{\infty}_{-\infty} k \rmd x  = \frac{1}{6\pi}\int \limits^{2h}_0 F(a)\rmd a.   
\end{equation}    
Substituting (\ref{karpman1}) into (\ref{total_N}) we obtain, upon
reversing the order of integration,
   \begin{equation} 
N  
    \sim  \frac{1}{\pi \epsilon}  \int \limits ^{\infty}_{-\infty}
    [u_0(x)]^{1/2} \rmd x, 
\end{equation}
which is the standard quantum-mechanical  result for the total numer of discrete eigenvalues in the semi-classical approximation and hence, the  total number of solitons found via the IST \cite{whitham_linear_1974}.

Thus, remarkably, modulation theory gives the correct analytical
description of the long-time asymptotic solitary wavetrain, even
though formally the underlying modulation description is based on the
asymptotic equivalence of spatial averaging over a large number of
oscillations to averaging over one period
\cite{whitham_non-linear_1965,whitham_linear_1974}, which
is not applicable here as the period of the traveling wave tends to
infinity in the soliton limit. We note that the direct description of
the solitary wavetrain as an esnsemble of non-interacting solitons
readily yields the formula (\ref{karpman0}) \cite{whitham_linear_1974}
but does not provide information about the distribution function
$F(a)$, which is related to the KdV initial condition.  In conclusion,
we note that the modulation description of the evolution of a
localized initial elevation without the restriction $u_0(x>0) \equiv
0$ imposed here on the initial data was obtaibed in
\cite{krylov_evolution_1992} and leads to the same asymptotic formulae
(\ref{x_soliton}), (\ref{karpman0}).  \medskip

{\it Localized depression}

We now consider the KdV evolution of a localized depression
$u_0(x)=u_0^-(x) \le 0$, where we assume $u_0^-(x)=-u_0^+(-x)$, the
function $u_0^+(x)$ being the localized elevation described
above. This problem was studied in the modulation theory framework in
\cite{gurevich_evolution_1992,gurevich_supersonic_1996}.  Similar to
the already considered elevation case, wavebreaking of a localized
depression with initial form $u_0^-(x)$ occurs at the boundary
with a uniform state implying now $r_3=0$ and $-h \le r_1 \le r_2 \le
0$.  Wavebreaking is regularized by the generation of a DSW, and
the corresponding modulation solution for the function $q(r_1,
r_2)$ is constructed in a completely analogous way to the one for the
elevation case.  In fact, it can be obtained from solution
(\ref{sheet_I}), (\ref{sheet_II}) by replacing $r_2 \to -r_2$, $r_3
\to -r_1$, $D(r) \to -D(r)$.  Remarkably, the ``depression'' DSW
exhibits solitons near its leading edge, while globally, there is no
discrete spectrum in the associated scattering problem. At the level
of modulation theory, this is a direct consequence of hyperbolicity
of the modulation equations so that the DSW solution is confined to
the finite region $(x_-(t), x_+(t))$ and is not affected, for $t
=O(1)$, by the behavior of the ``external'' solution at $|x| \to
\infty$.  The intermediate solitonic asymptotics in the
``solitonless'' KdV evolution were studied in
\cite{claeys_solitonic_2010} using the IST.

While the intermediate asymptotic behaviors of initial elevation and
depression evolution reveal a great deal of similarity, the
respective long-time asymptotic solutions are drastically different.
As was shown in \cite{el_evolution_1993}, the long-time asymptotic of
the modulation solution for the initial depression is given by:
\begin{equation}\label{lin_x}
t \gg 1: \quad x= 2 r_1 t  - \frac{1}{2}at + o (t^{1/2}),
\end{equation}
\begin{equation}\label{ampl}
 a(x,t) = 2(r_2-r_1) \simeq \frac{4}{t^{1/2}} \left( \tau  \right)^{1/4} A^{1/2} \left(-\tau \right).  
\end{equation}
Here
\begin{equation*}
  \tau=-{x}/{2t}, \quad A(y) =  \frac{2}{\pi} \int \limits ^{x_2}
  _{x_1}[y-u_0^-(x)]^{1/2} \rmd x, 
\end{equation*}
where $x_1(y) < x_2(y)$ are the roots of the equation $u_0^-(x) = y$.
Equation (\ref{ampl}) describes the asymptotic behavior of the
modulated wave's amplitude $a$. One can see that $a \to 0$ as $t \to
\infty$. Since $m=-a/2r_1$, formula (\ref{ampl}) also implies $m \to
0$ as $t \to \infty$, i.e., nonlinear modulation theory prescribes the
asymptotic transformation of the initial depression into a modulated
linear wave packet in full agreement with results from IST
\cite{ablowitz_solitons_1981}.  We note that the function $A(y)$ is
related to the logarithm of the transmission coefficient for the
potential barrier $-u_0^-(x)$.  The behavior $a \sim t^{-1/2}$
prescribed by (\ref{ampl}) is consistent with linear wavepacket energy
conservation and can be obtained as an exact solution of the
modulation equations for linear dispersive waves
\cite{whitham_linear_1974}. Linear modulation theory, however, does
not provide a coefficient of $t^{-1/2}$ in (\ref{ampl}), which is
related to the KdV initial conditions.

The wave number distribution in the wave packet is found to be  
\begin{equation}\label{lin_k}
\varepsilon^2 k^2 = - \frac{x}{3t} - \frac{1}{6} a(x,t) + O(1/t), \quad x<0 \, .
\end{equation}
The first term in the expansion (\ref{lin_k}) corresponds to the
motion of the linear wave packet to the left with the group velocity
$\omega'(k) = - 3 \varepsilon k^2 = x/t$, again in full agreement with
linear modulation theory.

We note that the evolution of an initial depression on shallow water
leading to the generation of a DSW has been observed in water tank
experiments \cite{hammack_korteweg-vries_1978,trillo_observation_2016}
that showed very good agreement with the KdV predicted dynamics.  Note
that in \cite{trillo_observation_2016}, an unconventional ``T-KdV''
version of the KdV equation was used where the roles of time and space
were swapped.

%
\smallskip
{\it Dispersive evolution of $N$-wave}

\smallskip
\begin{figure}
  \centering
  \includegraphics[scale=0.125] {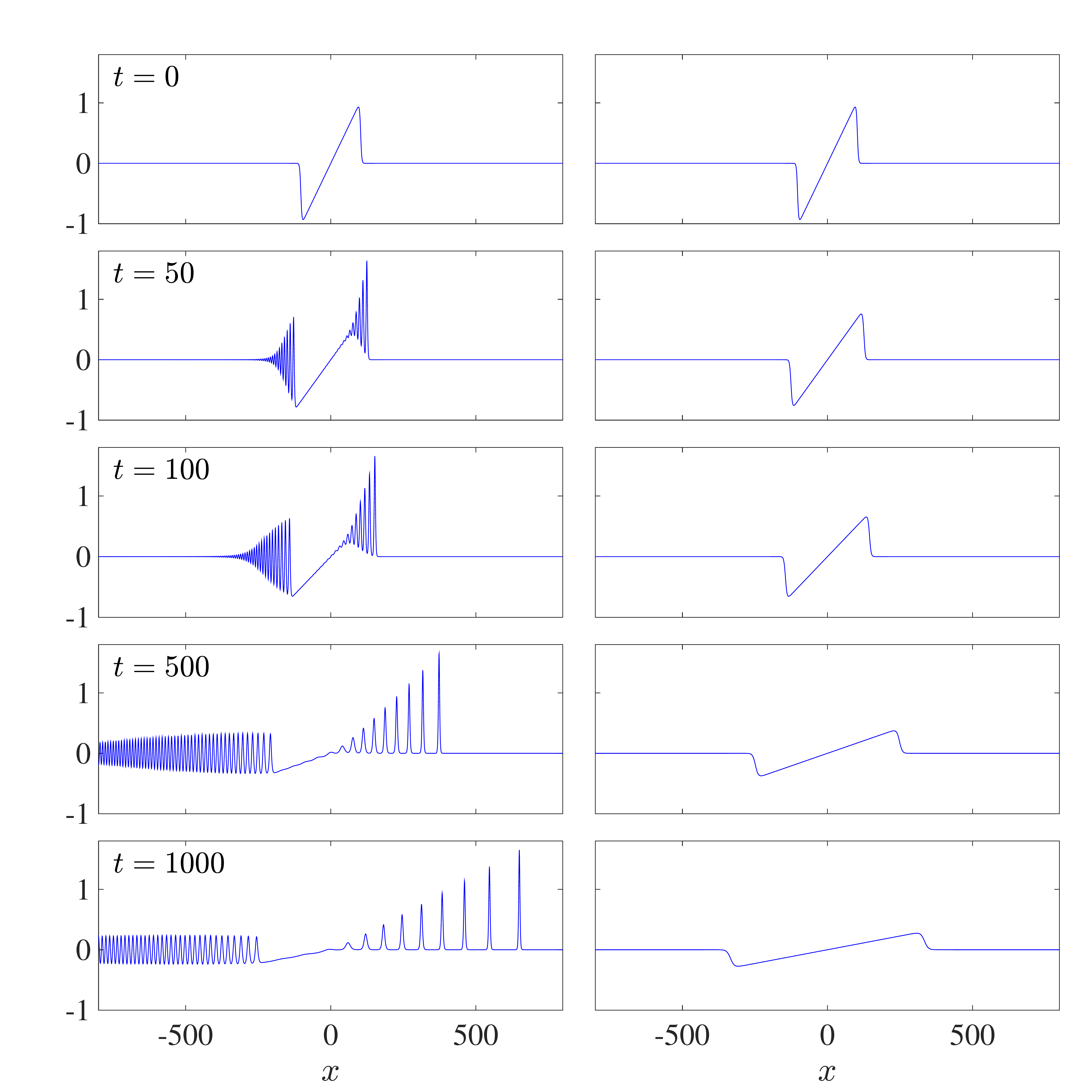}
  \caption{$N$-wave evolution in dispersive (KdV, left) and
    dissipative (Burger's, right) hydrodynamics.}
  \label{fig:N-wave}
\end{figure}
Finally, we consider the KdV evolution of an $N$-wave. The $N$ wave
represents the universal waveform occuring in the asymptotic evolution
of bipolar profiles in viscous fluid dynamics, described, e.g., by
Burger's equation \cite{whitham_linear_1974}. Such density profiles
leading to $N$ waves naturally occur in supersonic flows around
finite-length objects, e.g., an airfoil. The classical $N$-wave
represents a combination of two viscous shocks of opposite polarity
propagating in opposite directions into the same equilibrium state,
and connected by a rarefaction wave (see the initial state in
Fig.~\ref{fig:N-wave}). Importantly, the two shock waves in the
viscous $N$-wave are of the same nature and differ only in polarity,
both yielding similar decay with time.  See the viscous Burger's
equation evolution of the $N$-wave in Fig.~\ref{fig:N-wave}, right. The
evolution of the $N$-wave in dispersive hydrodynamics is drastically
different, yielding two contrasting types of long-time behavior shown
in Fig.~\ref{fig:N-wave}, left. The elevation part of the initial
profile evolves into a solitary wavetrain described by the asymptotic
formulae (\ref{x_soliton}), (\ref{karpman0}), while the depression
part evolves into a linear wave packet whose amplitude and frequency
modulation is described by (\ref{ampl}), (\ref{lin_k}).  However, the
intermediate asymptotic state of the dispersive-hydrodynamic evolution
of the $N$ wave consists of two DSWs connected by a rarefaction wave
as shown in Fig.~\ref{fig:N-wave}, left, $t=50, 100$.  This is a
dispersive counterpart of viscous $N$-wave evolution.

\subsection{Defocusing NLS} 
\label{sec:defocusing-nls}

Wavebreaking in the KdV equation represents the canonical example of
DSW formation for integrable, uni-directional dispersive hydrodynamics.
For the bi-directional case, we turn to the integrable, defocusing NLS
equation \eqref{eq:NLS} with $\sigma = 1$.  The NLS equation written
in dispersive hydrodynamic form \eqref{eq:NLS_disp_hydro} is an
example of a dispersive regularization of the Euler or shallow water
equations for the density $\rho$ and velocity $u$.  Applying the
approach of GP for KdV, Gurevich and Krylov constructed NLS DSWs in
\cite{gurevich_dissipationless_1987}.  The full Riemann problem was
later classified in \cite{el_decay_1995}.  It turns out that NLS DSWs
exhibit more subtle physical features than KdV DSWs, such as
cavitation or points of zero density.  This leads us to consider,
after reviewing the Riemann problem solution, an application of DSW
theory to the NLS piston problem, studied in \cite{hoefer_piston_2008}.

\subsubsection{Rarefactions}
\label{sec:rarefactions}

We seek the left $(\rho_{\rm l},u_{\rm l})$ and right $(\rho_{\rm
  r},u_{\rm r})$ states that can be connected by either a
1-rarefaction or 2-rarefaction solution of the dispersionless
hydrodynamic equations \eqref{eq:sw}.  These equations are equivalent
to the well-studied shallow water equations (e.g.,
\cite{leveque_finite_2002}).  Because this strictly hyperbolic system
admits the Riemann invariants $r_\pm$ \eqref{eq20}, rarefactions can
be characterized by the constancy of one of $r_\pm$.  A 1-rarefaction
requires $r_+$ be constant across the wave.  In particular, $r_+ =
\frac{1}{2} u_{\rm l} + \sqrt{\rho_{\rm l}} = \frac{1}{2} u_{\rm r} +
\sqrt{\rho_{\rm r}} = $ constant.  This relation and the admissibility
requirement that the left rarefaction edge move slower than the right
edge, $V_-|_{\text{left}} < V_-|_{\text{right}}$, give the 1-wave
curve
\begin{equation}
  \label{eq:51}
  \text{1-wave curve:}\quad u_{\rm l} - u_{\rm r} = -2( \sqrt{\rho_{\rm l}} -
  \sqrt{\rho_{\rm r}}), \quad \rho_{\rm r} < \rho_{\rm l} .
\end{equation}
The wave speeds are
\begin{equation*}
  \begin{split}
    V_-|_{\text{left}} &= u_{\rm l} - \sqrt{\rho_{\rm l}} \\
    V_-|_{\text{right}} &= u_{\rm r} - \sqrt{\rho_{\rm r}} .
  \end{split}
\end{equation*}
Similarly, the constancy of $r_-$ and admissibility
$V_+|_{\text{left}} < V_+|_{\text{right}}$ give the 2-wave curve
\begin{equation}
  \label{eq:103}
  \text{2-wave curve:}\quad u_{\rm l} - u_{\rm r} = 2( \sqrt{\rho_{\rm l}} -
  \sqrt{\rho_{\rm r}}) , \quad \rho_{\rm l} < \rho_{\rm r},
\end{equation}
with edge speeds
\begin{equation*}
  \begin{split}
    V_+|_{\text{left}} &= u_{\rm l} + \sqrt{\rho_{\rm l}}\\
    V_+|_{\text{right}} &= u_{\rm r} + \sqrt{\rho_{\rm r}} .
  \end{split}
\end{equation*}

The spatial structure of a centered 1-rarefaction can be determined by
seeking a self-similar solution in the form $r_- = r_-(x/t)$.  Then
the rarefaction is determined by $V_-(u,\rho) = x/t$ and $r_+ =
\frac{1}{2}u + \sqrt{\rho} = \frac{1}{2} u_{\rm l} + \sqrt{\rho_{\rm l}}$.  A
similar construction holds for the 2-rarefaction.  See
\cite{leveque_finite_2002} for details.

Note that if $\rho_{\rm l}$ or $\rho_{\rm r}$ is zero, i.e., one side
is vacuum, the solution to the Riemann problem is always a rarefaction
wave.

Rarefaction waves are weak solutions of the dispersionless NLS
equations that exhibit weak discontinuities.  Similar to the
uni-directional KdV case, the inclusion of dispersion serves to
regularize the weak discontinuities with decaying, linear dispersive
waves.

\subsubsection{DSWs}
\label{sec:dsws}

The main difference between the KdV and NLS DSW constructions is the
bi-directionality of the latter.  This manifests as a system of two
hyperbolic equations in the dispersionless limit \eqref{er} rather
than one for KdV \eqref{eq:37}.  Also, because the dispersive
hydrodynamic form of NLS \eqref{eq:NLS_disp_hydro} is a fourth order
system of equations, NLS modulations are described by a system of four
hyperbolic equations \eqref{eq18} as opposed to three for KdV
\eqref{eq:31}.  These differences lead to the possibility of two NLS
DSW families we label as 1-DSW and 2-DSW, depending upon which
characteristic family they are associated.  Otherwise, the NLS DSW
construction closely parallels the KdV DSW construction.

Motivated by our understanding of the KdV DSW, we seek a self-similar,
simple wave solution of the NLS Whitham equations \eqref{eq18} that
connects the soliton ($k = 0$) and harmonic ($a = 0$) edges for the
left $(\rho_{\rm l},u_{\rm l})$ and right $(\rho_{\rm r},u_{\rm r})$
states.  From Sec.~\ref{sec:NLS_Whitham_prop}, we observe that $a \to
0$ when either $r_2 \to r_1$ or $r_3 \to r_4$ and $k \to 0$ when $r_2
\to r_3$.  This implies that we can connect, from left to right, the
harmonic edge to the soliton edge via the 2-wave curve or the soliton
edge to the harmonic edge via the 3-wave curve of the Whitham system.
There are two distinct wave curves giving rise to DSW modulation
solutions.  We identify the slower 2-rarefaction modulation solution
with a 1-DSW and the faster 3-rarefaction modulation with a 2-DSW.

First we construct the 1-DSW.  The 2-wave curve of the Whitham
equations requires $r_j = $ constant for $j = 1,3,4$.  The GP boundary
conditions are obtained as follows.  At the left, harmonic edge where
$r_2 \to r_1$, the Whitham Riemann invariants limit to the
dispersionless Riemann invariants $r_\pm$ according to $r_3 \to r_-$
and $r_4 \to r_+$, by virtue of the ordering $r_- < r_+$, $r_3 < r_4$.
On the right, soliton edge where $r_2 \to r_3$, we still have $r_4 \to
r_+$ but now $r_1 \to r_-$.  These observations imply
\begin{equation}
  \label{eq:55}
  \begin{split}
    r_1 &= r_-|_{\text{right}}, \quad r_3 = r_-|_{\text{left}}, \\
    r_4 &= r_+|_{\text{left}} = r_+|_{\text{right}} .
  \end{split}
\end{equation}
Thus, $r_+$ is constant across the entire 1-DSW.  Utilizing
\eqref{eq:55}, the admissibility constraint for the existence of a
2-rarefaction solution of the Whitham equations $V_2|_{\text{left}} <
V_2|_{\text{right}}$ can be written
\begin{equation*}
  V_2|_{\text{left}} - V_2|_{\text{right}} = \frac{(\sqrt{\rho_{\rm l}} -
    6\sqrt{\rho_{\rm r}})(\sqrt{\rho_{\rm l}} - \sqrt{\rho_{\rm r}})}{\sqrt{\rho_{\rm l}} -
    2\sqrt{\rho_{\rm r}}} < 0 .
\end{equation*}
A sufficient condition is $\rho_{\rm l} < \rho_{\rm r}$ 
Utilizing the constancy of $r_+$ and the admissibility constraint, we
are now able to state the locus of states that can be connected by a
1-DSW
\begin{equation}
  \label{eq:54}
  \text{1-DSW locus:}\quad u_{\rm l} - u_{\rm r} = -2(\sqrt{\rho_{\rm l}} -
  \sqrt{\rho_{\rm r}}), \quad \rho_{\rm l} < \rho_{\rm r} .
\end{equation}
This is the dispersive analogue of the classical Hugoniot shock locus.
The spatial structure of the modulation is a centered 2-rarefaction
solution of the Whitham equations determined implicitly according to
\begin{equation*}
  V_2\left (r_1,r_2(s),r_3, r_4 \right ) = s = \frac{x}{t}.
\end{equation*}
The macroscopic properties of a 1-DSW can now be determined by
evaluating the characteristic speeds of the 2-rarefaction solution to
the Whitham equations at the left and right edges, yielding the speeds
at the left and right edges.  Additionally, evaluating the wave
amplitude \eqref{amp} and modulation wavenumber $k = 2\pi/L$
\eqref{eq017}, we obtain
\begin{equation}
  \label{eq:61}
  \begin{split}
    s_l &= V_2(r_1,r_1,r_3,r_4) = u_{\rm r} + \frac{\rho_{\rm l} +
      2\sqrt{\rho_{\rm r}\rho_{\rm l}}- 4\rho_{\rm r}}{2\sqrt{\rho_{\rm r}}-\sqrt{\rho_{\rm l}}}, \\
    s_r &= V_2(r_1,r_3,r_3,r_4) = u_{\rm r} + \sqrt{\rho_{\rm r}} - 2
    \sqrt{\rho_{\rm l}} , \\
    a_l &= 0, \quad a_r = 4\sqrt{\rho_{\rm l}}(\sqrt{\rho_{\rm r}} -
    \sqrt{\rho_{\rm l}}), \\
    k_l &= 4\sqrt{\rho_{\rm r} - \sqrt{\rho_{\rm r} \rho_{\rm l}}} , \quad k_r = 0 .
  \end{split}
\end{equation}
Because NLS admits dark solitons, the 1-DSW has polarity $p = -1$ in
the density $\rho$.  The orientation of the 1-DSW is $d = 1$.

An analogous calculation for the 2-DSW yields the constant Riemann invariants
\begin{equation}
  \label{eq:59}
  \begin{split}
    r_1 &= r_-|_{\text{left}} = r_-|_{\text{right}}, \\
    r_2 &= r_+|_{\text{right}}, \quad r_4 = r_+|_{\text{left}},
  \end{split}
\end{equation}
and the 3-rarefaction solution
\begin{equation}
  \label{eq:58}
  V_3(r_1,r_2,r_3(s),r_4) = s = \frac{x}{t} .
\end{equation}
The constancy of $r_1$ across the entire 2-DSW and the admissibility
criterion $V_3|_{\text{left}} < V_3|_{\text{right}}$ yield the locus
of left and right states that can be connected by a 2-DSW
\begin{equation}
  \label{eq:57}
  \text{2-DSW locus:}\quad u_{\rm l} - u_{\rm r} = 2(\sqrt{\rho_{\rm l}} -
  \sqrt{\rho_{\rm r}}), \quad \rho_{\rm r} < \rho_{\rm l} .
\end{equation}
The macroscopic 2-DSW properties are
\begin{equation}
  \label{eq:64}
  \begin{split}
    s_l &= V_3(r_1,r_2,r_2,r_4) = u_{\rm r} + \sqrt{\rho_{\rm l}} \\
    s_r &= V_3(r1,r_2,r_4,r_4) = u_{\rm r} + \frac{8\rho_{\rm l} - 8\sqrt{\rho_{\rm l}
        \rho_{\rm r}} + \rho_{\rm r}}{2\sqrt{\rho_{\rm l}} - \sqrt{\rho_{\rm r}}} \\
    a_l &= 4\sqrt{\rho_{\rm r}}(\sqrt{\rho_{\rm l}} - \sqrt{\rho_{\rm r}}), \quad a_r =
    0 , \\
    k_l &= 0, \quad k_r = 4\sqrt{\rho_{\rm l} - \sqrt{\rho_{\rm l} \rho_{\rm r}}} .
  \end{split}
\end{equation}
While the polarity of the 2-DSW is the same as that for the 1-DSW $p =
-1$, the orientation is the opposite $d = -1$.

Comparison of the DSW loci with eqs.~\eqref{eq:51} and \eqref{eq:103}
show that 1- and 2-DSW loci are \textit{inadmissible 1- and 2-wave
  curves} of the dispersionless equations \eqref{eq:sw}, respectively.
This seemingly peculiar feature of DSWs also occurs in classical shock
theory for a restricted class of hyperbolic equations when the
Hugoniot loci and wave curves coincide \cite{temple_systems_1983}.

\begin{figure}
  \centering
  \includegraphics[scale=0.3333]{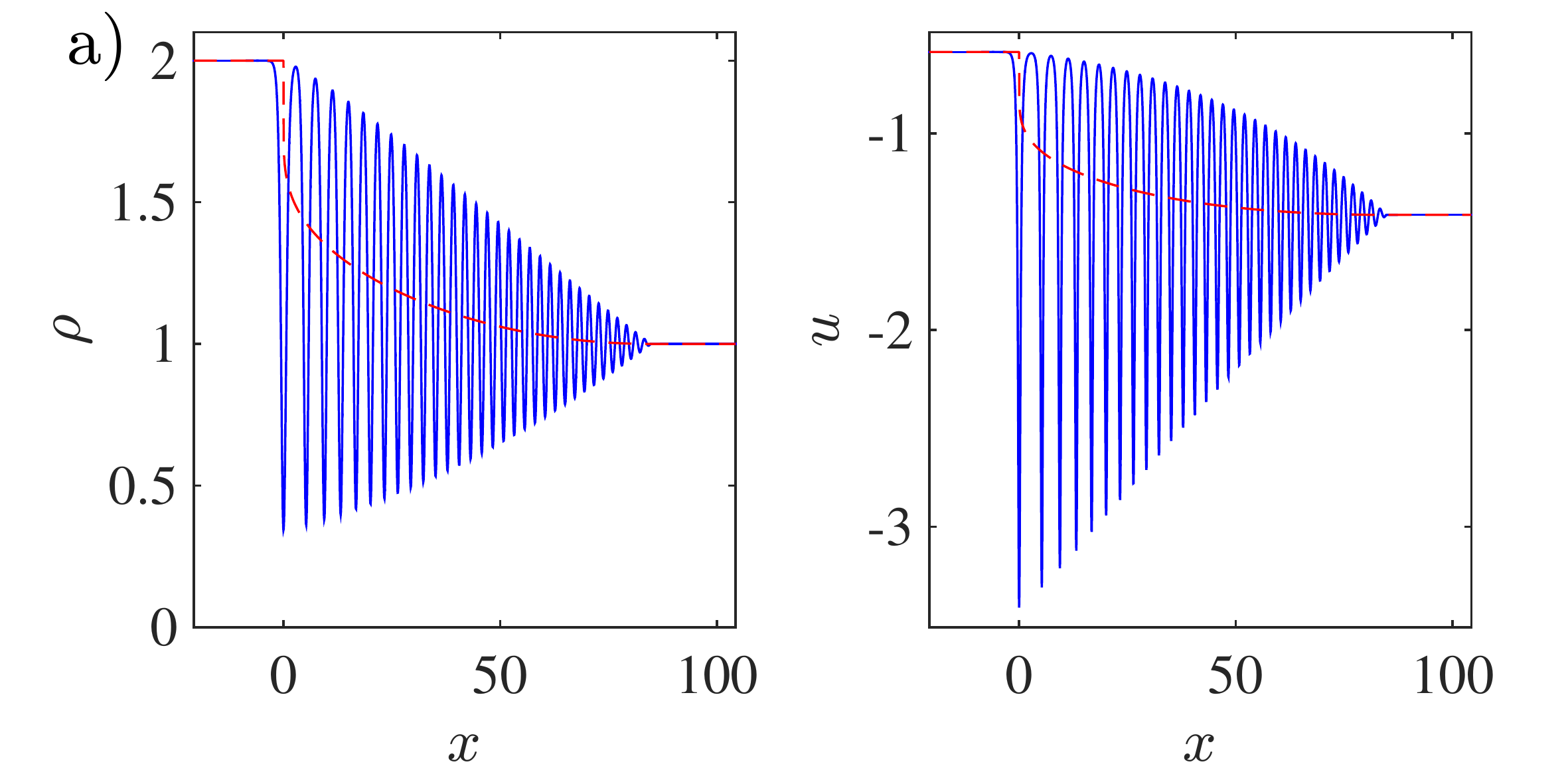} \\
  \includegraphics[scale=0.3333]{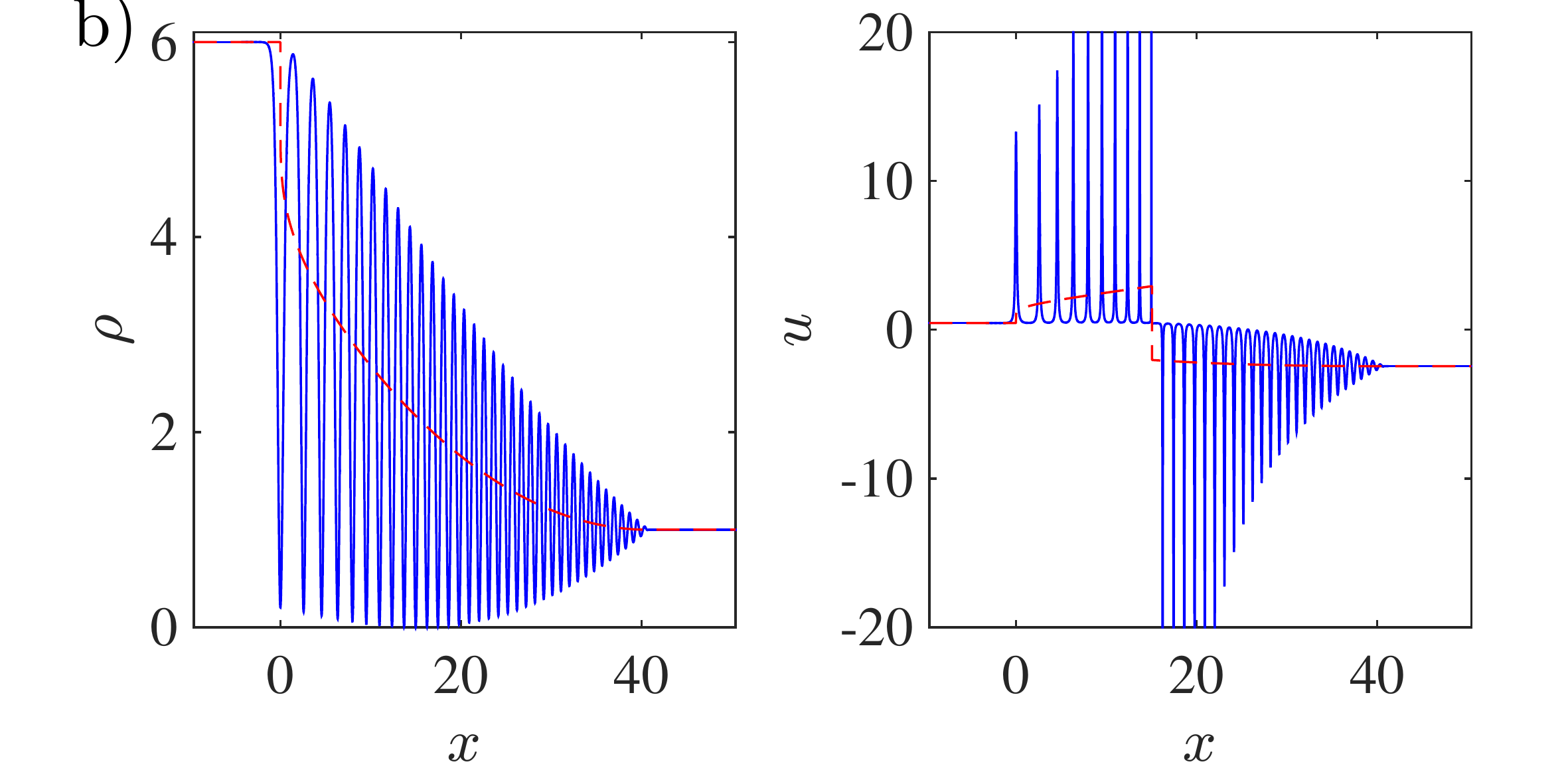}
  \caption{NLS 2-DSWs (solid) and their mean values (dashed) in the
    soliton reference frame with (a) $\rho_0 = 2$ at $t = 50$ and (b)
    $\rho_0 = 6$ at $t = 8$ with an interior vacuum point.}
  \label{fig:nls_2dsw}
\end{figure}
NLS DSWs exhibit curious physical behavior.  For example, the density
at the soliton edge goes to zero (vacuum) for sufficiently large
jumps.  The minimum density of the traveling wave solution
\eqref{eq013} for the 2-DSW modulation \eqref{eq:59}, \eqref{eq:58} is
\begin{equation}
  \label{eq:71}
  \rho_{\text{min}} = \frac{1}{4} \left ( \frac{1}{2}u_{\rm l} +
    \sqrt{\rho_{\rm l}} - 
  2\sqrt{\rho_{\rm r}} - r_3 \right )^2 .
\end{equation}
When evaluated at the soliton edge $r_3 \to r_2 = u_{\rm r} +
2\sqrt{\rho_{\rm r}}$, $\rho_{\text{min}} = (\sqrt{\rho_{\rm l}} - 2
\sqrt{\rho_{\rm r}})^2$, which is zero when $\rho_{\rm l} = 4\rho_{\rm
  r}$.  The existence of a vacuum or cavitation point persists for
larger density jumps and its location is embedded in the interior of
the wave \cite{el_general_1995}.  Associated with the development of a
vacuum point is a change in the direction of fluid flow relative to
the soliton edge.  We can see this by choosing a reference frame in
which the soliton edge is stationary, e.g., $u_{\rm r} = -
\sqrt{\rho_{\rm l}} < 0$ for the 2-DSW.  We can define a Mach number
for the flow by forming the ratio of the local flow speed to the
dispersionless speed of sound
\begin{equation*}
  M = \frac{|u|}{\sqrt{\rho}} .
\end{equation*}
In the 2-DSW soliton reference frame, the downstream flow $u_{\rm r}$ is
supersonic $M_{\rm r} = \sqrt{\rho_{\rm l}}/\sqrt{\rho_{\rm r}} > 1$.  The upstream flow
$u_{\rm l} = \sqrt{\rho_{\rm l}} - 2\sqrt{\rho_{\rm r}}$ is subsonic $M_l = |1 -
2\sqrt{\rho_{\rm r}/\rho_{\rm l}}| < 1$, comparable to classical shock wave flows.
However, the upstream flow \textit{changes direction} when $\rho_{\rm l} = 4
\rho_{\rm r}$, precisely when a vacuum point forms.  For $\rho_{\rm l} > 4
\rho_{\rm r}$, the fluid is flowing into the DSW from both the downstream
and upstream directions.  Example 2-DSWs with and without a vacuum
point are shown in Fig.~\ref{fig:nls_2dsw}.  The local fluid speed
gets very large in the vicinity of a vacuum point and is undefined at
the vacuum point.  However, the fluid momentum $\rho u$ is defined at
the vacuum point \cite{hoefer_dispersive_2006}.

\subsubsection{Riemann problem}
\label{sec:riemann-poblem}

Now that we have constructed the two basic wave types, rarefactions
and DSWs, we can consider the general Riemann initial conditions
\begin{equation}
  \label{eq:50}
  \rho(x,0) =
  \begin{cases}
    \rho_0 & x < 0 \\
    1 & x > 0
  \end{cases}, \quad u(x,0) =
  \begin{cases}
    u_0 & x < 0 \\
    0 & x > 0
  \end{cases}.
\end{equation}
This is a two-parameter family of initial conditions $\rho_0 > 1$,
$u_0 \in \R$.  The results for more general left and right states can
be found by utilizing the Galilean $u \to u + v$, $x \to x - vt$,
scaling $\rho \to A \rho$, $u \to A^{3/2} u$, $x \to A x$, $t \to
A^{-1/2} t$, and reflection $x \to -x$, $u \to -u$ symmetries of the
dispersive hydrodynamic form for the NLS equation
\eqref{eq:NLS_disp_hydro} with $v \in \R$, $A > 0$.

Due to its bi-directionality, the NLS Riemann problem will generically
result in the generation of two waves, a 1-wave and a 2-wave, each of
which is either a rarefaction or DSW with an intermediate, constant
state connecting them \cite{el_decay_1995}.  A peculiar feature of
dispersive hydrodynamic systems is that the constancy of the
intermediate state is required of the dispersionless \textit{or}
Whitham modulation equations.  This means that in the latter case, the
intermediate state can be oscillatory for the NLS solution itself.

\begin{figure}
  \centering
  \includegraphics{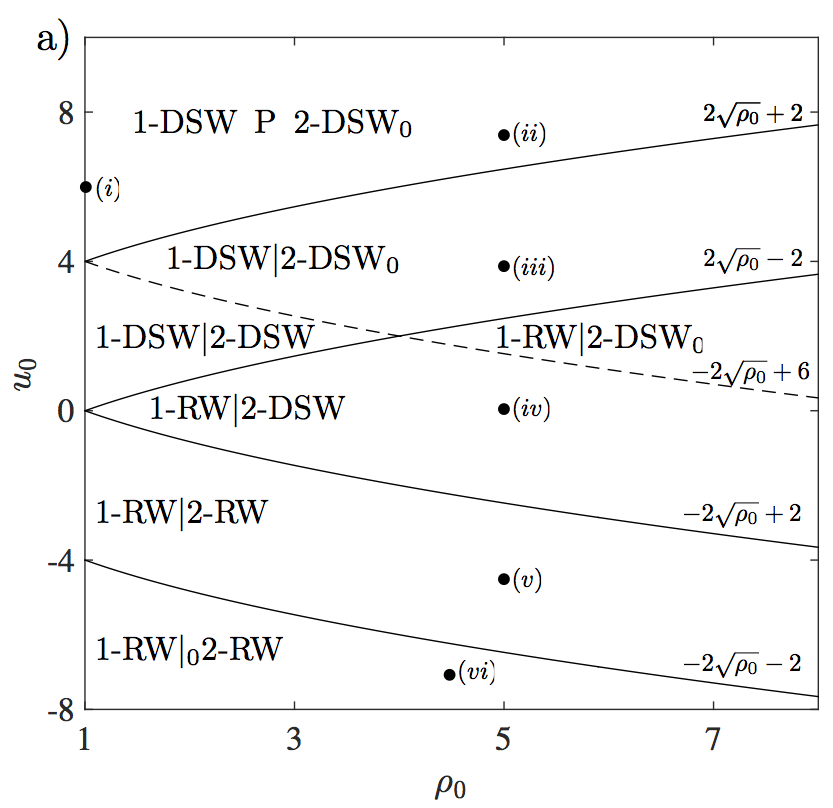}\\
  \includegraphics{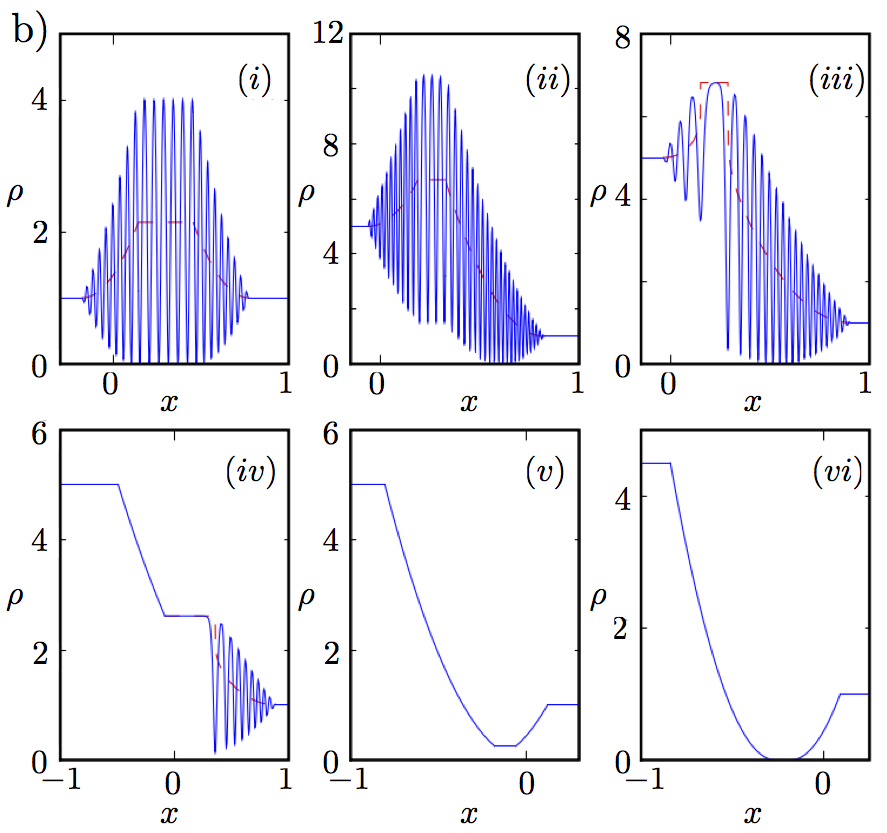}
  \caption{a) Classification of the NLS Riemann problem \eqref{eq:50}.
    (RW) - rarefaction wave; ($|$) - intermediate constant state; (P)
    - intermediate periodic state; zero subscript implies vacuum
    point. b) Modulated density (solid) and mean (dashed) for Riemann
    problems $(i-vi)$ identified in (a).}
  \label{fig:nls_riemann_classification}
\end{figure}
\begin{figure}
  \centering
  \includegraphics{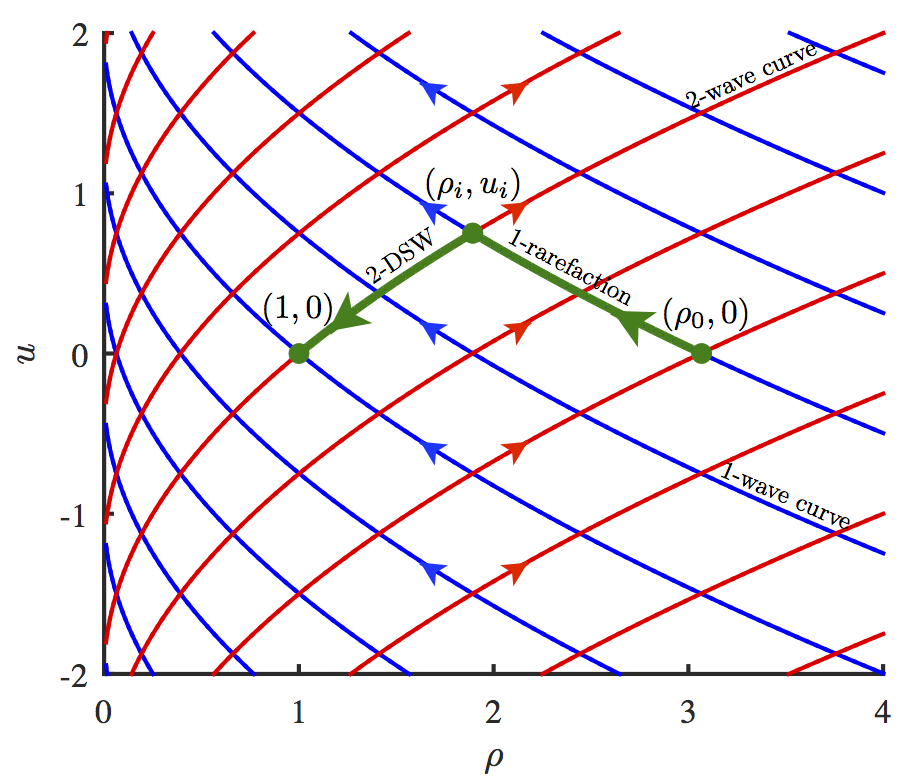}
  \caption{Traversal of 1- and 2-wave curves in solving the shock tube problem.}
  \label{fig:wave_dsw_loci}
\end{figure}

To determine the intermediate state $(\rho_{\rm i},u_{\rm i})$ connecting the 1-
and 2-waves, we simply equate the 1-wave curve/1-DSW locus
\eqref{eq:51} and \eqref{eq:54} to the
2-wave curve/2-DSW locus \eqref{eq:103} and \eqref{eq:57} according to
\begin{equation}
  \label{eq:72}
  \underbrace{u_0 + 2(\sqrt{\rho_0} - \sqrt{\rho_{\rm i}})}_{\text{1-wave}}
  = u_{\rm i} = \underbrace{2(\sqrt{\rho_{\rm i}} - 1)}_{\text{2-wave}} .
\end{equation}
Solving for the intermediate state, we obtain
\begin{equation}
  \label{eq:73}
  \begin{split}
    \rho_{\rm i} &= \frac{1}{16} \left ( u_0 + 2\sqrt{\rho_0} + 2 \right )^2,
    \\
    u_{\rm i} &= \frac{1}{2} u_0 + \sqrt{\rho_0} - 1 .
  \end{split}
\end{equation}
Comparing the values of $\rho_0$ with $\rho_{\rm i}$ and $\rho_{\rm
  i}$ with 1 determines the type of 1- and 2-wave, respectively.  For
example, the dam break \cite{leveque_finite_2002} or, equivalently,
the shock tube \cite{liepmann_elements_1957} problem where $u_0 = 0$
and $\rho_0 > 1$ results in
\begin{equation*}
  \rho_0 > \rho_{\rm i} = \frac{1}{4}(\sqrt{\rho_0} + 1)^2 > 1 ,
\end{equation*}
implying a 1-rarefaction connected to a 2-DSW shown in
Fig.~\ref{fig:nls_riemann_classification}b$iv$.  The construction of
this solution is depicted geometrically in
Fig.~\ref{fig:wave_dsw_loci}.  Blue (red) curves oriented to the left
(right) are 1-wave (2-wave) curves.  Since the rarefaction wave curves
and DSW loci coincide, it is the direction that they are traversed
that determines the wave type.  The arrows point in the direction of
increasing dispersionless characteristic speed for each wave family.
Tracing a wave curve in the direction of increasing characteristic
speed corresponds to an admissible rarefaction wave.  The decreasing
characteristic speed direction corresponds to an admissible DSW.  We
identify the solution to the Riemann problem by tracing an admissible
1-rarefaction wave curve from the left state $(\rho_0,0)$ to the
intermediate, constant state $(\rho_{\rm i},u_{\rm i})$ determined by
the intersection of the 1- and 2-wave curves \eqref{eq:72}.  Then, we
follow the 2-DSW locus to the right state $(1,0)$.  For more details,
see \cite{hoefer_shock_2014}.


The classification of solutions to the NLS Riemann problem depends
upon the relative ordering of the four values
$r_\pm|_{\text{left,right}}$ and was carried out in
\cite{el_decay_1995}.  There are five ways in which these values can
be ordered, leading to five qualitatively distinct types of solutions
shown in Fig.~\ref{fig:nls_riemann_classification}a,b.  For sufficiently
negative initial velocities, $u_0 < -2\sqrt{\rho_0} + 2$, there are no
DSWs generated (Figs.~\ref{fig:nls_riemann_classification}b$v$,$vi$).
The intermediate state \eqref{eq:73} exceeds the threshold for DSW
cavitation, $\rho_{\rm i} > 4$, when $u_0 > -2\sqrt{\rho_0} + 6$.  For
sufficiently large initial velocities $u_0 > 2\sqrt{\rho_0} -2$, there
are no rarefactions, only DSWs.  In this regime, the 1-DSW and 2-DSW
soliton edges propagate with velocities $s_1 = u_{\rm i} + \sqrt{\rho_{\rm i}} -
2\sqrt{\rho_0}$ (eq.~\eqref{eq:61}) and $s_2 = \sqrt{\rho_{\rm i}}$
(eq.~\eqref{eq:64}), respectively.  These velocities coincide when
$u_0 = 2\sqrt{\rho_0} + 2$.  When the initial velocity exceeds this
value, the intermediate state is a non-modulated, nonlinear periodic
wave (Fig.~\ref{fig:nls_riemann_classification}b$ii$).  The case
$u_0 > 4$, $\rho_0 = 1$ leads to two counterpropagating DSWs connected
by a standing periodic wave as shown in
Fig.~\ref{fig:nls_riemann_classification}b$i$.

\subsubsection{Piston problem}
\label{sec:piston-problem} 

Just as the shock tube problem is a fundamental, textbook problem of
classical gas dynamics, the piston problem has similar importance.
Here we consider the analogous piston problem for NLS dispersive
hydrodynamics.  The piston problem was studied in
\cite{hoefer_piston_2008,kamchatnov_flow_2010} with application to
Bose-Einstein condensates where the action of a piston can be achieved by
a repulsive laser modeled by a linear potential added to the NLS
equation \eqref{eq:NLS}.  This results in the Gross-Pitaevskii
equation \cite{gross__1961,pitaevskii__1961}
\begin{equation}
  \label{eq:75}
  i \psi_t + \frac{1}{2} \psi_{xx} + V(x,t)\psi - |\psi|^2 \psi = 0 .
\end{equation}
The piston problem also arises in the hypersonic regime of steady,
two-dimensional dispersive hydrodynamic flows past an obstacle
\cite{el_two-dimensional_2009}.

By assuming that the linear potential in the Gross-Pitaevskii equation
\eqref{eq:75} acts as a sharp barrier, a piston, with position $x =
p(t)$, through which fluid cannot pass, it's effect can be
encapsulated by the boundary condition
\begin{equation}
  \label{eq:76}
  u(p(t),t) = \dot{p}(t) .
\end{equation}
The local fluid velocity moves with the piston velocity.  We follow
\cite{hoefer_piston_2008} where an impulsively accelerated piston is
analyzed, i.e., we assume
\begin{equation*}
  p(t) =
  \begin{cases}
    0 & t < 0 \\
    v_p t & t > 0
  \end{cases} ,
\end{equation*}
where $v_p > 0$ is the piston velocity.  For a retracted piston $v_p <
0$, the solution is a rarefaction wave, the same as in classical gas
dynamics \cite{courant_supersonic_1948}.  The effect of a gradually
accelerated piston was described in \cite{kamchatnov_flow_2010}.

\begin{figure}
  \centering
  \includegraphics{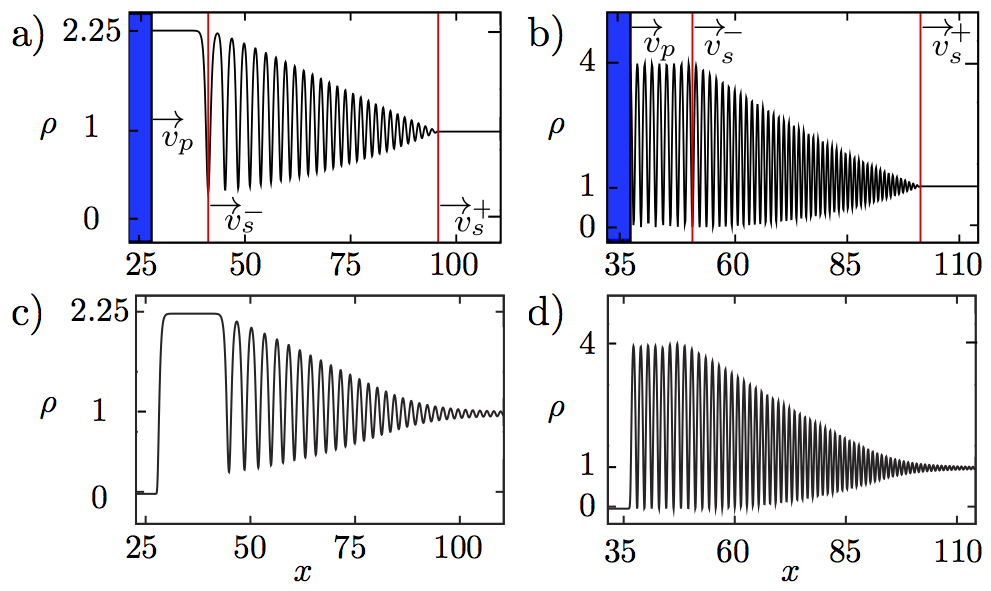}
  \caption{Modulation solutions (a,b) and numerical solutions (c,d) of
    the piston problem with $v_p = 1$ (a,c) at $t = 27$ and $v_p =
    2.5$ (b,d) at $t = 15$.  Figure adapted with permission from
    \cite{hoefer_piston_2008}, copyrighted by the American Physical
    Society.}
  \label{fig:piston_dsw}
\end{figure}

The initial fluid profile is assumed quiescent, with normalized
density
\begin{equation*}
  \rho(x,0^-) = 1, \quad u(x,0^-) = 0, \quad x > 0 .
\end{equation*}
Although we have only one boundary condition at the piston
\eqref{eq:76}, we can formulate a uniquely solvable problem by
assuming that only right-going waves in the second dispersionless
characteristic field $V_+$, eq.~\eqref{V}, are generated, i.e., that
the solution consists of a simple wave.  As remarked earlier in
Sec.~\ref{sec:cubic_breaking}, wave propagation of a quasi-linear
hyperbolic system into a constant state or quiescent ``gas'' must
occur via a simple wave \cite{courant_supersonic_1948}.  Thus, we seek
a 2-DSW that connects the known fluid velocity $u = v_p$ and some
unknown density $\rho = \rho_p$ at the piston to the quiescent state
$\rho = 1$, $u = 0$.  The 2-DSW locus \eqref{eq:57} yields the density
\begin{equation*}
  \rho_p = \left ( \frac{1}{2} v_p + 1 \right )^2.
\end{equation*}
The velocity of the 2-DSW soliton edge $v_s^-$ and harmonic edge
$v_s^+$ are therefore \eqref{eq:64}
\begin{equation*}
  v_s^- = \frac{1}{2} v_p + 1, \quad v_s^+ = \frac{2 v_p^2 + 4 v_p +
    1}{v_p + 1} .
\end{equation*}
This modulation solution and corresponding numerical solution are
shown in Fig.~\ref{fig:piston_dsw}a,c.  The main quantitative
difference between the modulation and numerical solutions is an
overall phase shift that we have not captured in this leading order
description.  The slowly varying phase shift could be recovered as
described for KdV modulations in Sec.~\ref{gen_sol_kdv-whitham}.

The behavior observed is analogous to a piston compressing a classical
gas where a shock wave is generated ahead of the piston.  However, the
DSW velocity exceeds the piston velocity only until $v_p = 2$,
precisely when a vacuum point appears in the DSW \eqref{eq:71}.  Now,
the oscillatory region is in contact with the piston.  The boundary
condition \eqref{eq:76} neglected the effects of dispersion.  We
introduce a modification to the 2-DSW solution by imposing an
alternate boundary condition to \eqref{eq:76}, namely that the
periodic traveling wave velocity $V$ \eqref{eq016} equal the piston
velocity
\begin{equation}
  \label{eq:81}
  V(p(t),t) = \frac{1}{2}(r_1 + r_2 + r_3 + r_4) = v_p .
\end{equation}
We continue to utilize the 2-DSW modulation \eqref{eq:58} but do not
require it to terminate at the soliton edge, i.e., $r_3 > r_2$.
Inserting the values of the known Riemann invariants \eqref{eq:59}
into the boundary condition \eqref{eq:81}, we obtain 
\begin{equation*}
  r_3(p(t),t) = v_p - 1, \quad v_p > 2 .
\end{equation*}
Since $r_2 = 1$, we observe that the solution consists of a
non-modulated nonlinear, periodic traveling wave moving with the
piston velocity adjacent to a partial 2-DSW.  Moreover, the density
minimum of the non-modulated traveling wave \eqref{eq013} is 
\begin{equation*}
  \rho_{\text{min}} = \frac{1}{4}(r_1 - r_2 - r_3 + r_4)^2 =
  \frac{1}{4} (v_p - 1 - r_3)^2 = 0.
\end{equation*}
The amplitude of the wavetrain is
\begin{equation*}
  a = (r_4 - r_3)(r_2 - r_1) = 4,
\end{equation*}
hence is independent of the piston velocity, once the critical value
$v_p = 2$ is exceeded.  The interface between the non-modulated
traveling wave and the DSW moves with speed
\begin{equation*}
  v_s^- = v_p + (v_p + 3) \left [ \frac{v_p E(m)}{(v_p - 2)K(m)}
  \right ]^{-1}, \quad m = \frac{4}{v_p},
\end{equation*}
where $E(m)$ is the complete elliptic integral of the second kind.
This modulation solution for $v_p = 2.5$ and an associated numerical
solution of \eqref{eq:75} are depicted in
Fig.~\ref{fig:piston_dsw}b,d.  It was shown that the predictions of
modulation theory agree remarkably well with direct numerical
simulations \cite{hoefer_piston_2008}.

In summary, there are two dynamical regimes.  The first, when $0 < v_p
< 2$, is analogous to the action of a piston compressing a classical
gas.  A dispersive shock wave propagates ahead of the piston.  The
second regime, when $v_p > 2$, has no classical correlate.
Sufficiently large piston speeds compressing a dispersive fluid
generate a uniform, cavitating nonlinear wavetrain at a fixed
amplitude adjacent to the piston.  Further downstream, this
wavetrain's amplitude becomes modulated via a 2-DSW.

\subsubsection{NLS DSWs: concluding remarks}

\label{sec:NLS_concl_remarks}
In the previous sections dedicated to defocusing NLS DSWs, we have
only considered DSWs described by simple wave solutions of the
modulation equations that arise due to a dispersive regularization of
Riemann initial data or due to the uniform motion of a piston.  Since
the NLS-Whitham system \eqref{eq18}, \eqref{eq019} is integrable (see
Sec.~\ref{sec:gener-hodogr-meth}), generalized hodograph solutions are
available for it so that one can describe DSWs generated by the
evolution of more general initial profiles as we did for KdV DSWs in
Sec.~\ref{sec:hod_phase_funct}.  The general hodograph solution to the
NLS-Whitham equations for the evolution of initially monotone profiles
in $\rho$ and $u$ with a single point of wavebreaking was constructed
in \cite{el_general_1995} using the method established in
\cite{gurevich_evolution_1992}, connecting the NLS-Whitham system with
the EPD system \eqref{EPD} in four-dimensional hodograph space with
coordinates $r_1, r_2, r_3, r_4$. An alternative form of this solution
was obtained in \cite{tian_whitham_1999}, where a detailed study of
its properties has been performed.   A particular solution of the EPD system was used in  \cite{kamchatnov_asymptotic_2002} to  describe regularization of a ``cubic'' wavebreaking
in the NLS dispersive hydrodynamics (see  Sec. \ref{sec:cubic_breaking} for the counterpart  KdV problem).  Also, in \cite{kamchatnov_asymptotic_2002}  a long-time asymptotic solution in the form of a pair of modulated dark soliton trains propagating on a constant background was 
obtained  for  the evolution of  ``large and smooth''  initial data  whose IST (Zakharov-Shabat) spectrum is domonated by a discrete component.
The amplitude distributions in the soliton trains were found using a Bohr-Sommerfeld type semi-classical quantization rule for the linear spectral problem associated with the defocusing NLS equation --- an analogue of the KdV soliton train distribution  described in Sec. \ref{sec:evolution_decaying_profiles} (localized elevation case). 
A  full modulation solution describing the evolution of a class of  ``localized'' (sufficiently rapidly approaching the  same non-zero constant  state at $x \to \pm \infty$) initial profles was obtained in \cite{el_two-dimensional_2009} in
the context of hypersonic 2D superfluid flow past an airfoil (see
Sec.~\ref{sec:hypersonic_NLS_piston}).  

Application of the  NLS DSW modulation theory in the context of   Bose-Einstein condensates was first considered  in \cite{kamchatnov_dissipationless_2004}. In \cite{hoefer_dispersive_2006} the NLS modulation solutions were used to identify the experimentally observed BEC blast waves (see Fig.\ref{fig:dsw_expt_super_bec_optics}a,b) with DSWs. Applications of the NLS DSW modulation theory to fiber optics were  discussed in \cite{kodama_whitham_1999},  \cite{biondini_whitham_2006}.

Concluding, we note that since the NLS equation is a generic model
for the evolution of an envelope of a weakly nonlinear,
quasi-monochromatic wave train, the DSWs described by the defocusing
NLS equation generically represent {\it envelope DSWs}. The notion of
an envelope shock (both monotone and oscillatory) was first introduced
by Ostrovsky in \cite{ostrovsky_envelope_1968} (see also
\cite{ostrovsky_modulated_1999}) in the context of weakly dissipative,
dispersive systems. Such an envelope shock is described by a
heteroclinic TW solution, analogous to the diffusive-dispersive shocks
of the KdVB equation described in Sec.~\ref{sec:disp-vs-diff}.

\subsection{Focusing NLS and dispersive regularization of gradient catastrophe}

\label{sec:focusing_NLS}

The general framework of dispersive hydrodynamics \eqref{eq:1} is
intimately related to hyperbolicity of the dispersionless limit
equation ($\mathbf{D}=0$). In the case of the defocusing NLS equation,
the dispersionless limit is given by the hyperbolic shallow water
equations (eq.~\eqref{eq:sw} with $\sigma=1$).  The case of the {\it
  focusing} NLS (fNLS) equation (eq.~\eqref{eq:NLS} with $\sigma=-1$)
is very different as the dispersionless limit \eqref{eq:sw} is now an
elliptic system, implying ill-posedness of the Cauchy problem for all
but analytic initial data as discussed in
Sec.~\ref{sec:NLS_disp_hyd}. Ellipticity of the dispersionless limit
of the focusing NLS equation is the mathematical manifestation of {\it
  modulational instability} of plane wave solutions, $\rho=\rho_0,
u=u_0$, both constants, with respect to infinitesimal perturbations.
The introduction of dispersion stabilizes the propagation of short
waves, as follows from the fNLS linear dispersion relation $\omega=k
u_0 \pm k(k^2/4 - \rho_0)^{1/2}$, which yields real values of the
frequency for $k>2\sqrt{\rho_0}$. This observation provides a glimpse
of the ``partial hyperbolicity'' of the modulation equations, allowing
for nonlinear dispersive saturation of modulational instability via
the generation of a focusing analogue of a DSW in some situations.

\begin{figure}
  \centering
  \includegraphics[height=2.5in]{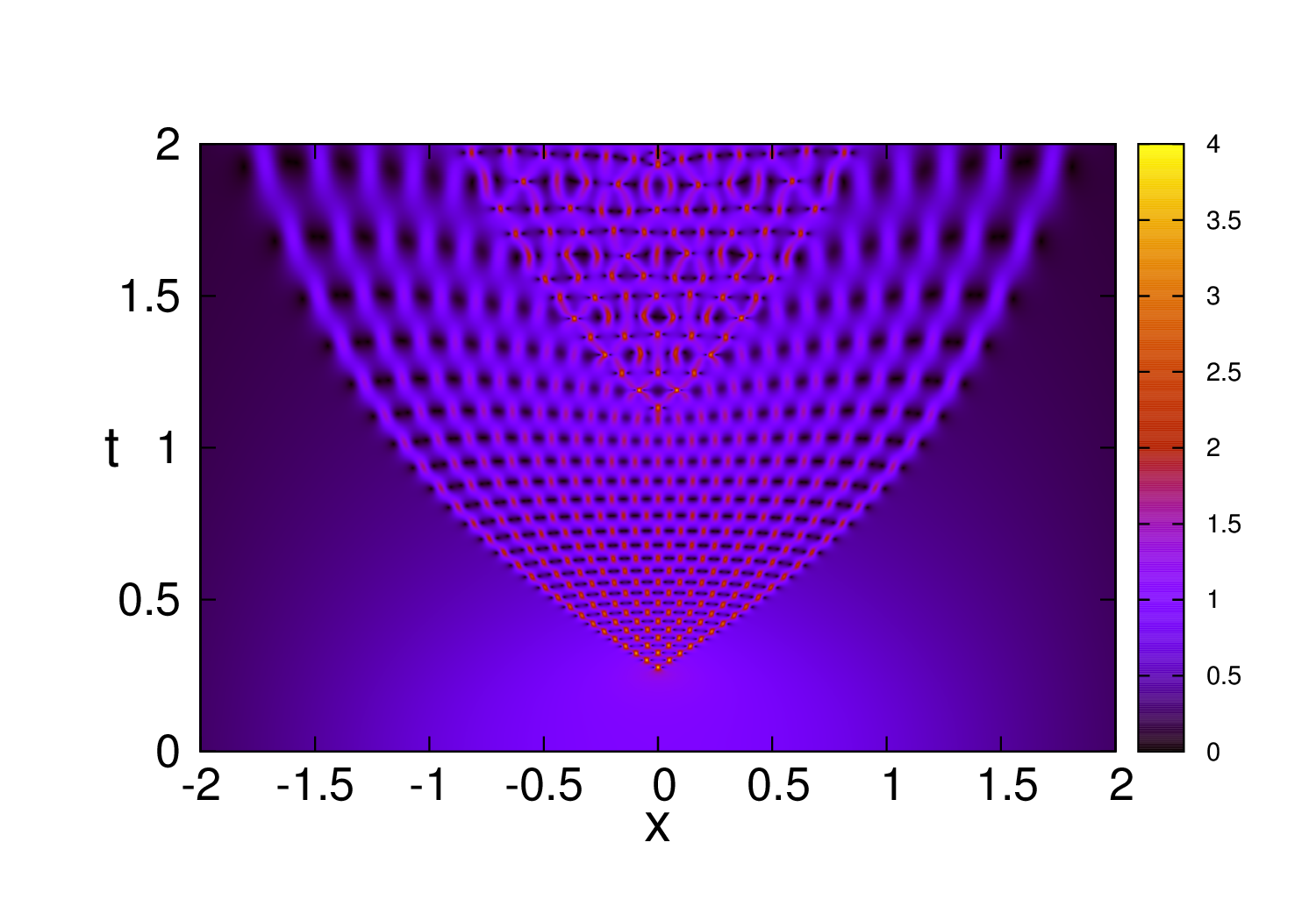}
  \caption{Contour plot of $|\psi|$ for the dispersive regularization
    of gradient catastrophe in the focusing NLS equation
    \eqref{eq:fNLS} with analytic initial condition
    $\psi(x,0)=\hbox{sech} \ x$ and small dispersion parameter
    $\varepsilon=1/33$ -- numerical solution. Figure adapted, with
    permission, from \cite{el_dam_2015}.}
  \label{fig:fNLS_sech}
\end{figure}

The periodic solutions for the focusing NLS equation (see, e.g.,
\cite{kamchatnov_nonlinear_2000}), similar to the KdV and defocusing
NLS cases, are expressed in terms of Jacobi elliptic functions with
harmonic and soliton limits.  Note that unlike solitons of the
defocusing NLS equation, fNLS solitons have positive, ``bright''
polarity and exist only on a zero density, or vacuum, background. The
fNLS-Whitham modulation equations have the same form as in the
defocusing case (see Sec.~\ref{sec:NLS_Whitham_prop}) with the crucial
difference that the Riemann invariants are now complex, satisfying the
relations $r_2=r_1^*$, $r_4=r_3^*$, leading to complex conjugate
characteristc velocities \eqref{vi}, implying the ellipticity of the
modulation system. The elliptic structure of the modulation equations
is the signature of {\it nonlinear} modulational instability
\cite{whitham_linear_1974,zakharov_modulation_2009}. It is remarkable
that, despite the elliptic structure of modulations, focusing NLS
evolution can exibit highly ordered wave structures on a
spatio-temporal scale $O(1)$. Some of these structures can be viewed
as focusing counterparts of DSWs. The appropriate framework for the
analysis of the emergence and evolution of such structures is the
small-dispersion fNLS equation,
\begin{equation}
  \label{eq:fNLS}
  i \varepsilon \psi_t+\frac{\varepsilon^2}{2}\psi_{xx}  +
  |\psi|^2\psi=0, \quad 0 < \varepsilon \ll 1.
\end{equation} 
The evolution of \eqref{eq:fNLS} as $\varepsilon \to 0$ with analytic
initial data
leads to the occurrence of gradient catastrophe at some point $(x_0,
t_0>0)$, which represents a focusing counterpart of hyperbolic
wavebreaking. Dispersive regularization of this gradient catastrophe
results in the generation of rapid oscillations, which are
asymptotically described by modulated solutions of the fNLS-Whitham
equations. Unlike typical single-phase DSW evolution in KdV or
defocusing NLS equations, the oscillations regularizing gradient
catastrophe in fNLS generically have a two-phase structure and may
undergo further bifurcations, giving rise to multiphase regions with
the number of phases greater than two. An example of such an evolution
is shown in Fig.~\ref{fig:fNLS_sech}.  Analysis of these
semi-classical structures heavily relies on the integrability of the
NLS equation and typically involves a Riemann-Hilbert problem
formulation of the IST (see
\cite{kamvissis_semiclassical_2003,tovbis_semiclassical_2004} and
references therein; also the next section for the discussion of
various relations between IST and Whitham theory).

\begin{figure}
  \centering
  \includegraphics{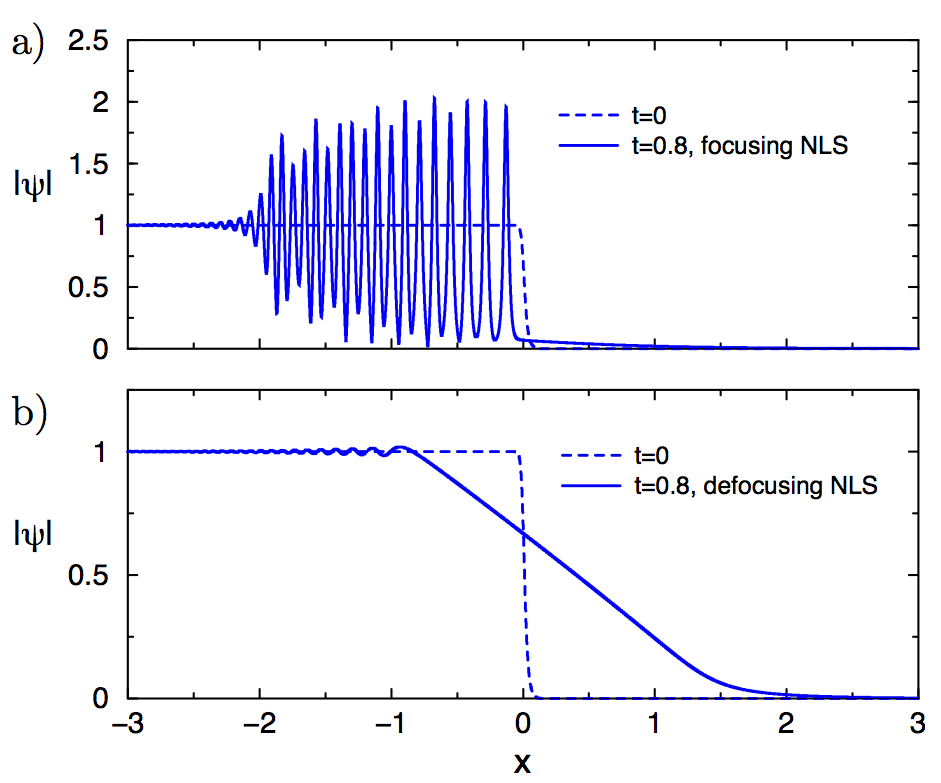}
  \caption{Dispersive regularization of the dam break problem
    with $\psi=0$ for $x>0$ (dashed line) in the NLS equation ---
    numerical solution. a) focusing NLS; b) defocusing NLS.}
  \label{fig:dam_break}
\end{figure}

There is a special case of focusing NLS evolution accompanied by the
generation of a single-phase, modulated structure that strongly
resembles a DSW in the sense that it connects two different,
non-oscillating states and expands in time.  This is the case of the
dam break Riemann problem with a vacuum state on one side: at $t=0$,
$\rho=0, u=0$ for $x>0$ and $\rho>0, u=0$ for $x<0$. An example of
focusing dam break evolution into vacuum is shown in
Fig.~\ref{fig:dam_break}a. Figure \ref{fig:dam_break}b shows the
evolution of the same dam break problem for the defocusing NLS
equation, which, quite similar to the classical shallow water case,
exhibits a smooth rarefaction wave \cite{whitham_linear_1974}, amended
by small amplitude, dispersive oscillations of the weak
discontinuities at the corners.

Dispersive regularization of the ``dry bottom'' dam break flow in
focusing NLS is asymptotically described by modulated {\it
  single-phase} periodic fNLS solutions with the modulation being given
by the similarity solution of the Whitham-NLS system, an analogue of
the simple-wave modulation solutions of the GP problem for the KdV and
defocusing NLS equations. This solution was obtained in
\cite{el_modulational_1993} (see also \cite{el_dam_2015}) and has the
form
\begin{equation}
  \label{eq:fNLS_DSW1}
  \begin{split}
    r_1=r_2^*=i q \\
    \mathrm{Re} [V_3(iq, -iq, r_3, r_4)]=\frac{x}{t}, \quad
    \mathrm{Im} [V_3 (iq, -iq, r_3, r_4)]=0,
  \end{split}
\end{equation}
where $q=\sqrt{\rho_0}$ and $V_3$ is given in \eqref{vi}. Since
$V_3=V_4^*$, the solution \eqref{eq:fNLS_DSW1} represents a {\it
  double characteristic fan}, a counterpart of the GP similarity
solutions for the KdV- and defocusing NLS-Whitham equations (see
Secs.~\ref{sec:case-u_-}, \ref{sec:dsws}).

If we let $r_3=a+ib$, $r_4=a-ib$, the solution \eqref{eq:fNLS_DSW1} is
represented in a particularly simple form \cite{bikbaev_whitham_1994,kamchatnov_new_1997,kamchatnov_nonlinear_2000}
\begin{equation*}
  \begin{split}
    \frac{x}{t}=2a + \frac{q^2 - b^2}{a},\\
    \frac{a^2+(q-b)^2}{a^2 -b^2+ q^2}=\frac{\E(m)}{\K(m)}\, , \quad
    m=\frac{4qb}{a^2+(q+b)^2}. 
  \end{split}
\end{equation*}
This solution describes modulations of a single-phase wavetrain
approximating the structure shown in Fig.~\ref{fig:dam_break}a.  A
small amplitude two-phase modulation correction is visible but the
dominant behavior is one-phase.  Within the wavetrain, the modulus $m$
varies from $m=0$ at the zero-amplitude trailing edge, which
propagates with the velocity $s_-=-2\sqrt{2\rho_0}$. The leading edge
is given by a non-propagating soliton, $s_+=0$, which has amplitude
$\rho_m=4\rho_0$ (note that Fig.~\ref{fig:dam_break} depicts
$\sqrt{\rho}$).

We note that the existence of the centered, double characteristic fan
solution \eqref{eq:fNLS_DSW1} for an {\it elliptic} modulation system
is highly non-trivial and is possible due to the special structure of
the NLS-Whitham equations whose complex conjugate characteristic
velocities can merge -- and thus become real -- on certain surfaces in
the space of Riemann invariants.  Recently, the modulation solution
\eqref{eq:fNLS_DSW1} has been recovered by a rigorous,
Riemann-Hilbert problem analysis of the semi-classical NLS square
barrier problem \cite{jenkins_semiclassical_2014}.  The direct IST
analysis of the asymptotic development of a rather generic localized
perturbation of a plane wave in \cite{biondini_universal_2016} showed
that this self-similar modulation of a single-phase, elliptic fNLS
solution provides a universal asymptotic description of the nonlinear
stage of modulational instability.


 \subsection{Modulation theory vs IST}
\label{sec:dsw-ist}

The Inverse Scattering Transform and Whitham modulation theory are two
major mathematical approaches for describing nonlinear dispersive
waves. Both methods were formulated in the mid 1960s and have since
been extensively used in various applications. Each of these methods
has its own merits and limitations. DSW theory for KdV and other
integrable equations is the common territory for both methods, hence
in this section we briefly outline some prominent interrelations
between them.

As we have already seen, there are several important connections
between IST and Whitham theory. Firstly, the Whitham equations were
shown to describe the slow evolution of finite-band IST spectrum of
traveling waves (Sec.~\ref{sec:integr-riem-invar}). Secondly, the
long-time asymptotic evolution of decaying initial profiles described
by modulation solutions were shown in
Sec.~\ref{sec:evolution_decaying_profiles} to be consistent with
results from IST.  However, there still remains a conceptual gap
between the initial value problem for the small-dispersion KdV
equation (\ref{kdv_eps}) and the GP matching problem (\ref{GP_match})
for the Whitham equations. Indeed, despite convincing evidence of the
validity of the GP approach (mostly numerical, see, e.g.,
\cite{fornberg_numerical_1978}), the ascertainment of its rigorous
justification remained open for a good number of years. The decisive
step was made in the seminal works of Lax and Levermore
\cite{lax_small_1983} and Venakides
\cite{venakides_zero-dispersion_1985} who used the IST to study the
{\it zero dispersion limit} of the solution to the initial value
problem (\ref{kdv_eps}).  By considering an approximation of the
initial data $u_0(x)$ by a $N$-soliton solution, with $N \sim
\varepsilon^{-1} \gg 1$, then passing to the limit as $N \to \infty$,
Lax and Levermore derived the Whitham equations as equations
describing the {\it weak limits} of rapidly oscillating KdV solutions
generated beyond the wavebreaking point of the dispersionless limit
(Hopf equation) solution (\ref{hs}).  Venakides
\cite{venakides_zero-dispersion_1985} extended the Lax-Levermore
results to solitonless potentials, the localized depression case in
terms of the initial value problem (\ref{kdv_eps}), and further, to the
periodic case \cite{venakides_zero_1986}.  Additionally, Venakides
bridged the Lax-Levermore \cite{lax_small_1983} and FFM
\cite{flaschka_multiphase_1980} results by revealing the generation
mechanism of rapid oscillations in small-dispersion solutions
\cite{venakides_generation_1985} and later showing in
\cite{venakides_korteweg-vries_1990} that higher-order corrections to
the Lax-Levermore weak limits describe locally periodic or
quasi-periodic oscillatory structure of the nonlinear wavepackets {\it
  assumed} in the FFM work.  Finally, a comprehensive approach to KdV
with decaying initial conditions in the small dispersion regime was
developed by Deift, Venakides and Zhou (DVZ) \cite{deift_new_1997} who
employed the powerful nonlinear steepest descent method for the oscillatory
Riemann-Hilbert problem associated with semi-classical inverse
scattering introduced earlier by Deift and Zhou
\cite{deift_steepest_1992}. The DVZ analysis provides strong,
pointwise asymptotics of the small-dispersion KdV solution and yields
the modulation solution as part of the full information about the
evolution.

The above works on rigorous IST analysis of KdV evolution have the
inherent constraint of rapidly decaying initial data as $|x| \to
\infty$.  The IST approach was extended to non-decaying step-like
initial conditions by Khruslov \cite{khruslov_asymptotics_1976} who
explained the apparent contradiction between the absence of discrete
spectrum in the KdV step problem and the appearance of solitons at the
DSW leading edge in the GP solution. He showed that the soliton state
at the leading edge is realized only asymptotically as $t \to
\infty$. Further development in the application of IST to the KdV step
problem can be found in \cite{venakides_long_1986}.  These results
have recently been refined \cite{egorova_long-time_2013}.  Ablowitz
and Baldwin \cite{ablowitz_interactions_2013} employed a combination
of IST with matched asymptotic expansions to study the KdV equation
with ``multi-step'' initial data. Apart from providing the detailed
structure of the solution at all times, the important conceptual
result of \cite{ablowitz_interactions_2013} is that the long-time
asymptotic state in the multistep KdV problem is a single DSW
connecting the asymptotic leftmost and rightmost states. A similar
result was obtained within the modulation theory approach in
\cite{el_generating_1996} and \cite{grava_generation_2002}, for smooth
initial data with several wavebreaking points. This result also
correlates with the classical result of viscous shock merger in the
overtaking interaction of two shock waves \cite{whitham_linear_1974}.
See Sec.~\ref{sec:dsw-interactions} for further details.

The development of IST theory for the defocusing NLS equation with
non-zero boundary conditions has been the subject of several recent
works including \cite{demontis_inverse_2013} and
\cite{jenkins_regularization_2015}. The latter paper utilizes the
Deift-Zhou Riemann-Hilbert problem approach to study the
regularization of initial sharp step data for the semi-classical
defocusing NLS equation and, in particular, provides rigorous
justification for some of the modulation theory solutions obtained
earlier in \cite{el_decay_1995} (see Section
\ref{sec:riemann-poblem}).  The connection of IST for the
semi-classical focusing NLS with modulation theory has been briefly
touched upon in Sec.~\ref{sec:focusing_NLS}.

We conclude this section by highlighting the key advantages,
particularly for physical applications, of the modulation theory
approach to the description of a DSW over fully rigorous, but also
more restrictive, IST related techniques.  DSW modulation theory: (i)
does not impose any global limitations (rapid decay as $|x|\to \infty$
or periodicity in $x$) on the solutions; ii) provides a universal
mathematical framework for DSW study within the well developed theory
of quasilinear hyperbolic systems, which is greatly enhanced by the
availability of Riemann invariants and the generalized hodograph
transform \cite{dubrovin_hydrodynamics_1989}; (iii) directly yields
all the key physical macroscopic DSW parameters (e.g., the amplitude
and the wavelength distributions, the speeds of the edges) not readily
available via IST. Most importantly, modulation theory can be
applied to non-integrable systems \cite{whitham_non-linear_1965},
although without the advantage of having Riemann invariant
structure for the modulation system, which is of crucial importance
for applications. The development of DSW theory for non-integrable
dispersive hydrodynamic systems will be described in the next section.

\section{DSWs in non-integrable systems}
\label{sec:dsw-non-integrable}

So far, the emphasis in this review has been upon completely
integrable equations, which predictably yield a very detailed
mathematical description.  As is often the case, integrable systems
provide insight into fundamental processes that carry over to other
systems, which are not integrable but have greater physical relevance.
In particular, there has been strong numerical evidence that such
fundamental features of dispersive hydrodynamics as the formation of
oscillatory regions and weak convergence are true for systems that are
not completely integrable but are structurally similar
\cite{hou_dispersive_1991,lax_generation_1994}.  In particular, a
detailed numerical study of the Riemann problem for a system
describing the dynamics of a two-temperature collisionless plasma
\cite{gurevich_expanding_1984} revealed that the qualitative features
of KdV DSWs hold true for ion-acoustic DSWs of finite amplitude, while
quantitatively there are significant differences.  In particular, the
qualitative DSW structure has been numerically observed to be a
modulated nonlinear wavetrain with a solitary wave at one edge and a
harmonic, small-amplitude wave packet at the other.


In contrast to IST, which is applicable only to integrable
systems, the Whitham method of slow modulations is much more general,
applicable to equations satisfying modest conditions
\cite{whitham_non-linear_1965}.  In what follows, we utilize Whitham
theory in order to describe DSWs in a number of apparently
non-integrable equations.  One important benefit of this
generalization is the ability to accurately model more physical
systems.  We demonstrate this with several modern applications.

\subsection{DSW fitting method}
\label{sec:dsw-fitting-method}

As described in Sec.~\ref{sec:whith-modul-theory}, the
applicability of the Whitham method \cite{whitham_non-linear_1965} to
a dispersive hydrodynamic system of $N$th order requires the existence
of a $N$-parameter family of periodic solutions and the availability
of $N-1$ local conservation laws. These requirements are satisfied by
many physically relevant dispersive hydrodynamic systems. Prominent
examples include systems describing ion-acoustic and magnetoacoustic
waves of finite amplitude in collisionless plasma
\cite{karpman_non-linear_1974,whitham_non-linear_1965}, the
Serre (Green-Naghdi, Su-Gardner) equations for fully nonlinear shallow
water waves
\cite{serre_contribution_1953,green_derivation_1976,su_korteweg-vries_1969}
and various versions of the generalized NLS equation, e.g., the NLS
equation with saturable nonlinearity describing light propagation in
photorefractive crystals \cite{boyd_nonlinear_2013}.

The modulation systems for ion-acoustic and magnetoacoustic waves in
collisionless plasma were derived in \cite{gurevich_nonlinear_1990}
and were shown to share a number of basic properties with KdV and NLS
modulation systems, despite the apparent absence of Riemann invariant
structure.  Specifically, these systems were shown to admit {\it exact
  reductions} to the dispersionless limit of the original
equations. Similar to the KdV and NLS cases, these reductions could be
achieved either via the zero-amplitude $a \to 0$ or zero-wavenumber $k
\to 0$ limit.  As these two limiting regimes correspond to the DSW
edges, the availability of such reductions opens up the possibility of
formulating the GP matching problem.

It was shown in \cite{el_resolution_2005} (see also
\cite{el_determination_2003,el_undular_2005}) that the generic
structure of modulation systems in the linear and soliton limits
prescribes the {\it DSW fitting relations} for a broad class of
dispersive hydrodynamic systems. These DSW fitting relations are
dispersive hydrodynamic counterparts of the Rankine-Hugoniot
conditions of viscous shock theory and hence enable one to classify
solutions of dispersive Riemann problems. In the general bi-directional
case, the DSW fitting relations include: (i) the DSW locus relating
admissible upstream and downstream flow states; (ii) the speeds of
the DSW edges; (iii) the causality conditions determining the DSW
orientation.  The causality conditions are analogues of the Lax
entropy inequalities in viscous shock theory. In the uni-directional
case, the DSW locus does not arise. Remarkably, the DSW fitting method
does not require knowledge of the full simple-wave modulation
solution--although it assumes its existence--and thus circumvents a
major difficulty in the analytical treatment of non-integrable
systems.
 
The mechanisms and conditions for the breakdown of the DSW fitting
construction were identified in
\cite{lowman_dispersive_2013,hoefer_shock_2014}.  In what follows, we
shall provide a brief exposition of the results of
\cite{el_resolution_2005} and \cite{hoefer_shock_2014}, which will
then be applied to several physically relevant systems.

\subsubsection{Scalar equations}
\label{sec:scalar-equations}

We start with a general scalar, model equation
\begin{equation}
  \label{uni_dh}
  u_t+ f(u)_x+D_2[u]_x=0
\end{equation}
with $f''(u) \ne 0$ and $D_2$ a conservative dispersive
operator of the second order so that equation (\ref{uni_dh}) has a
real-valued linear dispersion relation $\omega=\omega_0(k)$ with the
long-wave expansion
\begin{equation}
  \omega_0(k)=c_0 k + \mu k^3 +o(k^3), \quad k \ll 1, \quad \mu \ne 0,
\end{equation}
for small-amplitude waves $\sim e^{i(kx - \omega t)}$, propagating on
the background $u=u_0$ so that $c_0=f'(u_0)$.  We shall be assuming
that the dispersion relation is purely convex or concave, so that
$\partial_{kk}\omega_0 \ne 0$. Equation (\ref{uni_dh}) fits into the
general category of uni-directional dispersive-hydrodynamic systems
(\ref{eq:105}).

We shall consider the evolution of Riemann step initial data $u(x,0) =
u_\pm$, $\pm x > 0$ (\ref{eq:36}) for uni-directional dispersive
hydrodynamics (\ref{uni_dh}).  Our consideration will be based upon
the fundamental assumption that the initial step is resolved into a
DSW that has the structure of a modulated single-phase wavetrain. This
assumption is exactly the one already employed in the GP modulation
regularization for the KdV and NLS equations (see Secs.~\ref{sec:
  dsw-riemann} and \ref{sec:defocusing-nls}). Unlike in the integrable
case, there is no method at present to rigorously confirm this
assumption, so direct numerical confirmation is essential. The
assumption is rooted in Whitham's fundamental proposition that, by
analogy with linear dispersive wave theory, modulated nonlinear
wavetrains represent the generic, long-time outcome of a broad range
of dispersive-hydrodynamic problems \cite{whitham_linear_1974}.  Our
aim is to determine analytically the macroscopic DSW properties.  In
particular, we seek the speeds of the trailing and leading edges
$s_{\pm}$ in a general form, using only basic information provided
by equation \eqref{uni_dh}.  The speeds $s_\pm$ are the dispersive
counterpart of the classical Rankine-Hugoniot relation determining the
viscous shock speed.  In contrast to classical shock theory, where the
shock speed is determined by the hyperbolic structure alone and
viscosity only modifies the transition width/shape, in dispersive
hydrodynamics, the DSW speeds are determined by both the hyperbolic
{\it and} dispersive structure of the equation.  With the DSW speeds
$s_\pm$ in hand, additional macroscopic DSW properties can be deduced
by identifying the edge speeds with the soliton amplitude-speed
relation and the group velocity, yielding the DSW soliton edge
amplitude and harmonic edge wavenumber.

Suppose equation (\ref{uni_dh}) has a three-parameter periodic
traveling wave solution and at least two local conservation
laws. These requirements are quite generic and are satified for a
majority of dispersive hydrodynamic equations (\ref{uni_dh}) arising
in applications.  Then equation (\ref{uni_dh}) is amenable to Whitham
averaging \eqref{eq:6}, \eqref{eq:7}, which yields a system of three
quasilinear equations for slow variations of the traveling wave
parameters. The fundamental difference from the KdV and NLS cases is
that now the modulation system is unlikely to have Riemann invariants
so the expressions for simple-wave solutions involve integration of
the full Whitham equations.  Anyway, they are not of much use for our
task of identifying a generic DSW fitting description.  In the absence
of Riemann invariants, it is convenient to choose physical parameters
as modulation variables, e.g., the period-mean $\overline u$, the
wavenumber $k$, and the amplitude $a$.

Let the periodic traveling wave solution to (\ref{uni_dh}), $u=\tilde
u(\theta)$, $\theta_x=k$,  $\theta_t=-\omega$, be specified by an ODE
\begin{equation}
  \label{ODE11}
 k^2 (u_\theta)^2 = G(u), \quad u(\theta+ 2\pi) = u(\theta),
\end{equation}
where we have omitted the tilde for $u$,
assuming from now on that $u$ is the traveling wave solution $u(\theta)$,
rather than the general field $u(x,t)$.  Let the ``potential''
function $G(u)$ have exactly three real roots, say $b_1 \le b_2 \le
b_3$ where
\begin{equation}
  \label{eq:129}
  G(u) = (b_1 -
  u)(b_2 - u)(b_3 - u)H(u), 
\end{equation}
where $H(b_i) \ne 0$, $i = 1,2,3$.  To be definite, we assume that
$H(u)>0$ so that $G(u)$ has a graph similar to the KdV cubic potential
depicted in Fig.~\ref{fig:kdv_potential} and $b_2 \le u \le b_3$. Then
the wavenumber $k$ and the wave amplitude $a$ are expressed in terms
of ${\mathbf b}$ by (cf.~(\ref{eq:2pi_per}), (\ref{eq:33}))
\begin{equation}
  \label{eq:127}
  k=\pi \left(  \int \limits_{b_2}^{b_3}\frac{\rmd u}{\sqrt{G(u)}}
  \right)^{-1}, \quad a=b_3-b_2.
\end{equation} 
The period average of $\mathcal {F}[u]$ is defined by (cf.~(\ref{eq:aver}))
\begin{equation}
  \label{aver_func}
  \overline{\mathcal{F} [u] } = \frac{k}{\pi} \int \limits^{b_3}_{b_2}
  \frac{\mathcal{F} [u]}{\sqrt{G(u)}} \rmd u,
\end{equation}
In particular, the wave mean $\overline{u}$ is defined by
eq.~\eqref{aver_func} with $\mathcal{F}[u] = u$.  We also define the
{\it conjugate wavenumber}
\begin{equation}
  \label{eq:11}
  \tilde k=\pi \left(  \int \limits_{b_1}^{b_2}\frac{\rmd
      u}{\sqrt{-G(u)}} \right)^{-1}. 
\end{equation}
When $b_2 \to b_1$, the potential $G(u)$ develops a double root and
the wavenumber integral in eq.~\eqref{eq:127} is logarithmically
singular so that $k \to 0$.  In this limit, $a = b_3 - b_1 = O(1)$ and
\begin{equation*}
  \begin{split}
    \lim_{b_2 \to b_1} \int\limits_{b_1}^{b_2} \frac{\rmd
      u}{\sqrt{-G(u)}} = \frac{\pi}{\sqrt{(b_3 - b_1)H(b_1)}},
  \end{split}
\end{equation*}
so that $\tilde{k} = O(1)$.  For the same reason, when $a \to 0$,
$\tilde k \to 0$, and $k = O(1)$.  Therefore, $\tilde k$ is an
amplitude type variable.  The introduction of $\tilde k$ instead of
$a$ as a modulation variable is advantageous because $\tilde k$
provides symmetry in the description of the linear and soliton
regimes.  Now, $k=0$ at the DSW soliton edge and $\tilde k=0$ at the
harmonic edge.  We also note that if $b_1 \le u \le b_2$, i.e., $H(u)
< 0$ in \eqref{eq:129}, then the definitions of $k$ and $\tilde k$
should be simply swapped.  

The full nonlinear dispersion relation
$\omega(k,\overline{u},\tilde{k})$ can be determined as follows.
Given $k$, $\overline{u}$, $\tilde{k}$, the three roots
$\{b_j\}_{j=1}^3$ can be determined from \eqref{eq:127},
\eqref{aver_func}, \eqref{eq:11} assuming a nonzero Jacobian.  The
phase velocity $V$, as any other parameter in the traveling wave, is
determined by $\{b_j\}_{j=1}^3$, hence, in principle, the dependence
of $V$ on $k, \overline{u}, \tilde{k}$ is available. The nonlinear
dispersion relation is then found from $\omega = k V$.
 
Let us now formulate the GP matching conditions for a DSW modulation
solution of equation (\ref{uni_dh}) (cf.~(\ref{GP_match})):
\begin{equation}
  \begin{array}{l}
    x=x_{\rm h}(t):\qquad  \tilde k =0\, , \ \ \overline {u} = u_{\rm
      h} \, , \\ 
    x=x_{\rm s}(t):\qquad  k=0\, , \ \  \overline{u} = u_{\rm s}  \, ,
  \end{array}
  \label{GP_match1}
\end{equation} 
where the subscripts `h' (harmonic) and `s' (soliton) are associated
with the DSW edges. One has either $({\rm h}, {\rm s}) = (-, +)$ or
$({\rm h}, {\rm s}) = (+, -)$, depending on the DSW orientation $d$,
which is unknown at the outset and will be determined by an analogue
of the entropy condition (see Sec.~\ref{sec: dsw-riemann}).  If
$d=1$, then $({\rm h}, {\rm s})=(-,+)$, i.e., the trailing edge is associated with
harmonic vanishing amplitude waves and the leading edge with a solitary
wave. If $d=-1$, the edge association is opposite.

The edges $x_{{\rm h}, {\rm s}}(t)$ are determined by the kinematic
conditions (cf.~conditions \eqref{kinematic_conds} for the KdV case)
\begin{equation}\label{kinematic1}
  \frac{dx_{\rm h}}{dt} = \partial_k \omega (k_{\rm h}, u_{\rm h}),
  \quad \frac{dx_{\rm s}}{dt} = \tilde V_s(\tilde k_{\rm s}, u_{\rm
    s}), 
\end{equation}
where $\tilde V_s(\tilde k, \bar u)= V_s(a_s(\tilde k, \bar u), \bar
u)$ is the soliton amplitude-speed relation re-written in terms of
$\tilde k$.  We note that for each specific case, the identification
of the harmonic edge velocity with the linear group velocity and the
soliton edge with the soliton amplitude-speed relation can be deduced
from an asymptotic analysis of the characteristic velocities of the
modulation equations, which can be done for small amplitudes or small
wavenumbers.  Alternatively, conditions (\ref{kinematic1}) can be
generically postulated using the properties of linear and solitary
waves \cite{whitham_linear_1974}.
 
We now represent the modulation system, obtained by averaging two
conservation laws and complemented by wave conservation
$k_t+\omega_x=0$, in the general form of a quasilinear system
(\ref{hyp1}) (c.f.~\eqref{eq:8})
\begin{equation}\label{hyp2}
   {\bf U}_t +  \mathrm{A}({\bf U}){\bf U}_x   = 0\, ,
\end{equation}
where ${\bf U}({\bf x},t) = ( \bar u, k, \tilde k)^T$ and
$\mathrm{A}(\mathbf U)$ is the Jacobian matrix. Our major assumption
is that the modulation system (\ref{hyp2}) is strictly hyperbolic and
genuinely nonlinear, which guarantees the existence of self-similar
solutions compatible with the Riemann data (\ref{eq:36}).  Later, we
shall present a simple analytic criterion for the breakdown of genuine
nonlinearity of the modulation system in the context of DSW theory.
 
Due to the form of the potential $G(u)$ \eqref{eq:129}, an asymptotic
calculation of (\ref{aver_func}) in both harmonic ($\tilde k=0$) and
soliton ($k=0$) limits, we have $\overline{\mathcal{F} [u] } =
\mathcal{F} [\overline{u}]$. Then, for example, in the harmonic limit the
modulation system assumes the form
\begin{equation}
  \label{h_red}
  \begin{split}
    \tilde k =0, \ \ {\overline u}_t + V({\overline u}) {\overline
      u}_x =0, \ \ k_t+ \omega_0(k, {\overline u})_x =0,
  \end{split}
\end{equation}
where $V(u)= f'(u)$ and $\omega_0(k, {\overline u})$ is the linear
dispersion relation obtained by the linearization of (\ref{uni_dh})
against the background state $u= {\overline u}$.  The exact reduction
(\ref{h_red}) has been deduced directly from the full modulation system
of (\ref{uni_dh}) but it can be convincingly postulated on the
``physical'' grounds that in the harmonic, zero amplitude limit, the
modulation equation for the mean must agree with the dispersionless
limit equation, while the wave conservation equation is generic.
 
A similar reduction of the modulation system to just two equations
exists in the opposite, zero wavenumber case. The derivation of this
reduction, however, is more subtle than in the zero-amplitude case as
the limit $k \to 0$ is a singular one for the wavenumber conservation
law (see \cite{el_determination_2003,el_resolution_2005} for details).
 
We shall now look at how the reduced system (\ref{h_red}) and its
zero-wavenumber counterpart can characterize the similarity solution
of the full modulation system (\ref{hyp2}), describing DSW evolution
for the Riemann problem. Let the characteristic velocities of
(\ref{hyp2}) be ordered as $V_1 \le V_2 \le V_3$.
It can then be concluded that the required modulation solution is
associated with the 2-wave integral curve, (cf.~(\ref{eq:43}) in the
KdV case)
\begin{equation}
  \label{int_curve1}
  V_2 = s = \frac{x}{t}, \quad I_1(\overline u, k, \tilde k)=0,
  \quad I_2(\overline u, k, \tilde k)=0,
\end{equation}
where $I_1, I_2$ are the characteristic integrals of (\ref{hyp2})
associated with the simple-wave solution and correspond to fixed
Riemann invariants for diagonalizable systems.  The integral curve
(\ref{int_curve1}) connects two constant states of the solution vector
$(\overline u, k, \tilde k)$: the state $(u_{\rm h}, k_{\rm h}, 0)$ at
the harmonic edge and $(u_{\rm s}, 0, \tilde k_{\rm s})$ at the
soliton edge.  At these endpoints of the integral curve, the merger of
two of the characteristic velocities, as well as of the integrals
$I_1$ and $I_2$, occurs.

While $u_{\rm h}$ and $u_{\rm s}$ are given by the initial Riemann
data (modulo the DSW orientation), the parameters $k_{\rm h}$ and
$\tilde k_{\rm s}$ are unknown. These can be determined from the
condition that the DSW edges be characteristics of the modulation
system (\ref{hyp2}), so for each of them, there is a characteristic
relation (\ref{charf}) connecting the differentials of all modulation
variables: $d \overline u$, $dk$ and $d \tilde k$ (see
(\ref{diff_form})). For the limiting cases $\tilde k=0$ and $k=0$,
these characteristic relations, and therefore the integrals $I_1$ and
$I_2$ in (\ref{int_curve1}), can be found directly from the reduced
modulation systems. For example, along the harmonic edge $d \tilde k
=0$ and so the relevant characteristic relation connects only $dk$ and
$d \overline u$, implying $k=k(\overline u)$.  Substituting this into
the zero-amplitude reduction system (\ref{h_red}) immediately yields
the characteristic ODE for the harmonic edge
\begin{equation}
  \label{harm_ODE}
  \frac{\rmd k}{\rmd \overline u} = \frac{\partial \omega_0/\partial {\bar
      u}}{V({\bar u}) - \partial \omega_0/{\partial k}}, \quad
  k(u_{\rm s})=0\, . 
\end{equation}
The initial condition for the harmonic edge ODE (\ref{harm_ODE}) is
found from the GP matching condition (\ref{GP_match1}) at the
opposite, soliton edge. Indeed, this GP condition does not contain
$\tilde k$ and thus must be valid for all $\tilde k$ including $\tilde
k =0$.  Said differently, the characteristic integrals
\eqref{int_curve1} must hold for all $\tilde k$.  One of them, say
$I_1$, determines the integration constant for the simple wave $k =
k(\bar u)$ along the harmonic edge via $I_1(u_{\rm s},0,\kt_{\rm s}) =
I_1(u_{\rm s},0,0) = 0$.  
The solution $k({\overline u})$ of
(\ref{harm_ODE}) determines the value of $k_{\rm h}=k(u_{\rm h})$ at
the harmonic edge. Then, since the harmonic edge propagates with the
linear group velocity (see \eqref{kinematic1}), one has $s_{\rm h} = \partial_k \omega_0
(k_{\rm h}, u_{\rm h})$.  

An analogous consideration of the soliton edge leads to the ODE for
$\tilde k$ (the derivation, however, is less straightforward than for
the harmonic edge, see \cite{el_resolution_2005}),
\begin{equation}
  \label{sol_ODE}
  \frac{\rmd \tilde k}{\rmd \overline u} = \frac{\partial \tilde \omega_0/\partial
    {\overline u}}{V({\bar u})- \partial \tilde \omega_0/{\partial
      \tilde k}}, \quad \tilde k(u_{\rm h})=0, 
\end{equation}
where $\tilde \omega_0(\tilde k, \overline u)= -i \omega_0(i \tilde k,
\bar u)$ is the conjugate dispersion relation. Integrating
(\ref{sol_ODE}), we determine the value of $\tilde k$ at the soliton
edge.  The velocity of the soliton edge coincides with the soliton
velocity, defined as $V_{\rm s}=\lim_{k \to 0} \omega/k$, where
$\omega$ is the nonlinear traveling wave frequency. It can then be
shown \cite{el_resolution_2005} that $V_{\rm s}=\tilde \omega_0/\tilde
k$. Then the soliton edge velocity, see the second expression in
\eqref{kinematic1}, is found as $s_{\rm s}= \tilde \omega_0(\tilde
k_{\rm s}, u_{\rm s})/\tilde k_{\rm s}$, where $\tilde k_{\rm s} =
\tilde k(u_{\rm s})$.

 
Now we will identify the harmonic (h) and soliton (s) edges with the
leading or trailing edge propagation with the speeds $s_+$ and $s_-$
respectively. This is done by choosing a DSW orientation, i.e., by
identifying $s_{\rm s}$ and $s_{\rm h}$ with $s_+$ and $s_-$ and
verifying the causality conditions,
\begin{equation}
  \label{causality}
  s_-< V(u_-), \ \ s_+ > V(u_+), \ \ s_+ > s_- ,
\end{equation}
ensuring that the external characteristic family $\rmd x/\rmd t= V(u)$
carries data {\it into} the DSW region
(cf.~Fig.~\ref{fig:kdv_characteristics}).  If the inequalities are not
satisfied, it is necessary to change the DSW orientation.  We stress
that the correct choice of the DSW orientation is a necessary, but
generally not sufficient condition for the fulfillment of the
causality conditions.
 
\smallskip {\it Example}
\smallskip
 
As an example of the application of the DSW fitting method, we recover
the results of the modulation analysis of the KdV Riemann problem
without invoking the Riemann invariant structure.  The KdV
\eqref{kdv1} dispersion relation is $\omega_0=k \bar u - k^3$, and the
hyperbolic velocity $V(\bar u) = \bar u$. Then the ODE
(\ref{harm_ODE}) readily yields $k= \sqrt{\tfrac{2}{3}(\bar u - u_{\rm
    s})}$. Similarly, from (\ref{sol_ODE}) we obtain $\tilde k =
\sqrt{\tfrac{2}{3}(u_{\rm h}-\bar u)}$.
 
Now, $k_{\rm h}=\tilde k_{\rm s} = \sqrt{\tfrac{2}{3}(u_{\rm h} -
  u_{\rm s})}$ implying $u_{\rm h} > u_{\rm s}$.  Let $\Delta = u_{\rm
  h} - u_{\rm s}>0$. Then, for the harmonic edge we have $s_{\rm h} =
\omega_k(k_{\rm h}, u_{\rm h})= u_{\rm h} - 3 k_{\rm h}^2= u_{\rm s} -
\Delta$. For the soliton edge we use $\tilde \omega_0=\tilde k \bar u
+ \tilde k ^3$, giving $s_{\rm s} = u_{\rm s}+ \tilde k_{\rm s}^2=
u_{\rm s}+\frac{2}{3}\Delta$. Then the condition $u_{\rm h}>u_{\rm s}$
implies $s_{\rm s} > s_{\rm h}$, which yields the orientation $d=1$.
Hence $u_{\rm s} \equiv u_+$ and $u_{\rm h} \equiv u_-$ and $s_+ =
s_{\rm s}$, $s_- = s_{\rm h}$ as in \eqref{eq:125},
\eqref{eq:126}. One can see that the causality conditions
(\ref{causality}) are satisfied, being, in the KdV case, equivalent to
the single condition $u_->u_+$.
 
\medskip

In order for this simple wave DSW construction to hold, we require
monotonicity of the characteristic velocity $V_2$ as the 2-wave curve
is traversed (recall Sec.~\ref{sec:simple_waves}) or, equivalently,
genuine nonlinearity of the Whitham modulation system \eqref{hyp2}.
This requirement places restrictions on the dispersive and nonlinear
properties of the governing equation \eqref{uni_dh}.  In particular,
the assumption of the existence of a simple wave modulation solution
implies 
\begin{equation}
  \label{eq:104}
  \frac{\rmd V_2}{\rmd \bar u} = \frac{\partial V_2}{\partial \bar u}
  + \frac{\partial V_2}{\partial \bar k} \frac{\partial k}{\partial
    \bar u} + \frac{\partial V_2}{\partial \tilde k} \frac{\partial
    \tilde k}{\partial \bar u} \ne 0 .
\end{equation}
Without integrability, the direct verification of \eqref{eq:104} is in
general quite difficult.  However, the exact reduction \eqref{h_red}
at the harmonic edge and the corresponding reduction at the soliton
edge enable the verification of genuine nonlinearity in the vicinity
of the DSW trailing and leading edges \cite{lowman_dispersive_2013}.
Near the harmonic edge, the characteristic velocity is $V_2 = \partial_k
\omega_0(k_{\rm h},u_{\rm h}) + \mathcal{O}(\tilde{k}^2)$.
This and \eqref{harm_ODE} simplify \eqref{eq:104} to
\begin{equation}
  \label{eq:117}
  \left . \partial_{k \bar u} \omega_0 \left ( V -
      \partial_k \omega_0 \right ) + \partial_{kk} \omega_0
    \partial_{\bar u} \omega_0 \right |_{k_{\rm h},u_{\rm
      h}}\ne 0 .
\end{equation}
Similarly, for the soliton edge
\begin{equation}
  \label{eq:118}
  \left . \partial_{\bar u} \left ( \frac{\tilde \omega_0}{\tilde k}
    \right ) \left ( V -
      \partial_{\tilde k} \tilde \omega_0 \right ) + \partial_{\tilde
      k} \left ( \frac{\tilde \omega_0}{\tilde k} \right )
    \partial_{\bar u} \tilde \omega_0 \right |_{\tilde k_{\rm
      s},u_{\rm s}}\ne 0 .
\end{equation}
Inequalities \eqref{eq:117} and \eqref{eq:118} are necessary
conditions for the existence of a simple wave modulation solution.
The point at which either of these criteria do not hold corresponds to
an extrema in $s_{\rm h}$ or $s_{\rm s}$ as the right or left state is
varied.

It has been observed that an additional restriction on the DSW fitting
procedure is the convexity (or concavity) of the linear dispersion
relation, $\partial_{kk} \omega_0 \ne 0$
\cite{lowman_dispersive_2013,conforti_2014,el_radiating_2015}.  There
are two known implications of a zero dispersion point $\partial_{kk}
\omega_0 = 0$.  Due to the well-known Benjamin-Feir-Lighthill
criterion \cite{zakharov_modulation_2009}, the change in sign of
dispersion $\sgn (\partial_{kk} \omega_0)$ can lead to modulational
instability.  At the level of the Whitham equations, this implies a
change of type from hyperbolic to elliptic \cite{whitham_linear_1974}.
Physically, this can be understood as a lack of well-ordered waves
within the DSW.  At the soliton edge, $k = 0$, whereas at the harmonic
edge $k = k_{\rm h} > 0$.  If the group velocity exhibits an extremum
for some nonzero $k$ between 0 and $k_{\rm h}$, then the DSW can
experience an internal collision of waves or ``implosion'' as has been
observed in a model of magma dynamics \cite{lowman_dispersive_2013}
(see also Sec.~\ref{sec:visc-fluid-cond}).  This manifests as a
two-phase modulation region adjacent to the DSW.  Another example
accompanying the existence of a zero dispersion point is a linear
resonance where small amplitude radiation is shed from the large
amplitude soliton edge whose velocity is coincident with the linear
phase velocity \cite{conforti_2014,el_radiating_2015}.
 
Concluding this section, we make two important remarks.  The DSW
orientation is determined by the sign of dispersion, $d =-\sgn
(\partial_{kk}\omega_0)$.  Indeed, the sign of dispersion determines
the oscillation ordering in the DSW.  For example, in media with
negative dispersion, waves with smaller $k$ propagate faster, hence
the DSW structure exhibits a leading soliton edge and trailing
harmonic edge. Thus DSW orientation can be assumed at the very
beginning of the DSW fitting construction and verified a
posteriori. The second remark concerns DSW polarity and the amplitude
of the edge soliton.  Let's assume $d=1$, so that we are describing
the leading edge soliton. The DSW polarity coincides with the polarity
of the lead soliton and is readily inferred from a qualitative
analysis of the traveling wave equation (\ref{ODE11}).  For example,
if the potential function has the form \eqref{eq:129} and $G(u) > 0$
for $b_2 < u < b_3$, then the polarity is $p = 1$
(cf.~Fig.~\ref{fig:kdv_potential}b).  If $G(u) > 0$ for $b_1 < u <
b_2$, then $p = -1$.  For $p = 1$, the soliton amplitude $a_+$ is
determined from the condition $s_+ = V_{\rm s}(u_+, a_+)$, where
$V_{\rm s}(\bar u, a_{\rm s})$ is the speed-amplitude relation for
solitons propagating on the background $u=\bar u$, available via
analysis of the separatrix solution of the ODE (\ref{ODE11}) for the
soliton configuration.

\subsubsection{Dispersive Eulerian systems}
\label{sec:disp-euler-syst}

In the previous section, insight gleaned from the DSW GP problem for
KdV was utilized in order to build a general DSW fitting method,
applicable to uni-directional, dispersive hydrodynamic equations
\eqref{uni_dh}.  In this section, we reflect upon DSWs for the NLS
equation, presented in Sec.~\ref{sec:defocusing-nls}, in order to
generalize the DSW fitting method to bi-directional dispersive Euler
equations \eqref{eq:110}, rewritten here for convenience
\begin{equation}
  \tag{\ref{eq:110}}
  \begin{split}
    \rho_t + (\rho u)_x &= D_1[\rho,u], \\
    (\rho u)_t + \left ( \rho u^2 + P(\rho) \right )_x &= D_2[\rho,u] .
  \end{split}
\end{equation}
The general DSW fitting procedure conveyed here is based upon
\cite{el_resolution_2005,el_undular_2005,hoefer_shock_2014}.

We begin the analysis by identifying some basic
properties of the dispersionless $P$-system
\begin{equation}
  \label{eq:114}
  \begin{split}
    \rho_t + (\rho u)_x &= 0 \\
    (\rho u)_t + \left ( \rho u^2 + P(\rho) \right )_x &= 0 .
  \end{split}
\end{equation}
The $P$-system is genuinely nonlinear when
\begin{equation}
  \label{eq:115}
  \left ( \rho^2 P'(\rho) \right )' \ne 0 ,
\end{equation}
its characteristic velocities are
\begin{equation*}
  V_\pm = u \pm c(\rho) = u \pm \sqrt{P'(\rho)} ,
\end{equation*}
and it is diagonalized by the Riemann invariants
\begin{equation}
  \label{eq:131}
  r_\pm = u \pm \int \frac{c(\rho)}{\rho} \rmd
  \rho ,
\end{equation}
so that
\begin{equation}
  \label{eq:132}
  \frac{\p r_\pm}{\p t} + V_\pm \frac{\p r_\pm}{\p x} = 0 .
\end{equation}
Note that the definition of the Riemann invariants (\ref{eq:131})
differs by the factor $2$ from the definition of the analogous set
(\ref{eq20}) for the dispersionless cubic NLS equation. In what
follows, we assume strict hyperbolicity $V_- < V_+$ and genuine
nonlinearity \eqref{eq:115}.  These are the bi-directional
generalizations of nonzero flux curvature for scalar hyperbolic
equations.

As is the case for the genuinely nonlinear $P$-system
\cite{lax_hyperbolic_1973,smoller_shock_1994} and defocusing NLS
hydrodynamics (Sec.~\ref{sec:defocusing-nls}), the Riemann problem for
the dispersive Euler equations \eqref{eq:110} generically results in
two waves separated by a constant, intermediate state.  The leftmost
(rightmost) wave is associated with the slower (faster) characteristic
velocity $V_-$ ($V_+$) and is identified as a 1-wave (2-wave).  Each
wave is either a rarefaction or DSW, whose left and right states are
related via a wave curve or a DSW locus.  

We will make several further simplifying assumptions.  We assume that
the dispersive operators $D_{1,2}$ are second order and that the linear
dispersion relation $\omega_\pm$ for the dispersive Eulerian system
\eqref{eq:110} has the form
\begin{equation}
  \label{eq:134}
  \begin{split}
    \omega_\pm(k,\rho_0,u_0) = V_\pm(\rho_0,u_0)k\, \pm\, \mu k^3 +
    o(k^3), \\
    k  \to 0, \quad \mu > 0.
  \end{split}
\end{equation}
For simplicity, we have assumed positive dispersion.  The focus here
will be on the ``+'' branch $\omega_+$ of the dispersion relation
associated with 2-waves.  In order to implement Whitham averaging on
this fourth order system, we require an additional conservation law to
the ``mass'' and ``momentum'' conservations associated with
\eqref{eq:110}.  Finally, we assume the existence of a four-parameter
family of periodic traveling waves.

\smallskip

\textit{Rarefactions}

\smallskip

Rarefaction waves are self-similar, simple wave solutions of the
dispersionless $P$-system \eqref{eq:114}.  They are integral curves
(recall Sec.~\ref{sec:simple_waves}), efficiently determined from the
diagonal equations \eqref{eq:132}.  For example, a 2-rarefaction
corresponds to variation in the second characteristic family
associated with $V_+$ so that $r_-$ is constant throughout.  The 1-
and 2-wave curves are (c.f.~\eqref{eq:51}, \eqref{eq:103})
\begin{align}
  \label{eq:130}
  &\textrm{1-wave curve:}\quad u_{\rm l} - u_{\rm r} = \int_{\rho_{\rm l}}^{\rho_{\rm r}}
  \frac{c(\rho)}{\rho} \rmd \rho , \quad \rho_{\rm l} > \rho_{\rm r} , \\
  \label{eq:133}
  &\textrm{2-wave curve:}\quad u_{\rm l} - u_{\rm r} = 
  -\int_{\rho_{\rm l}}^{\rho_{\rm r}} \frac{c(\rho)}{\rho} \rmd \rho ,
  \quad \rho_{\rm l} < \rho_{\rm r} ,
\end{align}
where $(\rho_{\rm{l,r}},u_{\rm{l,r}})$ are the left and right states.
Admissibility follows from the requirement that $V_-(\rho_{\rm
  l},u_{\rm l}) < V_-(\rho_{\rm r},u_{\rm r})$ (1-rarefaction) and
$V_+(\rho_{\rm l},u_{\rm l}) < V_+(\rho_{\rm r},u_{\rm r})$
(2-rarefaction).  The rarefaction wave's spatiotemporal profile is
determined by inverting, for example, $V_+(r_-,r_+(s)) = s$, $s = x/t$
for the 2-rarefaction.

\smallskip

\textit{DSWs}

\smallskip

The DSW fitting method applied to the dispersive Euler system
\eqref{eq:110} closely follows the prescription for scalar dispersive
hydrodynamic equations \eqref{uni_dh}.  The four parameter family of
periodic traveling waves $\rho(x,t) = \tilde{\rho}(\theta)$, $u(x,t) =
\tilde{u}(\theta)$, $\theta_x = k$, $\theta_t = - \omega$, is assumed
to satisfy an ODE $k^2 (\rho')^2 = G(\rho)$ where the potential $G$
has three real roots, and we have dropped the tilde for notational
simplicity.  The dependence of $k$ on these roots is, as usual, fixed
by the condition of $2\pi$-periodicity in $\theta$
(cf. \eqref{eq:2pi_per}). We invoke the ``symmetric'' physical
parametrization $\mathbf{U} = (\rhob,\ub,k,\kt)^T$ consisting of the
wave's average density $\rhob$ and velocity $\ub$, as well as the
wavenumber $k$ and conjugate $\kt$.  The four Whitham equations
\begin{equation}
  \label{eq:137}
  \mathbf{U}_t + \mathrm{A} \mathbf{U}_x = 0
\end{equation}
are obtained by averaging three local conservation laws over a wave
period and adjoining the conservation of waves $k_t + \omega_x = 0$.

By either similar reasoning to the scalar case or hypothesis, we have
that the Whitham equations \eqref{eq:137} admit exact reductions to
the dispersionless $P$-system \eqref{eq:114} when $\kt \to 0$ (zero
amplitude) and $k \to 0$.  The zero amplitude reduction has the form
\begin{equation}
  \label{eq:138}
  \begin{split}
    \rhob_t + (\rhob\, \ub)_x &= 0 \\
    (\rhob\, \ub)_t + \left ( \rhob\, \ub^2 + P(\rhob) \right )_x &= 0 \\
    \kt = 0, \quad k_t + \left ( \omega_+(\rhob,\ub,k) \right )_x &= 0,
  \end{split}
\end{equation}
where we have chosen the ``+'' branch of the linear dispersion
relation and have therefore assumed a 2-DSW.  Since we have assumed
positive dispersion \eqref{eq:134}, we expect a 2-DSW with orientation
$d = -1$.  Then the GP matching conditions are formulated according to
\begin{equation*}
  \begin{split}
    x &= s_{\rm l} t: \quad k = 0, \quad \rhob = \rho_{\rm l}, \quad
    \ub = u_{\rm l}, \\
    x &= s_{\rm r} t: \quad \kt = 0, \quad \rhob = \rho_{\rm r}, \quad
    \ub = u_{\rm r} ,
  \end{split}
\end{equation*}
and the edge speeds $s_{\rm{l,r}}$ are to be determined.

The DSW locus can be determined by invoking a backward characteristic
argument \cite{el_resolution_2005}, where the ``dual'' of a 2-DSW,
resulting from negative time propagation of an initial jump from
$(\rho_{\rm l},u_{\rm l})$ to $(\rho_{\rm r},u_{\rm r})$, is a
2-rarefaction \eqref{eq:133}.  Because time has been reversed, the
2-rarefaction admissibility criterion $\rho_{\rm l} < \rho_{\rm r}$
must also be reversed.  Therefore, the 2-DSW locus is
\begin{equation}
  \label{eq:135}
  \textrm{2-DSW locus:}\quad u_{\rm l} - u_{\rm r} = 
  -\int_{\rho_{\rm l}}^{\rho_{\rm r}} \frac{c(\rho)}{\rho} \rmd \rho ,
  \quad \rho_{\rm l} > \rho_{\rm r} .
\end{equation}
The causality conditions \eqref{causality} generalize to the 2-DSW
here according to
\begin{equation}
  \label{eq:136}
  \begin{split}
    \textrm{2-DSW causality:}\quad V_-|_{\rm left} < s_{\rm l} <
    V_+|_{\rm left}, \\
    V_+|_{\rm right} < s_{\rm r}, \quad s_{\rm
      r} > s_{\rm l} .
  \end{split}
\end{equation}
Repeating these arguments for the 1-DSW yields
\begin{align}
  \label{eq:140}
  &\textrm{1-DSW locus:}\quad u_{\rm l} - u_{\rm r} = 
  \int_{\rho_{\rm l}}^{\rho_{\rm r}} \frac{c(\rho)}{\rho} \rmd \rho ,
  \quad \rho_{\rm l} < \rho_{\rm r} . \\
  \label{eq:141}
  &\textrm{1-DSW causality:}\quad V_-|_{\rm right} < s_{\rm r} <
  V_+|_{\rm right}, \\
  \nonumber
  &\qquad\qquad\qquad\qquad s_{\rm l} < V_-|_{\rm left}, \quad s_{\rm
    r} > s_{\rm l} .
\end{align}
As in the NLS case, the DSW loci correspond to inadmissible wave
curves.

Returning to the 2-DSW construction, we now seek a simple wave
solution of the zero amplitude reduction \eqref{eq:138}.  This is a
system of three hyperbolic equations so it is not immediately clear
that we will be able to integrate it.  The key is to assume a local
relationship $F(\rhob,\ub) = C_0$ \cite{el_resolution_2005}.  Because
it is independent of both $\kt$ and $k$, it must be valid along both
the zero amplitude reduction \eqref{eq:138} and the analogous zero
wavenumber reduction so that $F(\rho_{\rm l}, u_{\rm l}) = F(\rho_{\rm
  r},u_{\rm r})$.  In order to be consistent with the 2-DSW locus
\eqref{eq:135}, we must have the local relation
\begin{equation}
  \label{eq:142}
  \ub = u_{\rm l} + \int_{\rho_{\rm l}}^{\rhob} \frac{c(\rho)}{\rho}
  \rmd \rho ,
\end{equation}
valid along both the zero amplitude and zero wavenumber reductions.
Using \eqref{eq:142}, we can now eliminate one variable from the
reduction \eqref{eq:138} and obtain
\begin{equation*}
  \begin{split}
    \rhob_t + V_+(\rhob,\ub(\rhob)) \rhob_x &= 0, \quad \kt = 0, \\
    k_t + \left (\omega_+(\rhob,\ub(\rhob),k) \right )_x &=0 ,
  \end{split}
\end{equation*}
which is the same as the scalar, zero amplitude reduction
\eqref{h_red}.  The simple wave ODE is obtained in the same manner
\begin{equation}
  \label{eq:144}
  \frac{\rmd k}{\rmd \rhob} = \frac{\p \omega_+/\p \rhob}{V_+ - \p
    \omega_+/\p k}, \quad k(\rho_{\rm l}) = 0 .
\end{equation}
Upon integration of \eqref{eq:144}, and evaluating $k_{\rm r} =
k(\rho_{\rm r})$, we can obtain the harmonic, leading edge 2-DSW speed via
the group velocity $s_{\rm r} = \partial_k \omega_+(\rho_{\rm
  r},u_{\rm r},k_{\rm r})$.

The zero wavenumber, soliton edge is analyzed in a similar fashion
(see \cite{el_resolution_2005} for details), leading to the simple
wave ODE
\begin{equation}
  \label{eq:145}
  \frac{\rmd \kt}{\rmd \rhob} = \frac{\p \tilde\omega_+/\p \rhob}{V_+ - \p
    \tilde \omega_+/\p \kt}, \quad \kt(\rho_{\rm r}) = 0 ,
\end{equation}
with the conjugate frequency $\tilde\omega_+ = -i
\omega_+(\rhob,\ub,i\kt)$.  Integrating \eqref{eq:145} and determining
$\kt_{\rm l} = \kt(\rho_{\rm l})$ yields the soliton, trailing edge
2-DSW speed $s_{\rm l} = \tilde \omega_+(\rho_{\rm l},u_{\rm
  l},\kt_{\rm l})/\kt_{\rm l}$.  As in the scalar case, the soliton
edge amplitude $a_{\rm l}$ can be determined from the soliton
amplitude-speed relation.

The DSW fitting carried out here was for a 2-DSW in the presence of
positive dispersion so that $d = -1$.  Appropriate modifications are
required for other cases, e.g., $+ \to -$ in eqs.~\eqref{eq:144},
\eqref{eq:145} and usage of the 1-DSW locus \eqref{eq:140} apply to a
1-DSW with negative dispersion.

Similar to the scalar case, this DSW fitting method requires
additional admissibility criteria such as genuine nonlinearity of the
Whitham equations and convex dispersion \cite{hoefer_shock_2014}.
For the 2-DSW, we require monotonicity in the third characteristic
Whitham velocity $V_3$, which amounts to the requirement
\begin{equation}
  \label{eq:119}
  \begin{split}
    &\partial_{kk} \omega_+ \partial_{\rhob} \omega_+ \\
    &+ (V_+ - \partial_k \omega_+) \left ( \partial_{k\rhob} \omega_+
    + \partial_{k\ub} \omega_+ \frac{c}{\rhob} \right ) \Big |_{k_{\rm h}, \rho_{\rm h},
    u_{\rm h} } \ne 0 ,
  \end{split}
\end{equation}
where $(k_{\rm h}, \rho_{\rm h}, u_{\rm h})$ are the modulation
parameters at the harmonic edge.  At the soliton edge, genuine
nonlinearity implies
\begin{equation*}
  \begin{split}
    &\partial_{\tilde k} \left (\frac{\tilde \omega_+}{\tilde k} \right
    )  \partial_{\rhob} {\tilde \omega_+} \\
    &+ (V_+ - \partial_{\tilde k} {\tilde \omega_+})\left
      ( \partial_{\rhob} \left ( \frac{\tilde \omega_+}{\tilde k}
      \right )
    + \partial_{\ub} \left ( \frac{\tilde \omega_+}{\tilde k} \right )
    \frac{c}{\rhob} \right ) \bigg |_{k_{\rm s}, \rho_{\rm s},
    u_{\rm s} } \ne 0 ,
  \end{split}
\end{equation*}
where $(k_{\rm s}, \rho_{\rm s}, u_{\rm s})$ are the modulation
parameters at the soliton edge.

In summary, the DSW fitting method applied to the bi-directional
dispersive Euler equations \eqref{eq:110} reduces to the same simple
wave ODEs as in the scalar dispersive hydrodynamic case \eqref{uni_dh}
with the additional, local application of the DSW locus. 

\subsection{Applications of the DSW fitting method }
\label{sec:appl-dsw-fitt}

We now provide several examples of the DSW fitting method applied to
non-integrable dispersive hydrodynamic equations.

\subsubsection{Viscous fluid conduits}
\label{sec:visc-fluid-cond}

When a viscous fluid is injected from below into a tall reservoir of
miscible, heavier, more viscous fluid, the injected fluid buoyantly
rises and eventually establishes a steady flow configuration
consisting of a fluid conduit or fluid-filled pipe
\cite{whitehead_dynamics_1975}.  Due to high viscosity contrast, there
is negligible drag at the interface between the two fluids.  As such,
the driven fluid conduit can be modeled as a deformable pipe
\cite{olson_solitary_1986}.  A multiple scales, perturbation
derivation leads to the conduit equation
\cite{lowman_dispersive_2013-2}
\begin{equation}
  \label{eq:84}
  A_t + (A^2)_z - (A^2(A^{-1}A_t)_z)_z = 0,
\end{equation}
describing long interfacial waves along the fluid conduit.  The
dimensionless cross-sectional area of the conduit $A$ satisfies a
nonlinear, dispersive wave equation that expresses conservation of the
injected fluid's mass.  The nonlinear convection term expresses the
self-steepening effect of buoyancy while the nonlinear, nonlocal
dispersive term expresses the effect of normal and tangential
interfacial stresses.

The conduit equation is a long wave model with no amplitude assumption
in its derivation.  This combined with the form of the nonlinear
dispersion term makes the conduit equation a scalar analogue of the
Serre equations for long, shallow water waves of arbitrary height,
studied in the next section.  The conduit equation \eqref{eq:84}
happens to be one of a class of equations modeling magma migration in
the Earth's mantle
\cite{mckenzie_generation_1984,simpson_multiscale_2010}.  The viscous
fluid conduit system has been used as a model system to investigate
solitons experimentally
\cite{olson_solitary_1986,scott_observations_1986,whitehead_wave_1988,lowman_interactions_2014}.
Mathematically, it falls into the class of dispersive hydrodynamic equations (\ref{uni_dh}) amenable to the DSW fitting method.

\begin{figure}
  \centering
  \includegraphics{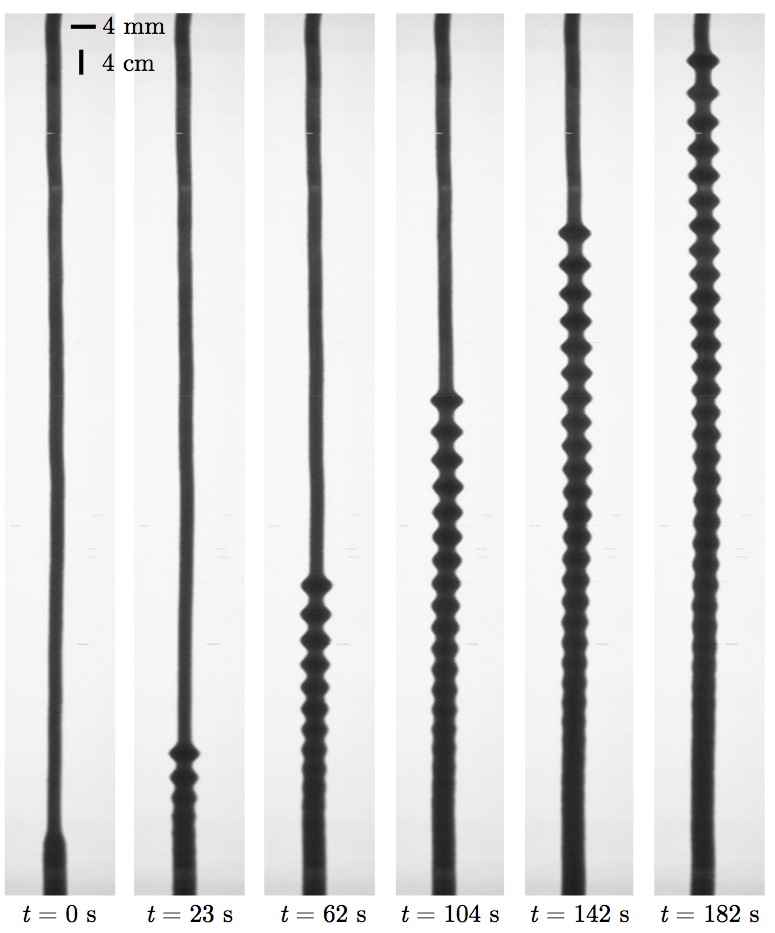}
  \caption{Viscous fluid conduit DSW formation and propagation
    \cite{maiden_wavebreaking_2015}.  Images have a 1:10 horizontal to
    vertical aspect ratio.}
  \label{fig:conduit_dsw}
\end{figure}
An example conduit DSW is pictured in Fig.~\ref{fig:conduit_dsw}
\cite{maiden_wavebreaking_2015}.  A narrower conduit, supported by
injection at a fixed rate, is followed by a wider conduit, generated
by a gradual ramp-up to a larger injection rate.  This leads to a
sharp transition in conduit diameter, gradient catastrophe, at $t = 0$
for simplicity.  The nominal jump ratio in Fig.~\ref{fig:conduit_dsw}
is $A_-/A_+ \approx 2.9$, where $A_\pm$ are the leading and trailing
conduit cross sectional areas, respectively. Due to interfacial
dispersion, this approximate Riemann problem is resolved by the
generation of a DSW.

Previous studies have described conduit DSWs in the small amplitude,
long wave regime of \eqref{eq:84} where the KdV equation is valid
\cite{elperin_nondissipative_1994}.  Here we follow
\cite{lowman_dispersive_2013} and apply the DSW fitting method to
solutions of the non-integrable \cite{harris_painleve_2006} conduit
equation \eqref{eq:84}.

We require the linear dispersion relation for $A(z,t) = \overline{A} +
a e^{i(kz-\omega t)} + c.c.$, $a \to 0$
\begin{equation*}
  \omega_0(k,\overline{A}) = \frac{2 \overline{A} k}{1 + \overline{A} k^2},
\end{equation*}
and the long wave sound speed
\begin{equation*}
  c(\overline{A}) = \lim_{k \to 0} \frac{\omega(k,\overline{A})}{k} =
  2 \overline{A} .
\end{equation*}
The conduit equation itself is one conservation law for the system.
It can be shown that the only other one is
\cite{barcilon_nonlinear_1986,harris_conservation_1996}
\begin{equation*}
  \left ( \frac{1}{A} + \frac{A_z^2}{A^2} \right )_t + \left (
    \frac{A_{tz}}{A} - \frac{A_z A_t}{A^2} - 2 \ln A \right )_z = 0 .
\end{equation*}
With this information in hand, we can apply the DSW fitting method.
We consider the Riemann initial conditions
\begin{equation*}
  A(z,0) =
  \begin{cases}
    A_0 & z < 0 \\
    1 & z > 0
  \end{cases}, \quad A_0 > 1 .
\end{equation*}
The scaling symmetry $A \to \overline{A}^{-1} A$, $z \to \overline{A}^{-1/2}
z$, $t \to \overline{A}^{1/2} t$ satisfied by solutions of
\eqref{eq:84} enables more general initial jumps.

\begin{figure}
  \centering
  \includegraphics{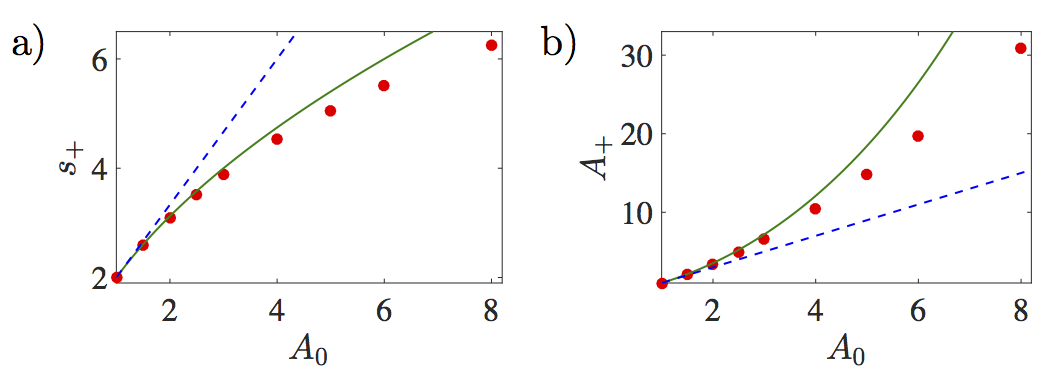}
  \caption{Conduit DSW macroscopic properties.  a) Soliton edge speed
    versus jump height.  b) Soliton edge amplitude.  Numerics (circles),
    DSW fitting (solid), KdV approximation (dashed).}
  \label{fig:conduit_dsw_numerics}
\end{figure}
From eq.~\eqref{harm_ODE}, the harmonic, trailing edge of the DSW is
described by integrating
\begin{equation*}
  \frac{\rmd k}{\rmd \overline{A}} = \frac{1}{\overline{A}^2
    k(\overline{A}k^2 + 3)} , \quad k(1) = 0.
\end{equation*}
The substitution $\alpha = \overline{A}k^2$ gives the separable ODE
\begin{equation*}
  \frac{\rmd \alpha}{\rmd \overline{A}} = \frac{(\alpha + 2)(\alpha +
    1)}{\overline{A}(\alpha + 3)} , \quad \alpha(1) = 0 .
\end{equation*}
Integration and evaluation at the trailing, harmonic edge $\overline{A} = A_0$ yields
the wavenumber
\begin{equation}
  \label{eq:93}
  k_-^2 = \frac{1}{4 A_0} \left( A_0 - 4 + \sqrt{A_0(8 +
    A_0)} \right ) .
\end{equation}
The harmonic edge speed is determined by evaluating the group velocity
\begin{equation}
  \label{eq:94}
  s_- = \frac{\partial \omega_0}{\partial k}(k_-,A_0) = 8 + 3 A_0 - 3
  \sqrt{A_0(8 + A_0)} .
\end{equation}

The soliton, leading edge of the DSW is analyzed by introducing the
conjugate wavenumber $\kt$ and frequency $\wt$
\begin{equation*}
  \wt_0(\kt,\oA) = -i \omega_0(i\kt,\oA) = \frac{2\oA \kt}{1 - \oA
    \kt^2} .
\end{equation*}
Then the soliton edge speed is found by integrating \eqref{sol_ODE}
\begin{equation*}
  \frac{\rmd \kt}{\rmd \oA} =\frac{1}{\oA^2 \kt(\oA\kt^2 - 3)}, \quad
  \kt(A_0) = 0 . 
\end{equation*}
Introducing the change of variable $\at = \oA \kt^2$ leads to
\begin{equation*}
  \frac{\rmd \at}{\rmd \oA} = \frac{(\at - 2)(\at -1)}{\oA(\at - 3)},
  \quad \at(A_0) = 0 .
\end{equation*}
Integration and evaluation at the soliton edge $\oA = 1$ identifies
the conjugate wavenumber
\begin{equation*}
  \kt_+^2 = 1 + \frac{\sqrt{1 + 8 A_0} - 1}{4 A_0} .
\end{equation*}
Evaluating the phase speed gives the leading edge soliton speed
\begin{equation}
  \label{eq:98}
  s_+ = \frac{\wt_0(\kt_+,1)}{\kt_+} = \sqrt{1 + 8 A_0} - 1 .
\end{equation}
The leading edge soliton amplitude $A_+$ can be determined by
inverting the soliton amplitude-speed relation
\begin{equation}
  \label{eq:99}
  \sqrt{1 + 8 A_0} - 1 = \frac{A_+^2(2 \ln A_+ - 1) + 1}{(A_+ - 1)^2}.
\end{equation}
Small jump $0 < \Delta = A_0-1 \ll 1$ expansions of \eqref{eq:98} and
\eqref{eq:99} yield the approximate results $s_+ \sim 2 +
\frac{4}{3}\Delta$, $A_+ \sim 2 \Delta$, in agreement with the KdV
approximation \cite{whitehead_korteweg_1986,lowman_dispersive_2013}.

\begin{figure}
  \centering
  \includegraphics{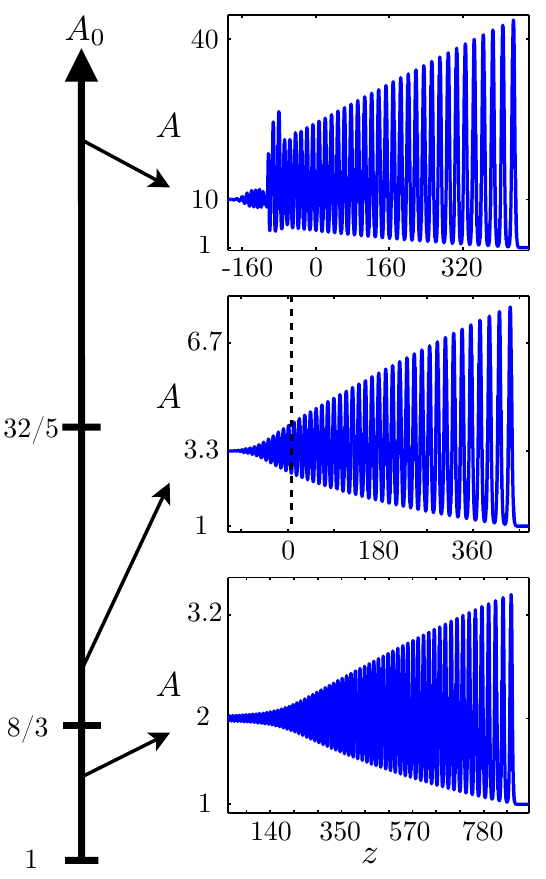}
  \caption{Numerical computation of conduit DSWs with differing jump
    heights.  Small jump ($A_0 = 2$), the DSW is KdV-like.  Moderate
    jump ($A_0 = 3.3$), the DSW exhibits backflow.  Large jump ($A_0 =
    10$), the DSW implodes.  See \cite{lowman_dispersive_2013}.}
  \label{fig:conduit_profiles}
\end{figure}
Thus we have arrived at the relatively simple expressions
\eqref{eq:93}, \eqref{eq:94} for the harmonic edge and \eqref{eq:98},
\eqref{eq:99} for the soliton edge macroscopic DSW properties.  A
comparison of macroscopic DSW properties at the soliton edge with
numerical simulations of eq.~\eqref{eq:84} for a smoothed step are
shown in Fig.~\ref{fig:conduit_dsw_numerics}.  The KdV approximation
only holds in the extremely small jump regime, demonstrating the
importance of this non-integrable, DSW fitting approach. The deviation
between modulation theory and numerics for large jump heights could be
due to a loss of genuine nonlinearity in the full Whitham system or
other causes.  We consider one such possibility below.

The trailing edge speed in \eqref{eq:94} can be negative when $A_0 >
8/3$.  This corresponds to a negative group velocity and the
development of backflow.  Although the injected fluid is buoyantly
rising in the column, interfacial waves are backpropagating downward.
Such behavior has been observed experimentally
\cite{maiden_wavebreaking_2015} and is shown numerically in
Fig.~\ref{fig:conduit_profiles}.

For sufficiently long waves, the dispersion 
\begin{equation}
  \label{eq:100}
  \frac{\partial^2 \omega_0}{\partial k^2}(k,\oA) = -\frac{4 \oA^2 k(3
    - \oA k^2)}{(1 + \oA k^2)^3} ,
\end{equation}
is negative, concave.  There is an inflection point at the critical
wavenumber $k^2 = 3/\oA$.  Inserting the expression for $k_-$
\eqref{eq:93} into \eqref{eq:100}, we find that the DSW trailing edge
crosses the zero dispersion point when $A_0 = 32/5$.  This change in
dispersion sign leads to what was termed implosion, because the waves
at the trailing edge move faster than the waves in the interior of the
DSW, leading to an internal two-phase interaction
\cite{lowman_dispersive_2013}.  This behavior is due to a change of
type for the Whitham equations from hyperbolic to elliptic.  See
Sec.~\ref{sec:scalar-equations}.

The three DSW behaviors are shown in Fig.~\ref{fig:conduit_profiles}
as numerical simulations of a smoothed step initial condition for the
conduit equation \eqref{eq:84}.

\subsubsection{Shallow water DSWs --- undular bores}
\label{sec:shallow-water}

\begin{figure}
  \centering
  \includegraphics{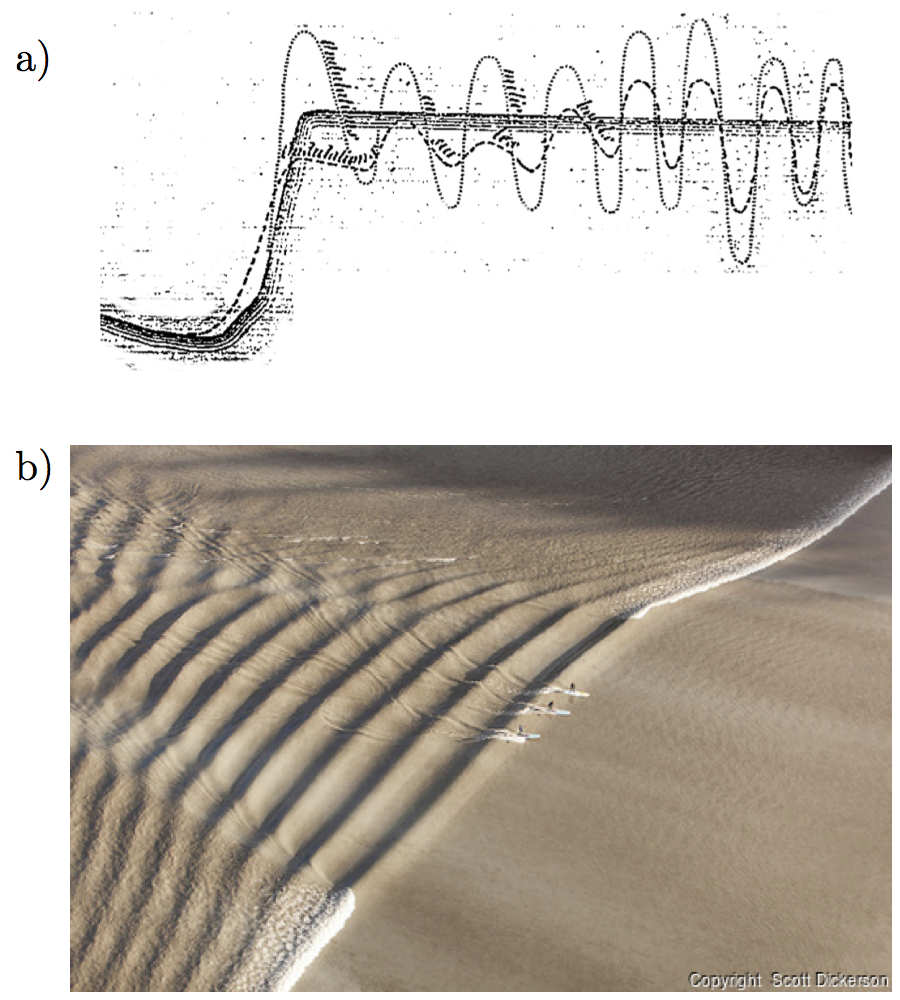}
  \caption{Undular bores.  a) Measurements of a DSW generated by a
    hydraulic jump on Pont-aqueduc de Crau, Canal de Craponne, France
    in 1865 \cite{bazin_recherches_1865,darcy_recherches_1865}.  b)
    Three standing paddle board riders on a river undular bore in
    Turnagain Arm, Alaska (copyright Scott Dickerson, 2013).}
  \label{fig:dsw_sw}
\end{figure}

One of the earliest experiments on DSWs occurred in the context of
channel flow where an undular hydraulic jump or undular bore was
created in 1865 by Darcy and Bazin
\cite{bazin_recherches_1865,darcy_recherches_1865}.  Some measurements
from these experiments are depicted in Fig.~\ref{fig:dsw_sw}a.  A
tidal bore propagating upstream in certain rivers can form when the
incoming tide is strong enough.  If the Froude number (ratio of the
vertically averaged horizontal flow speed to the long wave speed) is
larger than unity but less than 1.4 to 1.7, then the bore takes on an
undular character.  An example from Turnagain Arm, Alaska is shown in
Fig.~\ref{fig:dsw_sw}b, revealing a combined undular, DSW character in
the center and a dissipative, non-oscillatory behavior at the edges.

As a first physical example of DSWs in non-integrable Eulerian
dispersive hydrodynamics (\ref{eq:110}), we consider the Serre
(Green-Naghdi, Su-Gardner) system describing fully nonlinear,
unsteady, shallow water waves
\cite{serre_contribution_1953,green_derivation_1976,su_korteweg-vries_1969} (see also \cite{whitham_variational_1967})
\begin{align}
  &\eta_t+(\eta u)_x = 0\, ,  \label{SG}\\
  &u_t+uu_x+\eta_x=\frac{1}{\eta}\left[\frac{1}{3}\eta^3(u_{xt}+uu_{xx}-
    (u_x)^2)\right]_x \, . \nonumber 
\end{align} 
Here $\eta$ is the total depth and $u$ is the layer-mean horizontal
velocity; all variables are non-dimensionalized by their typical
values. The first equation is the exact equation for conservation of
mass and the second equation can be regarded as an approximation to
the equation for conservation of horizontal momentum.  The system
(\ref{SG}) has the structure of well-known Boussinesq-type systems for
shallow water waves, but differs from them in retaining full
nonlinearity in the leading-order dispersive term.  The system
(\ref{SG}) can be consistently obtained from the full, irrotational
Euler equations using an asymptotic expansion in the small dispersion
parameter $\varepsilon=h_0/L \ll 1$, where $h_0$ is the (dimensional)
equilibrium depth and $L$ is a typical wavelength
\cite{el_unsteady_2006}.  We stress, however, that there is no
limitation on the amplitude.

Equations \eqref{SG} can be put in dispersive Eulerian form
\eqref{eq:110} by identifying the total depth $\eta$ with $\rho$ and
taking
\begin{equation*}
  P(\eta) = \frac{1}{2} \eta^2, \quad D_1 \equiv 0, \quad D_2 =
  \frac{1}{3} \eta^3 \left ( u_{tx} + u u_{xx} - u_x^2 \right ) .
\end{equation*}
The mathematical similarity between the dispersionless Serre equations
($D_2 = 0$) and the isentropic Euler equations of gas dynamics lead to
a significant body of work on the \textit{hydraulic analogy} (see,
e.g., \cite{harleman_studies_1950,hoyt_hydraulic_1962}), which sought
to study supersonic gas flows with shallow water.  Note, however, that
the role of dispersion was completely neglected.

The Serre system has attracted significant attention due to its
superiority, compared to other Boussinesq type systems, in capturing
fully nonlinear dispersive shallow-water dynamics (see,
e.g., \cite{nadiga_different_1996}).  DSWs in the Serre system have
been studied numerically in a number of papers, including some 
recent works
\cite{le_metayer_numerical_2010,mitsotakis_galerkin/finite-element_2014}. The
modulation description of Serre system DSWs via the DSW fitting method
was done in \cite{el_unsteady_2006}. Below we outline the results of
this latter work.

\begin{figure}[ht]
  \centerline{\includegraphics[scale=0.25]{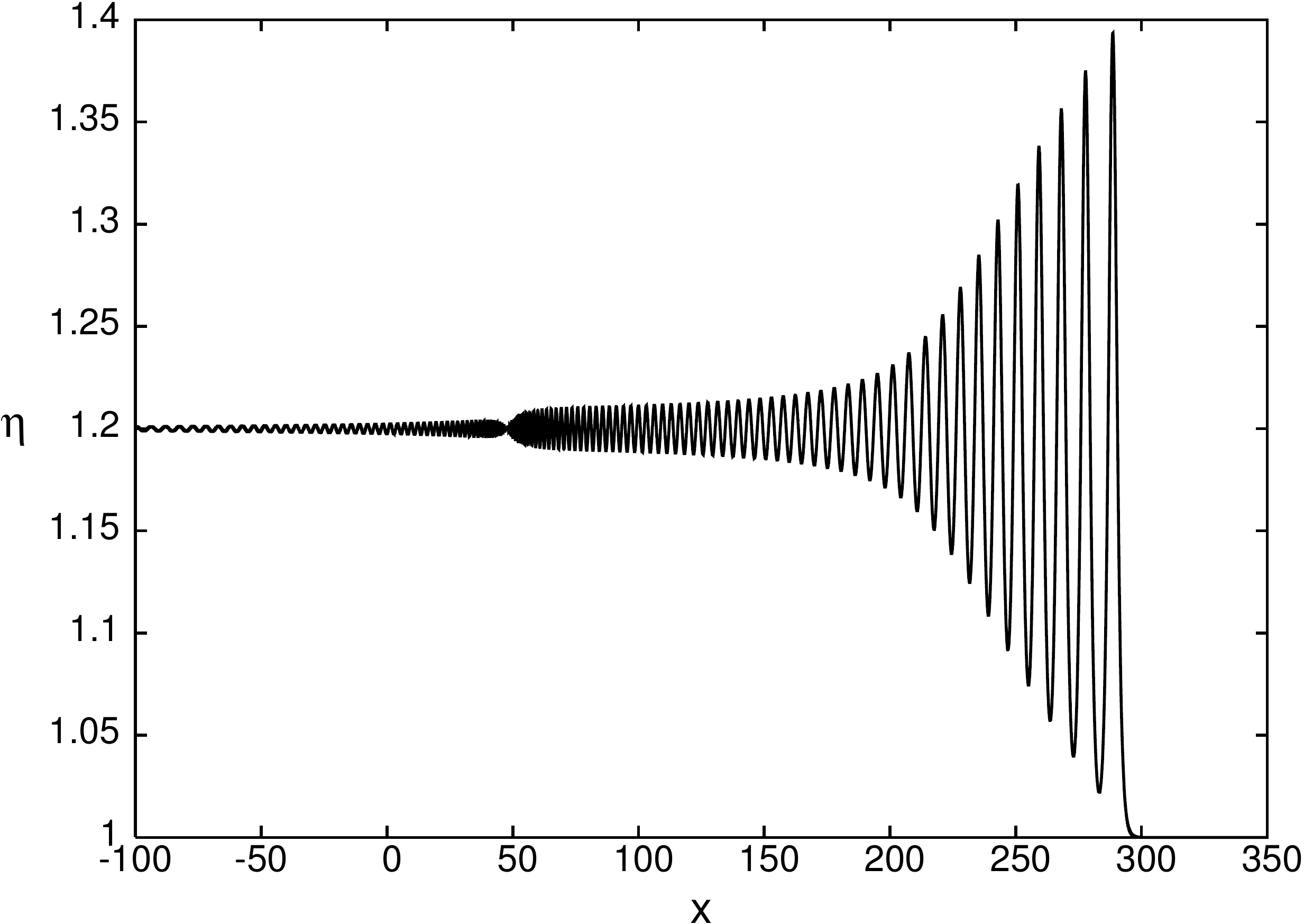}} \caption{The
    numerical solution of the Serre equations (\ref{SG}) for smoothed 
    Riemann initial data producing a single 2-DSW at $t=250$. Initial
    Riemann data: $\eta_-=\Delta=1.2, \ \eta_+=1, u_-=2(\sqrt{\eta_-}
    -1), \ u_+=0$.  Reprinted with permission from
    \cite{el_unsteady_2006}, copyright 2006, American Institute
    of Physics.} \label{fig:serre_DSW}
\end{figure}

The dispersion relation of (\ref{SG}) for linear waves propagating on
the background $\eta = \bar \eta$, $u=\bar u$ has the form
\begin{equation}
  \label{ldr}
  \omega_\pm(k;  \bar \eta, \bar u)=k \bar u \pm 
  \frac{k \bar \eta^{1/2}}{(1+ \bar \eta^2k^2/3)^{1/2}}\, .
\end{equation}
One can readily see that $\sgn[\partial_{kk} \omega_+]=-1$, implying
the orientation $d=1$ for 2-DSWs. The solitary wave solutions of
(\ref{SG}) are waves of elevation, hence the polarity $p=1$ of DSWs in
terms of $\eta$, propagating on the background $\eta=\bar \eta, u=\bar
u$, and are characterized by the following speed-amplitude relation
\cite{el_unsteady_2006}
\begin{equation} \label{serre_speed-amp}
  c_s(a; \bar  \eta, \bar u)=\bar u + \sqrt{\bar \eta 
    +a} .
\end{equation} 
The dispersionless limit of the Serre equations is the ideal
shallow-water system. Thus the locus of faster, 2-DSWs connecting the
left state $\eta=\eta_-, \ u=u_-$ and the right state $\eta=\eta_+, \
u=u_+$ is given by the dispersionless limit 2-wave curve:
\begin{equation}
  \label{serre_locus}
  u_+ - 2 \eta_+^{1/2} = u_- - 2 \eta_-^{1/2}.
\end{equation}
Without loss of generality, we assume $\eta_+=1$, $u_+=0$.  Numerical
solution of the Serre equations with the Riemann data satisfying
(\ref{serre_locus}) is shown in Fig.~\ref{fig:serre_DSW} and indeed
exhibits a single, 2-DSW with $d=1$, $p=1$.

Incorporating the DSW locus (\ref{serre_locus}) into the linear
dispersion relation (\ref{ldr}) by imposing the constraint $\bar u
(\bar \eta)=2(\bar \eta^{1/2}-1)$, we obtain for the frequency of
the fast linear waves
\begin{equation}\label{OM}
  \begin{split}
    \Omega (k, \bar \eta)= \omega_+(k; \bar \eta, \bar u(\bar \eta)) \\
    = 2k(\bar \eta^{1/2}-1)+ \frac{k\bar
      \eta^{1/2}}{(1+\bar \eta^2k^2/3)^{1/2}} \, .
  \end{split}
\end{equation}
The second ingredient for DSW fitting analysis is the simple 2-wave's
characteristic speed for the dispersionless limit equations of $\bar
\eta, \bar u$,
\begin{equation}
  V(\bar \eta)=\bar u(\bar \eta)+ \bar \eta^{1/2}=3\bar
  \eta^{1/2}-2. 
\end{equation}
Then the characteristic ODEs \eqref{eq:144}, \eqref{eq:145} for the
trailing and leading edge become
\begin{equation}\label{ode1}
  \frac{dk}{d\bar \eta}=\frac{\partial \Omega / \partial \bar
    \eta}{V(\bar \eta) - \partial \Omega / \partial k} \, , \qquad
  k(1)=0\, ,
\end{equation}
\begin{equation}\label{ode2}
  \frac{d \tilde k}{d\bar \eta}=\frac{\partial
    \tilde  \Omega /
    \partial \bar \eta}{V(\bar \eta) - \partial  \tilde \Omega / \partial \tilde k} \,
,\qquad \tilde k(\eta_-)=0 \, ,
\end{equation}
where $\tilde \Omega (\tilde k, \bar \eta)=-i \Omega (i \tilde k, \bar
\eta)$.  Given solutions of (\ref{ode1}), (\ref{ode2}), the velocities
of the trailing and leading edges are found as
\begin{equation}\label{spm}
  s_-=\frac{\partial \Omega }{\partial k}( k_-, \eta_-)\, , \qquad
  s_+=\frac{\Omega(\tilde k_+, 1 )}{\tilde k_+} \, ,
\end{equation}
where $k_-=k(\eta_-)$ and $\tilde k_+=\tilde k(1)$.

Equations (\ref{ode1}), (\ref{ode2}) assume a separable form by
introducing $\alpha=(1+ k^2 \bar \eta^2/3)^{-1/2}$ and $\tilde
\alpha=(1- \tilde k^2 \bar \eta^2/3)^{-1/2}$ instead of $k$ and
$\tilde k$ respectively. Then (\ref{ode1}) becomes
\begin{equation}\label{sep1}
  \frac{d\bar \eta}{\bar
    \eta}=\frac{2(1+\alpha+\alpha^2)}{\alpha(1+\alpha)(\alpha-4)}d\alpha\,
  , \quad \alpha(1)=1 \, .
\end{equation}
Equation (\ref{sep1}) is readily integrated to give
\begin{equation}\label{res}
\bar\eta
=\frac{1}{\sqrt{\alpha}}\left(\frac{4-\alpha}{3}\right)^{21/10}
\left(\frac{1+\alpha}{2}\right )^{2/5}\, .
\end{equation}
Next, using (\ref{spm}), (\ref{OM}) we obtain an implicit expression
for the trailing edge $s_-$ in terms of the total depth ratio across
the DSW $\Delta=\eta_-/\eta_+=\eta_-$:
\begin{equation}\label{tr_serre}
\begin{split}
    \sqrt{\beta}\Delta-\left(
    \frac{4-\beta}{3}\right)^{21/10}\left(\frac{1+\beta}{2}\right)^{2/5}=0\,, \\
\hbox{where} \ \ \beta = \left(
    \frac{2+s_-}{\sqrt{\Delta}}-2\right)^{1/3} \, .
    \end{split}
\end{equation}
Similarly, integrating the ODE (\ref{ode2}) and using the definition
of the leading edge velocity (\ref{spm}), we obtain an implicit
equation for $s_+$,
\begin{equation}\label{lead_serre}
  \frac{\Delta}{\sqrt{s_+}}-\left(\frac{3}{4-s_+}\right)^{21/10}
  \left(\frac{2}{1+s_+}\right)^{2/5}=0 \, .
\end{equation}
\begin{figure}[ht]
  \centerline{ \includegraphics[scale=0.25] {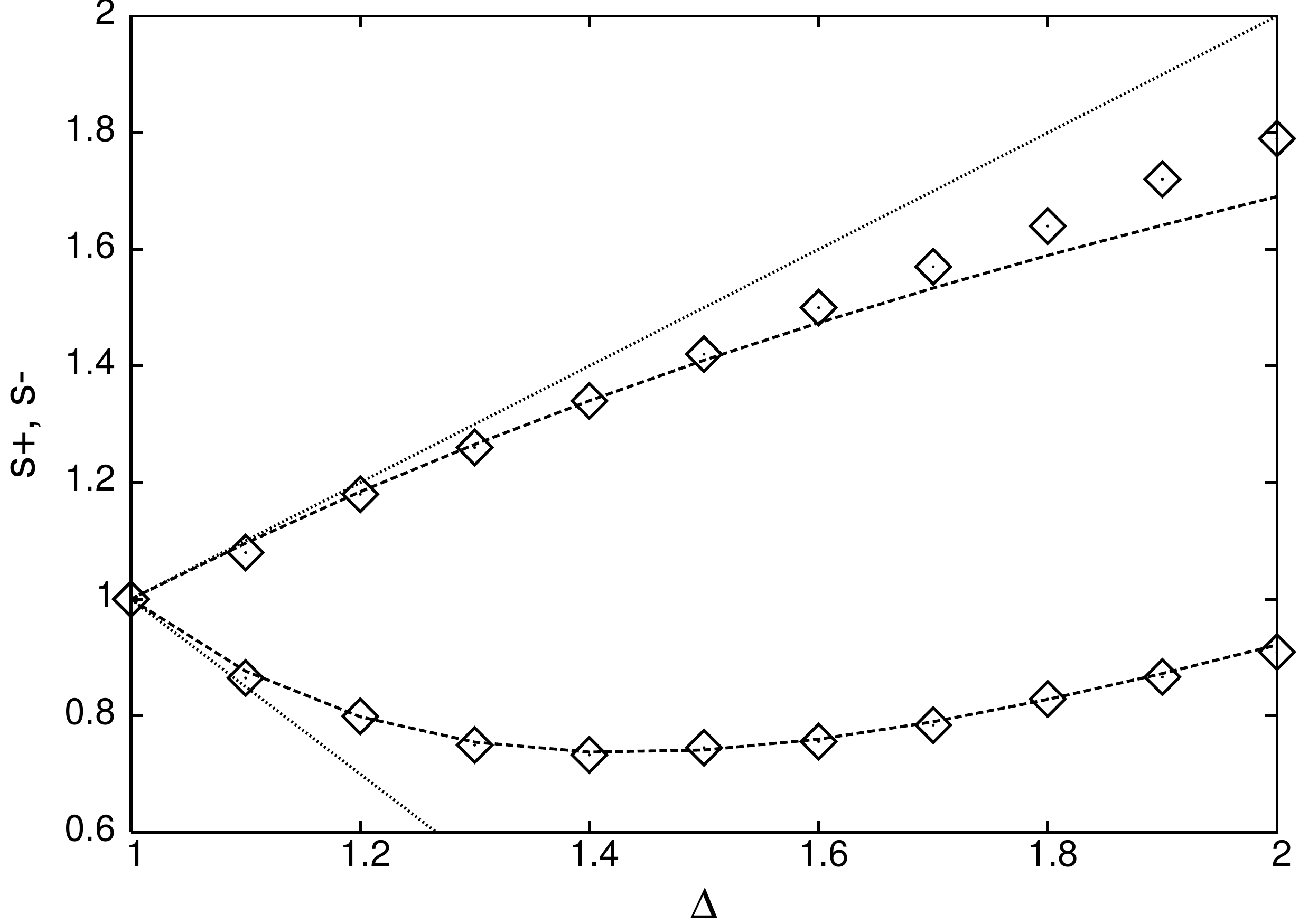}}
  \caption{The leading $s_+$ (upper curve) and trailing $s_-$
    (lower curve) edge speeds vs the depth ratio $\Delta$ across the
    2-DSW of the Serre equations. Dashed line: modulation solution
    (\ref{tr_serre}), (\ref{lead_serre}); Diamonds: values of $s_-$
    extracted from the full numerical solution; Dotted line: the KdV
    DSW boundaries.  Reprinted with permission from
    \cite{el_unsteady_2006}, copyright 2006, American Institute
    of Physics. }\label{fig:serre_edges}
\end{figure}
From the expression (\ref{serre_speed-amp}) relating the speed and the
amplitude of the soliton, and the identification of the leading edge
speed with the soliton speed $s_+=c_s(a_+; 1, 0)$, we derive the
equation for the amplitude $a_+$ of the leading solitary wave for the
free surface elevation,
\begin{equation}\label{leada}
  \begin{split}
    \frac{\Delta}{(a_++1)^{1/4}}-\left(\frac{3}{4-\sqrt{a_++1}}\right)^{21/10} \\
    \times \left(\frac{2}{1+\sqrt{a_++1}}\right)^{2/5}=0 \, .
  \end{split}
\end{equation}
Expansion of (\ref{leada}) for small jumps $\delta=\Delta-1$ yields
$a^+=2\delta + O(\delta^2)$ which agrees to leading order with the KdV
result eq.~\eqref{eq:9}.

Now we need to check the admissibility conditions. The causality
inequalities \eqref{eq:136} assume the form
\begin{equation} \label{in2}
  \begin{split}
    \Delta^{1/2} - 2 <s_-< 3 \Delta^{1/2} -2 \, , \\
    s_+>1\, , \quad s_+>s_-\, ,
  \end{split}
\end{equation}
and can be shown to be satisfied for all $\Delta$
\cite{el_unsteady_2006}.  However, the monotonicity condition for
$s_-(\Delta)$ (eq.~\eqref{eq:119}) can be shown to exhibit breakdown
at $\Delta=\Delta_c \approx 1.43$. This is the indication of linear
degeneracy of the modulation system occuring for $\Delta> \Delta_c$
and thus, inapplicability of the simple-wave integral curve for such
jumps. The comparison of the modulation solution and numerical curves
for the DSW speeds as function of $\Delta$ is shown in
Fig.~\ref{fig:serre_edges} and demonstrates excellent agreement for a
broad range of $\Delta$. One can see that the analytical and numerical
curves for the leading edge start to diverge for $\Delta > \Delta_c$
but, surprisingly, the trailing edge curve still agrees with numerics
even for values of $\Delta$ well beyond $\Delta_c$ where, formally,
the simple wave DSW fitting is not applicable. Also shown in
Fig.~\ref{fig:serre_edges} are appropriately scaled speeds of the KdV
DSW. One can see that for small jumps the KdV and Serre DSW speeds
agree quite well but quickly start to diverge as $\Delta$ increases.

An interesting observation made in \cite{el_unsteady_2006} is that for
large values of $\Delta$, the DSW apparently satisfies the classical
Rankine-Hugoniot relations rather than the DSW locus
(\ref{serre_locus}). A similar observation has been made for DSWs in
the photo-refractive NLS equation \cite{el_theory_2007} where
simple-wave DSW theory also experiences linear degeneracy breakdown,
although the nonmonotonicity is now exhibited by the leading edge due
to a different dispersion sign \cite{hoefer_shock_2014}, see Section \ref{sec:nonlinear-optics} below. This apparent
effect of the Rankine-Hugoniot locus appearing in conservative
dispersive systems still requires explanation.

Concluding this section, we note that the DSW fitting method can be
extended to estimate the number of solitary waves and their
distribution over amplitude in solitary wave trains developing from
large-scale, decaying initial profiles $\eta_0(x)$, $u_0(x)$
satisfying the simple-wave condition $u_0(x)-2\sqrt{\eta_0(x)}=const$
for fast waves \cite{el_asymptotic_2008}. The results of
\cite{el_asymptotic_2008} are ``non-integrable'' analogues of the
Karpman results \cite{karpman_asymptotic_1967,karpman_non-linear_1974}
for the KdV equation based on semi-classical Bohr-Sommerfeld
quantization, see Section \ref{sec:evolution_decaying_profiles}.

\subsubsection{DSWs in nonlinear optics}
\label{sec:nonlinear-optics}

Intense laser beam propagation through defocusing (normal dispersion)
optical media can lead to the generation of DSWs.  Observations in
optical fibers
\cite{rothenberg_observation_1989,liu_wave-breaking-extended_2012,fatome_observation_2014}
and in spatial optics
\cite{wan_dispersive_2007,jia_dispersive_2007,ghofraniha_shocks_2007,conti_observation_2009,ghofraniha_measurement_2012,barsi_spatially_2012}
support the interpretation of light as a dispersive hydrodynamic
medium.

\begin{figure}
  \centering
  \includegraphics[width=0.7\columnwidth]{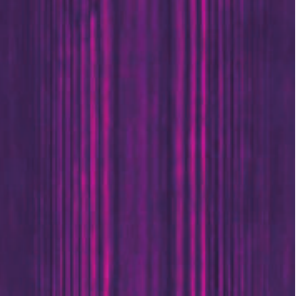}
  \caption{Image of the output face of a defocusing, photorefractive
    crystal resulting from a pulse at the input face.  Two
    counter-propagating, one-dimensional DSWs resulting from
    wavebreaking.  Reprinted by permission from Macmillan Publishers
    Ltd: Nature Physics \cite{wan_dispersive_2007}, copyright
    2007.}
  \label{fig:photorefractive_dsw}
\end{figure}

The dispersive hydrodynamics of light can be modeled with the gNLS
equation (recall \eqref{eq:111})
\begin{equation}
  \label{eq:101}
  i \psi_z + \frac{1}{2} \nabla^2\psi - f(|\psi|^2) \psi = 0 ,
\end{equation}
where $z$ is the propagation direction, playing the role of time, and
the Laplacian term represents either diffraction in the transverse
direction $\nabla^2 = \partial_{xx} + \partial_{yy}$ for spatial
optics or temporal dispersion $\nabla^2 = \partial_{tt}$ in fibers.
The nonlinearity $f$, assumed positive and an increasing function of
its argument, is due to an intensity dependent refractive index.  The
simplest example is the Kerr electrooptic effect where $f(\rho) =
\rho$ \cite{boyd_nonlinear_2013}, and we obtain the integrable,
defocusing NLS equation \eqref{eq:NLS}, $\sigma = 1$ assuming one
transverse direction.  This form is often used to model fibers.
Another example is photorefractive, spatial media where the
nonlinearity takes the form \cite{boyd_nonlinear_2013}
\begin{equation}
  \label{eq:102}
  f(\rho) = \frac{\rho}{1 + \gamma \rho} , \quad \gamma > 0.
\end{equation}
Two-dimensional DSWs imaged from a photorefractive crystal are shown
in Fig.~\ref{fig:dsw_expt_super_bec_optics}c,d and two
counter-propagating one-dimensional DSWs are shown in
Fig.~\ref{fig:photorefractive_dsw} \cite{wan_dispersive_2007}.  Other
examples include nonlocal, e.g., thermal \cite{ghofraniha_shocks_2007}
or nematic liquid crystal \cite{assanto_collisionless_2008}, media
where $f$ is found by inverting an elliptic operator, an additional
source of dispersion.  We will focus on the one-dimensional case,
taking $\nabla^2 = \partial_{xx}$ for simplicity.

The hydrodynamic form for the gNLS equation is given in
eqs.~\eqref{eq:110} and \eqref{eq:113} via the mapping of the complex
wavefunction $\psi \to \sqrt{\rho} \exp( i\int u \rmd x)$
(eq.~\eqref{eq:112}) and we take $t \to z$.  DSWs for gNLS were
constructed utilizing the DSW fitting method for photorefractive
nonlinearity \eqref{eq:102} in \cite{el_theory_2007} and for general
nonlinearities in \cite{hoefer_shock_2014}.  In what follows, we first
identify some properties of the gNLS equation and then apply the DSW
fitting method to the photorefractive nonlinearity case.

We verify the requisite ingredients to apply Whitham theory.  The
dispersion relation for the hydrodynamic form of the gNLS equation is
\begin{equation*}
  \begin{split}
    \omega_{\pm}(k,\rho_0,u_0) &= u_0 k \pm k\sqrt{c_0^2 + k^2/4} \\
    &\sim
    V_\pm(\rho_0,u_0) k \, \pm \, \frac{1}{8c_0} k^3 + O(k^5),
  \end{split}
\end{equation*}
where 
\begin{equation*}
  \begin{split}
    V_\pm(\rho_0,u_0) &= u_0 \pm c_0, \\
    c_0 &= \sqrt{\rho_0 f'(\rho_0)} ,
  \end{split}
\end{equation*}
so that the dispersion sign is positive $\mathrm{sgn} \, \partial_{kk}
\omega_+ > 0$ for 2-DSWs.  This suggests that the 2-DSW will have
orientation $d = -1$. For photorefractive nonlinearity \eqref{eq:102},
we obtain the long wavelength speed of sound
\begin{equation*}
  c(\rho) = \frac{\sqrt{\rho}}{1 + \gamma \rho} .
\end{equation*}
The dispersionless limit is strictly hyperbolic and genuinely
nonlinear so long as $(\rho f'(\rho))' > 0$.  This enforces the
restriction $\gamma \rho < 1$ in the case of photorefractive
nonlinearity.  The gNLS equation admits the local energy conservation
law \cite{jin_semiclassical_1999}
\begin{equation*}
  \begin{split}
    &\mathcal{E} = \frac{1}{2} \rho u^2 + \frac{\rho_x^2}{8\rho} +
    \int_0^\rho f(\tilde \rho) \rmd \tilde \rho, \\
    &\mathcal{E}_z + \big ( u( \mathcal{E} + P(\rho) ) \big )_x =
    \frac{1}{4} \left ( u \rho_{xx} - \frac{\rho_x (\rho u)_x}{\rho}
    \right )_x .
  \end{split}
\end{equation*}
A traveling wave solution $\rho(x,z) = \tilde \rho(x - Vz)$, $u(x,z) =
\tilde u(x-Vz)$ satisfies (dropping tildes)
\begin{equation*}
  \begin{split}
    u &= V + \frac{A}{\rho} \\
    (\rho')^2 &= 8 \rho \int_0^\rho f(\tilde \rho) \rmd \tilde \rho + B
    \rho^2 + C \rho - 4 A^2 \\
    &\equiv G(\rho),
  \end{split}
\end{equation*}
where $A$, $B$, and $C$ are integration constants.  For 
photorefractive nonlinearity \eqref{eq:102}, 
\begin{equation*}
  G(\rho) = -\frac{8}{\gamma^2} \rho \ln(1 + \gamma \rho) + B
  \rho^2 + C \rho - 4 A^2,
\end{equation*}
where we have absorbed an integration constant by redefining $B$.  We
assume the existence of three real, ordered roots $G(b_i) = 0$, $i =
1,2,3$.  Since $G(0) = -4A^2 < 0$, the density of the periodic
solution oscillates between $0 < b_1$ and $b_2$. The three parameters $A$,
$B$, and $C$ can be related to the three roots.  The roots
$\{b_i\}_{i=1}^3$ and $V$ can further be related to the physical
modulation variables $\rhob$, $\ub$, $k$, and $\kt$ per the standard
approach described in Sec.~\ref{sec:dsw-fitting-method}.  Finally, we
can identify the soliton amplitude-speed relation in the limit $b_2
\to b_3$
\begin{equation}
  \label{eq:152}
  \begin{split}
    (V-\ub)^2 = &\frac{2 \rho_{\rm min} f(\rhob)}{\rhob - \rho_{\rm
        min}} \\
    & \qquad -\frac{2 \rho_{\rm min}}{(\rhob - \rho_{\rm
        min})^2} \int_{\rho_{\rm
          min}}^{\rhob} f(\tilde \rho) \rmd \tilde \rho ,
  \end{split}
\end{equation}
where $\rho_{\rm min} \equiv \min_{\xi \in \R} \rho(\xi)$ is the
minimum of the ``dark'' or depression soliton implying the DSW
polarity $p = -1$.

The final ingredients we need to implement DSW fitting are the Riemann
invariants of the dispersionless limit \eqref{eq:131}
\begin{equation*}
  r_\pm = u \pm \int \sqrt{\frac{f'(\rho)}{\rho}}
  \rmd \rho .
\end{equation*}
For photorefractive nonlinearity, we have
\begin{equation*}
  r_\pm = u \pm \frac{2}{\sqrt{\gamma}} \mathrm{arctan}\,(\sqrt{\gamma
    \rho}) .
\end{equation*}

We now focus upon the case of photorefractive nonlinearity as in
\cite{el_theory_2007}.  We introduce the locus of left states
$(\rhob,\ub)$ that can be connected to the right state $(\rho_{\rm
  r},u_{\rm r})$ by a 2-DSW \eqref{eq:135}
\begin{equation}
  \label{eq:155}
  \ub(\rhob) = u_{\rm r} + \frac{2}{\sqrt{\gamma}} 
  \mathrm{arctan}\left (\frac{\sqrt{\gamma \rhob} -
      \sqrt{\gamma \rho_{\rm r}}}{1 + \gamma\sqrt{\rhob\rho_{\rm r}}}
  \right ) , \quad \rhob > \rho_{\rm r}. 
\end{equation}
This is the local relation \eqref{eq:142} that holds for a simple wave
solution along the exact reductions of the Whitham equations.

Because the DSW orientation $d = -1$, the DSW leading edge speed $s_{\rm r}$
is associated with the harmonic limit and the simple wave ODE
\eqref{eq:144} subject to the zero wavenumber boundary condition at
the left edge $k(\rho_{\rm l}) = 0$.  This ODE for $k(\rhob)$ is somewhat
complicated and can be simplified by the transformation
\begin{equation}
  \label{eq:156}
  \alpha(\rhob) = \frac{\omega_+ - \ub(\rhob) k}{c(\rhob) k} = \sqrt{1 +
    \frac{k^2(1+\gamma \rhob)^2}{4 \rhob}} ,
\end{equation}
to a scaled phase velocity.  Then $\alpha(\rhob)$ satisfies
\begin{equation}
  \label{eq:157}
  \begin{split}
    \frac{\rmd \alpha}{\rmd \rhob} &= - \frac{(1+\alpha)\left ( 1 +
        3\gamma \rhob + 2 \alpha (1 - \gamma \rhob) \right
      )}{2\rhob(1+\gamma \rhob)(1 + 2 \alpha)},
  \end{split}
\end{equation}
subject to the initial condition
\begin{equation*}
  \alpha(\rho_{\rm l}) = 1 .
\end{equation*}
Integration of \eqref{eq:157} results in $\alpha_{\rm r} =
\alpha(\rho_{\rm r})$, which via \eqref{eq:156} yields $k_{\rm r}$.
The leading edge speed is therefore the group velocity $s_{\rm r}
= \partial_k \omega_+(\rho_{\rm r},u_{\rm r},k_{\rm r})$.  

A similar analysis at the soliton, trailing edge can be performed in
order to determine the trailing edge speed $s_{\rm l}$.  The simple wave ODE
\eqref{eq:145}, subject now to $\kt(\rho_{\rm r}) = 0$, can again be
simplified by the transformation
\begin{equation}
  \label{eq:158}
  \tilde\alpha(\rhob) = \frac{\tilde\omega_+ - \ub(\rhob)
    \kt}{c(\rhob) \kt} = \sqrt{1 - 
    \frac{\kt^2(1+\gamma \rhob)^2}{4 \rhob}} ,
\end{equation}
to a scaled phase velocity.  Then $\tilde\alpha(\rhob)$ satisfies the
\textit{same equation} \eqref{eq:157} with $\alpha \to \tilde \alpha$
subject to the boundary condition at the leading edge
\begin{equation}
  \label{eq:12}
  \tilde \alpha(\rho_{\rm r}) = 1 .
\end{equation}
Integration of \eqref{eq:157} with $\alpha \to \tilde\alpha$ and
\eqref{eq:12} results in $\tilde\alpha_{\rm l} =
\tilde\alpha(\rho_{\rm l})$, which via \eqref{eq:158} yields $\kt_{\rm
  l}$.  The trailing edge speed is therefore the phase velocity
$s_{\rm l} = \tilde\omega_+(\rho_{\rm l},u_{\rm l},\kt_{\rm
  l})/\kt_{\rm l}$.

Here we reach an obstacle that sometimes arises in applications.  The
simple wave ODE \eqref{eq:157} at both the leading and trailing edges
cannot be integrated analytically.  Several options are available.
Recalling that photorefractive nonlinearity \eqref{eq:102} can be
considered a perturbation to the integrable NLS equation when $0 <
\gamma \ll 1$, one could appeal to an asymptotic analysis of
\eqref{eq:157}.  Alternatively, one can numerically integrate
\eqref{eq:157} to obtain predictions for $s_\pm$.  Both of these
strategies were carried out in \cite{el_theory_2007}.

\begin{figure}
  \centering
  \includegraphics{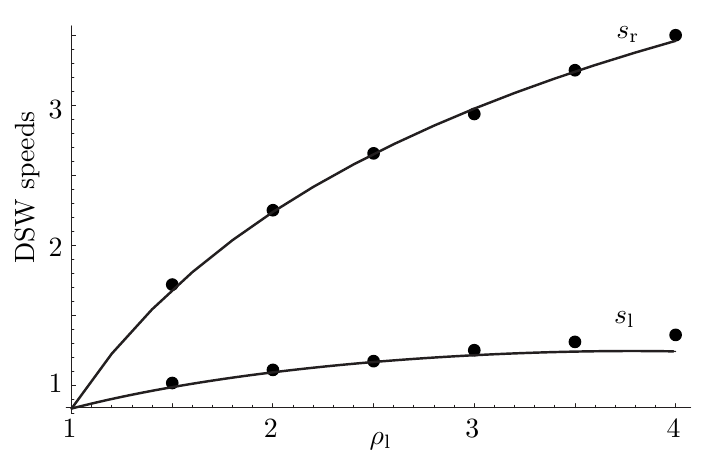}
  \caption{DSW leading $s_{\rm r}$ and trailing $s_{\rm l}$ speeds as
    a function of the left density $\rho_{\rm l}$ with $u_{\rm l} =
    \ub(\rho_{\rm l})$ (eq.~\eqref{eq:155}), $\rho_{\rm r} = 1$,
    $u_{\rm r} = 0$, and $\gamma = 0.2$. Reprinted figure with
    permission from \cite{el_theory_2007} copyright 2007 by the
    American Physical Society.}
  \label{fig:photorefractive_speeds}
\end{figure}
Figure \ref{fig:photorefractive_speeds} depicts the DSW leading and
trailing edge speeds for a 2-DSW computed from the DSW fitting
procedure (solid curves) and from direct numerical simulation of the
gNLS equation \eqref{eq:101} with photorefractive nonlinearity
\eqref{eq:102} \cite{el_theory_2007}.  The DSW fitting procedure
yields excellent agreement across a range of density jumps.  

\begin{figure}
  \centering
  \includegraphics{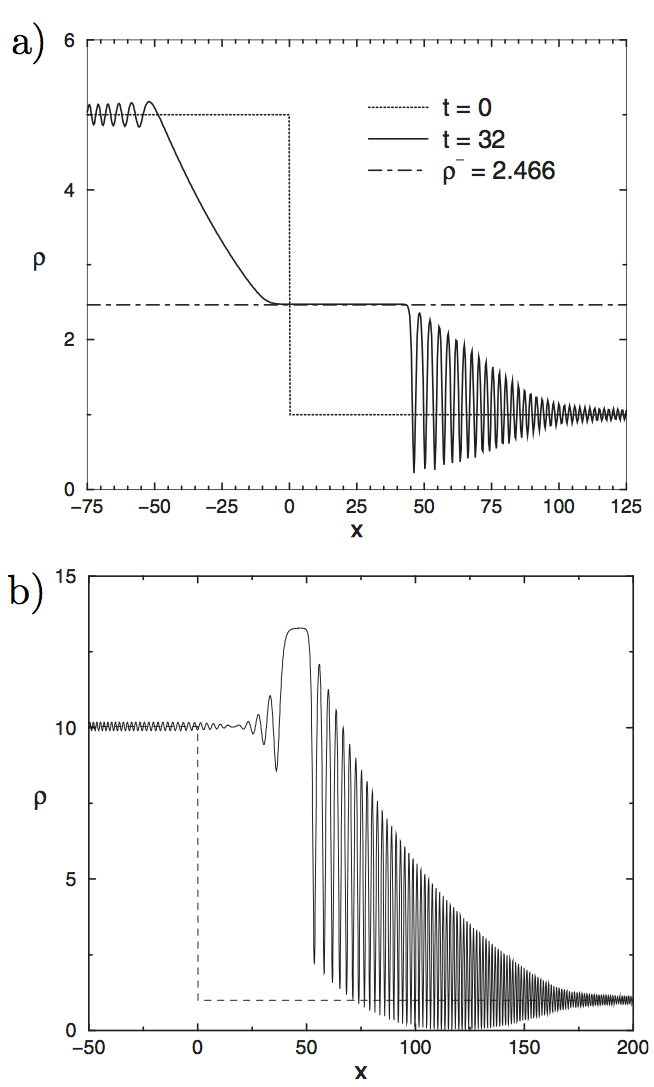}
  \caption{Numerically computed DSWs for the gNLS equation with
    photorefractive nonlinearity. a) Shock tube problem with
    $\rho_{\rm l } = 5$, $\rho_{\rm r} = 1$, $u_{\rm l} = u_{\rm r} =
    0$, and $\gamma = 0.1$.  b) 2-DSW resulting from $\rho_{\rm l} =
    10$, $\rho_{\rm r} = 1$, $u_{\rm l}= \ub(10)$
    (eq.~\eqref{eq:155}), $u_{\rm r} = 0$, $\gamma = 0.2$.  Reprinted
    figures with permission from \cite{el_theory_2007} copyright
    2007 by the American Physical Society.}
  \label{fig:photorefractive_dsws}
\end{figure}

An example DSW resulting from the shock tube problem with $u_{\rm l} =
u_{\rm r} = 0$ and $\rho_{\rm l} > \rho_{\rm r}$ is shown in
Fig.~\ref{fig:photorefractive_dsws}a.  The dispersive shock tube
results in the generation of a 1-rarefaction propagating upstream and
a 2-DSW downstream.  

It was observed in \cite{el_theory_2007} that for 2-DSW density jumps
$\rho_{\rm l}/\rho_{\rm r} \gtrsim 4$, the DSW structure is altered
and there is a significant discrepancy between the prediction for
$s_{\rm l}$ and the observed soliton edge as we can begin to see in
Fig.~\ref{fig:photorefractive_speeds}.  An example profile is shown in
Fig.~\ref{fig:photorefractive_dsws}b.  We see that $\gamma \rho$
significantly exceeds unity ($\gamma = 0.2$ in
Fig.~\ref{fig:photorefractive_dsws}b), therefore the dispersionless
system loses genuine nonlinearity.  As described in
Sec.~\ref{sec:non-classical-dsws}, new wave structures are possible
when scalar dispersive hydrodynamics exhibit a non-convex flux.  The
linearly degenerate point $\gamma \rho = 1$ is the hyperbolic system
analogue of an inflection point in the scalar flux.

Although the simple wave ODE \eqref{eq:157} is complex, additional
analysis is still available \cite{el_theory_2007,hoefer_shock_2014}.
For example, equating the soliton amplitude-speed relation
\eqref{eq:152} to the trailing edge speed $V = s_{\rm l}$, one can
determine the DSW trailing edge amplitude.  Moreover, one can see
directly from eq.~\eqref{eq:152} the possibility that $\rho_{\rm min}
= 0$, i.e., the existence of a vacuum point.  An internal vacuum point
is visible in Fig.~\ref{fig:photorefractive_dsws}b.  There are also
admissibility conditions \eqref{eq:136} to check.  Finally, an
extremum in either of the DSW speeds as a function of $\rho_{\rm l}$
\textit{or} $\rho_{\rm r}$ is associated with a breakdown of the
simple wave assumption \cite{hoefer_shock_2014}.  An extremum in
$s_{\rm l}$ for the 2-DSW constructed here was observed in
\cite{el_theory_2007} when $\rho_{\rm l} \approx 4$, $\rho_{\rm r} =
1$, and $\gamma = 0.2$.  This  critical jump height was exceeded for
the simulation in Fig.~\ref{fig:photorefractive_dsws}b.

In summary, the DSW fitting method applied to non-integrable
dispersive hydrodynamic problems yields practical results that agree
with numerical simulations across a wide range of parameters.  The
break down of the method can be identified by structural properties of
the zero amplitude or zero wavenumber reductions of the Whitham
modulation equations \cite{hoefer_shock_2014}.



\section{Non-classical DSWs}
\label{sec:non-classical-dsws}


The DSW theory described in the previous sections can be viewed as the
dispersive-hydrodynamic counterpart of classical viscous shock theory.
The admissibility criteria for such viscous shocks include the
causality conditions, termed Lax entropy conditions
\cite{lax_hyperbolic_1973}. An analogous set of causality conditions
has been introduced for DSWs (see
Sec.~\ref{sec:dsw-fitting-method}). In classical shock theory, the
notion of a Lax shock is closely related to convexity of the flux of
the underlying hyperbolic conservation law $u_t+f(u)_x=0$: in a system
with convex or concave flux, i.e., $f''(u) \ne 0$ for a scalar
equation, all admissible shock waves are Lax shocks.

When the flux $f(u)$ is non-convex, i.e. $f''(u)=0$ at some point
within the range of $u$, new wave features emerge in hyperbolic shock
theory including composite wave solutions and undercompressive shocks
that do not satisfy the Lax entropy condition
\cite{lefloch_hyperbolic_2002}.  
One of the standard approaches to shock admissibility analysis
involves regularization of the equation by higher order terms that are
dissipative and dispersive, representing more of the physics than is
contained in the hyperbolic conservation law alone.  For scalar
conservation laws with convex flux, the fundamental model describing a
diffusive-dispersive regularization is the KdV-Burger's equation
(\ref{eq:2}) considered in Sec.~\ref{sec:anat-disp-shock}. For {\it
  non-convex} conservation laws, such a model is the modified
KdV-Burger's (mKdVB) equation, \be \label{mkdvb} u_t+ (u^3)_x= \nu
u_{xx}+\mu u_{xxx}, \quad \nu \ge 0, \ \ \mu \ne 0.  \ee The
remarkable feature of the mKdVB equation is that it exhibits, for
$\mu>0$, traveling waves called {\it undercompressive shocks} because
the corresponding limiting ($\nu/\sqrt{|\mu|} \to 0$) shock wave
\be\label{shock1} u(x,t)=\left\{\begin{array}{ll}
    u_-, \quad & x<st\\[6pt]
    u_+, &x>st
\end{array}
\right.  \ee is subsonic both ahead of and behind the wave:
\be\label{ucw} |s|< V(u_\pm) = 3u^2_\pm , \ee
where $V$ is the characteristic velocity or sound speed.

Shock wave theory for the mKdVB equation was constructed in
\cite{jacobs_travelling_1995} by analyzing traveling wave solutions,
where it was shown that undercompressive shocks appear only for
$\mu>0$.  When $\mu<0$ there are no undercompressive shocks but there
are shock-rarefaction complexes exhibiting contact shocks propagating
with the sound speed.

What would be the counterparts of these ``non-classical'' shock
phenomena in the inviscid case when $\nu=0$ so that eq.~(\ref{mkdvb})
becomes the conservative mKdV equation?  Or, putting it the other way
around, what is the effect of a non-convex hyperbolic flux on the
solutions of dispersive Riemann problems, and, in particular, on the
modulation theory of DSWs?  This question is not only of significant
mathematical interest but is also important for applications, such as
oceanography and nonlinear optics where dispersive-hydrodynamic
systems with nonconvex hyperbolic flux are quite common.  A prominent
model is the Gardner equation, the combined KdV-mKdV equation, which
is the standard model for internal waves in stratified fluids
\cite{helfrich_long_2006} but also arises in plasma physics
\cite{ruderman_dynamics_2008} and BEC dynamics
\cite{kamchatnov_nonlinear_2014}. The Gardner equation is
mathematically equivalent to the mKdV equation, however, the
transformation between mKdV and Gardner involves a change of boundary
conditions at infinity, which is not always physically acceptable so
the use of Gardner, rather than mKdV, can be essential for
applications. The fully nonlinear bi-directional counterpart of the
Gardner equation, the Miyata-Choi-Camassa (MCC) system
\cite{choi_fully_1999} is a two-layer generalization of the Serre
system (\ref{SG}).  The dispersionless limit of the MCC system is the
two-layer ideal shallow water system exhibiting non-genuine
nonlinearity \cite{esler_dispersive_2011}. Other important non-convex
dispersive hydrodynamic models include the cubic-quintic NLS equation
\cite{crosta_crossover_2012} describing, in particular, the evolution
of the complex wavefunction $\psi$ of a Bose-Einstein condensate with
two- and three-body interactions, and the so-called derivative NLS
equation for nonlinear Alfven waves in dispersive magnetohydrodynamics
\cite{kennel_nonlinear_1988}.

\begin{figure}
  \centering
  \includegraphics{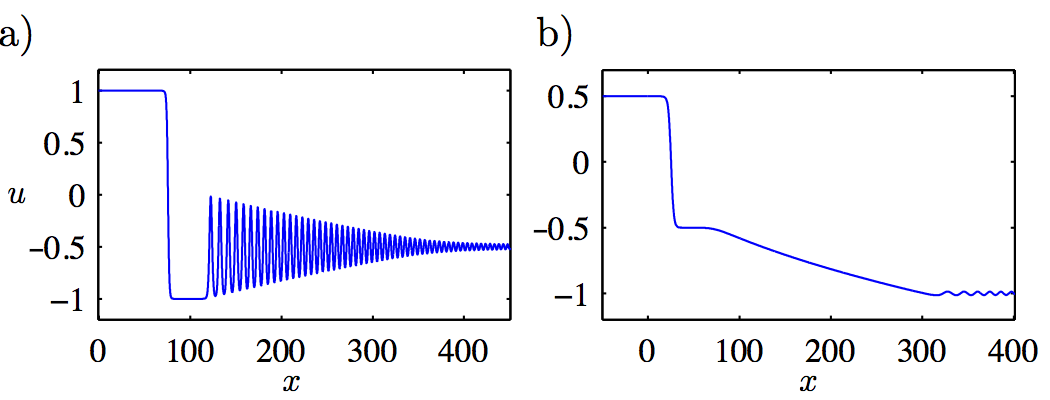}
  \caption{Numerical solutions of the mKdV equation (\ref{mkdvb})
    ($\nu=0$, $\mu = 1 > 0$) with step-like initial data. a) kink-DSW;
    b) kink-rarefaction.} \label{fig:kink_DSW}
\end{figure}

Analysis of the traveling wave solutions of the mKdV (or Gardner)
equation \cite{kamchatnov_undular_2012,el_dispersive_2015} shows that,
along with familiar KdV type cnoidal waves and solitons of both
polarities, due to the symmetry $u \to - u$ of the mKdV equation, two
new types of solutions emerge: for $\mu>0$ these new solutions are
termed {\it kinks} or {\it anti-kinks} while for $\mu<0$ there are
{\it nonlinear trigonometric solutions}. Kinks are smooth, propagating
fronts connecting conjugate states $u \to u_\pm$, $u_- = -u_+$ as $x
\to \pm\infty$ and propagate with the classical shock speed $s= u_+^2$
of the hyperbolic conservation law $u_t+(u^3)_x=0$. The kink solution
of the mKdV equation with $\mu>0$ is described by the formula
\begin{equation}
  \label{kink}
  \begin{split}
    u &= u_+ \tanh[ u_+ \eta] \, ,  \quad
    \eta =\frac{x-u_+^2 t}{\sqrt{2 \mu}} .
  \end{split}
\end{equation}
The kink speed satisfies \eqref{ucw} and thus the kink represents an
{\it undercompressive DSW}, the direct counterpart of the
undercompressive shock wave of the mKdVB equation.  Having the
restricted locus $u_++u_-=0$, kinks are usually generated in Riemann
problems as part of a double wave structure, the other part being
either a ``classical'', KdV-type DSW or a rarefaction wave (see
Fig.~\ref{fig:kink_DSW}). Depending on the initial step parameters,
DSWs of both polarities can be generated (separately or in
combinations with kinks/anti-kinks).

\begin{figure}
  \centering
  \includegraphics{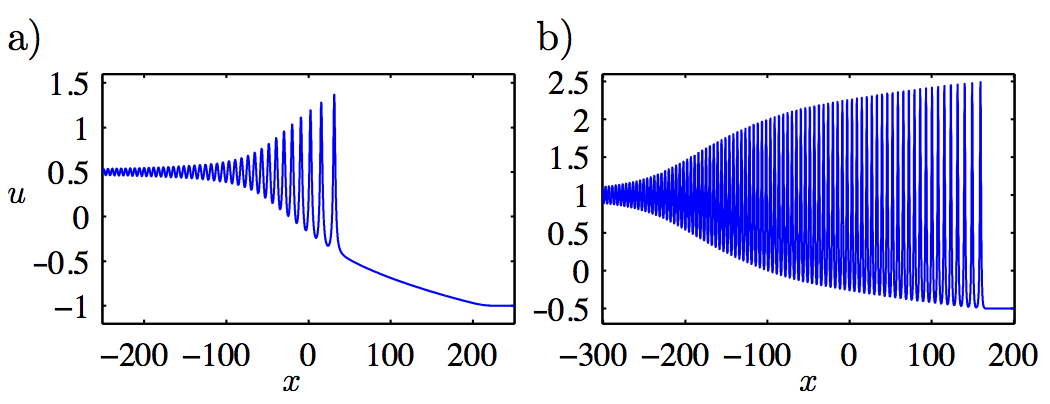}
  \caption{Numerical solutions of the mKdV equation (\ref{mkdvb})
    ($\nu=0$, $\mu = -1 <0$) with step-like initial data. a) CDSW-rarefaction;
    b) CDSW$|$DSW.} \label{fig:CDSW}
\end{figure}

The mKdV equation with $\mu<0$ supports another non-classical family
of traveling wave solutions, which are described in terms of
trigonometric functions with no linearization involved.  These
trigonometric solutions are the ``carrier waves'' for {\it contact
  DSWs} (CDSWs), some of which do not have counterparts in the
diffusive-dispersive theory of the mKdVB equation
\cite{el_dispersive_2015}. CDSWs, similar to kinks, have the
restricted locus $u_++u_-=0$ and so usually are generated in 
combination with a rarefaction (see Fig.~\ref{fig:CDSW}a) or are
realized only partially, forming part of a compound wave --- the
CDSW$|$DSW complex (Fig.~\ref{fig:CDSW}b). When fully realized, a CDSW
exhibits an algebraic soliton at its leading edge.  CDSWs always
exhibit a ``Bordeaux glass'' type envelope.
 
The modulation system for the mKdV equation was first derived in an
analogous fashion to Whitham's derivation of the KdV modulation system
\cite{driscoll_modulational_1975} (see Sec.~\ref{sec:kdv-mod-sys}) by
averaging the first three mKdV conservation laws.
It transpired later
\cite{pavlov_double_1995,kamchatnov_dissipationless_2004-1}, and more
generally in \cite{kamchatnov_undular_2012} using a reduced version of
finite-gap spectral theory \cite{kamchatnov_nonlinear_2000}, that the
modulation system for the mKdV equation with $\mu >0$ can be mapped
{\it onto} the KdV modulation system (\ref{eq:31}).  The mapping can
be viewed as the modulation theory counterpart of the Miura transform
\cite{miura_kortewegvries_1968}. It was also shown in
\cite{driscoll_modulational_1975} that the modulation system for the
mKdV equation with $\mu<0$ can be elliptic, which is the signature of
nonlinear modulational instabilty. However, this unstable regime is
not realized by solutions of the Riemann problem. Further, it was
shown in \cite{el_dispersive_2015} that within the hyperbolic region,
the mKdV-Whitam system is neither genuinely nonlinear nor strictly
hyperbolic, leading to the breakdown of the ``classical'' DSW
admissibility conditions and the occurrence of solutions that cannot
be realized in convex dispersive hydrodynamics. Other modulation
systems exhibiting non-strict hyperbolicity and yielding novel
dynamics not captured by classical DSW theory have been studied in
\cite{pierce_self-similar_2006} (the integrable fifth order KdV
equation), \cite{kodama_whitham_2008} (the complex modified mKdV
equation) and \cite{grava_initial_2009} (the Camassa-Holm equation).

The modulation solutions for DSWs in the mKdV equation with $\mu<0$
were first constructed in \cite{marchant_undular_2008}, where the new
DSW type realized by a modulated nonlinear trigonometric solution was
termed ``sinusoidal undular bore''.   The
analogous DSWs for the Gardner equation were studied in
\cite{kamchatnov_undular_2012}.  Their nature as contact DSWs was
realized in \cite{el_dispersive_2015} by an analysis of the modulation
solution's characteristics.  The composite CDSW-rarefaction exhibits a
\textit{triple characteristic} at the interface between the CDSW and
rarefaction, similar to the properties of a contact shock.

Elements of DSW theory for the mKdV equation with $\mu>0$ appear in
\cite{chanteur_formation_1987} in the context of collisionless plasma
physics.  The detailed description of kink or CDSW emergence in
Riemann problem solutions of the Gardner equation was carried out in
\cite{kamchatnov_undular_2012} for both signs of $\mu$.  These Gardner
equation results were applied to the resonant generation of internal
undular bores in stratified fluid flows over topography
\cite{kamchatnov_transcritical_2013} (see Sec.~\ref{sec:dsw_forced}).
The very recent paper \cite{el_dispersive_2015} contains detailed
analysis of the nontrivial connections between the Riemann problem
classifications of the mKdVB and mKdV equations.

DSW theory for non-integrable systems whose dispersionless limit lacks
genuine nonlinearity is only just emerging with few existing
contributions.  The identification of kinks in fully nonlinear,
near-critical, two-layer flows with undercompressive shocks was made
in \cite{kluwick_near-critical_2007}.  The paper
\cite{esler_dispersive_2011} uses the DSW fitting method to describe
the dispersive dam break and lock exchange flows in the framework of
the MCC system. The authors show that MCC lock exchange flows give
rise to kink-DSW compound wave complexes, similar to those found in
the Gardner and mKdV equations in
\cite{kamchatnov_undular_2012,el_dispersive_2015} (see
Fig.~\ref{fig:kink_DSW}). The occurrence of DSWs with different
polarities related to non-convex dispersive hydrodynamics in some
solutions of the Riemann problem for the cubic-quintic NLS equation
has been reported in \cite{crosta_crossover_2012}.

%

\section{DSW interactions}
\label{sec:dsw-interactions}

\begin{figure}
  \centering
  \includegraphics{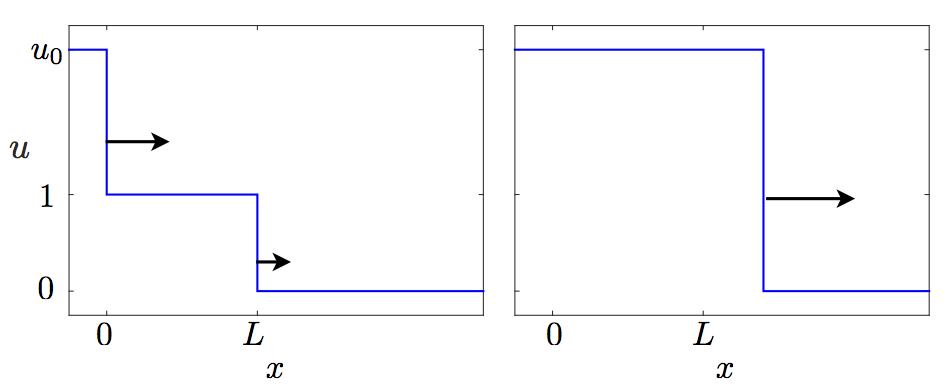}
  \caption{Schematic of an overtaking viscous shock interaction in
    uni-directional dispersionless hydrodynamics with time progressing
    to the right.}
  \label{fig:overtaking}
\end{figure}

\begin{figure}
  \centering
  \includegraphics{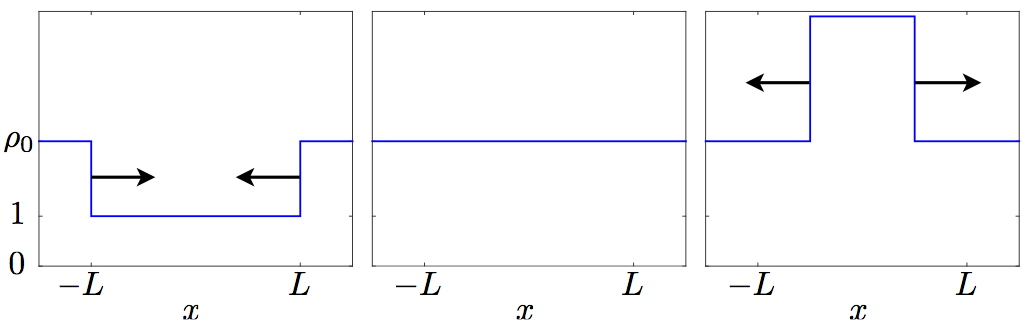}
  \caption{Schematic of the density for a head-on collision of two
    viscous shock waves in bi-directional dispersionless hydrodynamics
    with time progressing rightward.}
  \label{fig:headon}
\end{figure}

DSWs are coherent, long-lived structures.  It is therefore natural to
consider their interaction properties.  Fundamental shock interactions
include an overtaking interaction with two co-propagating shocks and a
head-on interaction with two counter-propagating shocks.  Typical
dynamics for these two interaction types when the hydrodynamics are
\textit{dispersionless} (e.g., for the dispersionless KdV or Hopf
equation \eqref{eq:37} and the dispersionless NLS or shallow water
equations \eqref{eq:sw}) are shown in Figs.~\ref{fig:overtaking} and
\ref{fig:headon}.  The overtaking viscous shock interaction results in
the eventual merger of the two shocks (speeds $s_1 > s_2$) into a
single shock (speed $s_{\rm m}$).  For nonconvex flux, the shock
speeds are guaranteed to be interleaved
\begin{equation}
  \label{eq:4}
  s_2 < s_{\rm m} < s_1 .
\end{equation}
The symmetric head-on collision of two viscous shocks depicted in
Fig.~\ref{fig:headon} shows that, post-interaction, two new shocks are
generated with a larger density and propagate away from one another.

In this section, we consider the canonical overtaking DSW interaction
in the context of uni-directional KdV dispersive hydrodynamics and the
head-on DSW interaction for bi-directional NLS dispersive
hydrodynamics.  We conclude with interactions involving dispersively
regularized rarefaction waves.

\subsection{Overtaking DSW interaction}
\label{sec:kdv-overtaking-dsw}

We consider the KdV equation \eqref{kdv1} with double-step initial
conditions
\begin{equation*}
  u(x,0) =
  \begin{cases}
    u_0 & x < 0 \\
    1 & 0 < x < L \\
    0 & L < x
  \end{cases},
\end{equation*}
where $u_0 > 1$.  We assume that $L$ is large enough so that each DSW
resulting from the initial two steps has sufficient distance to
develop.  Then the simple wave DSW analysis in Sec.~\ref{sec:
  dsw-riemann} applies to the initial two DSWs.  From
eq.~\eqref{eq:52}, the leftmost, faster DSW expands with speeds
$s_{1,-} = -u_0 + 2$, $s_{1,+} = \frac{2}{3} u_0 + \frac{1}{3}$ and
the slower DSW propagates with speeds $s_{2,-} = -1$, $s_{2,+} =
\frac{2}{3}$.  If we simply \textit{assume}, by motivation from the
viscous case, that the two DSWs merge into a single DSW connecting $u
= 0$ and $u = u_0$, then its speeds are $s_{\textrm{m},-} = -u_0$ and
$s_{\textrm{m},+} = \frac{2}{3} u_0$.  This would imply that the
soliton edge DSW speeds are interleaved
\begin{equation}
  \label{eq:164}
  s_{2,+} < s_{\textrm{m},+} < s_{1,+},
\end{equation}
as in the viscous shock case \eqref{eq:4}.

\begin{figure}
  \centering
  \includegraphics[width=\columnwidth]{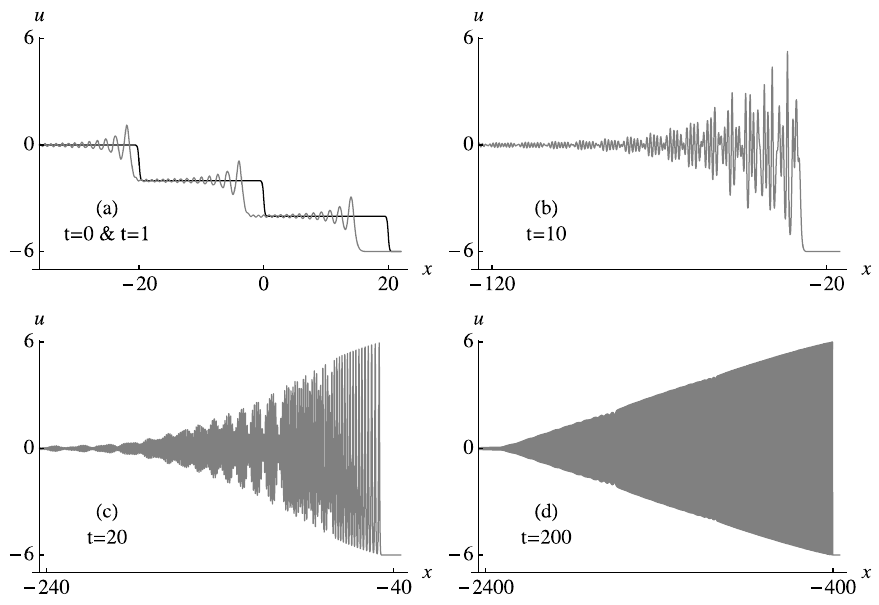}
  \caption{Numerical solution of the KdV equation \eqref{kdv_eps} with
    $\varepsilon^2 = 0.1$ for multi-step initial data.  The dynamics
    display a multiphase interaction region and the effective merger
    into a one-phase DSW after long time.  Reprinted from
    \cite{ablowitz_interactions_2013} with permission from Elsevier.}
  \label{fig:3phase}
\end{figure}

In fact, it has been shown using IST with matched asymptotic
expansions \cite{ablowitz_interactions_2013} and the rigorous IST
Riemann-Hilbert problem approach \cite{egorova_long-time_2013} that a
general class of initial data with boundary conditions $u \to u_\pm$
as $x \to \pm \infty$ and $u_+ > u_-$ result in a \textit{single} DSW,
dispersive radiation, and possibly solitons for large $t$. Thus,
multiple KdV DSWs will eventually merge into a single DSW.  An example
numerical solution of the KdV equation \eqref{kdv_eps} with multi-step
initial data is shown in Fig.~\ref{fig:3phase}
\cite{ablowitz_interactions_2013}.

Modulation theory provides additional information about the dynamics
of step-like initial data \cite{el_generating_1996}.  Recall that a
single DSW is described by KdV-Whitham modulations of a one-phase
wave.  When two DSWs collide, the one-phase equations exhibit a
singularity and there is an increase in the number of phases required
to describe the modulation solution.  The resultant DSW interaction
region is described by two-phase modulations, and can be viewed as the
nonlinear superposition of two one-phase objects.  The multiphase
interaction region is eventually extinguished in long time, leaving
just one-phase modulations (a single DSW).  A rigorous examination of
this was performed in \cite{grava_generation_2002}.

\begin{figure}
  \centering
  \includegraphics{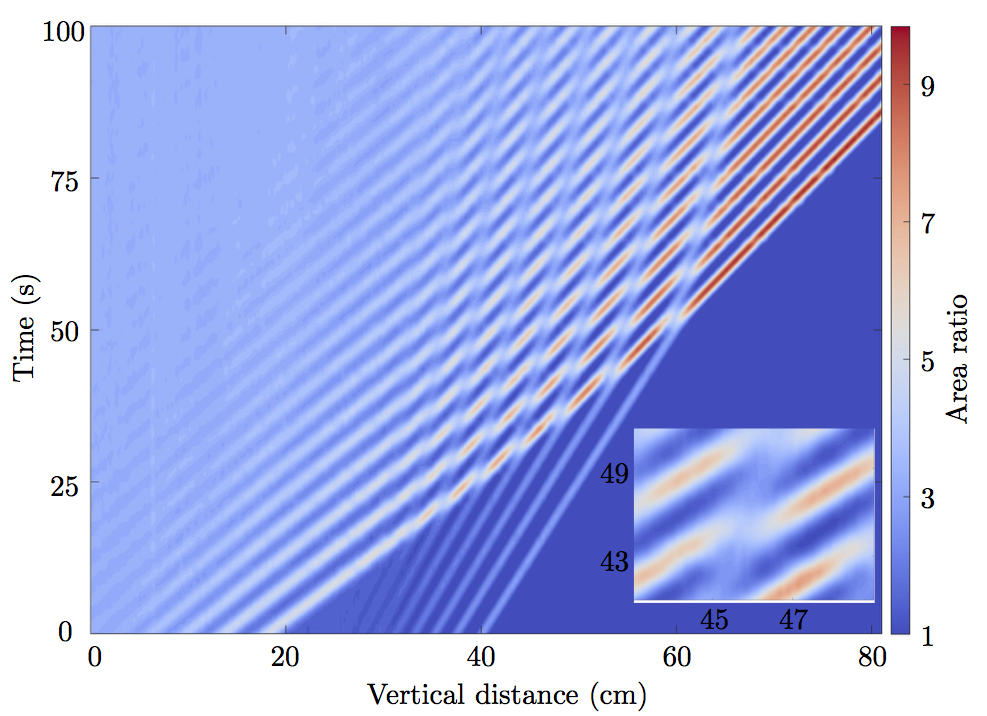}
  \caption{Contour plot in space-time depicting the normalized cross
    sectional area of an overtaking interaction experiment of viscous
    fluid conduit DSWs.  A two-phase interaction region appears
    (inset) and is eventually extinguished.  Figure provided with
    permission from \cite{maiden_wavebreaking_2015}.}
  \label{fig:conduit_overtaking_DSWs}
\end{figure}
The overtaking interaction of two DSWs is therefore quite similar to
the same involving two viscous shock waves.  The main difference is
that, due to the nonzero spatial extent of DSWs, their interaction
occurs over some time period, in contrast to the essentially
instantaneous interaction of viscous shock waves.  Using the
generalized hodograph transform, the KdV modulation solution for the
overtaking interaction of two DSWs was obtained in
\cite{el_generation_2002} and applied to the long-time evolution of
shelves formed behind solitary wave propagation through an
inhomogeneous medium.  Numerical computations of an overtaking DSW
interaction for both the KdV equation and the two-phase KdV-Whitham
equations were performed in \cite{ablowitz_soliton_2009} (see also
Fig.~\ref{fig:kdv_all_interactions}a).

Motivated by this KdV analysis, the merger of two co-propagating DSWs
can be postulated a-priori, therefore making it possible to determine
the macroscopic properties of an overtaking interaction of two DSWs in
other model equations using the DSW fitting approach from
Sec.~\ref{sec:dsw-fitting-method}.  The overtaking interaction of two
DSWs in uni-directional, dispersive hydrodynamics was observed in the
viscous fluid conduit system (Sec.~\ref{sec:visc-fluid-cond})
\cite{maiden_wavebreaking_2015} as shown in
Fig.~\ref{fig:conduit_overtaking_DSWs}.  The appearance and
annihilation of a modulated two-phase interaction region is apparent.
Moreover, the speed interleaving property \eqref{eq:164} is also
apparent and has been derived for conduit DSWs using the DSW fitting
method of Sec.~\ref{sec:visc-fluid-cond}
\cite{maiden_wavebreaking_2015}.

\subsection{Head-on DSW interaction}
\label{sec:nls-head-interaction}

\begin{figure}
  \centering
  \includegraphics{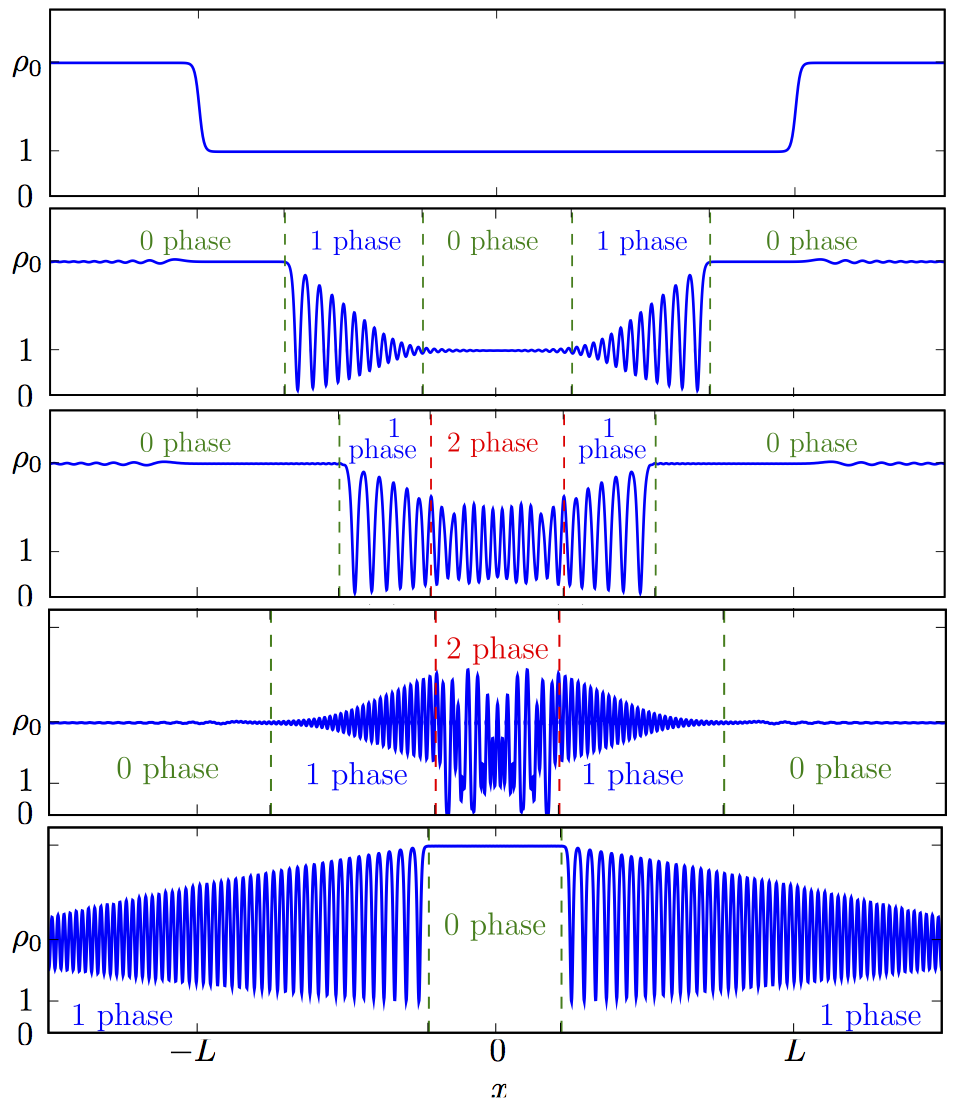}
  \caption{Density from the numerical solution of the NLS equation for a
    symmetric, head-on collision with time progressing downward.  The
    number of phases required to describe each modulation region is
    noted.  Figure adapted, with permission, from
    \cite{hoefer_interactions_2007}.}
  \label{fig:nls_headon}
\end{figure}

We now consider symmetric initial data for the defocusing NLS equation
\eqref{eq:NLS} ($\sigma = 1$) leading to the generation of two
counter-propagating NLS DSWs
\begin{equation}
  \label{eq:165}
  \rho(x,0) =
  \begin{cases}
    \rho_0 & |x| > L \\
    1 & |x| < L
  \end{cases}, \quad u(x,0) =
  \begin{cases}
    u_0, & x < -L \\
    0 & |x| < L \\
    -u_0 & x > L
  \end{cases} ,
\end{equation}
where $\rho_0 > 1$ to guarantee DSW generation and $u_0 =
2(\sqrt{\rho_0} - 1)$ lies on the 2-DSW locus \eqref{eq:57}.  The
result of numerical simulation of a smoothed version of this initial
value problem is depicted in Fig.~\ref{fig:nls_headon}.  Similar to
the KdV case, the interaction of two NLS DSWs results in a two-phase
interaction region.  This two-phase behavior eventually subsides,
leaving an unmodulated, zero phase region in its wake and two new DSWs
propagating away from each other.  These dynamics are, again,
analogous to the dispersionless case of Fig.~\ref{fig:headon}.  

\begin{figure}
  \centering
  \includegraphics{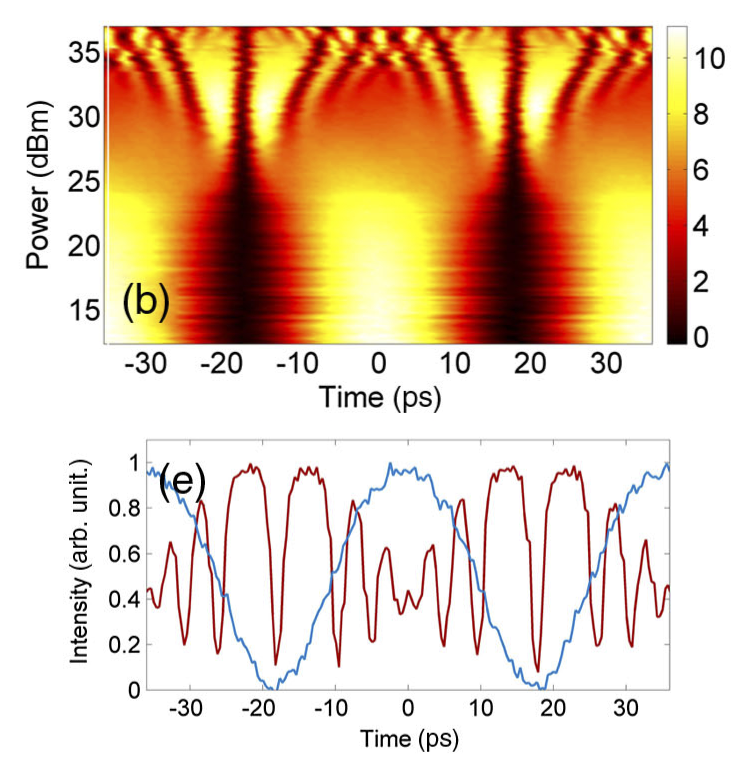}
  \caption{Head-on DSW collision observation in optical four-wave
    mixing.  Top (b): Contour plot of output power (color) versus
    input power (vertical axis) and time (horizontal axis).  Bottom
    (e) Output intensity (red) resulting from cosine input with power
    $P = 34.6$ dBm (blue).  Copyright \cite{fatome_observation_2014},
    DOI:10.1103/PhysRevX.4.021022, Creative Commons 3.0, 2014.}
  \label{fig:four_wave_mixing}
\end{figure}

The head-on collision of two DSWs in the context of optical four-wave
mixing was observed in \cite{fatome_observation_2014}.  Figure
\ref{fig:four_wave_mixing} shows the onset of the collision
process.

As for the multi-step KdV DSW interaction problem, NLS modulation
theory provides the means to identify the spontaneous creation and
annihilation of phases in the modulation solution
\cite{el_general_1995}.  A convenient method to determine the
dispersive regularization for piecewise constant initial conditions is
\textit{initial data regularization}
\cite{bloch_dispersive_1992,biondini_whitham_2006}.  Degenerate
initial conditions for the Whitham equations corresponding to a
minimum number of phases in the modulation solution are chosen so that
a global solution is obtained.  The overlap of two Riemann invariants
effectively nullifies the role of one phase in the modulation
solution.  As the modulation solution evolves, Riemann invariants may
separate (or merge), giving rise to the appearance (or annihilation)
of an additional phase in the solution.  Using this method, it was
shown in \cite{biondini_whitham_2006} that the long time evolution of
piecewise constant initial data for NLS can be described by expanding
zero-, one-, and two-phase regions but no higher phase regions.  In
\cite{hoefer_interactions_2007}, this method was utilized to determine
the space-time boundaries for the number of phases required to
describe the head-on and overtaking interaction of NLS DSWs.  For
example, dispersive regularization of the initial data \eqref{eq:165}
requires six Riemann invariants to guarantee a global solution of the
NLS-Whitham equations.  This implies that, at most, two phases will
appear in the solution.  As shown in Fig.~\ref{fig:nls_headon} for the
symmetric, head-on DSW collision, this two-phase region is
extinguished in finite time.  The overtaking interaction of two NLS
DSWs was also found to result in a single, merged DSW.  

We note that soliton-DSW interactions were also observed in
\cite{maiden_wavebreaking_2015} and shown to exhibit soliton
absorption or soliton refraction, depending on the initial ordering of
the soliton and DSW.  The theory of DSW-soliton interactions has not
been developed.

\subsection{DSW and rarefaction wave interactions}
\label{sec:dsw-rarefaction-wave}

At the level of hyperbolic, dispersionless equations, shocks are not
the only weak solutions available.  Rarefaction waves can also exist
and here we briefly examine their interaction properties in the
context of dispersive hydrodynamics.

\begin{figure}
  \centering
  \includegraphics[width=\columnwidth]{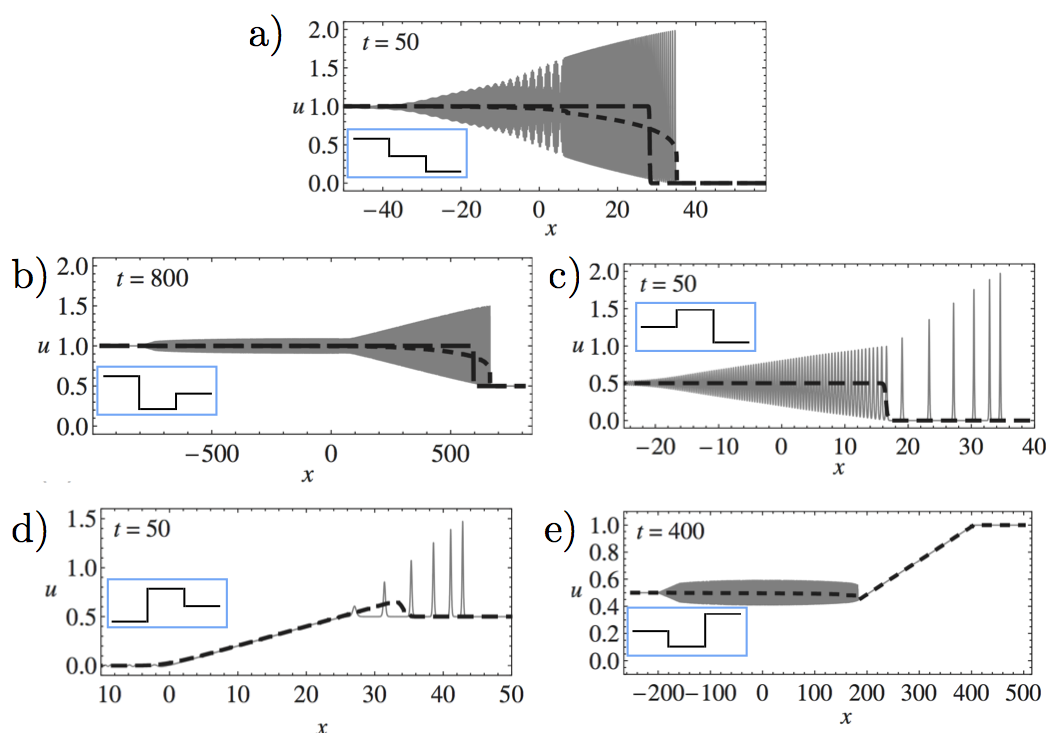}
  \caption{Numerical simulation (solid) and mean $\overline{u}$ from
    Whitham theory (short dash) for five of the six qualitatively
    distinct two-step initial conditions.  The initial condition for
    each panel is shown as an inset.  In a and b, the dissipative
    shock regularization of the Hopf equation is also shown (long
    dash).  Figure adapted, with permission, from
    \cite{ablowitz_soliton_2009}.}
  \label{fig:kdv_all_interactions}
\end{figure}
The six qualitatively distinct KdV wave interaction problems for
two-step initial data were examined in
\cite{el_generation_2002,ablowitz_soliton_2009}.  According to the
classification in Fig.~\ref{fig:gp_classification}, a depression step
gives rise to a DSW whereas an elevation step leads to a rarefaction
wave. These six interaction problems can be conveniently visualized
according to the relative magnitudes of $u$ adjacent to each step as
follows: \caseI, \caseII, \caseIII, \caseV, \caseIV, \caseVI.
Numerical solutions of the first five cases are depicted in
Fig.~\ref{fig:kdv_all_interactions}.  The final case \caseVI, not
shown, results in a single rarefaction.  Case \caseI\ is the
overtaking DSW interaction described in
Sec.~\ref{sec:kdv-overtaking-dsw}
(Fig.~\ref{fig:kdv_all_interactions}a).  Case \caseII\ results in a
DSW on the left that eventually overtakes and absorbs a small
rarefaction (Fig.~\ref{fig:kdv_all_interactions}b).  The long tail is
a remnant of the rarefaction-DSW interaction.  The generation of a
rarefaction on the left that interacts with a DSW, case \caseIII,
leads to a single DSW and a finite number of solitons
(Fig.~\ref{fig:kdv_all_interactions}c).  The case \caseIII\ is
qualitatively similar to the shoaling of an undular bore over a gentle
slope and will be described in Sec.~\ref{sec:dsw-perturbed} (see
Fig.~\ref{fig:shoaling}).  The large rarefaction in case \caseV\
interacts with a DSW that eventually turns into a finite train of
separated solitons (Fig.~\ref{fig:kdv_all_interactions}d).  A small
DSW cannot overtake the large rarefaction of case \caseIV, resulting
in a single rarefaction with a small amplitude oscillatory tail
(Fig.~\ref{fig:kdv_all_interactions}e).

In the NLS case, there are a number of qualitatively distinct
solutions resulting from two-step initial data.  Due to the large
number, they have not all been classified.  We consider two cases
here:  the refraction of a DSW by a rarefaction wave and the
interaction of two degenerate rarefaction waves.

\begin{figure}
  \centering
  \includegraphics[width=0.8\columnwidth]{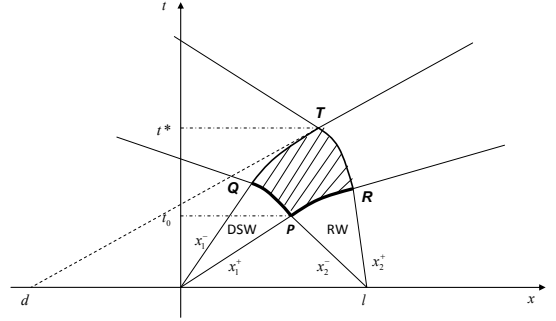}
  \caption{$x,t$-diagram of a 2-DSW interaction with a 1-rarefaction
    wave (RW). The interaction occurs within the region PQTR. The
    refraction shift $d$ of the DSW soliton edge quantifies the
    intensity of the interaction.  Reprinted from
    \cite{el_refraction_2012} Copyright (2012), with permission from
    Elsevier.}
  \label{fig:DSW_refraction}
\end{figure}
The head-on interaction of a NLS DSW with a centered rarefaction wave
has been considered in \cite{el_refraction_2012} as a dispersive
counterpart of the classical gas dynamics problem of shock wave
refraction (see, e.g.,
\cite{courant_supersonic_1948,rozhdestvenskii_systems_1983}). When a
one-dimensional viscous shock wave undergoes a head-on collision with
a rarefaction wave, the parameters of the two waves alter so that the
long-time output of such an interaction consists of a new pair of
shock and rarefaction waves propagating in opposite directions. Since
the shock wave speed undergoes a change from one constant value to
another as a result of its propagation through the finite rarefaction
region with varying density and velocity, the interaction diagram in
the $x,t$ plane is naturally interpreted as shock wave refraction on
the rarefaction wave.  The $x,t$ diagram of the interaction between a
2-DSW and a 1-rarefaction wave is shown in
Fig.~\ref{fig:DSW_refraction}. The interaction occurs within the
region PQTR and is described by the classical hodograph solution of
the NLS-Whitham equations in which only two Riemann invariants vary.
As a result, the interaction dynamics are expressed in terms of the
solution of a single EPD equation \eqref{EPD} in the plane of 
varying Riemann invariants.  The intensity of the interaction is
macroscopically quantified by the refraction shift $d$ of the DSW
soliton edge.

\begin{figure}
  \centering
  \includegraphics[width=0.6\columnwidth]{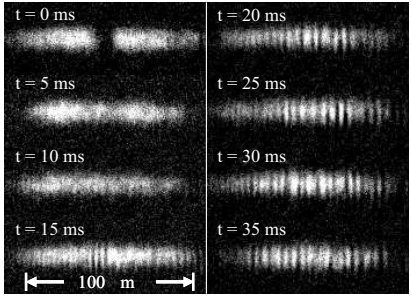}
  \caption{Density from a matter wave interference experiment in a
    Bose-Einstein condensate.  Reprinted from
    \cite{hoefer_matter-wave_2009} with permission from Elsevier.}
  \label{fig:bec_interference}
\end{figure}
\begin{figure}
  \centering
  \includegraphics{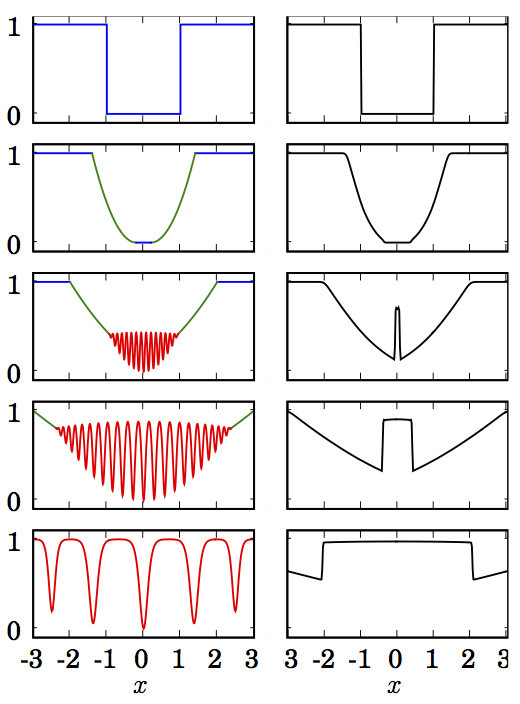}
  \caption{Density for degenerate rarefaction wave interaction in NLS
    (left) and dispersionless NLS (right) with time progressing
    downward.  The NLS solution was constructed using Whitham theory
    \cite{hoefer_matter-wave_2009}.  Figure adapted from
    \cite{hoefer_matter-wave_2009} with author permission.}
  \label{fig:nls_rarefaction_interaction}
\end{figure}

The second type of interaction we consider is that of two degenerate
rarefaction waves.  As noted in Sec.~\ref{sec:rarefactions}, a sharp
transition into the vacuum state $\rho = 0$, will always result in a
degenerate rarefaction wave.  Much like the interaction of two DSWs
generates two-phase modulations, the collision of two
counter-propagating, degenerate rarefaction waves leads to one phase
modulations.  This model problem has significance in BEC when two
initially separated clouds are allowed to merge together.  The result
is matter wave interference, a macroscopic realization of the quantum,
wave nature of matter \cite{andrews_observation_1997}.  An
experimental realization of this is shown in
Fig.~\ref{fig:bec_interference} \cite{hoefer_matter-wave_2009}.  The
modulation solution corresponding to the interaction of two degenerate
rarefaction waves is shown in
Fig.~\ref{fig:nls_rarefaction_interaction}.  Nonlinear matter wave
interference can be interpreted as the generation of two back-to-back
DSWs, analogous to the dispersionless, hyperbolic case, also shown in
Fig.~\ref{fig:nls_rarefaction_interaction}.  Note that the experiment
pictured in Fig.~\ref{fig:bec_interference} was also performed with
weaker transverse confinement, giving rise to fully three-dimensional
dynamics \cite{chang_formation_2008}.

In concluding this section, we note that the similarity between shock
and rarefaction interactions for both dispersive and dissipative
regularization of conservation laws is quite striking.  The principle
differences include the generation of finite time interaction regions
and, possibly, solitons for the dispersive case.

\section{DSWs in perturbed  systems}
\label{sec:dsw-perturbed}

\begin{figure}
  \centering
  \includegraphics[width=\columnwidth]{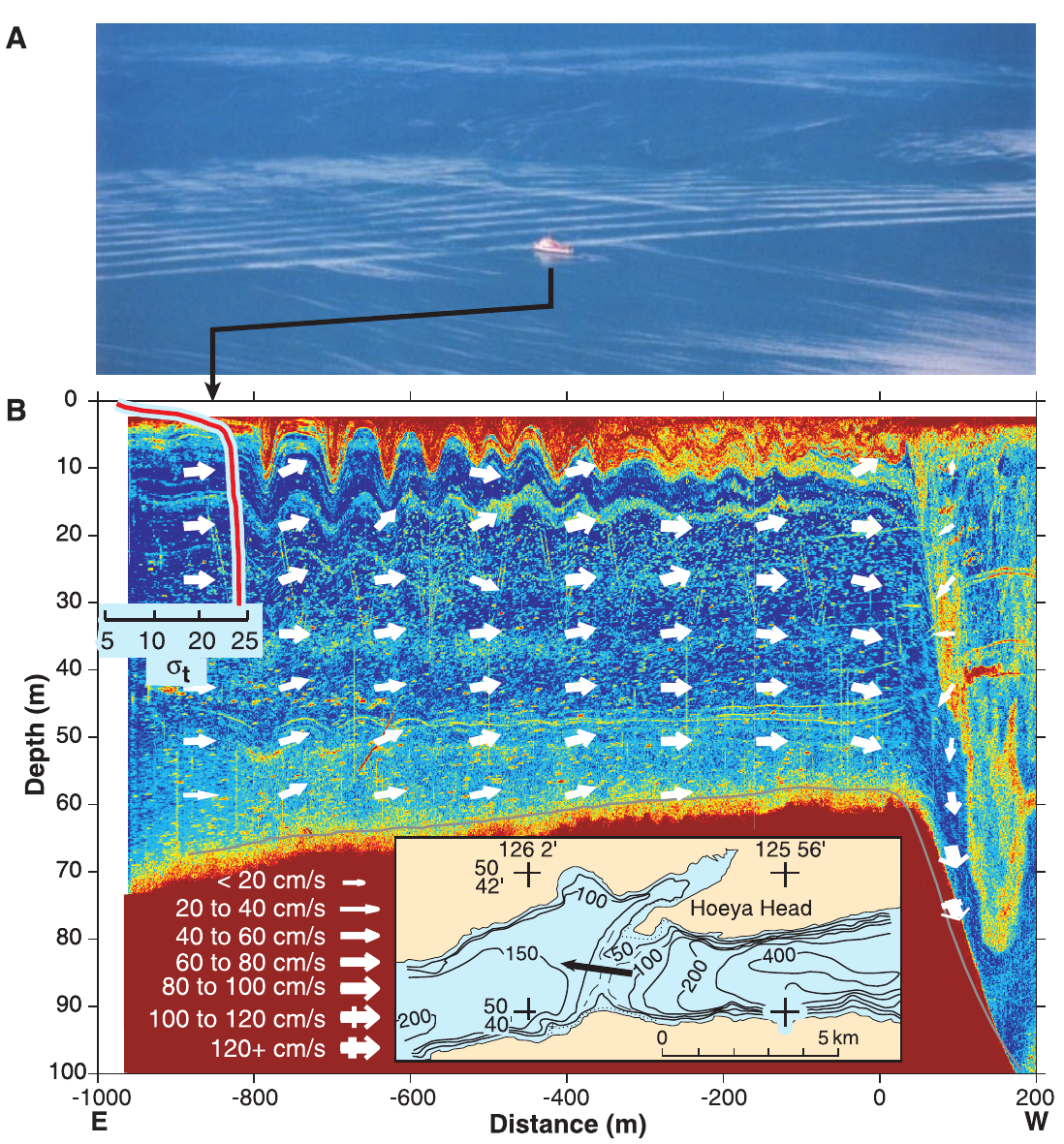}
  \caption{A shoaling internal undular bore in Knight Inlet, British
    Columbia, Canada.  (a) Ship traversing internal waves with (b)
    associated current vectors and echo sounder profile, revealing the
    internal DSW structure.  From \cite{farmer_generation_1999}.
    Reprinted with permission from AAAS.}
  \label{fig:internal_dsw}
\end{figure}

\begin{figure}
  \centering
  \includegraphics[width=\columnwidth]{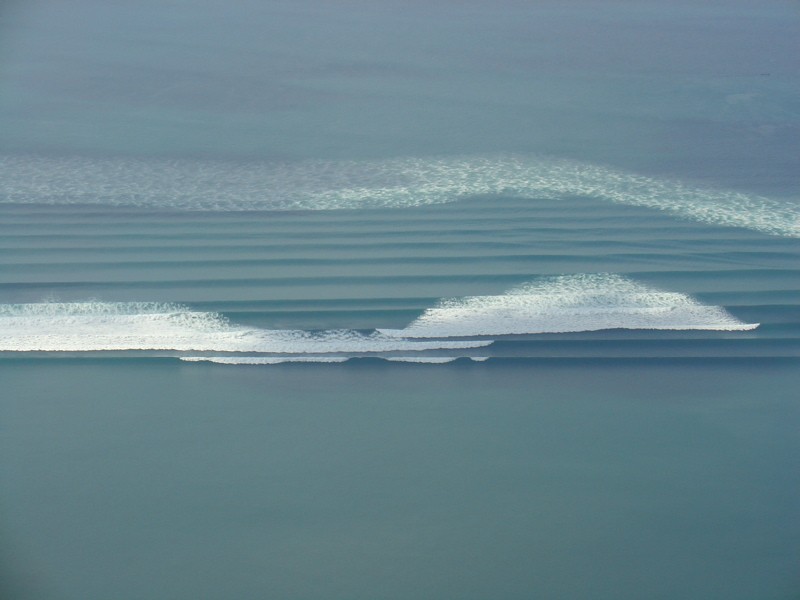}
  \caption{Shoaling of a tsunami off the island Koh Jum, Thailand
    (copyright Anders Grawin, 2006).}
  \label{fig:tsunami_dsw}
\end{figure}

All examples of DSWs considered so far have been described by
solutions of constant coefficient, dispersive-hydrodynamic equations
without perturbative and/or forcing terms.  However, the presence of
small perturbations in the system reflecting, e.g., the medium's
inhomogeneity or weak dissipation can strongly impact DSW dynamics.  A
fluid mechanics example of particular applied interest is the
propagation of a shallow water DSW (surface or internal undular bore)
over slowly varying bottom topography, termed shoaling.  This example
is of broader significance in the context of DSW propagation in weakly
inhomogeneous media.  We note that shoaling undular bore propagation
has been observed in many locations \cite{apel_oceanic_2002}, e.g.,
internal undular bores in Knight Inlet, Canada
\cite{farmer_generation_1999} (Fig.~\ref{fig:internal_dsw}) and
surface undular bores resulting from the 2004 Indian Ocean tsunami in
the Strait of Malacca \cite{grue_formation_2008} and elsewhere
(Fig.~\ref{fig:tsunami_dsw}).  See
\cite{madsen_solitary_2008,tissier_nearshore_2011,glimsdal_dispersion_2013}
regarding the role of undular bores in nearshore tsunami propagation.

The canonical model for the description of weakly nonlinear long
dispersive wave propagation over slowly varying topography is the
variable-coefficient KdV equation (see, e.g.,
\cite{grimshaw_internal_2007} and references therein),
\begin{equation}\label{varkdv}
u_\tau +  \mu(\tau) uu_\xi +  \lambda (\tau) u_{\xi \xi \xi} = 0 \, .
\end{equation}
Here, $u$ is the normalized amplitude of the dominant mode in the
waveguide, $\tau$ is related to the physical spatial variable $x$ and
measures time along the spatial path of the wave, while $\xi$ is the
temporal variable measuring the wave phase.  The slowly varying
coefficients $\mu(\tau)$ and $\lambda(\tau)$ are both expressed in
terms of the variable topography profile $h(\tau)$. The equation
(\ref{varkdv}) can be re-cast in an equivalent form of the
constant-coefficient KdV equation with a small perturbation
\cite{el_evolution_2007}, 
\begin{equation}
  \label{perturbed_kdv}
  U_t+UU_x+U_{xxx}=\delta R(U,t), \quad \delta \ll 1,
\end{equation}
where $U, x, t$ are new variables related to $u, \tau, \xi$ in
(\ref{varkdv}), and the value of the perturbation parameter $\delta$
is determined by the typical inhomogeneity gradient (variable
topography), $h'(\tau)$.  The presence of small dissipative terms in
the original equation describing, e.g., volume viscosity or turbulent
bottom friction, leads to the same perturbed KdV equation
(\ref{perturbed_kdv}) but with an appropriately modified perturbation
term, possibly depending upon derivatives of $U$. Thus one arrives at
the prototypical problem of describing a DSW in the perturbed KdV
equation and in perturbed dispersive-hyrodynamic models in general.

The modulation equations for perturbed dispersive hydrodynamic
equations can be derived by averaging the balance laws of the
perturbed equation, e.g., the KdVB equation, over the periodic family
of the unperturbed equation, e.g., the KdV equation. Modulation
systems for the perturbed KdV equation with different perturbation
types, typically dissipative, were derived by various authors using
either Whitham's original prescription
\cite{gurevich_averaged_1987,avilov_evolution_1987} or direct multiple
scales analysis \cite{myint_modulation_1995}. In all cases, the result
of the derivation is a non-homogeneous modulation system of the form
\begin{equation}
  \label{perturbed-whitham}
  \frac{\partial r_j}{\partial t} + V_j({\bf r})\frac{\partial
    r_j}{\partial x} = \delta R_j({\bf r}, t), \quad  j=1,2,3,   
\end{equation}
where $V_j({\bf r})$ are the KdV-Whitham characteristic velocities
eq.~\eqref{kdv_char_vel} and $R_j({\bf r}, t)$ arise from averaging the
perturbations in the governing equation's balance laws.  An effective
general method for the derivation of modulation systems for perturbed
integrable equations whose perturbation-free part belongs to the AKNS
hierarchy was developed by Kamchatnov in
\cite{kamchatnov_whitham_2004}.  See also
\cite{kamchatnov_whitham_2015} for a very recent extension of the
original method.
\begin{figure}
  \centering
  \includegraphics{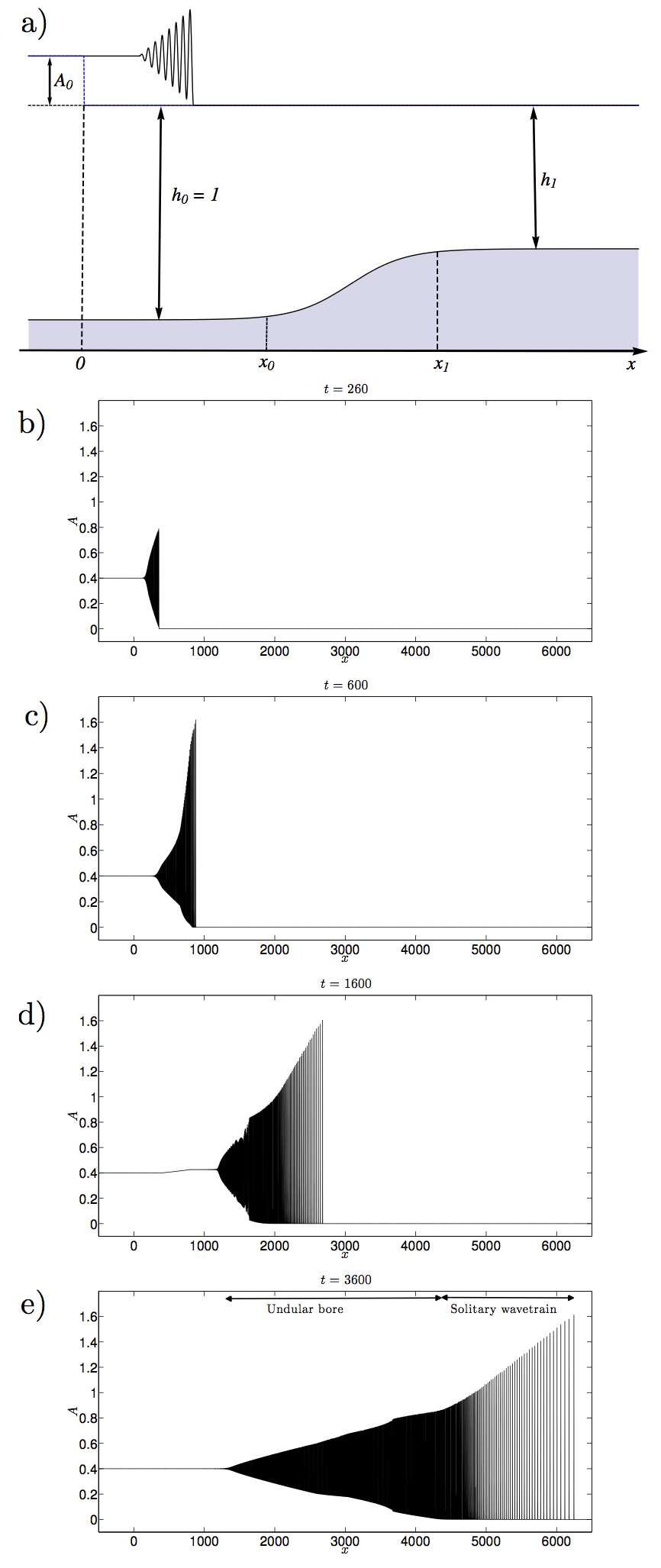}
  \caption{Transformation of a shallow water DSW (undular bore)
    propagating over a gentle bottom slope (Reprinted with permission
    from \cite{el_transformation_2012}, copyright Cambridge University
    Press, 2012).  a) schematic of the setting; b) - e) Numerical
    simulation of the variable coefficient KdV equation with DSW
    initial condition generated by an initial step in the flat bottom
    region. DSW interaction with shoaling topography leads to the
    generation of a large-amplitude solitary wavetrain propagating
    ahead of the DSW.} \label{fig:shoaling}
\end{figure}

Unlike the unperturbed modulation system (\ref{eq:31}), the perturbed
system (\ref{perturbed-whitham}) can possess TW solutions
$\mathbf{r}(x-st)$ if the perturbation term does not depend on time,
e.g., the KdVB equation (\ref{eq:2}).  It was shown in
\cite{avilov_evolution_1987,gurevich_averaged_1987} that the TW
solution of the KdVB-Whitham system connects two constant, disparate
states: $u \to u_\pm$ as $x \to \pm \infty$, $u_->u_+$, and satisfies
the Rankine-Hugoniot condition $s=(u_- + u_+)/2$.  Thus, this solution
describes slow modulations of a classical diffusive-dispersive shock
wave in the small diffusion regime.  As a matter of fact, this
outlined construction of the TW modulation solution is equivalent to a
direct multiple scales perturbation analysis of the ODE (\ref{eq:3})
with $\delta \ll 1$ for the TW solutions of the KdVB equation itself
(see \cite{johnson_non-linear_1970} and
Sec.~\ref{fig:kdvbPhasePlane}).  However, modulation theory provides a
much broader platform, enabling one to describe different stages of
diffusive-dispersive shock development (see
\cite{dubrovin_hydrodynamics_1989,el_dispersive_2015}).

The TW solutions of the perturbed, integrable Whitham equations
describing modulations of stationary, diffusive-dispersive shocks were
obtained for the Kaup-Boussinesq-Burger's equations
\cite{el_analytic_2005}, Benjamin-Ono-Burger's equation
\cite{matsuno_whitham_2007}, and the defocusing NLS equation with
linear damping \cite{larre_wave_2012}.  The unsteady dynamics of DSWs
in perturbed systems, particularly in the variable-coefficient KdV
equation (\ref{varkdv}), has only begun to be understood
\cite{el_evolution_2007,el_transformation_2012}.  In particular, the
study of a shoaling shallow water DSW in \cite{el_transformation_2012}
showed that, along with an adiabatic deformation of the DSW due to 
slowly varying topography, there is a non-adiabatic response (the
generation of a sequence of isolated solitons): an expanding,
large-amplitude, modulated solitary wavetrain propagating ahead of the
DSW (see Fig.~\ref{fig:shoaling}). Interestingly, a similar effect was
observed in the overtaking interaction of a DSW and a rarefaction
wave in \cite{ablowitz_soliton_2009} in the constant-coefficient KdV
equation (cf. Fig.~\ref{fig:kdv_all_interactions}).  These qualitative similarities between two apparently
different problems are not surprising when one observes that, in our
present formulation, the shallow water DSW essentially propagates
through a ``rarefaction region'' of decreasing depth.

\section{Resonant generation of DSWs in forced systems}
\label{sec:dsw_forced}

\begin{figure}
  \centering
  \includegraphics{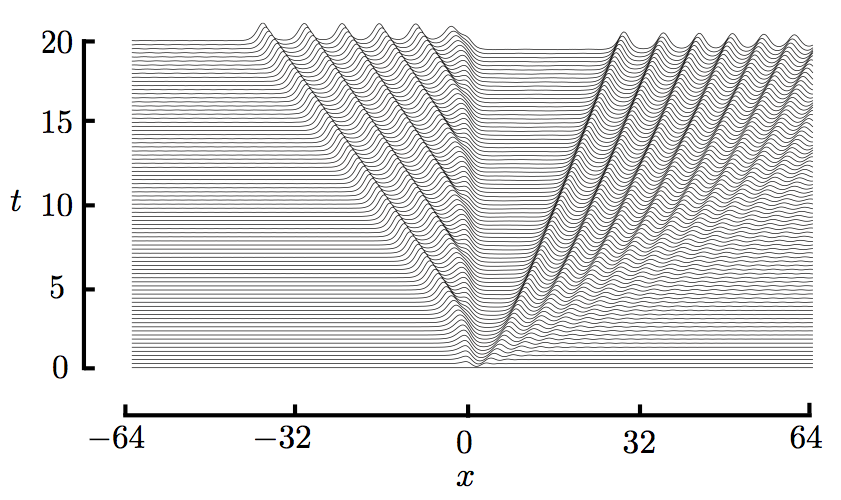}
  \caption{Transcritical flow past localized topography: a numerical
    solution of the forced KdV equation (\ref{forced_kdv}) at exact
    criticality $\Delta=0$. The partial, attached DSW is generated
    upstream (left) and a fully realized, detached DSW downstream
    (right).} \label{fig:transcritical}
\end{figure}

We have described DSWs forming as a result of dispersive
regularization of wavebreaking or Riemann initial data.  Another DSW
generation mechanism due to a moving piston was described in
Sec.~\ref{sec:piston-problem} where the DSW is realized by introducing
a forcing term in the equation as a moving, impenetrable, sharp
potential barrier \cite{hoefer_piston_2008}. If the forcing term is
instead a moving potential barrier of finite width, a penetrable
barrier, DSWs are generally not generated unless the barrier velocity
is sufficiently close to the local speed of linear long waves in the
reference frame of the barrier. In this case, the possibility arises
for a new, resonant, mechanism of DSW generation, which is realized in
a broad range of applications.  This resonant mechanism was first
identified and described in the framework of near-critical dispersive
shallow water flows past broad, localized obstacles by Grimshaw and
Smyth (GS) \cite{grimshaw_resonant_1986} and Smyth
\cite{smyth_modulation_1987}.  An appropriate model is the
forced KdV equation, which can be obtained in the canonical form:
\begin{equation}
  \label{forced_kdv}
  -A_t - \Delta A_x + 6AA_x + A_{xxx}+ F_x(x)=0.
\end{equation}
Here $A$ denotes an appropriate field variable (e.g., the free-surface
or interfacial displacement), $\Delta$ is the deviation of the
incident flow velocity from the long wave phase speed and $F(x) \ge 0$
represents localized topographic forcing, that is we assume that
\begin{equation}
  \label{1-2}
  F(x)\to0\quad \mathrm{as} \quad x\to\pm\infty.
\end{equation}
In practice, it is sufficient to assume that $F(x) \ge 0$ for $x \le l$
and is zero otherwise, with a single maximum at $x=0$. It is also
assumed that $A(x,0)=0$ and $A \to 0$ as $|x| \to \infty$.  A typical
solution of the forced KdV equation (\ref{forced_kdv}) at exact
criticality $\Delta=0$ is shown in Fig.~\ref{fig:transcritical}.

The prominent feature of a near-critical flow past broad ($l \gg 1$)
obstacles is the formation of a smooth, steady hydraulic transition
over the obstacle region so that the flow is subcritical, $A \to
A_->0$, in front of the obstacle and supercritical, $A \to A_+ < 0$,
after the obstacle.  The transition between $A_-$ upstream and $A_+$
downstream is described by the steady, dispersionless hydraulic limit
of (\ref{forced_kdv}), 
\begin{equation}
  \label{hydraulic} - \Delta A_x + 6AA_x + F_x(x)=0,
\end{equation}
augmented by the condition of exact criticality at the maxumum of the
forcing, expressed as the requirement that $A_x \ne 0$ when $F_x=0$
\cite{grimshaw_resonant_1986}.  One can establish that the required
heteroclinic solution of (\ref{hydraulic}) exists only if
\begin{equation}
  \label{transcrit}
  - (12 F_m)^{1/2}\le \Delta \le (12 F_m)^{1/2},
\end{equation}
where $F_m = \max [F(x)]=F(0)$, and 
$$
6 A_{\pm} = \Delta \mp (12 F_m)^{1/2}.
$$
Outside the interval (\ref{transcrit}), the flow is either purely
subcritical or purely supercritical, and in both cases is described by
a homoclinic solution of (\ref{hydraulic}) smoothly connecting the
equilibrium states $A=0$ as $x \to\pm \infty$.

In the transcritical regime (\ref{transcrit}), the non-equilibrium
upstream and downstream states $A_{\pm}$ are resolved back into the
equilibrium state $A=0$ away from the obstacle with the aid of DSWs,
propagating upstream and downstream. These are asymptotically
described by the appropriate GP similarity modulation solutions of the
{\it unforced} KdV equation, as these DSWs are formed away from the
obstacle. Upstream DSW propagation is a remarkable feature of
dispersive transcritical flows which does not have a counterpart in
hyperbolic conservation law theory. Another important feature is that
the asymptotic description of transcritical flow is defined by
just two parameters, $\Delta$ and $F_m$, and does not depend on the
detailed form of the obstacle.

\begin{figure}
  \centering
  \includegraphics[scale= 0.3]{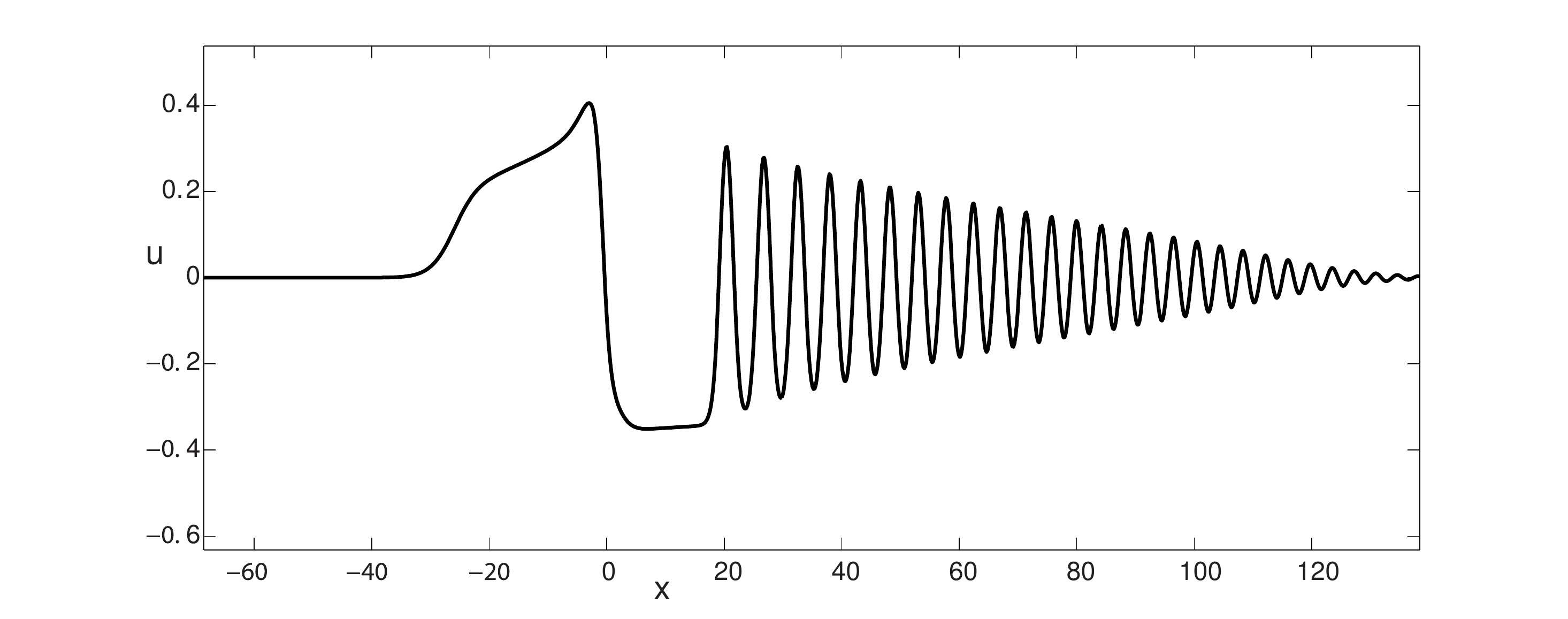}
  \caption{Transcritical internal wave flow past localized topography.
    Numerical solution of the forced Gardner equation with a
    kink-rarefaction generated upstream and a detached DSW downstream.
    Figure reproduced with permission from
    \cite{kamchatnov_transcritical_2013}.}
  \label{fig:transcritical_g}
\end{figure}

The outlined GS analytical framework for the study of transcritical
dispersive hydrodynamic flows proved to be quite general (see, e.g.,
\cite{baines_topographic_1995,madsen_transient_2012}). In the fluid
mechanics context, it was successfully applied to the description of
resonant generation of atmospheric internal waves (the Morning Glory,
see Fig.~\ref{fig:atmospheric_dsws}) using the forced Benjamin-Ono
equation \cite{porter_modeling_2002}.  Note that the modulation theory
for Benjamin-Ono DSWs was developed in \cite{jorge_modulation_1999}.
Resonant generation of internal waves in the framework of the forced
Gardner equation was studied in \cite{kamchatnov_transcritical_2013},
where, due to non-convexity of the hyperbolic flux (see
Sec.~\ref{sec:non-classical-dsws}), a rich variety of transcritical
regimes were shown to be possible, including non-classical DSWs
(kinks, contact DSWs) along with standard, KdV type, DSWs.  One such
regime with a kink-rarefaction double wave generated upstream and a
classical DSW propagating downstream is shown in
Fig.~\ref{fig:transcritical_g}. The resonant generation of
finite-amplitude shallow water DSWs in the framework of the forced
Serre system was studied in \cite{el_transcritical_2009} where the
upstream and downstream DSW closure was performed using the results of
the DSW fitting method applied to the unperturbed Serre equations
\cite{el_unsteady_2006} (see Sec.~\ref{sec:shallow-water}).
\begin{figure}
  \centering
  \includegraphics{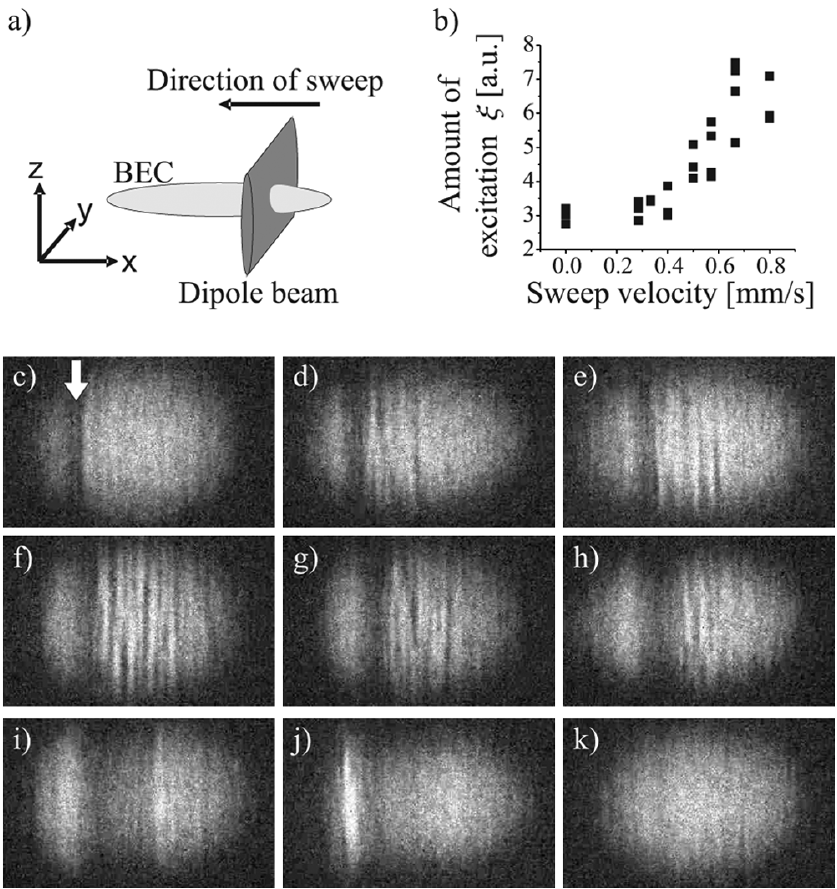}
  \caption{Demonstration of transcritical flow in a BEC induced by
    sweeping a repulsive laser beam. Reprinted figures with permission
    from \cite{engels_stationary_2007} copyright 2007 by the
    American Physical Society.}
  \label{fig:bec_transcritical}
\end{figure}

DSW generation by transcritical flows past obstacles has been observed
in superfluids and nonlinear optical media. The possibility of
resonant generation of unsteady wavetrains in superfluid flows through
broad, penetrable potential barriers was predicted by Hakim
\cite{hakim_nonlinear_1997}, who considered the forced, defocusing NLS
equation, i.e., the Gross-Pitaevskii equation (\ref{eq:75}).  The
subcritical, transcritical and supercritical regimes were identified
by analyzing the hydraulic solution obtained in the framework of
shallow-water theory. The generation of unsteady wavetrains (DSWs)
signifying the breakdown of superfluidity was studied numerically.

The resonant generation of dark soliton trains in 1D BEC flows past
penetrable barriers has been realized experimentally in
\cite{engels_stationary_2007}.  In the experiment, a wide, repulsive
penetrable barrier created by a laser beam was swept at a constant
speed through an elongated BEC confined to a cigar-shaped trap, as
illustrated schematically in Fig.\ref{fig:bec_transcritical}a.  The
panels c) - k) in Fig.~\ref{fig:bec_transcritical} show absorption
images of the BEC density for different speeds of the dipole barrier.
For very slow sweep speeds, such as in panel c), the part of the BEC
through which the dipole beam was swept appears essentially unaffected
(superfluidity maintained).  The only visible effect of the beam is a
density suppression that it leaves at its end position shown by the
white arrow.  At intermediate speeds, the shedding of dark solitons
was observed (panels d) -- g)) followed by a gradual decreasing of the
number of generated solitons (panels h), i)) and an apparent full
restoration of superfluidity at sufficiently large speeds (panel
k)). These dynamics are in qualitative agreement with the predictions
of the transcritical flow analysis in \cite{hakim_nonlinear_1997}.  A
detailed theory of resonant 1D NLS flows past wide penetrable barriers
was constructed in \cite{leszczyszyn_transcritical_2009} by extending
the GS analytical framework to the forced NLS equation. In
\cite{leszczyszyn_transcritical_2009}, the modulation solutions of the
defocusing NLS equation (see Sec.~\ref{sec:defocusing-nls}), the
potential-free Gross-Pitaevskii equation (\ref{eq:75}), were used for
the description of the downstream and upstream DSWs connected by a
smooth hydraulic solution. The downstream, almost stationary DSW
adjacent to the potential was identified as a counterpart of the dark
soliton train observed in the experiment
\cite{engels_stationary_2007}.

In the nonlinear optics context, DSW generation via the nonlinear
tunneling of a plane wave through a penetrable barrier potential was
observed in \cite{wan_wave_2010}.

%
%
%
%
%

\section{Supersonic dispersive flows past obstacles and 2D oblique
  DSWs}
\label{sec:dsw_oblique}

\begin{figure}
  \centering
  \includegraphics{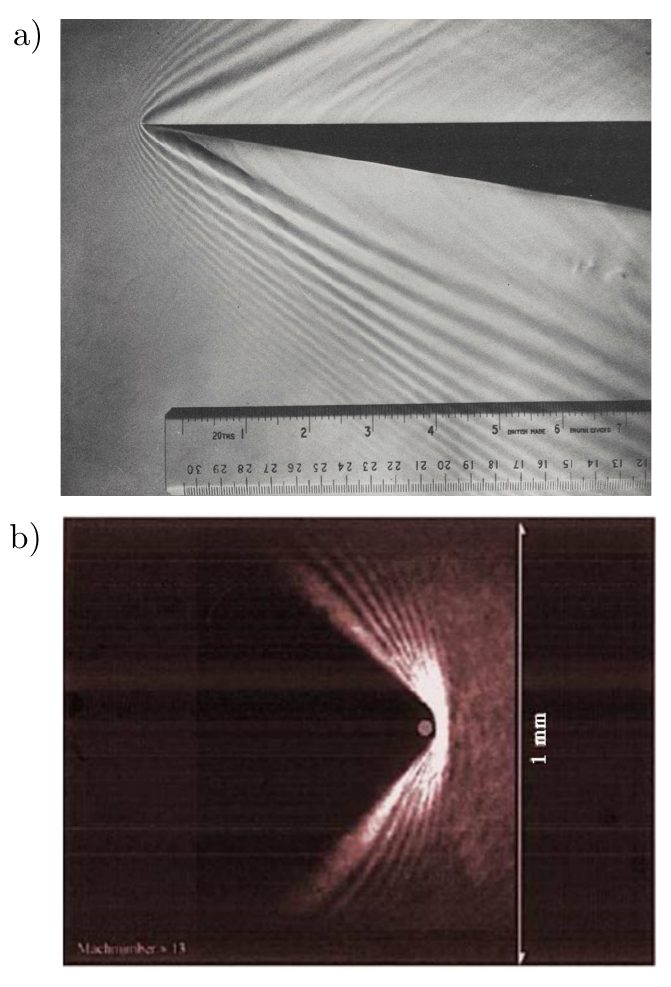}
  \caption{Experimental observations of 2D oblique DSWs. a) Oblique
    DSW in supercritical shallow water flow past obstacle
    \cite{merzkirch_flow_1974}.  Reprinted from
    \cite{merzkirch_flow_1974} with permission from Elsevier. b) BEC
    bow shock from expansion past an obstacle. Reprinted figure with
    permission from \cite{carusotto_2006} copyright 2006 by the
    American Physical Society.} \label{fig:oblique_DSW}
\end{figure}

Formation of shocks in multidimensional supersonic flows past
obstacles is a textbook problem in classical shock theory
\cite{courant_supersonic_1948,liepmann_elements_1957}.  It is
well-known that a two-dimensional (2D) supersonic flow, when turned
through a compression, e.g., by an obstacle, can lead to the formation
of a stationary viscous shock. Straight, oblique, and curved detached
shocks can form when the flow is supersonic or transonic,
respectively. These stationary shock patterns can exhibit stable, rich
dynamical behavior \cite{dyke_album_1982} so finding their
counterparts in dispersive hydrodynamics is a natural task with
applications in water waves, superfluids and optics. Two examples of
experimentally observed, 2D stationary oblique DSWs in supersonic
dispersive-hydrodynamic flows past obstacles are presented in
Fig.~\ref{fig:oblique_DSW}.

Assuming steady, irrotational, 2D dispersive hydrodynamic motion, the
governing Eulerian equations (\ref{eq:1}) become a PDE system for
three quantities: the density $\rho(x,y)$ and two velocity components
${\mathbf u} = (u(x,y),v(x,y))$.  For example, the
dispersive-hydrodynamic representation of the stationary, 2D defocusing
NLS equation has the form:
\begin{equation}
  \label{2D_NLS}
  \begin{split}
    (\rho u)_x+( \rho v)_y=0,\\
    uu_x+vu_y+\rho_x+\left(\frac{\rho_x^2+\rho_y^2}{8\rho^2}-
      \frac{\rho_{xx}+\rho_{yy}}{4\rho}\right)_x=0,\\
    uv_x+vv_y+\rho_y+\left(\frac{\rho_x^2+\rho_y^2}{8\rho^2}-
      \frac{\rho_{xx}+\rho_{yy}}{4\rho}\right)_y=0, \\
    u_y -v_x=0 \, .
   \end{split}
\end{equation}
We introduce the Mach number $M=|{\bf u}|/c$, where $c(\rho) =
\sqrt{P'(\rho)}$ is the speed of sound in the long-wave,
dispersionless limit. For NLS dispersive hydrodynamics,
$c=\sqrt{\rho}$.  The dispersionless limit of
(\ref{2D_NLS}) is hyperbolic if $M>1$ corresponding to supersonic
flow.  This hyperbolicity condition holds for general, 2D
dispersionless Euler systems under appropriate monotonicity conditions
on the pressure law $P(\rho)$ \cite{courant_supersonic_1948}.  In
certain application areas, e.g., shallow water waves, the condition $M
> 1$, e.g., Froude number above unity, is termed supercritical.

\begin{figure}
  \centering 
  \includegraphics[scale= 0.5]{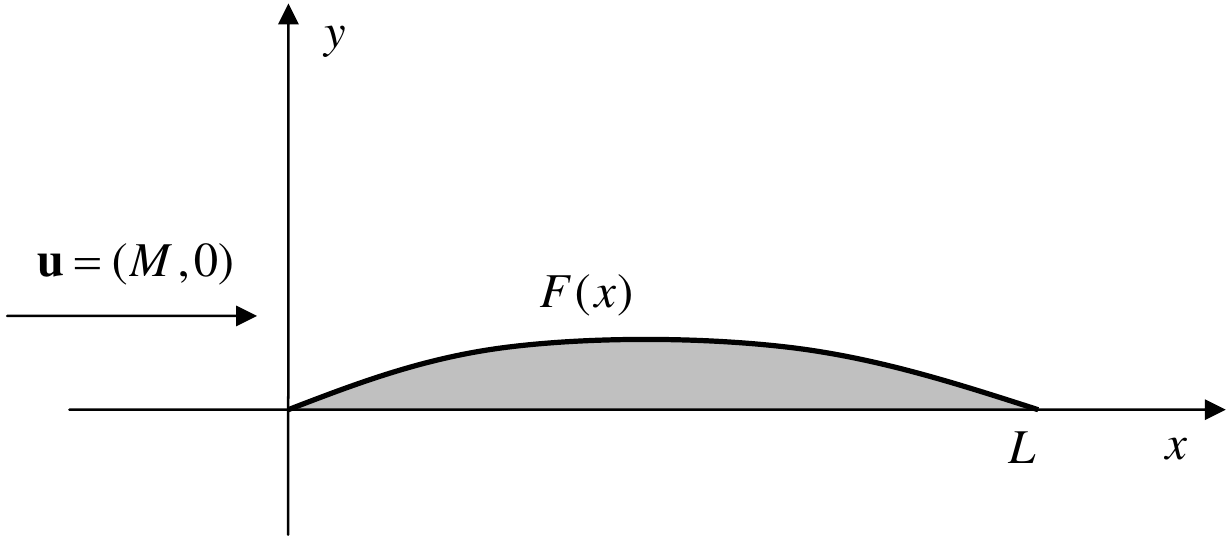}
  \caption{Schematic of 2D flow past an obstacle.} 
  \label{fig:2D_schematic}
\end{figure}

We now consider a stationary, supersonic flow past an impenetrable
obstacle in the uppper half-plane by introducing appropriate boundary
conditions for the 2D dispersive Eulerian systems. We shall be using
equations (\ref{2D_NLS}) as an instructive and representative
example. See Fig.~\ref{fig:2D_schematic} for a schematic of the
configuration considered.

Let the shape of the obstacle be given by $y=F(x) \ge 0$ for $x \ge 0
$, $F(0)=0$.  Assuming hyperbolicity of the problem, the flow in the
lower half-plane is isolated from the flow in the upper half plane so
can be considered independently.  The impenetrability condition ${\bf
  u} \cdot {\bf n}|_{S}=0$ at the obstacle's surface assumes the form
\begin{equation}\label{eq14b}
  v=uF'(x) \quad \text{at} \quad y=F(x).
\end{equation}
We also impose the condition of stationary equilibrium flow far away
from the obstacle
\begin{equation}
  \label{eq14a}
  \rho=1,\quad u=M,\quad v=0\quad\text{at}\quad  |x|, |y| \to \infty.
\end{equation}
The system (\ref{2D_NLS}) has a family of stationary, periodic
solutions \cite{el_two-dimensional_2007} and is amenable, in
principle, to a general modulation analysis using the DSW fitting
method. Care must be taken due to the presence of a 2D convective
instability \cite{kamchatnov_stabilization_2008,hoefer_dark_2012}, a
feature, not occuring in 1D flows.

Since the general problem (\ref{2D_NLS}) - (\ref{eq14a}) is quite
involved, it is instructive to consider first some more manageable
approximations by introducing reasonable small/large parameters.

\subsection{Supersonic flow past slender obstacles: weakly nonlinear
  KdV approximation}

\begin{figure}
  \centering
  \includegraphics{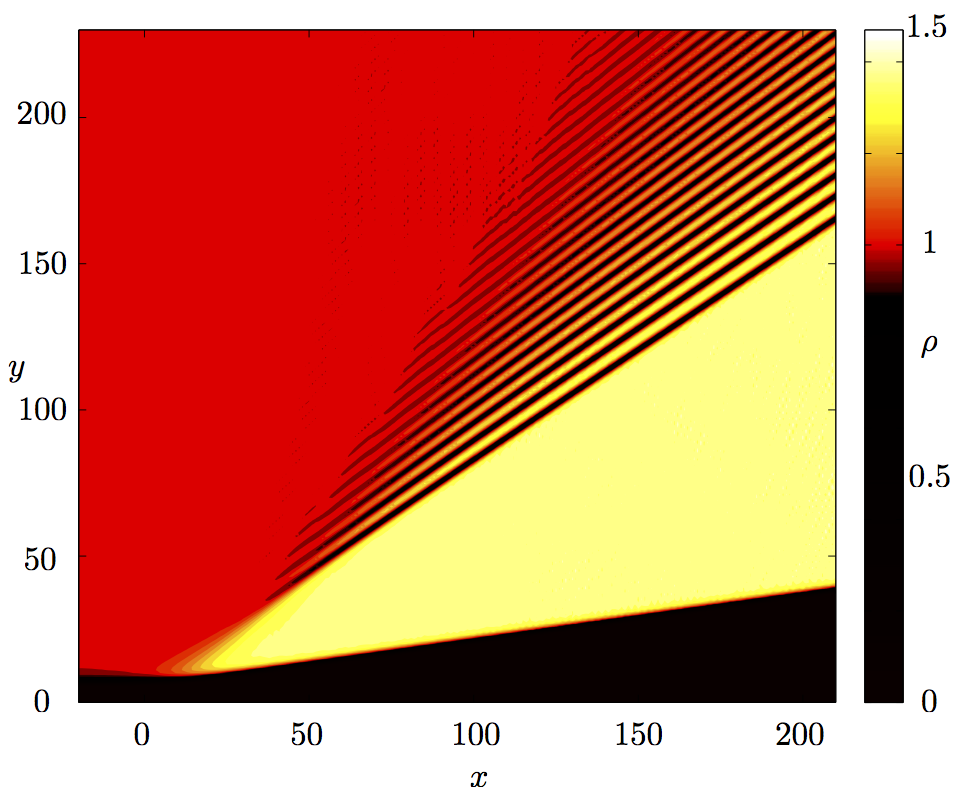}
\caption{Numerical solution of the 2D NLS equation yielding a
  steady, oblique DSW in supercritical flow past a
  corner.} \label{fig:DSW_corner}
\end{figure}

If the obstacle is slender, $\eps = \max |F'(x)| \ll 1$, one can use
multiple scales expansions for (\ref{2D_NLS}), reducing the boundary
value problem (\ref{2D_NLS}) - (\ref{eq14a}) to an initial value
problem for the KdV equation.  This has been done for a rather general
dispersive hydrodynamic setting in \cite{karpman_non-linear_1974} and,
later, in \cite{gurevich_supersonic_1995}. To this end, we introduce
the asymptotic expansions
\cite{el_two-dimensional_2007,hoefer_dark_2012},
\begin{equation}
  \label{eq:162}
  \begin{split}
    \rho &= 1+\eps \rho_1+\eps^2 \rho_2+\ldots,  \\
    u &= M+\eps u_1+\eps^2\rho_2+\ldots ,      \\
    v &= \eps v_1+\eps^2v_2+\ldots , 
  \end{split}
\end{equation}
$\rho_i, u_i, v_i \to 0$ as $x,y \to \infty$, along with new
independent variables
\begin{equation}
  \label{eq:163}
  \zeta= x- \sqrt{M^2-1}y,\quad \tau=\eps y.
\end{equation}
Substitution of (\ref{eq:162}), (\ref{eq:163}) into (\ref{2D_NLS})
yields, at the first order in $\eps$,
\begin{equation}\label{5-3}
    u_1=-\frac{\rho_1}M,\quad v_1=\frac{\sqrt{M^2-1}}M \rho_1,
\end{equation}
Continuing to the second order, we obtain  the KdV equation  for $\rho_1$:
\begin{equation}
  \label{5-4}
  \rho_{1,\tau}-\frac{3M^2}{2\sqrt{M^2-1}}\rho_1  \rho_{1,\zeta}+
  \frac{M^4}{8\sqrt{M^2-1}}\rho_{1,\zeta\zeta\zeta}=0.
\end{equation}
 The boundary condition (\ref{eq14b}) yields, at the
same order in $\eps$, the KdV initial condition \be \label{2D_kdv_ic}
\rho_1 (0, \zeta) = - \frac{M^2}{\sqrt{M^2-1}} F'(\zeta).  \ee Thus
the boundary value problem (\ref{2D_NLS}) - (\ref{eq14a}) for the 2D
stationary NLS equation is reduced to the KdV initial value problem
(\ref{5-4}), (\ref{2D_kdv_ic}).  One can see from (\ref{2D_kdv_ic})
that flow past a straight corner yields the Riemann step problem,
while flow past an airfoil corresponds to the evolution of a bi-polar
profile and $N$-wave type evolution.  See
Sec.~\ref{sec:evolution_decaying_profiles} and note the change of
dispersion sign in (\ref{5-4}) resulting in wave pattern inversion, in
both polarity and orientation.  The standard similarity DSW of the KdV
Riemann problem (Sec.~\ref{sec: dsw-riemann}) then maps to
the 2D {\it oblique DSW} in the $(x,y)$-plane, with an {\it oblique
  dark soliton} \cite{el_oblique_2006} at the edge facing the corner
and a stationary 2D linear wavepacket at the opposite, harmonic edge.
A typical solution of NLS flow past a corner is shown in
Fig.~\ref{fig:DSW_corner}.

The detailed study of the KdV approximation in supersonic
dispersive-hydrodynamic flow past obstacles has been undertaken in
\cite{gurevich_supersonic_1995,gurevich_supersonic_1996}.

\subsection{Hypersonic flow: NLS piston problem approximation}
\label{sec:hypersonic_NLS_piston}

\begin{figure}
  \centering
  \includegraphics[width=\columnwidth]{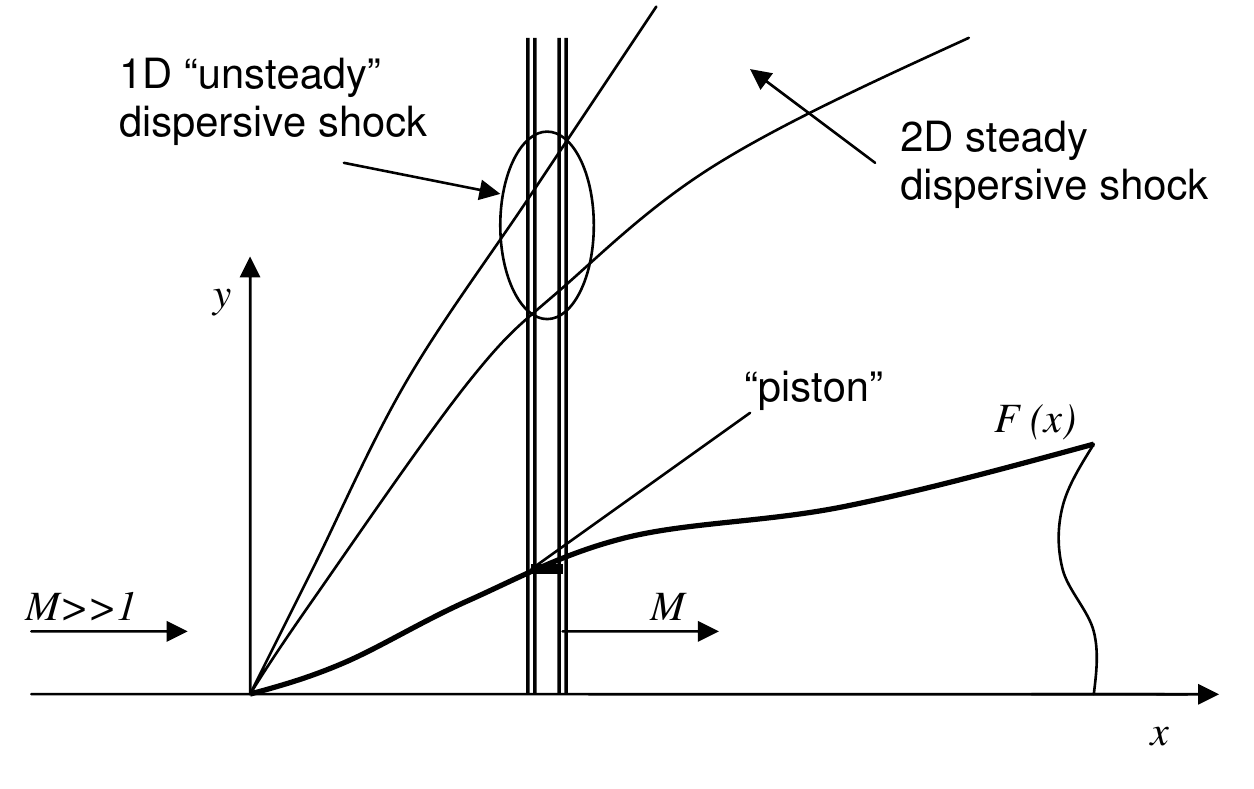}
  \caption{Piston analogy in hypersonic flow past an obstacle.
    Reprinted figure with permission from
    \cite{el_two-dimensional_2009} copyright 2009 by the American
    Physical Society.}
  \label{fig:piston_analogy}
\end{figure}

If the oncoming flow is highly supersonic, $M \gg 1$, the steady
problem of two-dimensional flow past a slender body can be
asymptotically transformed to the much simpler problem of 1D
``unsteady'' flow along the $y$ axis with the scaled $x$-coordinate
playing the role of time \cite{el_hypersonic_2004}.  This notion of
\textit{hypersonic similitude} is well-known in classical gas dynamics
(see, e.g., \cite{hayes_hypersonic_2004}). To this end, we substitute
into eqs.~(\ref{2D_NLS}) the expansion
\begin{equation}\label{eq15}
   u=M+u_1+O(1/M),\quad T=x/M,\quad Y=y,
\end{equation}
assuming $M^{-1} \ll 1$. Then to leading order we obtain
\begin{equation}\label{eq16}
\begin{split}
    \rho_T+(\rho v)_Y=0,\\
 v_T+vv_Y+\rho_Y+\left(\frac{\rho_Y^2}{8n^2}
   -\frac{\rho_{YY}}{4\rho}\right)_Y=0,
   \end{split}
\end{equation}
\begin{equation*}
 u_{1}=0.
\end{equation*}
Equations (\ref{eq16}) represent the hydrodynamic form of the 1D
defocusing NLS equation
\begin{equation}\label{nls-1D}
   i\Psi_T+\tfrac12\Psi_{YY}-|\Psi|^2\Psi=0
\end{equation}
for the complex field variable
\begin{equation}
\Psi=\sqrt{\rho}\exp\left(i\int^Y v(Y',T)dY'\right),
\end{equation}
and so we can apply the modulation theory developed in
Sec.~\ref{sec:defocusing-nls}. It is remarkable that, if $M \alpha
=O(1)$, where $\alpha=\mathrm{max}|F'(x)|$, the boundary condition
(\ref{eq14b}) asymptotically reduces to the piston problem with moving
boundary condition
\begin{equation}\label{eq19}
   v=v_p= df/dT\quad\text{at}\quad Y=f(T),
\end{equation}
where the piston motion is given by the function $f(T)=F(MT)$
describing the obstacle's profile.  Conditions (\ref{eq14a}) at
infinity transform into
\begin{equation}\label{eq19a}
  \rho=1,  \  \ v = 0 \quad \hbox{as} \ \  Y \to \infty \, .
\end{equation}
Thus the problem of hypersonic, steady 2D NLS flow past a slender
body has been reduced to the 1D unsteady piston problem, whose
solutions were obtained in Sec.~\ref{sec:piston-problem}.  The piston
analogy is illustrated in Fig.~\ref{fig:piston_analogy}.

The piston approximation, similar to the KdV approximation considered
in the previous subsection, is valid for slender obstacles, $\alpha
\ll 1$. The important difference, however, is that the piston
approximation does not contain a small amplitude assumption.
Indeed, there is no small-amplitude expansion for $\rho$ involved and
$u_1=O(1)$ in (\ref{eq15}), which enables one to capture
large-amplitude effects such as cavitation (see Sec.~\ref{sec:dsws}),
which are not present in the KdV approximation.
\begin{figure}
  \centering
  \includegraphics[scale= 0.4]{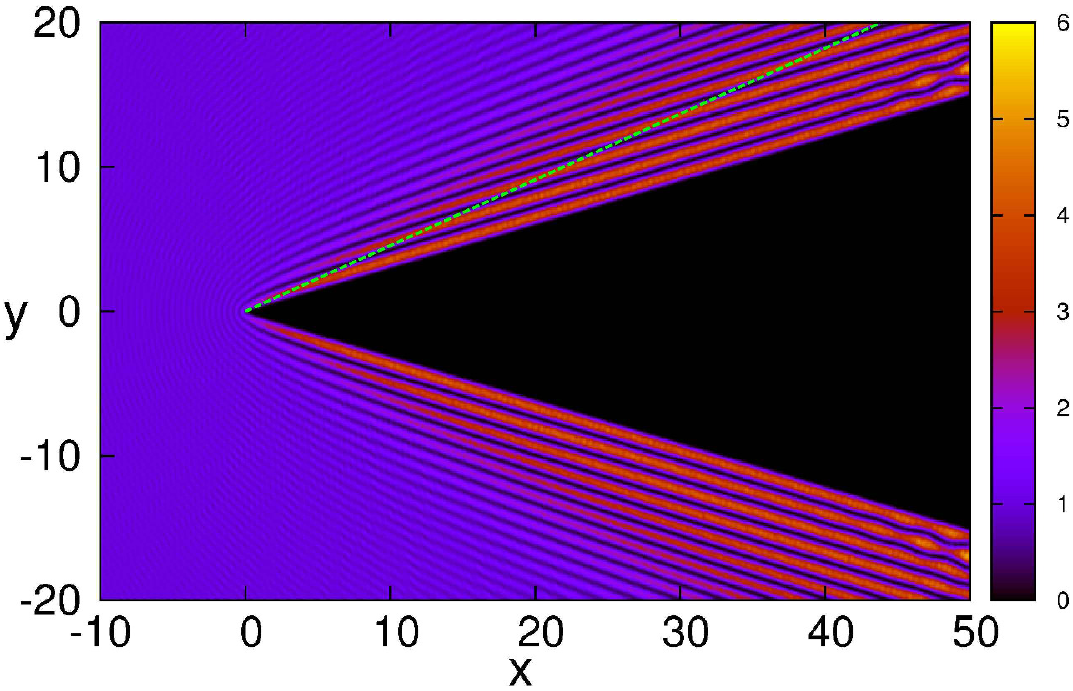} \includegraphics[scale=
  0.2]{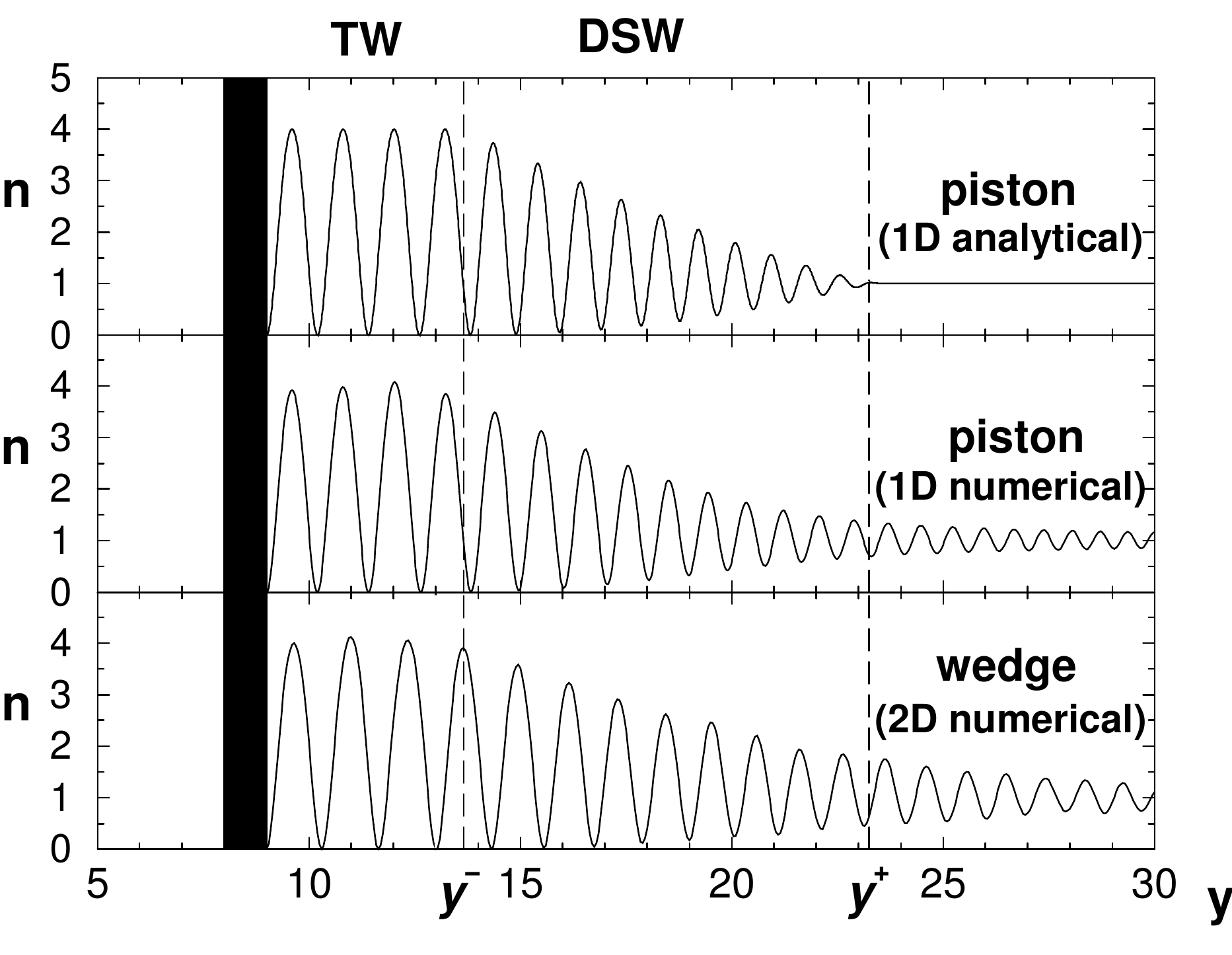}
\caption{Hypersonic ($M=10$) 2D NLS flow past a wedge with the opening
  angle $2 \alpha=0.6$: a) density plot. The green dashed line in the
  upper half-plane marks the end of the periodic transition wave
  predicted by the modulation theory.  b) profile $\rho(y)$ at $x=30$;
  top: analytical modulation theory solution; middle: numerical
  solution of the associated 1D piston problem; bottom: 1D cut of the
  full 2D solution at $t=15$. Points $y^-$ and $y^+$ mark the
  boundaries of the DSW specified by the modulation solution. The
  wedge surface in the upper half-plane (the piston) is located at
  about $y = 9$.  Reprinted figure with permission from
  \cite{el_two-dimensional_2009} copyright 2009 by the American
  Physical Society.} \label{fig:dsw_wedge}
\end{figure} 
 
\begin{figure}
  \centering
  \includegraphics[scale= 0.6]{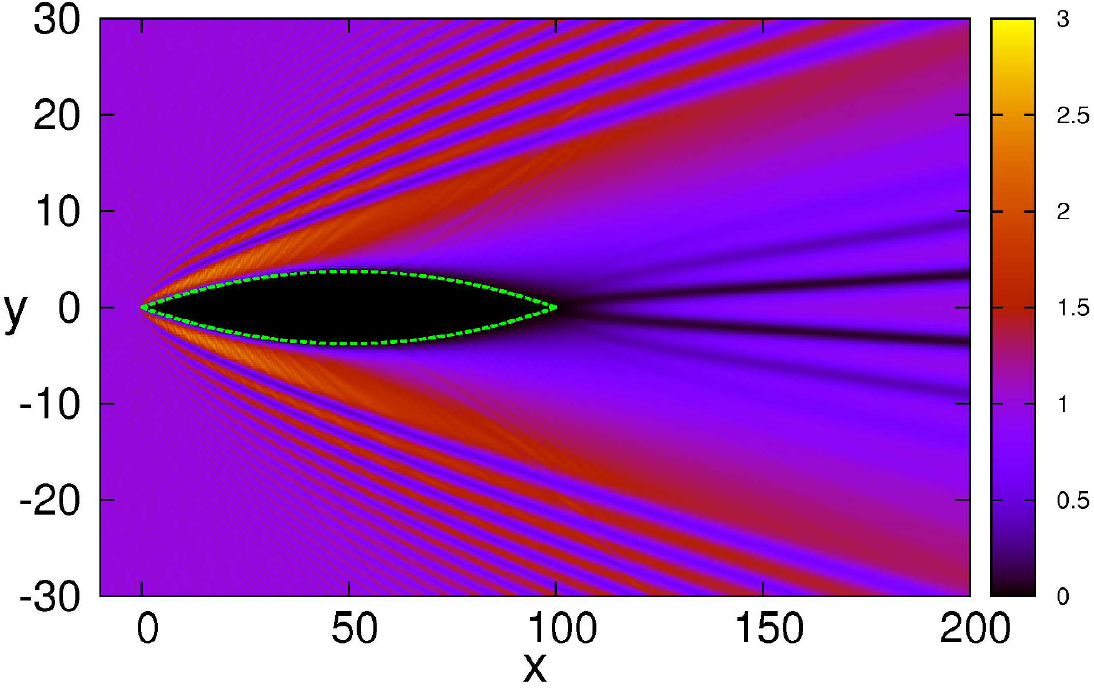}
  \caption{2D hypersonic NLS flow past an airfoil.  Reprinted figure
    with permission from \cite{el_two-dimensional_2009} copyright 2009
    by the American Physical Society.}
  \label{fig:airfoil}
\end{figure}

Modulation theory of NLS 2D hypersonic flow past obstacles was
developed in
\cite{el_spatial_2006,el_erratum_2006,el_two-dimensional_2009}.
Comparison of the modulation theory results for hypersonic flow past a
corner exhibiting large-amplitude cavitation effects is shown in
Fig.~\ref{fig:dsw_wedge}.  The numerical simulation of hypersonic flow
past an airfoil is shown in Fig.~\ref{fig:airfoil}. One can see the
formation of two oblique DSWs with contrasting asymptotic properties:
the rear DSW transforms into a train (fan) of oblique dark solitons,
while the front one asymptotically assumes a universal ``BEC
ship-wave'' pattern \cite{gladush_radiation_2007}, obtained by the
application of Kelvin's famous construction for linear deep water
waves \cite{whitham_linear_1974} to the 2D NLS equation. The analogy
with dispersive-hydrodynamic $N$-wave evolution is, again, helpful in
the interpretation of the global wave pattern arising in flow past an
airfoil.

In conclusion of this section, we mention that, along with steady
oblique DSWs, 2D dispersive hydrodynamics admits {\it unsteady oblique
  DSWs}, which result from an initial value problem and arise as an
initial pattern in the development of steady oblique DSWs. These
unsteady oblique DSWs were studied in \cite{hoefer_theory_2009}



%

\section{Conclusion}
\label{sec:conclusion}

Dispersive shock waves have recently become the subject of very active
research due to the growing recognition of their fundamental nature
and ubiquity in applications, which is comparable to the ubiquity of
solitons. This recognition, which had been steadily present in water
wave theory, has been recently facilitated by groundbreaking
experiments in superfluids and nonlinear optics.  The very recent
progress in dynamics of viscous fluid conduits has made DSWs readily
available for detailed quantitative measurements.

This review represents an attempt to summarize the results of the last
50 years of DSW research using the framework of nonlinear modulation
theory pioneered by Whitham in 1965.  The Whitham method of slow
modulations and the Gurevich-Pitaevskii matching regularization
problem have provided a powerful theoretical framework for the
development of DSW theory, encompassing both integrable and
non-integrable dispersive-hydrodynamic systems.

\section*{Acknowledgements}
\label{sec:acknowledgements}

This work was supported by the Royal Society International Exchanges
Scheme IE131353 (both authors) and NSF CAREER DMS-1255422 (MAH).
 
\bibliographystyle{elsarticle-num}

\end{document}